\documentclass[prd,preprint,preprintnumbers,nofootinbib,eqsecnum,superscriptaddress]{revtex4}

 \usepackage[dvips,final]{graphicx}
  \usepackage{amssymb}
   \usepackage{amsmath}
    \usepackage{amsfonts}
     \usepackage{epsfig}
      \usepackage{bm}% bold math
      
\usepackage{color}

\usepackage{mathpazo}
\usepackage[section]{placeins}

\input epsf.tex
\def\desepsf(#1 width #2){\epsfxsize=#2 \epsfbox{#1}}

\newcommand{\alphas}{\ensuremath{\alpha_\mathrm{s}}}

\newcommand{\PBM}{PB}

\usepackage[normalem]{ulem}

\usepackage{multirow}
\usepackage{ctable}
\usepackage{booktabs}
\usepackage{array}
\usepackage{tabularx}
\usepackage{xcolor}
\usepackage{pstricks}
\definecolor{fred}{rgb}{0.90053, 0.00369, 0.00159}  % ta3skyblue

\newcommand{\dif}{\mathrm{d}}
\newcommand{\diff}[1]{\frac{\mathrm{d}#1}{#1}}

\newcommand{\be}{\begin{eqnarray}}
\newcommand{\ee}{\end{eqnarray}}

\begin{document}

\author{Rafa{\l} Maciu{\l}a}
\email{rafal.maciula@ifj.edu.pl}
\affiliation{Institute of Nuclear
Physics, Polish Academy of Sciences, ul. Radzikowskiego 152, PL-31-342 Krak{\'o}w, Poland}

\author{Roman Pasechnik}
\email{roman.pasechnik@thep.lu.se}
\affiliation{Department of Astronomy and Theoretical Physics, Lund University, SE-223 62 Lund, Sweden}

\author{Antoni Szczurek}
\email{antoni.szczurek@ifj.edu.pl} 
\affiliation{Institute of Nuclear Physics, Polish Academy of Sciences, ul. Radzikowskiego 152, PL-31-342 Krak{\'o}w, Poland}
\affiliation{College of Natural Sciences, Institute of Physics, University of Rzesz{\'o}w, ul. Pigonia 1, PL-35-310 Rzesz{\'o}w, Poland} 
 
%\title{Impact of intrinsic charm concept on understanding \\the LHCb fixed-target open charm data}

\title{Production of forward heavy-flavour dijets at the LHCb\\ within $\bm{k_{T}}$-factorization approach}

\begin{abstract}
We calculate differential cross sections for $c \bar c$- and $b \bar b$-dijet production in $pp$-scattering at $\sqrt{s} = 13$ TeV
in the $k_T$-factorization and hybrid approaches with 
different unintegrated parton distribution functions (uPDFs). We present distributions 
in transverse momentum and pseudorapidity of the leading jet, rapidity difference between the jets and the dijet invariant mass. Our results are compared to recent LHCb data on forward production of heavy flavour dijets, measured recently for the first time individually for both, charm and bottom flavours. We found that an agreement between the predictions and the data within the full $k_T$-factorization is strongly related to the modelling of the large-$x$ behaviour of the gluon uPDFs which is usually not well constrained. The problem 
may be avoided following the hybrid factorization approach. Then a good description of the measured distributions is obtained with the Parton-Branching, the Kimber-Martin-Ryskin, the Kutak-Sapeta and the Jung setA0 CCFM gluon uPDFs.
%Only the most recent sets of the CCFM uPDFs are found to visibly overestimate the experimental data points. 
We calculate also differential distributions for the ratio of $c \bar c$ and $b \bar b$ cross sections. In all cases we obtain the ratio close to 1 which is caused by the condition on minimal 
jet transverse momentum ($p_{T}^{\mathrm{jet}} > 20$ GeV) introduced in the experiment, that
makes the role of heavy quark mass almost negligible. %This is in contrast to inclusive cross sections for $D$ and $B$ mesons where the ratio may be much larger than $1$.
The LHCb experimental ratio seems a bit larger.
We discuss potentially important for the ratio effect of $c$- or $b$-quark
gluon radiative corrections related to emission outside of the jet cone. The found effect seems rather small.
More refine calculation requires full simulation of $c$- and $b$-jets
which goes beyond of the scope of the present paper. 
\end{abstract} 

\maketitle

%----------------------------
\section{Introduction}
%----------------------------

The production of Heavy Flavours (HF) is a laboratory for studying
effects of perturbative Quantum Chromodynamics (QCD).
Very often it is used to "extract" parton distribution functions (PDFs) for gluon
in the nucleon. This is usually done within collinear-factorization 
approximation. There are a few methods for experimental investigation of charm and bottom quark production at hadron colliders.
One is the direct procedure based on full reconstruction of all decay products of open $D$ and $B$ mesons.
The corresponding hadronic decay products can be used to built invariant mass distributions, permitting direct observation of $D$ or $B$ meson as a peak in the experimental invariant mass spectrum. Different charm mesons $D^{\pm}$, $D^0 / {\bar D}^0$, $D_s^{\pm}$ were used in this context at the LHC, including ALICE, ATLAS and CMS measurements at mid-rapidities \cite{ALICE:2012mhc,ALICE:2012gkr,ALICE:2012inj,ALICE:2019nxm,ATLAS:2015igt,CMS:2016nrh} and the LHCb measurements at forward directions \cite{LHCb:2013xam,LHCb:2015swx}.
Similarly $B^{0}, B^{\pm}$ mesons were studied at ATLAS and CMS experiments \cite{ATLAS:2013sma,CMS:2011pdu,CMS:2011oft}.

Studies of open heavy flavour mesons require some additional modelling that takes into account effects related with a transition from quarks to hadrons. The transition is called hadronization or parton fragmentation and can be so far approached only through phenomenological models. In principle, in the case of multi-particle final states the Lund string model \cite{Andersson:1983ia} and the cluster fragmentation model \cite{Webber:1983if} are often used, however, those methods are originally devoted to mid-rapidities and their application in forward directions is an open question. The hadronization of heavy quarks in non-Monte-Carlo calculations is usually done with the help of fragmentation functions (FF). The latter are similar objects as PDFs and provide the probability for finding a hadron produced from a high energy quark or gluon. Unfortunately, also in this case the FF technique leads to some unambiquities \cite{Maciula:2019iak,Szczurek:2020vjn}, especially in forward directions. 

The inclusive open heavy flavour meson production at the LHC was studied differentially both in collinear
\cite{Cacciari:2012ny,Kniehl:2012ti,Klasen:2014dba,Bhattacharya:2016jce,Garzelli:2019kce} and 
$k_T$-factorization approaches \cite{Maciula:2013wg,Maciula:2019izq,Maciula:2020cfy,Maciula:2021orz,Jung:2011yt,Karpishkov:2016hnx,Guiot:2021vnp} as well as in the dipole model \cite{Goncalves:2017chx,Bhattacharya:2016jce}.
The inclusive distributions describe the experimental ones
within theoretical uncertainties (choice of the heavy quark mass,
choice of renormalization and factorization scales and choice of parton distributions).
Also correlation measurements (meson-antimeson or meson-meson pairs) were performed by the LHCb \cite{LHCb:2012aiv} and CMS \cite{CMS:2011yuk} collaborations. Those data were nicely explained in the framework of $k_T$-factorization approach \cite{Jung:2011yt,Maciula:2013wg,Maciula:2013kd,Karpishkov:2016hnx,Karpishkov:2017kph}.
Similar studies were also performed with the Next-to-Leading Order (NLO) collinear factorization \cite{Vogt:2018oje},
but there some problems with description of the data were identified that can be solved only by introducing a phenomenological intrinsic transverse momentum or by matching parton-level calculations and parton shower \cite{Souza:2015dgh}.
In any case, theoretical calculations of the open heavy flavour meson production are to some extent biased
by not fully understood fragmentation.

Another method is to measure heavy quark/antiquark jets. The reconstruction of jets containing heavy flavour hadrons provides more direct access to the primary heavy flavour parton kinematics than an inclusive measurement of heavy flavour hadrons and allows to
study separately production and fragmentation effects. This is
nowadays used almost routinely in the LHC era for $b$- or ${\bar b}$-jets. Corresponding measurements have been performed by the ATLAS \cite{ATLAS:2011ac}, the CMS \cite{CMS:2012pgw} and very recently by the ALICE \cite{ALICE:2021wct} experiments. The latter
can reconstruct the $b$-flavoured jets down to extremely low transverse momenta, i.e. $p_{T} \approx 10$ GeV \cite{ALICE:2021wct}. 
Recently, the CMS collaboration has measured for the first time jet shapes for $b$-jets in $pp$-collisions \cite{CMS:2020geg}.
Also $b \bar b$-dijets were measured by the ATLAS \cite{ATLAS:2016anw} and the CMS \cite{CMS:2018dqf} experiments.

Although $b$-jet identification algorithms have been deployed for several decades, a task of identification of $c$-jets is more challenging \cite{CMS:2021scf}. Some first trials for inclusive $c$-jet studies were initiated by the CMS \cite{CMS:2016wma}, ALICE \cite{ALICE:2019cbr} and LHCb \cite{LHCb:2015tna} collaborations. Very recently the LHCb collaboration has achieved the identification of charm jets in Run 2 \cite{LHCb:2021dlw}. Last year, the LHCb has reported very unique results of simultaneous measurement of $c\bar c $ and $b\bar b$-dijets in $pp$-scattering at $13$ TeV \cite{LHCb:2020frr}. This is the first $c\bar c$-dijet differential cross-section measurement at the hadron collider.

Here we wish to explore the new LHCb data  for $c\bar c $ and $b\bar b$-dijet production within
$k_T$-factorization approach. This data set is extremely interesting from the point of view 
of constraining unintegrated gluon densities in a proton. The data points were obtained for 2.2 $< \eta_{jet1,2} <$ 4.2, $p_{T}^{jet1,2} >$ 20 GeV, jet radius $R_{\mathrm{cone}} =$ 0.5 and for the difference in the azimuthal angle between the jets $\Delta \varphi > 1.5$. The specific kinematics set and characteristic mechanisms behind the considered reactions 
have many benefits in phenomenological treatment. In principle, because of large scales the pQCD methods are fully applicable and almost no heavy quark mass effects are expected. There is no need to use fragmentation functions and the production mechanism is fully dominated by gluon interactions. Additionally, as will be discussed in the following, the forward direction explored in the LHCb measurement allows for a direct probe of different models of the unitegrated parton distribution functions (uPDFs) for gluon, simultaneously in both small- and large-$x$ regions. All these aspects make the study very interesting from the context of present and future collider experiments on forward heavy flavour production, including those recently being proposed by the Forward Physics Facility (FPF) community \cite{Anchordoqui:2021ghd}. It might also be of a great importance in understanding the dynamics standing behind mechanisms of prompt atmospheric neutrino flux production at present IceCube, Bajkal GVD or other future neutrino observatories.      

%----------------------------------------------------
\section{Details of the model calculations}
%----------------------------------------------------

\subsubsection{The $k_{T}$-factorization framework}

%----------------------------------------------------------------------------
\begin{figure}[!h]
\centering
\begin{minipage}{0.4\textwidth}
  \centerline{\includegraphics[width=1.0\textwidth]{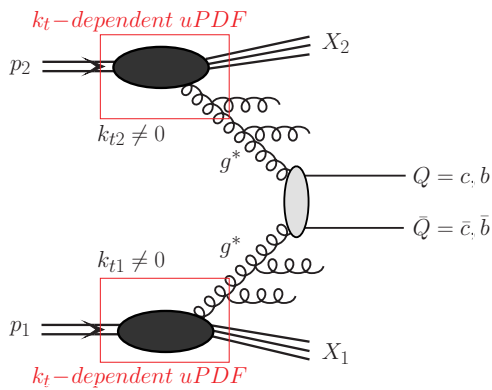}}
\end{minipage}
  \caption{
\small A diagramatic representation of the leading-order mechanism of heavy flavoured dijet production in the $k_{T}$-factorization approach.
}
\label{fig:diagram-kT}
\end{figure}
%----------------------------------------------------------------------------

We remind the theoretical formalism for the calculation of the inclusive $Q\bar{Q}$-pair production in the $k_{T}$-factorization approach \cite{kTfactorization}, where $Q = c,b$ stands for the charm and beauty quarks, respectively. In this framework the transverse momenta $k_{t}$'s (virtualities) of both partons entering the hard process are taken into account, both in the matrix elements and in the parton distribution functions. Emission of the off-mass-shell initial state partons is encoded in the transverse-momentum-dependent (unintegrated) parton distribution functions (uPDFs). In the case of heavy-flavoured dijet production, the parton-level cross section is usually calculated via the Leading-Order (LO) $g^*g^* \to Q\bar Q$ fusion mechanism of off-shell initial state gluons that is the dominant process at high energies. As will be shown when presenting our numerical results, the $q^*\bar q^* \to Q\bar Q $ mechanism remains subleading in the whole kinematical domain considered here. Then the hadron-level differential cross section for the $Q \bar Q$-dijet production, formally at leading-order, reads:
\begin{eqnarray}\label{LO_kt-factorization} 
\frac{d \sigma(p p \to Q \bar Q \, X)}{d y_1 d y_2 d^2p_{1,t} d^2p_{2,t}} &=&
\int \frac{d^2 k_{1,t}}{\pi} \frac{d^2 k_{2,t}}{\pi}
\frac{1}{16 \pi^2 (x_1 x_2 s)^2} \; \overline{ | {\cal M}^{\mathrm{off-shell}}_{g^* g^* \to Q \bar Q} |^2}
 \\  
&& \times  \; \delta^{2} \left( \vec{k}_{1,t} + \vec{k}_{2,t} 
                 - \vec{p}_{1,t} - \vec{p}_{2,t} \right) \;
{\cal F}_g(x_1,k_{1,t}^2,\mu_{F}^2) \; {\cal F}_g(x_2,k_{2,t}^2,\mu_{F}^2) \; \nonumber ,   
\end{eqnarray}
where ${\cal F}_g(x_1,k_{1,t}^2,\mu_{F}^2)$ and ${\cal F}_g(x_2,k_{2,t}^2,\mu_{F}^2)$
are the gluon uPDFs for both colliding hadrons and ${\cal M}^{\mathrm{off-shell}}_{g^* g^* \to Q \bar Q}$ is the off-shell matrix element for the hard subprocess.
The gluon uPDF depends on gluon longitudinal momentum fraction $x$, transverse momentum
squared $k_t^2$ of the gluons entering the hard process, and in general also on a (factorization) scale of the hard process $\mu_{F}^2$.
The extra integration is over transverse momenta of the initial
partons. Here, one keeps exact kinematics from the very beginning and additional hard dynamics coming from transverse momenta of incident partons. Explicit treatment of the transverse momenta makes the approach very efficient in studies of correlation observables. The two-dimensional Dirac delta function assures momentum conservation. The gluon uPDFs must be evaluated at longitudinal momentum fractions 
$x_1 = \frac{m_{1,t}}{\sqrt{s}}\exp( y_1) + \frac{m_{2,t}}{\sqrt{s}}\exp( y_2)$ and $x_2 = \frac{m_{1,t}}{\sqrt{s}}\exp(-y_1) + \frac{m_{2,t}}{\sqrt{s}}\exp(-y_2)$, where $m_{i,t} = \sqrt{p_{i,t}^2 + m_Q^2}$ is the quark/antiquark transverse mass.  

The off-shell matrix elements are known explicitly only in the LO and only for limited types of $2 \to 2$ QCD processes (see \textit{e.g.} heavy quark \cite{Catani:1990eg}, dijet \cite{Nefedov:2013ywa}, Drell-Yan \cite{Nefedov:2012cq} mechanisms). Recently, higher multiplicity processes ($2 \to 3$ and $2 \to 4$) with off-shell partons were also calculated at tree level in the context of $c\bar c \textrm{+ jets}$ \cite{Maciula:2019izq,Maciula:2016kkx,Maciula:2017egq}, $\gamma + \textrm{c-jet}$ \cite{Bednyakov:2017vck}, $Z^0 + \textrm{c/b-jet}$ \cite{Lipatov:2018oxm}, $4$-jets \cite{Kutak:2016mik} and double charm \cite{vanHameren:2015wva} production.
Some first steps to calculate NLO corrections in the $k_{T}$-factorization framework have been made for diphoton production \cite{Nefedov:2015ara,Nefedov:2016clr}.
There are ongoing intensive works on construction of the full NLO Monte Carlo generator for off-shell initial state partons that are expected to be finished in near future \cite{private-Hameren}. Another method for calculation of higher multiplicity final states is to supplement the QCD $2 \to 2$ processes with parton shower. For the off-shell initial state partons it was done only with the help of full hadron level Monte Carlo generator CASCADE \cite{Jung:2010si}. There, dedicated transverse-momentum dependent initial-state parton showers were introduced using backward evolution, that is not unique and needs to be matched to a given model of uPDFs. %However, in the moment this method can be consistently used only with the uPDFs that have a steep drop of the parton densities at $k_{t}^{2} > \mu_{F}^{2}$.

Technically, there is a direct relation between a resummation present in uPDFs in the transverse momentum dependent factorization and a parton shower in the collinear framework. The popular statement is that actually in the $k_{T}$-factorization approach already at leading-order some part of radiative higher-order corrections can be effectively included via uPDFs, without any additional showering procedure. However, it depends strictly on a theoretical construction of different uPDF models, in which extra emissions of soft and even hard partons can be encoded. In some uPDF models the off-shell gluon can be produced either from a gluon or quark, therefore, in the $k_{T}$-factorization all channels driven by $gg, q\bar q$ and even by $qg$ initial states are open already at leading-order (in contrast to the collinear factorization). Then, when calculating the heavy flavour production cross section via the $g^* g^* \to Q \bar Q$ mechanism one could expect to effectively include contributions related to an additional one or two (or even more) extra partonic emissions which in some sense plays a role of the initial state parton shower.

%------------------------------------
\subsection{The hybrid model}
%------------------------------------

%----------------------------------------------------------------------------
\begin{figure}[!h]
\centering
\begin{minipage}{0.4\textwidth}
  \centerline{\includegraphics[width=1.0\textwidth]{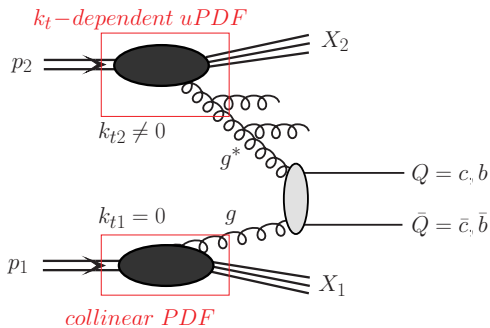}}
\end{minipage}
  \caption{
\small A diagramatic representation of the leading-order mechanism of heavy flavoured dijet production in the hybrid model.
}
\label{fig:diagram-kT}
\end{figure}
%----------------------------------------------------------------------------

The LHCb measurement of the $c\bar c$- and $b\bar b$-dijets is performed within the asymmetric kinematical configuration.
Under a general assumption that $x_1 \gg x_2$ the cross section for the processes under consideration can be also
expressed in the so-called hybrid factorization approach motivated by the
works in Refs.~\cite{Deak:2009xt,Kutak:2012rf}. 
In this framework the small-$x$ gluon is taken to be off mass shell, the large-$x$ gluon is treated as collinear and 
the differential cross section e.g. for $pp \to Q \bar Q X$ via $g^* g \to Q \bar Q$ mechanism reads:
\begin{eqnarray}
d \sigma_{pp \to Q\bar Q X} = \int d^ 2 k_{t} \int \frac{dx_1}{x_1} \int dx_2 \;
g(x_1, \mu^2) \; {\cal F}_{g^{*}}(x_2, k_{t}^{2}, \mu^2) \; d\hat{\sigma}_{g^{*}g \to Q\bar Q} \; ,
\end{eqnarray}
where $g(x_1, \mu^2)$ is a collinear PDF in one proton and ${\cal F}_{g^{*}}(x_2, k_{t}^{2}, \mu^2)$ is the unintegrated gluon distribution in the second one. The $d\hat{\sigma}_{g^{*}g \to Q\bar Q}$ is the hard partonic cross section obtained from a gauge invariant tree-level off-shell amplitude. In the present paper we shall not discuss the validity of the hybrid model on the theoretical level and concentrate only on its phenomenological application in forward production. A derivation of the hybrid factorization from the dilute limit of the Color Glass Condensate approach can be found in Ref.~\cite{Kotko:2015ura}.

%----------------------------------------------------------
\subsection{Unintegrated parton distribution functions} 
%----------------------------------------------------------

%---------------------------
\subsubsection{The CCFM uPDFs}
%---------------------------

The CCFM evolution equation for gluon, in the limits of high and low energies (small- and large-$x$ values),
is almost equivalent to the BFKL and very similar to the DGLAP evolution, respectively \cite{CCFM}.
In order to correctly treat gluon coherence effects it introduces the so-called angular-ordering which is commonly considered as a great advantage of this framework.

In the leading logarthmic approximation, the CCFM equation 
for unintegrated gluon density ${\cal F}_{g}(x,k_t^2,\mu^2)$
can be written as
\begin{equation}
  \displaystyle {\cal F}_{g}(x,k_t^2,\mu^2) = {\cal F}_{g}^{(0)}(x,k_t^2,\mu_0^2) \Delta_s(\mu,\mu_0) + \atop { 
  \displaystyle + \int\frac{dz}{z}\int\frac{dq^2}{q^2}\Theta(\mu-zq)\Delta_s(\mu,zq) \tilde P_{gg}(z,k_t^2, q^2) {\cal F}_{g}\left(\frac{x}{z},k^{\prime \, 2}_t,q^2\right) },
\end{equation}

\noindent
where $\mu^2$ is the evolution (factorization) scale which is further defined by the maximal allowed
angle for any gluon emission, $k_t^\prime = q(1 - z) + k_T$
and $\tilde P_{gg}(z,k^2_t,q^2)$ is the CCFM
splitting function:
\begin{equation}
  \displaystyle \tilde P_{gg}(z,k^2_t,q^2) = \bar\alpha_s(q^2(1-z)^2) \left[\frac{1}{1-z}+\frac{z(1-z)}{2}\right] + \atop {
  \displaystyle + \bar\alpha_s(k_t^2)\left[\frac{1}{z}-1+\frac{z(1-z)}{2}\right]\Delta_{ns}(z,k^2_t,q^2) }.
\end{equation}

\noindent
The Sudakov and non-Sudakov form factors read:
\begin{equation}
 \ln \Delta_s(\mu,\mu_0)= - \int\limits_{\mu_0^2}^{\mu^2}\frac{d\mu^{\prime \, 2}}{\mu^{\prime \, 2}}\int\limits_0^{z_M=1-\mu_0/\mu^\prime}dz\,\frac{\bar\alpha_s(\mu^{\prime \, 2}(1-z)^2)}{1-z},
\end{equation}
\begin{equation}
\ln \Delta_{ns}(z,k_t^2, q_t^2) = -\bar\alpha_s(k_t^2)\int\limits_0^1\frac{dz^\prime}{z^\prime}\int\frac{dq^2}{q^2}\Theta(k_t^2-q^2)\Theta(q^2-z^{\prime\,2} q^2_t).
\end{equation}

\noindent
where $\bar \alpha_s = 3 \alpha_s/\pi$. 

The first term in the CCFM equation is the initial 
unintegrated gluon density multiplied by the Sudakov form factor.
It corresponds to the contribution of non-resolvable branchings between 
the starting scale $\mu_0^2$ and the current running scale $\mu^2$.
The second term describes the details of the QCD evolution 
expressed by the convolution of the CCFM gluon splitting function
with the gluon density and the Sudakov form factor. The 
theta function introduces the angular ordering condition.

The CCFM equation can be solved numerically using 
the \textsc{updfevolv} program \cite{Hautmann:2014uua},
and the uPDFs for gluon and valence quarks can be 
obtained for any $x$, $k_t^2$ and $\mu^2$ values. Since only valence quarks are included in the evolution, the resulting CCFM gluon uPDF
can be characterized as nearly $0$-flavour scheme density. Therefore, it might be expected to be larger with respect to DGLAP-based unintegrated distributions that represent a variable-flavour-number-scheme (VFNS) framework.  

Within the CCFM approach the parton transverse momentum is allowed to be larger than the scale $\mu^2$.
This useful feature enables to effectively take into account
%translates into the ease of effective taking into account 
higher-order radiative corrections, that correspond to the initial-state real gluon
emissions which are resummed into the uPDFs.

The CCFM approach for the gluon uPDF was successfully used in the past to describe $B$-meson and $b\bar b$-dijet production data taken by the D0 and CDF at Tevatron \cite{Jung:2010ey} as well as by the CMS at the LHC \cite{Jung:2011yt}.

% Thus, for any phenomenological studies in the $k_{T}$-factorization approach
%the standard scheme with leading-order matrix elements is recommended as long as the CCFM uPDFs are used.   

%-----------------------
\subsubsection{The PB uPDFs}
%-----------------------

The Parton Branching (PB) method, introduced in Refs.~\cite{Hautmann:2017fcj,Martinez:2018jxt}, provides an iterative solution for the evolution of both collinear and transverse momentum dependent parton distributions. Within this novel method the splitting kinematics at each branching vertex stays under full control during the QCD evolution.
Here, soft-gluon emission in the region $z\to 1$ and transverse momentum recoils in the parton branchings along the QCD cascade are taken into account simultaneously. Therefore, the PB approach allows for a natural determination of the uPDFs, as the transverse momentum at every branching vertex is known. It agrees with the usual methods to solve the DGLAP equations, but provides in addition a possibility to apply angular ordering instead of the standard ordering in virtuality.

Within the PB method, a soft-gluon resolution scale parameter $z_M$ is introduced into the QCD evolution equations that distinguish between non-resolvable and resolvable emissions. These two types of emissions are further treated with the help of the Sudakov form factors 
\begin{equation}
\label{sud-def}
 \Delta_a ( z_M, \mu^2 , \mu^2_0 ) = 
\exp \left(  -  \sum_b  
\int^{\mu^2}_{\mu^2_0} 
{{d \mu^{\prime 2} } 
\over \mu^{\prime 2} } 
 \int_0^{z_M} dz \  z 
\ P_{ba}^{(R)}\left(\alphas , 
 z \right) 
\right) 
  \;\; ,   
\end{equation}
and with the help of resolvable splitting probabilities $P_{ba}^{(R)} (\alphas,z)$, respectively.
Here $a , b$ are flavor indices, 
$\alphas$ is the strong coupling at a scale being a function of ${\mu}^{\prime 2}$,   
$z$ is 
the longitudinal momentum 
splitting variable, and   
$z_M < 1 $ is the soft-gluon resolution parameter.
Then, by connecting the evolution variable $\mu$ in the splitting process $b \to a c$
with the angle $\Theta$ of the momentum of particle $c$ with respect to the beam direction, 
the known angular ordering relation $\mu = | q_{t,c} | / (1 - z)$ is obtained, that ensures quantum coherence of softly radiated partons.

The \PBM\ evolution  equations with angular ordering condition for unintegrated parton densities    
$ {\cal F}_a ( x , k_{t} , \mu^2) $ 
are given 
by~\cite{Hautmann:2017fcj}   
\begin{eqnarray}
\label{evoleqforA}
   { {\cal F}}_a(x,k_t, \mu^2) 
 &=&  
 \Delta_a (  \mu^2  ) \ 
 { {\cal F}}_a(x,k_t,\mu^2_0)  
 + \sum_b 
\int
{{d^2 q_{t}^{\prime } } 
\over {\pi q_{t}^{\prime 2} } }
 \ 
{
{\Delta_a (  \mu^2  )} 
 \over 
{\Delta_a (  q_{t}^{\prime 2}  
 ) }
}
\ \Theta(\mu^2-q_{t}^{\prime 2}) \  
\Theta(q_{t}^{\prime 2} - \mu^2_0)
 \nonumber\\ 
&\times&  
\int_x^{z_M} {{dz}\over z} \;
P_{ab}^{(R)} (\alphas 
,z) 
\;{ {\cal F}}_b\left({x \over z}, k_{t}+(1-z) q_{t}^\prime , 
q_{t}^{\prime 2}\right)  
  \;\;  .     
\end{eqnarray}
Here, the starting disitribution for the uPDF evolution is taken in the factorized form as a product of collinear PDF fitted to the precise DIS data and an intrinsic transverse momentum distribution in a simple gaussian form. Unlike the CCFM parton distributions, the PB densities have the strong normalization property:
\be
\int { {\cal F}}_a(x,k_t, \mu^2)\; d k_t = f_{a}(x,\mu^2), 
\ee
and therefore well reproduce modern collinear PDFs after integrating out their $k_{t}$-dependence.
The PB uPDFs, including gluon and quarks, can be calculated by an iterative Monte-Carlo method and are characterized by a steep drop of the parton densities at $k_{t}^2 > \mu^2$, again in contrast to the CCFM unintegrated distributions\footnote{Very recently, only a first attempt to incorporate CCFM effects into the PB method has been done \cite{Monfared:2019uaj}.}.

There are two basic sets of the parton-branching uPDFs: PB-NLO-2018-set1 and PB-NLO-2018-set2, that correspond to different choice of the parameters of the initial distributions \cite{Martinez:2018jxt}. Both of them, including uncertainties are available in TMDLIB \cite{Hautmann:2014kza} and both of them were used previously with success to describe the LHCb data on forward production of open charm meson \cite{Maciula:2019izq,Maciula:2020cfy}.

%-------------------------
\subsubsection{The KMR/MRW uPDFs}
%-------------------------

Another DGLAP-based and frequently used in phenomenological studies prescription for unintegrated parton densities is the Kimber-Martin-Ryskin (KMR) approach \cite{Kimber:1999xc,Kimber:2001sc,Watt:2003mx}. It has been successfully used especially for heavy flavour production at the LHC, including inclusive charm meson and baryon \cite{Maciula:2013wg,Maciula:2018iuh}, charm-anticharm meson pairs \cite{Maciula:2013wg,Karpishkov:2016hnx}, double and triple charm meson \cite{Maciula:2013kd,Cazaroto:2013fua,Maciula:2017meb,dEnterria:2016ids}, charm meson associated with jets \cite{Maciula:2016kkx,Maciula:2017egq} as well as $B$-meson and $b\bar b$-dijets \cite{Jung:2011yt}.

According to this approach the unintegrated gluon distribution is given
by the following formula
\begin{eqnarray} \label{eq:UPDF}
  f_g(x,k_t^2,\mu^2) &\equiv& \frac{\partial}{\partial \log k_t^2}\left[\,g(x,k_t^2)\,T_g(k_t^2,\mu^2)\,\right]\nonumber \\ &=& T_g(k_t^2,\mu^2)\,\frac{\alpha_S(k_t^2)}{2\pi}\,\sum_{b }\,\int_x^1\! d z\,P_{gb}(z)\,b \left (\frac{x}{z}, k_t^2 \right).
\end{eqnarray}
This formula makes sense for $k_t > \mu_0$, where $\mu_0\sim 1$ GeV
is the minimum scale for which DGLAP evolution of the conventional
collinear gluon PDF, $g(x,\mu^2)$, is valid. A similar expression can be written also for quarks.

The virtual (loop) contributions may be resummed to all orders by the Sudakov form factor
\begin{equation} \label{eq:Sudakov}
  T_g (k_t^2,\mu^2) \equiv \exp \left (-\int_{k_t^2}^{\mu^2}\!\diff{\kappa_t^2}\,\frac{\alpha_S(\kappa_t^2)}{2\pi}\,\sum_{b}\,\int_0^1\!\dif{z}\;z \,P_{b g}(z) \right ),
\end{equation}
which gives the probability of evolving from a scale $k_t$ to a scale $\mu$ without parton emission.
The exponent of the gluon Sudakov form factor can be simplified using the following identity: $P_{qg}(1-z)=P_{qg}(z)$.  Then the gluon Sudakov form factor reads
\begin{equation}
  T_g(k_t^2,\mu^2) = \exp\left(-\int_{k_t^2}^{\mu^2}\!\diff{\kappa_t^2}\,\frac{\alpha_S(\kappa_t^2)}{2\pi}\,\left( \int_{0}^{1-\Delta}\!\dif{z}\;z \,P_{gg}(z) + n_F\,\int_0^1\!\dif{z}\,P_{qg}(z)\right)\right),
\end{equation}
where $n_F$ is the quark--antiquark active number of flavours into which the gluon may split. Due to the presence of the Sudakov form factor in the KMR prescription only last emission generates transverse momentum of the gluons initiating hard scattering.

In the above equation the variable $\Delta$ introduces a restriction of the phase space for gluon emission and is crucial for the final shape and characteristics of the unintegrated density. In Ref.~\cite{Kimber:1999xc} the cutoff $\Delta$ was set in accordance with the strong ordering (SO) in transverse momenta of the real parton emission in the DGLAP evolution, 
\be
\label{eq:11}
\Delta =\frac{k_{t}}{\mu}\,.
\ee
This corresponds to the orginal KMR prescription where one always has $k_{t}^{2} < \mu_{F}^{2}$ restriction and the Sudakov form-factor always satisfies the $T_{g}(k_{t}^{2},\mu^{2})<1$ condition.

The prescription for the cutoff $\Delta$ was further modified in Ref.~\cite{Kimber:2001sc,Watt:2003mx} to account for the angular ordering 
(AO) in parton emissions in the spirit of the CCFM evolution,
\be
\label{eq:13}
\Delta=\frac{k_{t}}{k_{t}+\mu}\,.
\ee
This modification leads to a bigger upper limit for $k_{t}$ than in the DGLAP scheme and opens the $k_{t}^{2} > \mu_{F}^{2}$ region.
In this extra kinematical regime one gets $T_{g}(k_{t}^{2},\mu^{2})>1$, which
contradicts its interpretation as a probability of no real emission. Thus, the Sudakov form factor is usually set to be equal to one in that domain.
For transparency, here the modified KMR model will be referred to as the Martin-Ryskin-Watt (MRW) model \cite{Watt:2003mx}.

Different definitions of the ordering cut-off lead to significant differences between the two models. In the KMR model the $k_{t}^{2} > \mu_{F}^{2}$ region is forbidden while in the MRW case the $k_{t}^{2} > \mu_{F}^{2}$ contributions are directly allowed (see \textit{e.g.} a detailed discussion in Ref.~\cite{Golec-Biernat:2018hqo}). In the MRW model both in quark and gluon densities large $k_{t}$-tails appear, in contrast to the KMR case.
In the numerical calculations below we used the MMHT2014 \cite{Harland-Lang:2014zoa} 
%as well as the CT14 \cite{Dulat:2015mca} \textbf{[In the moment not shown in the text]}
collinear PDFs to calculate both, the KMR and the MRW unintegrated densities. 

%-------------------------
\subsubsection{The Kutak-Sapeta uPDFs}
%-------------------------

An alternative approach to those presented above has been applied in the Kutak-Sapeta (KS) model \cite{Kutak:2012rf,Kutak:2014wga} for the gluon uPDF. There the unintegrated gluon density is obtained from the unified framework of the Balitsky-Kovchegov (BK) \cite{Balitsky:1995ub,Kovchegov:1999yj,Balitsky:2007feb} and DGLAP evolution equations and then fitted to combined HERA data. This framework is a continuation of the model presented some time ago in Ref.~\cite{Kwiecinski:1997ee} for unified BFKL-DGLAP evolution and it extends the ideas behind by taking into account the nonlinear (or gluon saturation) effects in the QCD evolution. In order to account for some effects related to the saturation of gluons, the unified BFKL-DGLAP evolution equation
is supplemented with the nonlinear term in the BK form. Thus, the authors obtained the so-called modified BK equation in which the unified linear part deals with partial resummation of the NLLA corrections and the nonlinear term is taken in the basic LLA approach.

According to this approach the improved nonlinear equation for the unintegrated gluon density,
written in momentum space, reads as follows:
\begin{multline} 
{\cal F}_{g}(x,k^2) \; = \; {\cal F}_{g}^{(0)}(x,k^2) \\
+\,\frac{\alpha_s(k^2)N_c}{\pi}\int_x^1 \frac{dz}{z} \int_{k_0^2}^{\infty}
\frac{dl^2}{l^2} \,   \bigg\{ \, \frac{l^2{\cal F}(\frac{x}{z},l^2)\,   -\,
k^2{\cal F}(\frac{x}{z},k^2)}{|l^2-k^2|}   +\,
\frac{k^2{\cal F}(\frac{x}{z},k^2)}{|4l^4+k^4|^{\frac{1}{2}}} \,
\bigg\} \\
-\frac{2\alpha_s^2(k^2)}{\rho^2}\left[\left(\int_{k^2}^{\infty}\frac{dl^2}{l^2}{\cal F}(x,l^2)\right)^2
+{\cal F}(x,k^2)\int_{k^2}^{\infty}\frac{dl^2}{l^2}\ln\left(\frac{l^2}{k^2}\right){\cal F}(x,l^2)\right]\,, 
\label{eq:fkov} 
\end{multline} 
where $\rho$ is the radius of the hadronic target, and ${\cal F}_{g}^{(0)}(x,k^2)$ is the starting distribution.
The linear part of the equation is given by the BFKL kernel while the nonlinear part is proportional to the triple pomeron vertex which allows for the recombination of gluons.

The basic model of the KS gluon uPDF \cite{Kutak:2012rf} has been further extended by introducing a factorization scale dependence \cite{Kutak:2014wga} of the originally scale-independent density ${\cal F}_{g}(x,k^2)$. The scale dependence is obtained as follows:
\be
{\cal F}_{g}(x,k^2,\mu^2):=\theta(\mu^2-k^2)T_s(\mu^2,k^2)\frac{xg(x,\mu^2)}{xg_{hs}(x,\mu^2)}{\cal F}_{g}(x,k^2)+\theta(k^2-\mu^2){\cal F}_{g}(x,k^2).
\label{eq:hardgluon}
\ee

where 
\be
xg_{hs}(x,\mu^2)=\int^{\mu^2} dk^2 T_s(\mu^2,k^2){\cal F}_{g}(x,k^2),\,\, xg(x,\mu^2)=\int^{\mu^2} dk^2{\cal F}_{g}(x,k^2)
\label{eq:intglu2}
\ee
and the Sudakov form factor assumes the form:
\be
T_s(\mu^2,k^2)=\exp\left(-\int_{k^2}^{\mu^2}\frac{dk^{\prime 2}}{k^{\prime 2}}\frac{\alpha_s(k^{\prime 2})}
{2\pi}\sum_{a^\prime}\int_0^{1-\Delta}dz^{\prime}P_{a^\prime a}(z^\prime)\right) ,
\ee
where $\Delta=\frac{\mu}{\mu+k}$ and $P_{a^\prime a}$ is a splitting function with subscripts $a^\prime a$ specifying the type of transition.

In the numerical calculations below we applied both the KS-linear and KS-nonlinear gluon uPDF sets as implemented in the TMDlib. Phenomenological applications of these densities are limited to the $x < 10^{-2}$ kinematic domain and therefore they can be applied only when considering small-$x$ effects. The KS gluon uPDFs are frequently used within the hybrid approach and are found very useful especially in phenomenological studies of forward particle production that is taking place in highly asymmetric kinematical configurations. Recently, the densities have been examined in the context of forward jet, forward-forward dijet as well as central-forward dijet production at the LHC (see e.g. Refs.~\cite{Bury:2016cue,vanHameren:2016ftb,vanHameren:2020rqt,VanHaevermaet:2020rro}).   

%-----------------------------------
\section{Numerical results}
%-----------------------------------

In this section we present a variety of numerical results of the theoretical models described above for inclusive production of the $c\bar c$- and $b\bar b$-dijets at the LHC. The theoretical predictions are confronted with corresponding experimental data from the LHCb experiment \cite{LHCb:2020frr} collected recently at $\sqrt{s} = 13$ TeV. In the first step we show results obtained within the exact $k_{T}$-factorization framework and then we carefully discuss the kinematics behind the processes under consideration. Next, we focus on the predictions based on the hybrid model showing explicitly the role of the large-$x$ behaviour of gluon densities for a satisfactory description of the LHCb data and understanding the QCD dynamics behind. In the end we discuss the scale uncertainty of our predictions and $R =\frac{c\bar c}{b\bar b}$ dijet cross section ratio as a function of a few kinematical variables. For completeness of our study we also present a direct comparison to the corresponding results of calculations based on the collinear framework.     

\clearpage
\subsection{Predictions of the standard $\bm{k_{T}}$-factorization framework}

As a default set in the numerical calculations below we take the renormalization/factorization scales
$\mu^2 = \mu_{R}^{2} = \mu_{F}^{2} = \sum_{i=1}^{n} \frac{m^{2}_{it}}{n}$ (averaged transverse mass of the given final state) and the charm and bottom quark masses $m_{c}=1.5$ GeV and $m_{b}=4.75$, respectively. The strong-coupling constant $\alpha_{s}(\mu_{R}^{2})$ at NLO is taken from the MMHT2014 PDF routines.

For the CCFM uPDFs we always set the factorization scale to $\mu^2 = M_{Q\bar Q}^2 + p_{T}^{Q\bar Q}$, where $M_{Q\bar Q}$ is the invariant mass of the $Q\bar Q$ system (or energy of the scattering subprocess) and $p_{T}^{Q\bar Q}$ is the transverse momentum  of $Q \bar Q$-pair (or the incoming off-shell gluon pair). This has to be applied as a consequence of the CCFM evolution algorithm.

Here and in the following only the dominant at high energies pQCD gluon-gluon fusion $g^*g^* \to Q \bar Q$ mechanism is taken into account.
We have checked numerically that the annihilation $q^*\bar q^* \to Q \bar Q$ mechanism is negligible here and can be safely neglected.

\subsubsection{Charm dijets}

\begin{figure}[!h]
\begin{minipage}{0.47\textwidth}
  \centerline{\includegraphics[width=1.0\textwidth]{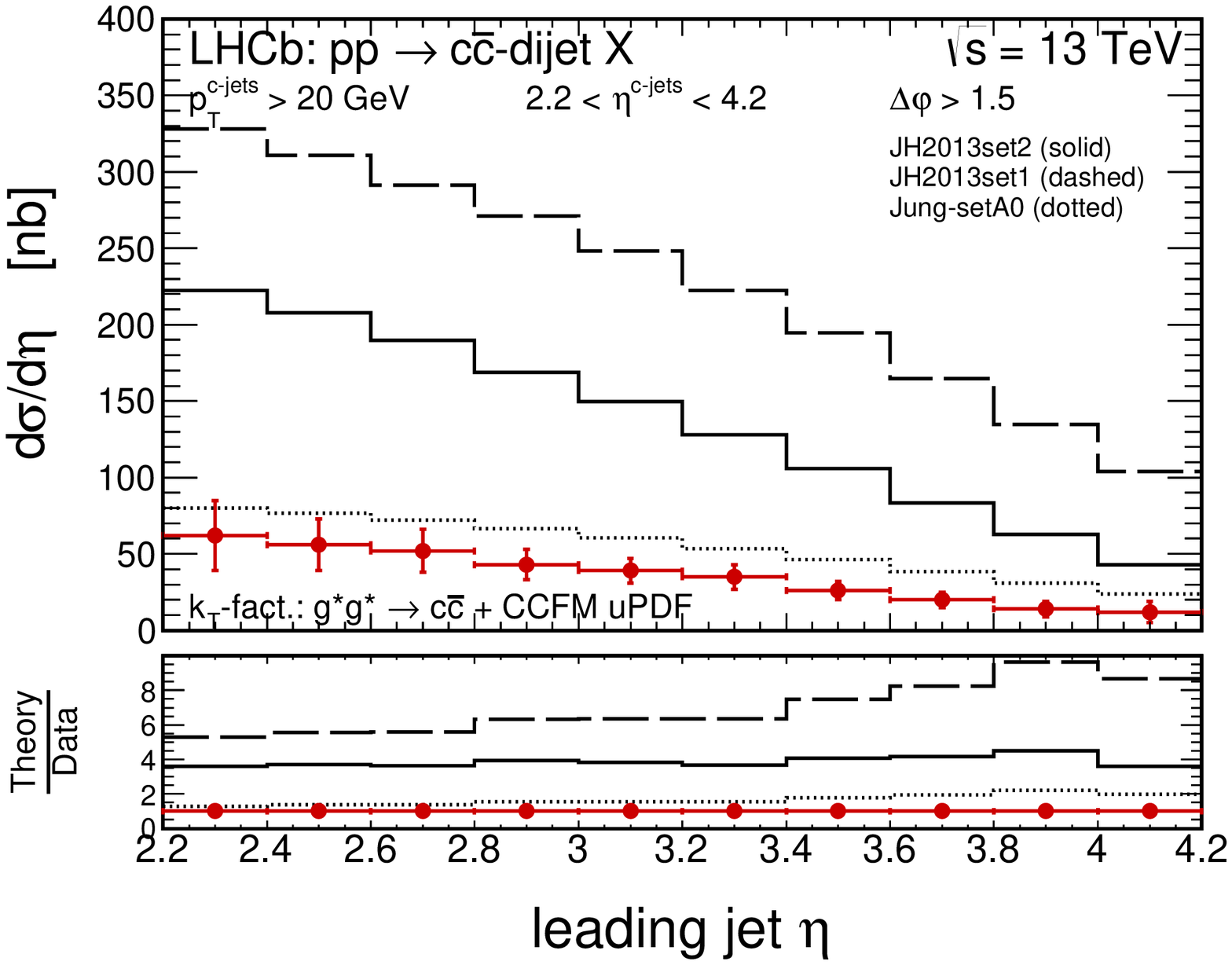}}
\end{minipage}
\begin{minipage}{0.47\textwidth}
  \centerline{\includegraphics[width=1.0\textwidth]{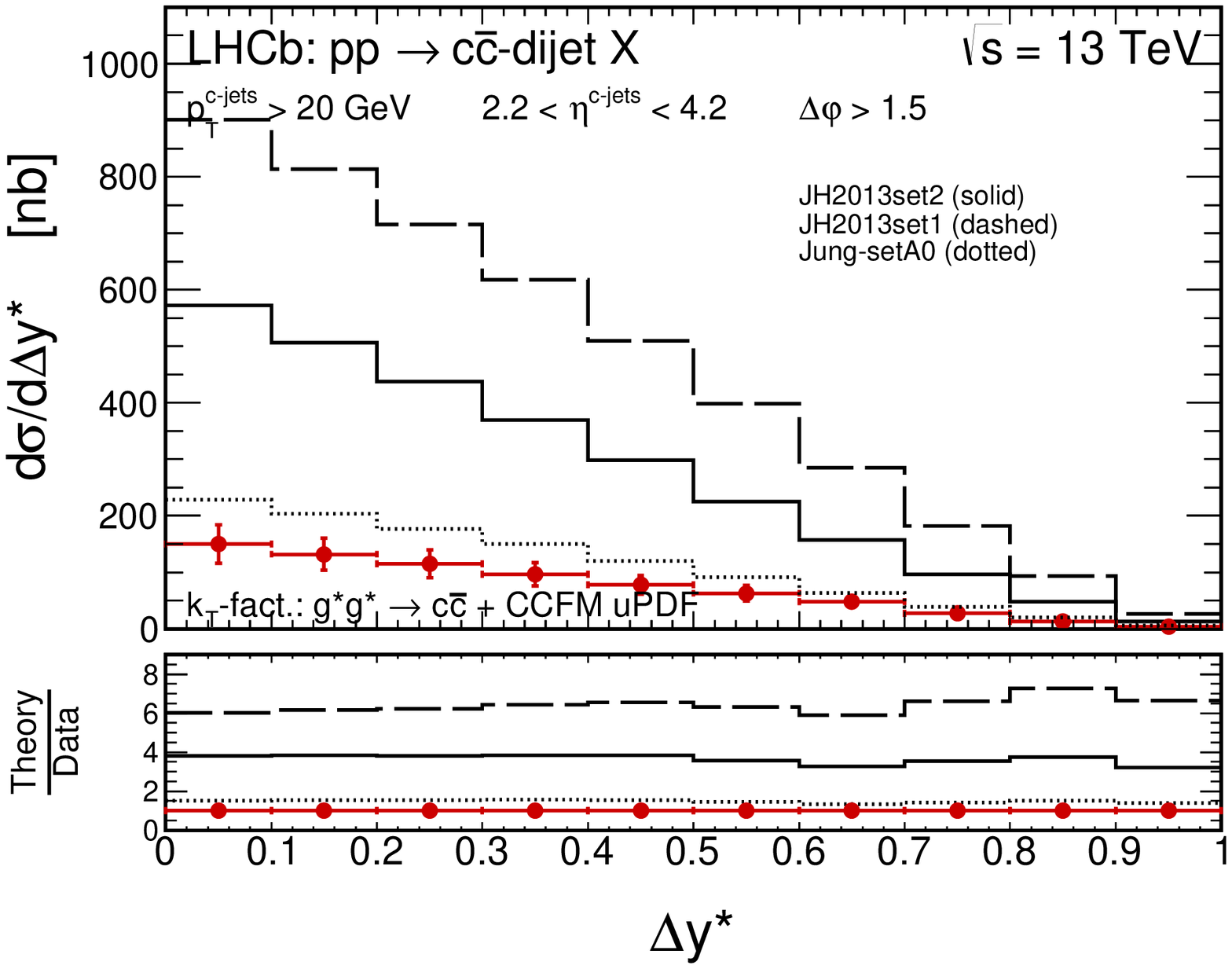}}
\end{minipage}
\begin{minipage}{0.47\textwidth}
  \centerline{\includegraphics[width=1.0\textwidth]{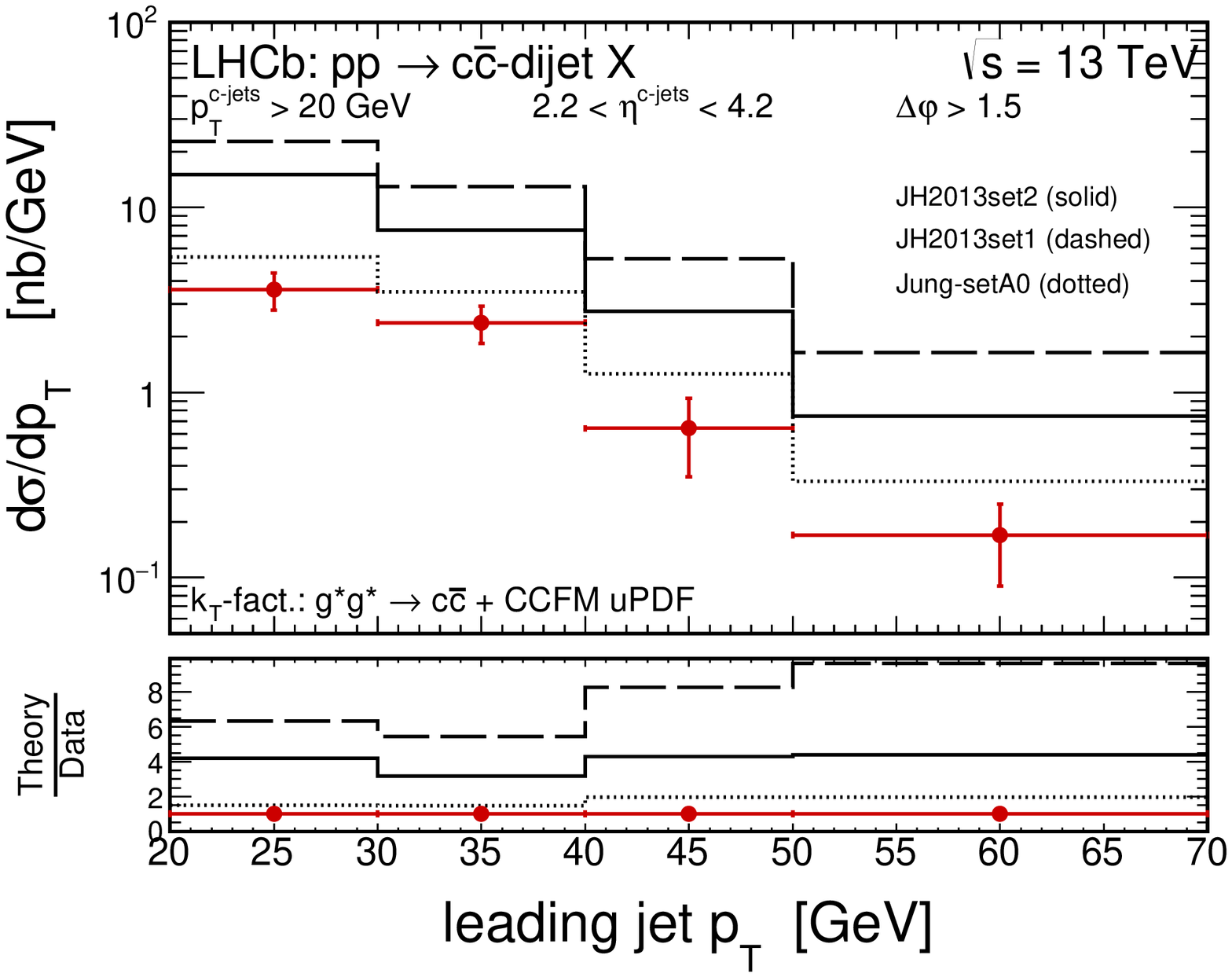}}
\end{minipage}
\begin{minipage}{0.47\textwidth}
  \centerline{\includegraphics[width=1.0\textwidth]{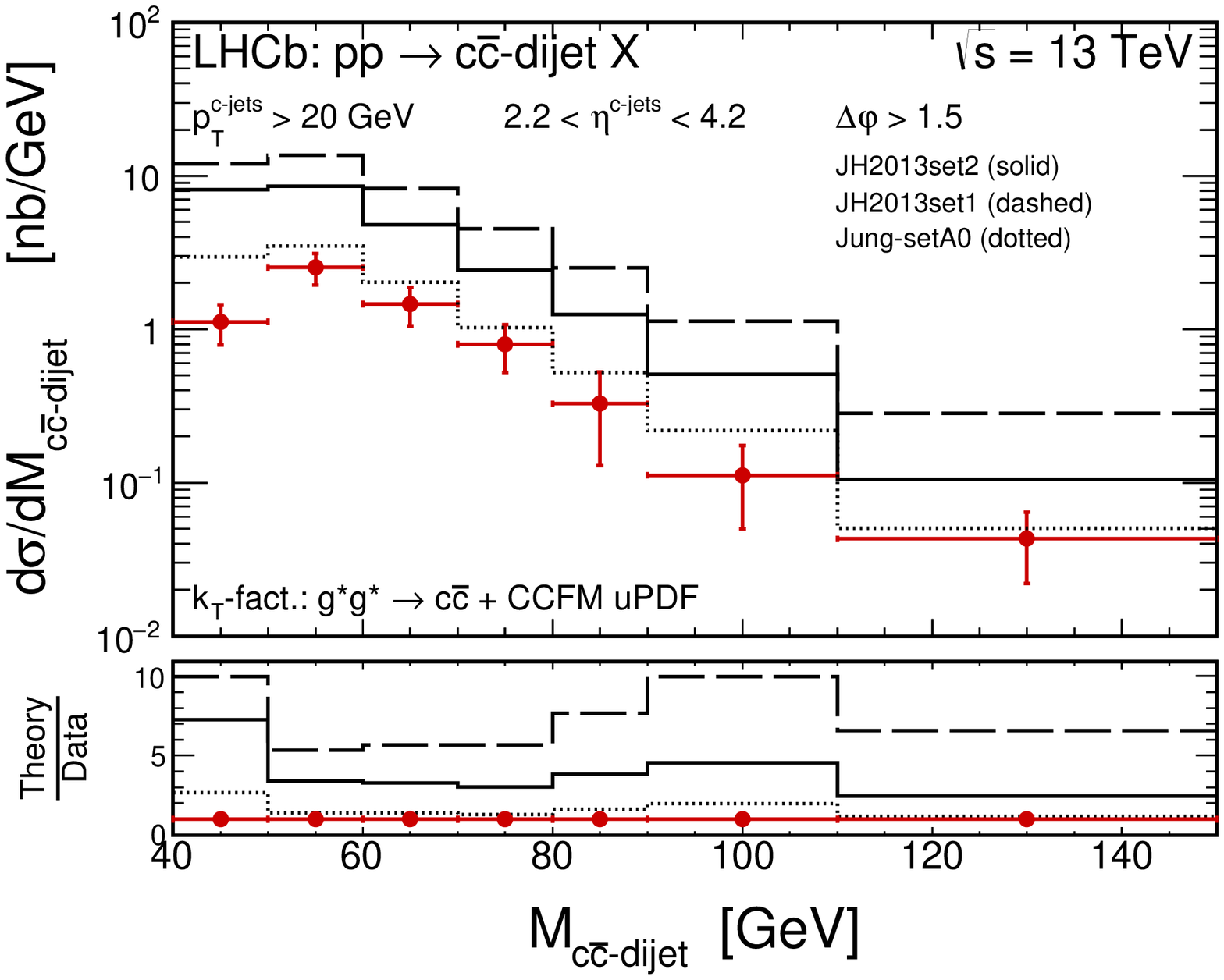}}
\end{minipage}
  \caption{The differential cross sections for forward production of $c\bar c$-dijets in $pp$-scattering at $\sqrt{s}=13$ TeV as a function of the leading jet $\eta$ (top left), the rapidity difference $\Delta y^{*}$ (top right), the leading jet $p_{T}$ (bottom left) and the dijet invariant mass $M_{c\bar c\text{-}\mathrm{dijet}}$ (bottom right). Here the dominant pQCD $g^*g^* \to c \bar c$ mechanism is taken into account. The theoretical histograms correspond to the $k_{T}$-factorization calculations obtained with the CCFM uPDFs.
\small 
}
\label{fig:3}
\end{figure}

We start with presentation of numerical results for forward production of the $c\bar c$-dijets in $pp$-scattering at $\sqrt{s}=13$ TeV. In Fig.~\ref{fig:3} we show the corresponding differential cross sections as a function of the leading jet $\eta$ (top left panels), the rapidity difference $\Delta y^{*}$ (top right panels), the leading jet $p_{T}$ (bottom left panels) and the dijet invariant mass $M_{c\bar c\text{-}\mathrm{dijet}}$ (bottom right panels).
The theoretical histograms are obtained with the $k_{T}$-factorization approach using three different sets of the CCFM uPDFs: JH2013set1 (dashed), JH2013set2 (solid), and Jung-setA0 (dotted). We observe that both of the most recent CCFM uPDF sets, i.e. JH2013set1 and set2 significantly overshoot the experimental data in the whole kinematic range probed by the LHCb experiment. The results of the $k_{T}$-factorization with the JH2013set2 uPDF are slightly closer to the data points than the one obtained with the JH2013set1 uPDF set. The JH-2013-set1 gluon density is determined from the fit to inclusive $F_2$ HERA data only while the JH-2013-set2 set is determined from the fit to both $F_{2}^{(charm)}$ and $F_{2}$ data, however, even for the latter case, here the discrepancy between the predictions and the LHCb measurement is huge. Surprisingly, the Jung setA0 gluon density that a bit older set of the CCFM gluon uPDF leads to a much better description of the experimental data and seems to only slightly overestimates the measured distributions.  

%----------------------------------------------------------------------------
%----------------------------------------------------------------------------

\begin{figure}[!h]
\begin{minipage}{0.47\textwidth}
  \centerline{\includegraphics[width=1.0\textwidth]{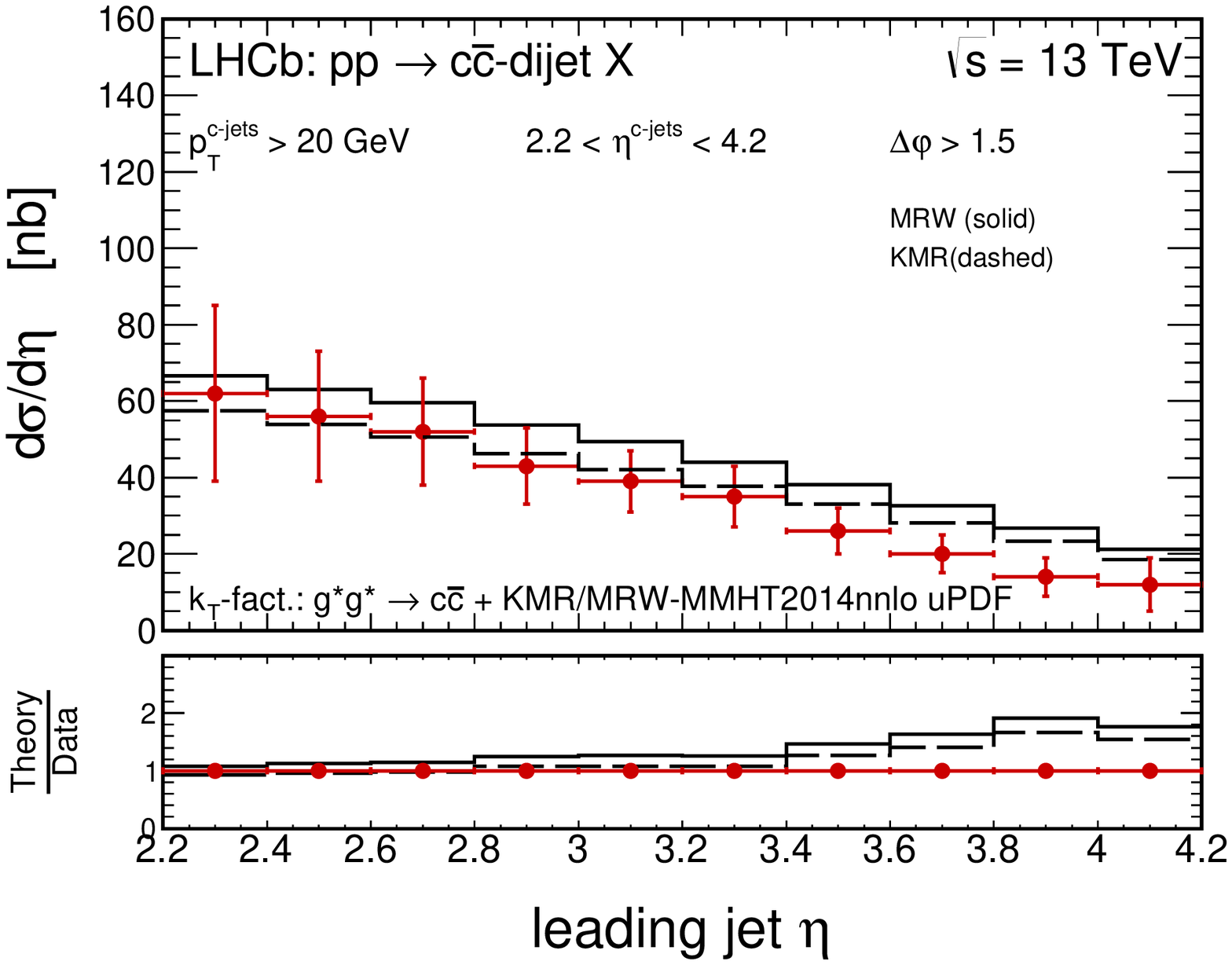}}
\end{minipage}
\begin{minipage}{0.47\textwidth}
  \centerline{\includegraphics[width=1.0\textwidth]{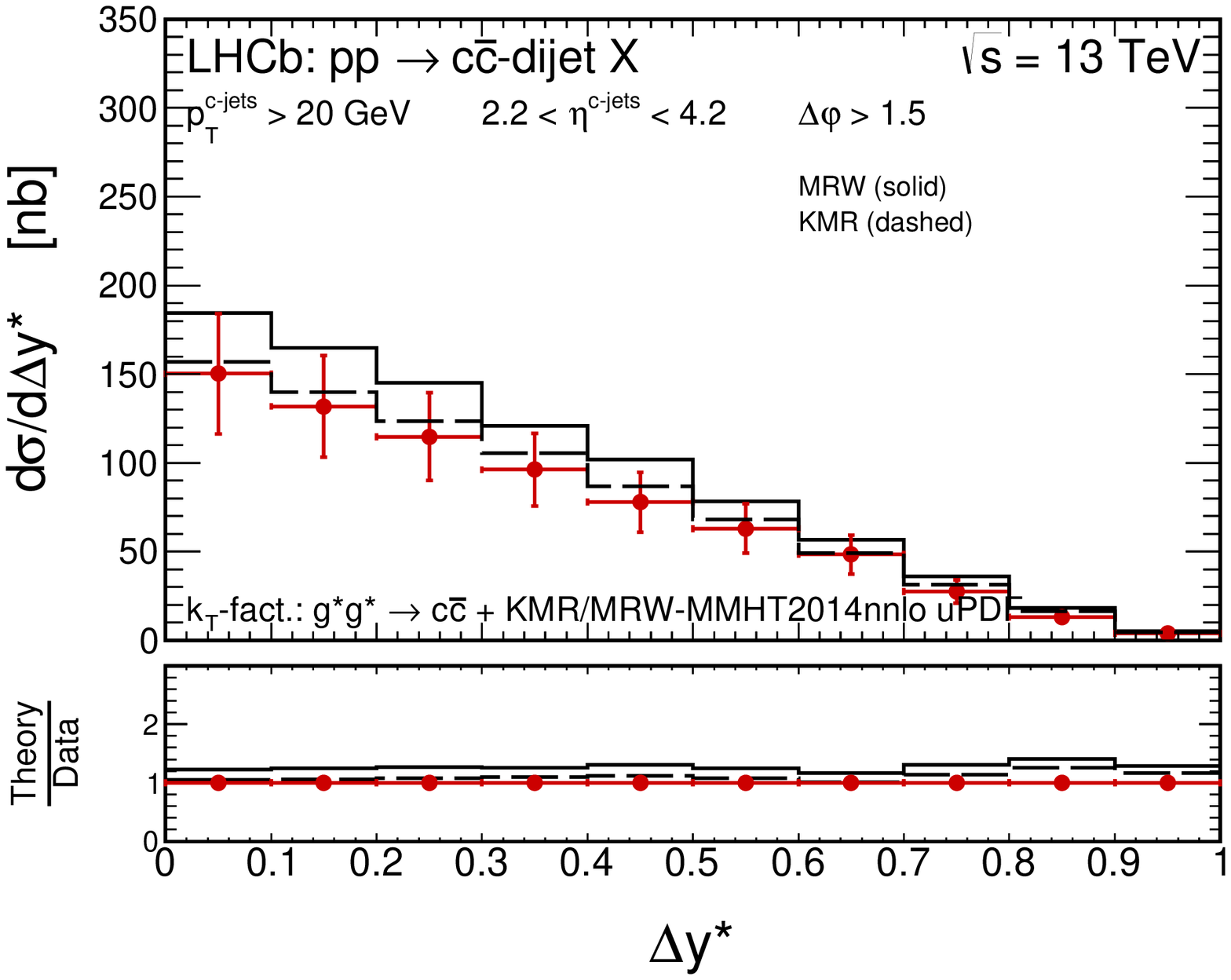}}
\end{minipage}
\begin{minipage}{0.47\textwidth}
  \centerline{\includegraphics[width=1.0\textwidth]{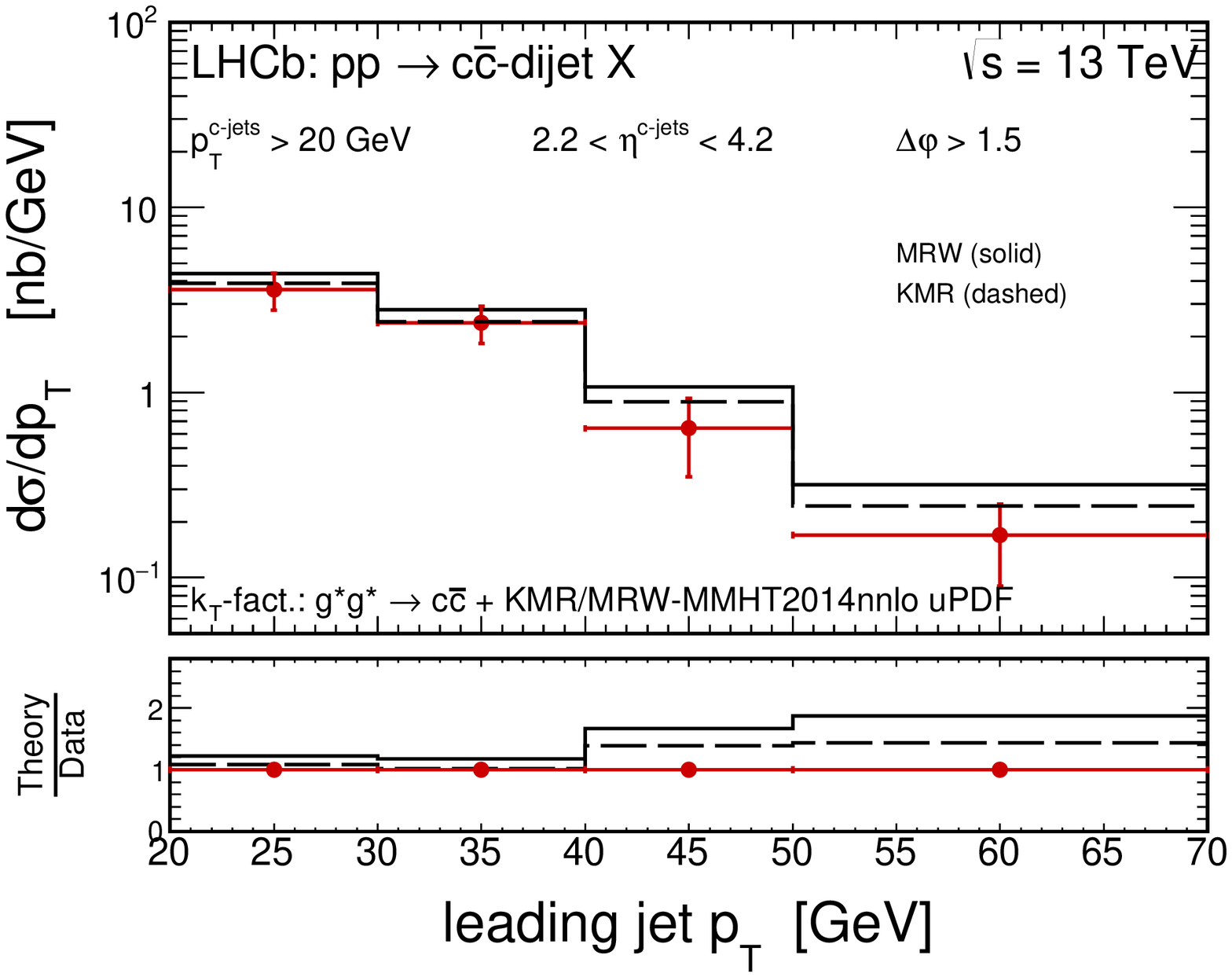}}
\end{minipage}
\begin{minipage}{0.47\textwidth}
  \centerline{\includegraphics[width=1.0\textwidth]{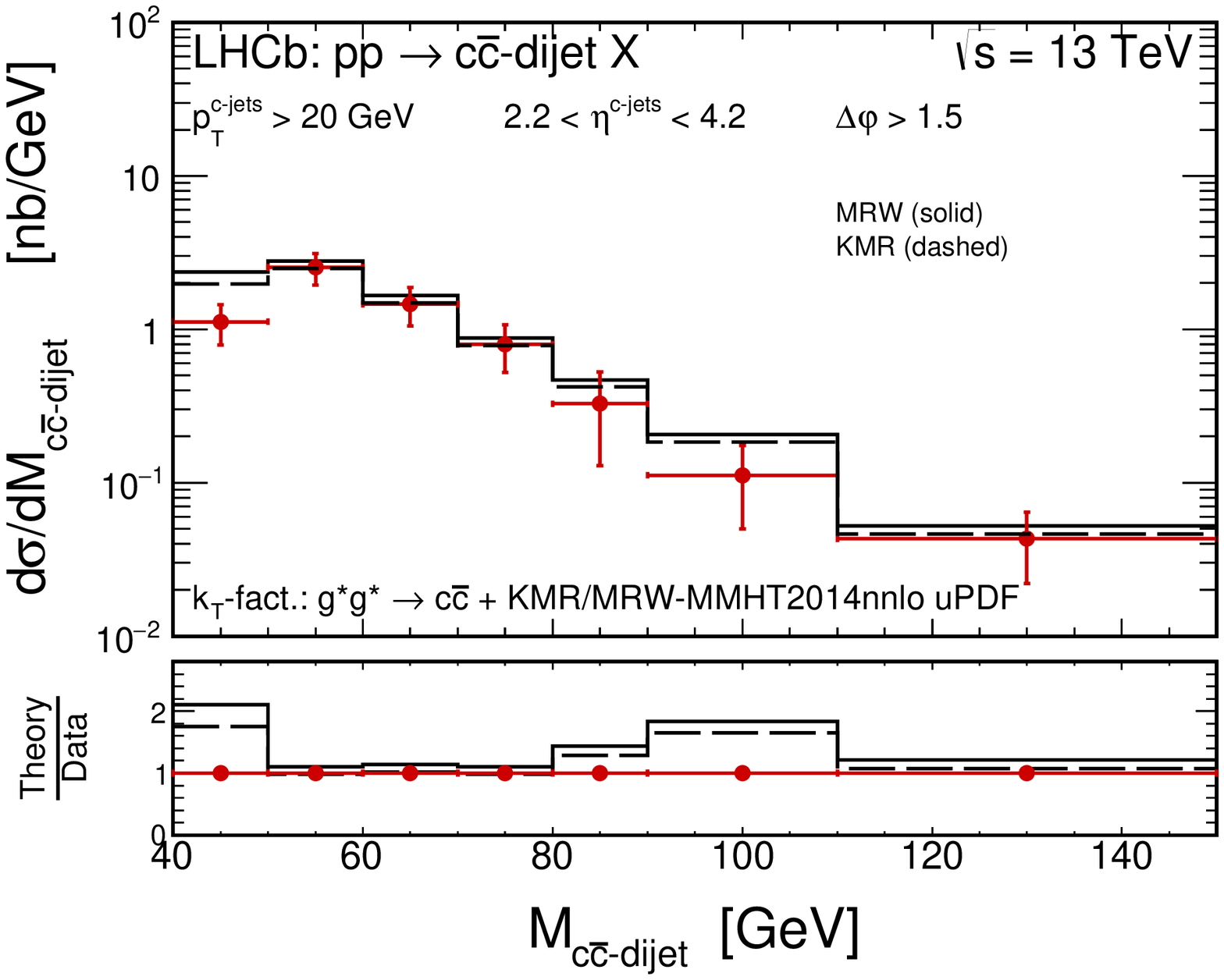}}
\end{minipage}
  \caption{The same as in Fig.~\ref{fig:3} but here the KMR and the MRW uPDFs are used.
\small 
}
\label{fig:4}
\end{figure}

In Fig.~\ref{fig:4} we show similar $k_{T}$-factorization results but here we use the KMR (dashed histograms) and the MRW (solid histograms) gluon uPDFs.
Both models seem to reasonably well describe the LHCb data. The histograms that represent the KMR predictions show a slightly better agreement with the data. It is visible especially for the leading jet $\eta$ and $\Delta y^*$ distributions which are displayed in linear scale. However, the effect is not significant for the overall picture and does not strongly disfavour the MRW gluon density here.  

%----------------------------------------------------------------------------
%----------------------------------------------------------------------------

\begin{figure}[!h]
\begin{minipage}{0.47\textwidth}
  \centerline{\includegraphics[width=1.0\textwidth]{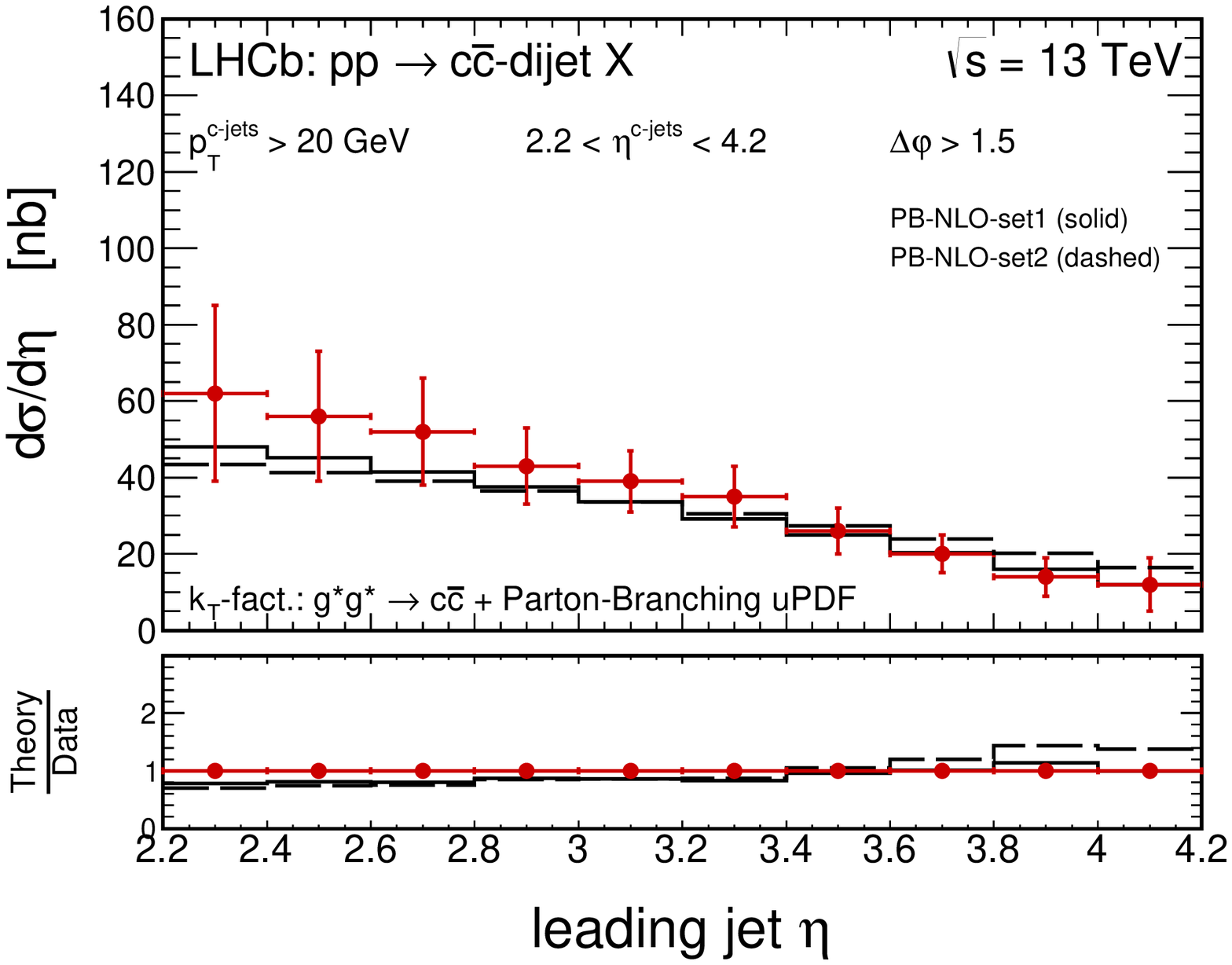}}
\end{minipage}
\begin{minipage}{0.47\textwidth}
  \centerline{\includegraphics[width=1.0\textwidth]{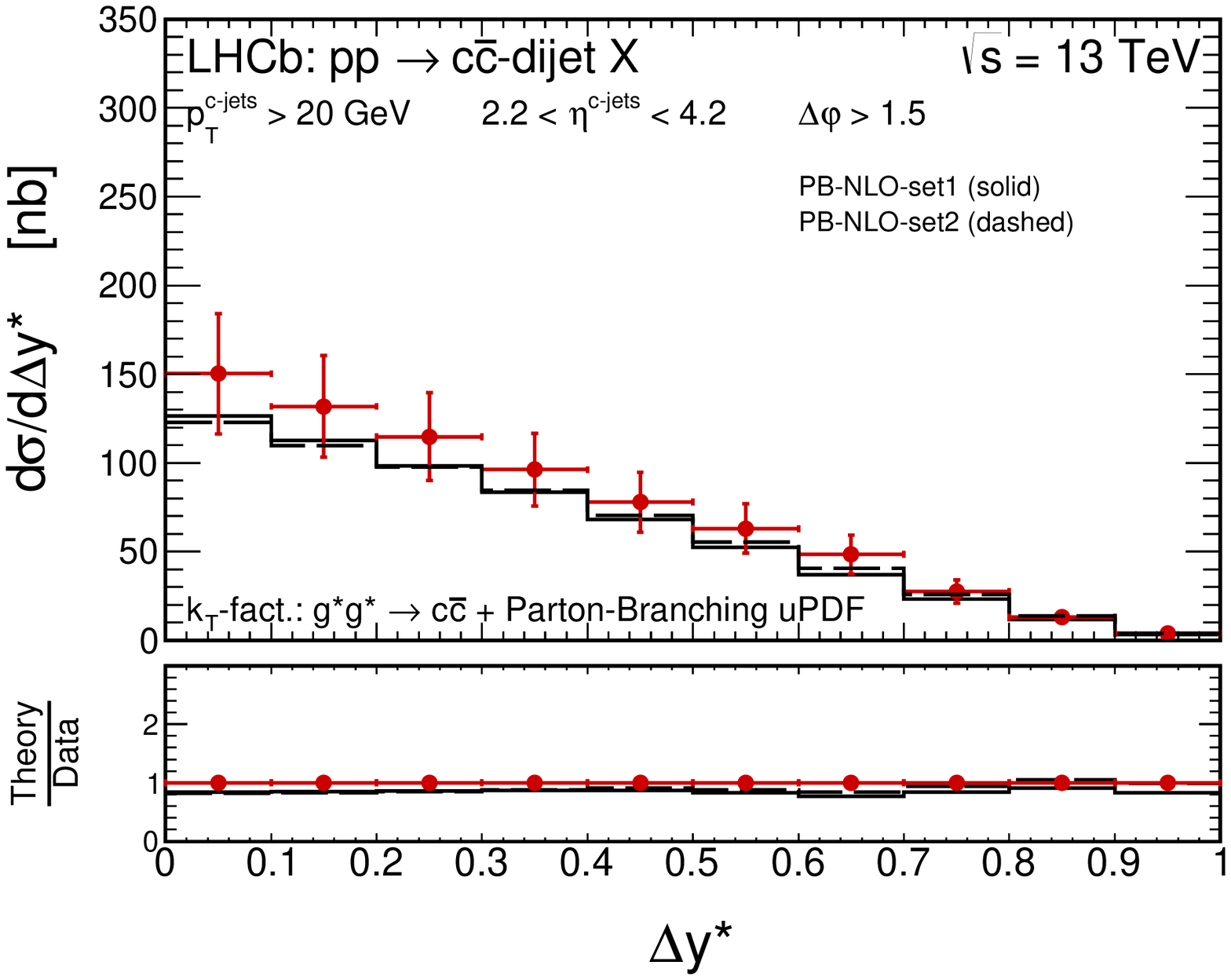}}
\end{minipage}
\begin{minipage}{0.47\textwidth}
  \centerline{\includegraphics[width=1.0\textwidth]{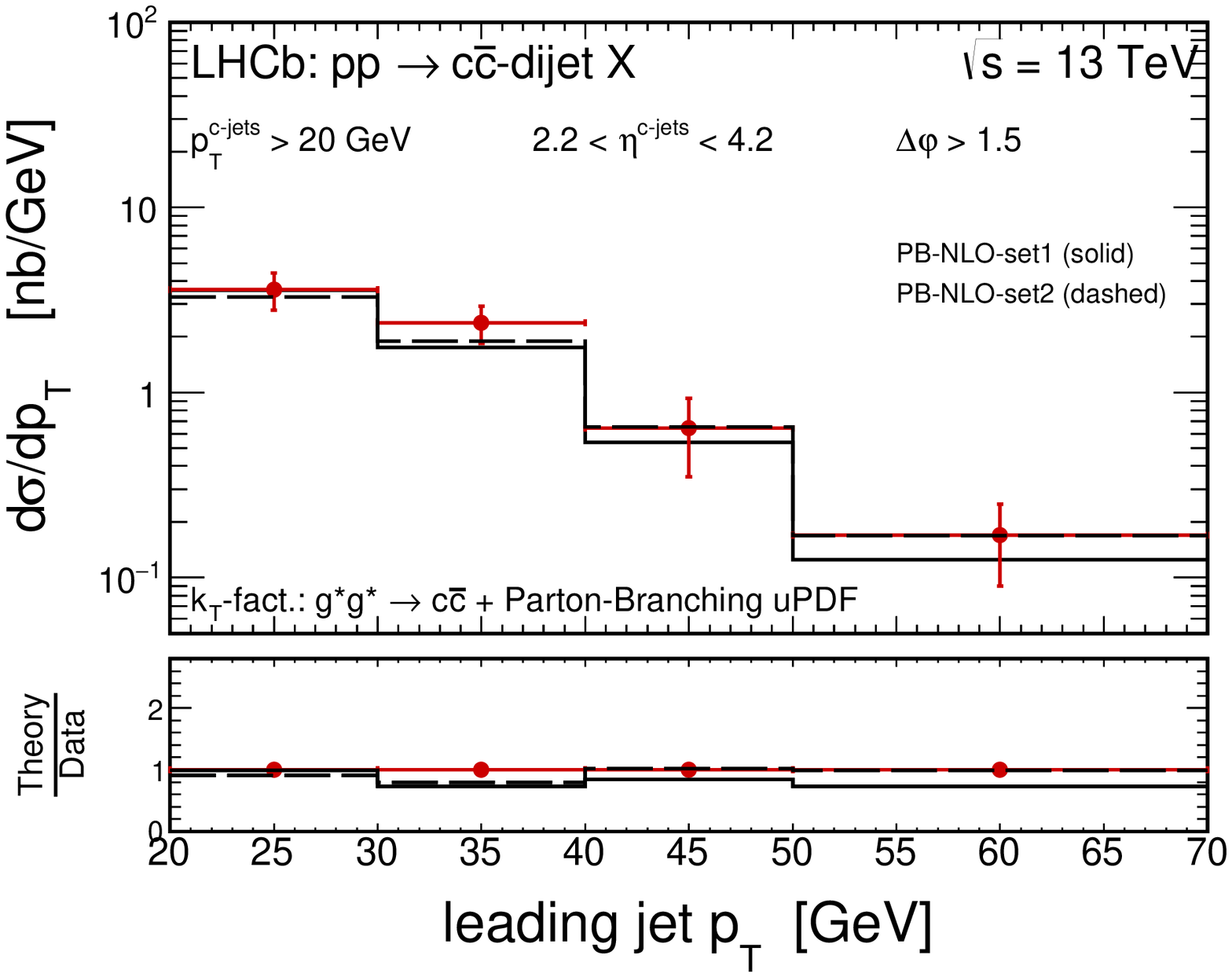}}
\end{minipage}
\begin{minipage}{0.47\textwidth}
  \centerline{\includegraphics[width=1.0\textwidth]{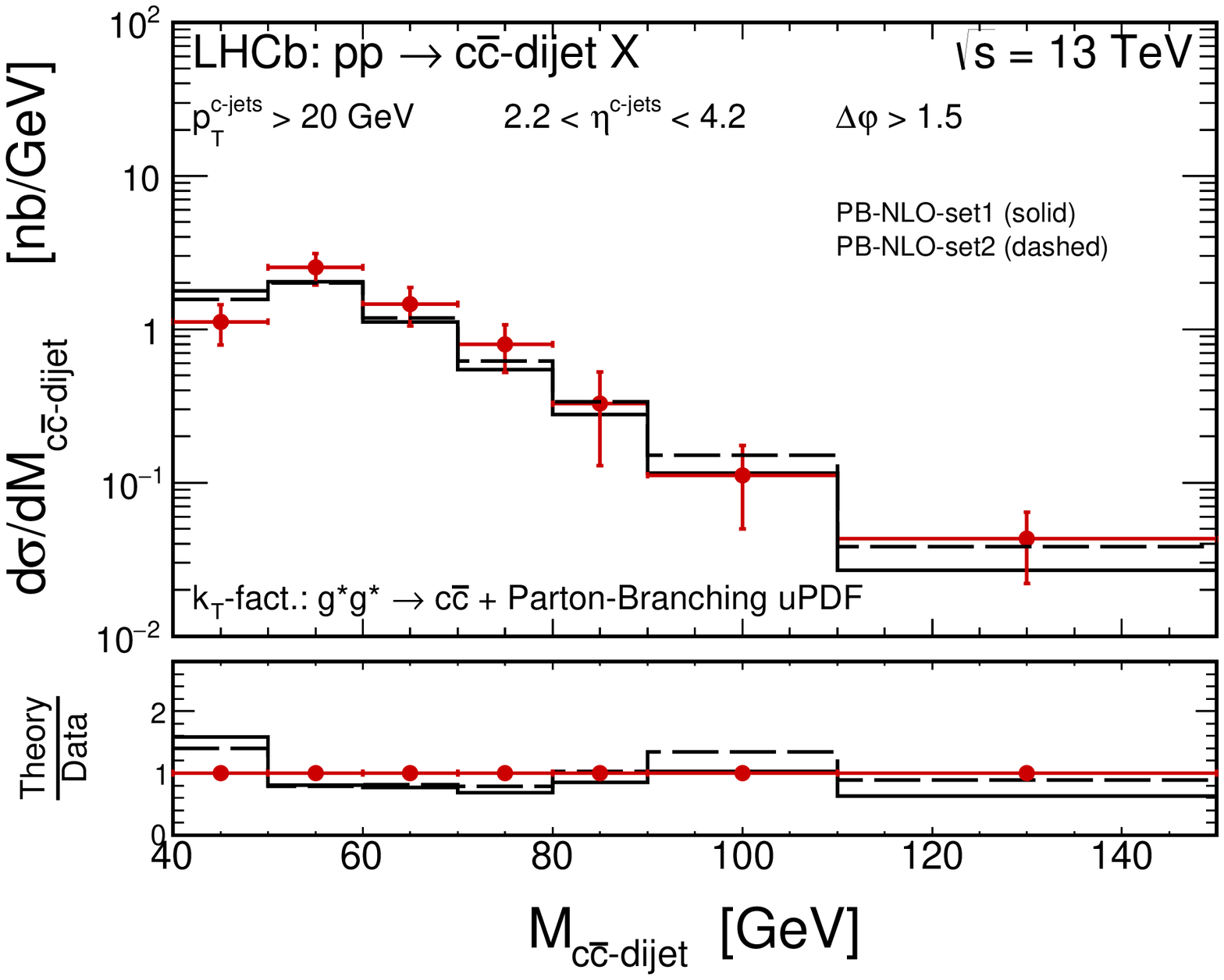}}
\end{minipage}
  \caption{The same as in Fig.~\ref{fig:3} but here the Parton Branching uPDFs are used.
\small 
}
\label{fig:5}
\end{figure}

Next, we also examined the Parton-Branching gluon uPDFs against the LHCb data. In Fig.~\ref{fig:5} we present our predictions for the PB-NLO-set1 (solid) and PB-NLO-set2 (dashed) gluon densities. In opposite to the CCFM and the KMR/MRW cases, here a slight tendency to underestimate the data points appears. However, both of the gluon densities provide a very good description of the measured distributions and both lead to a very similar results. A small missing strength is observed only at smaller values of the leading jet $\eta$ and some differences between the two predictions are found only at large leading jet $p_{T}$'s and large dijet invariant masses $M_{c\bar c\text{-}\mathrm{dijet}}$. The PB-NLO-set2 density seems to reproduce slopes of the distributions a bit better than the PB-NLO-set1 one. The two models of the gluon uPDF differ from each other in a value of the starting evolution scale as well as in a set of the argument of $\alpha_S$. In the set1 case, the integrated parton density and the initial parameters are the same as the ones obtained by HERAPDF2.0 PDF. In the set2 case, the parameters are slightly modified and thus the integrated parton distributions do not fully reproduce the collinear ones. However, in both cases a reasonably good fit to HERA data was obtained.

%----------------------------------------------------------------------------
%----------------------------------------------------------------------------

\subsubsection{Bottom dijets}

\begin{figure}[!h]
\begin{minipage}{0.47\textwidth}
  \centerline{\includegraphics[width=1.0\textwidth]{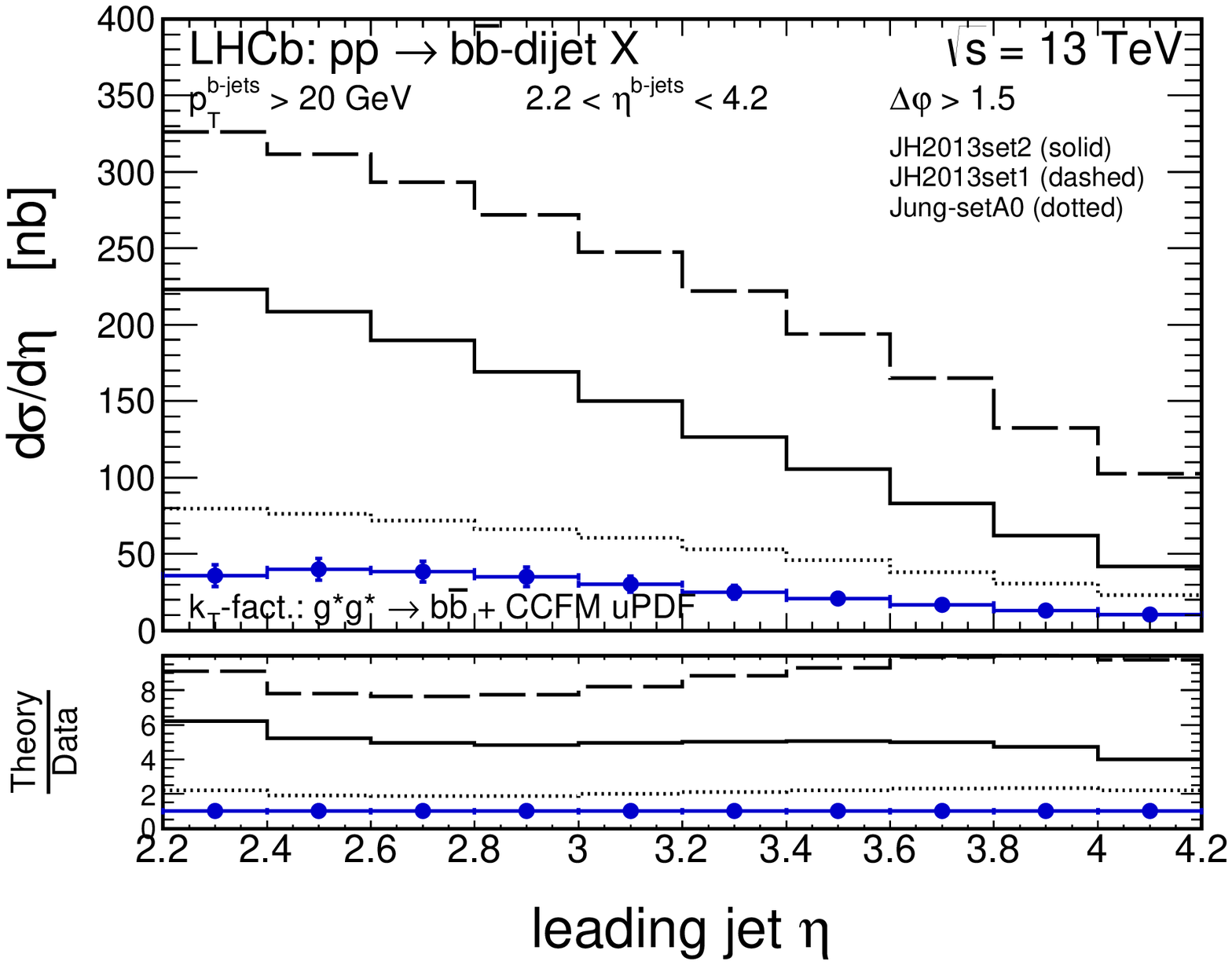}}
\end{minipage}
\begin{minipage}{0.47\textwidth}
  \centerline{\includegraphics[width=1.0\textwidth]{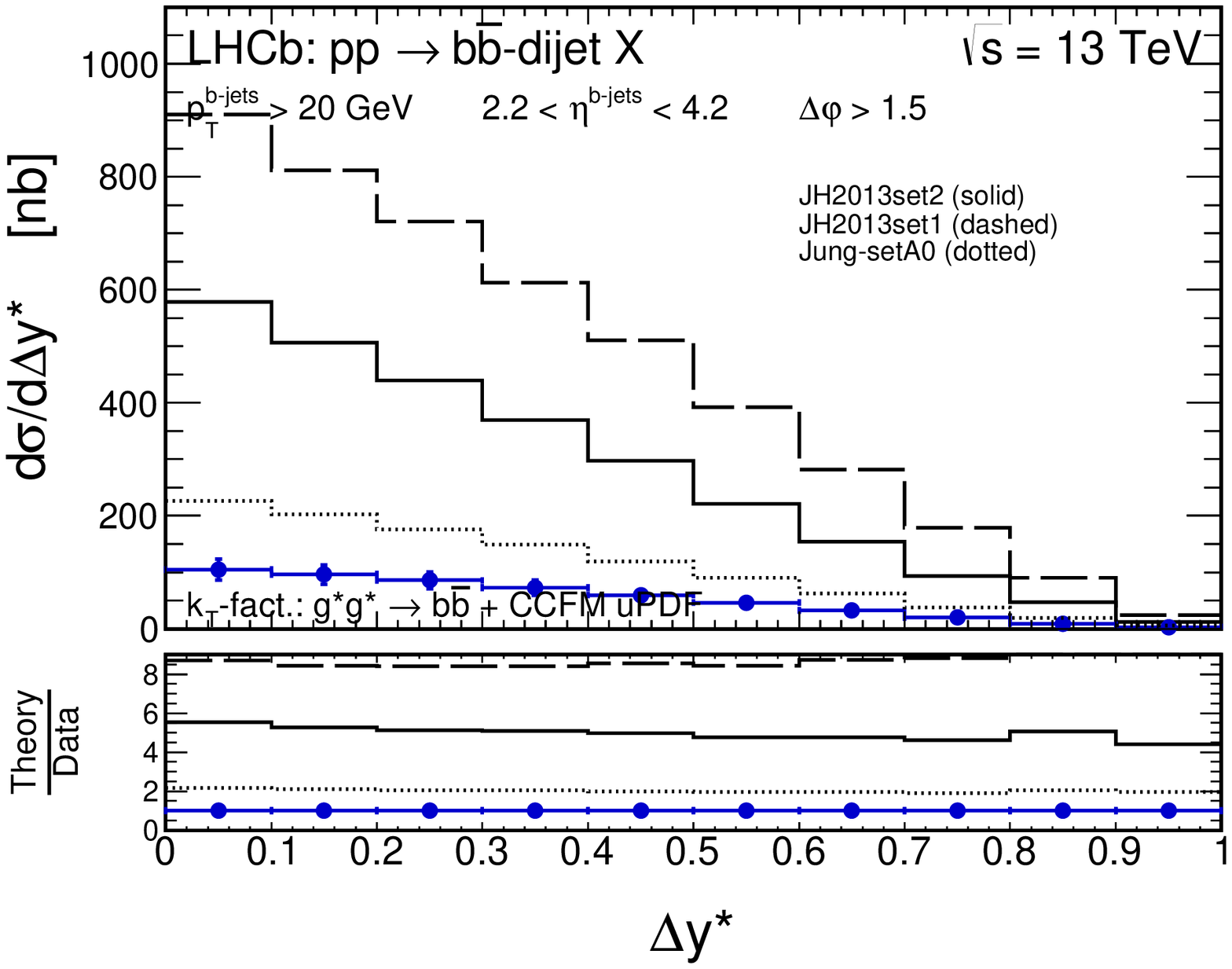}}
\end{minipage}
\begin{minipage}{0.47\textwidth}
  \centerline{\includegraphics[width=1.0\textwidth]{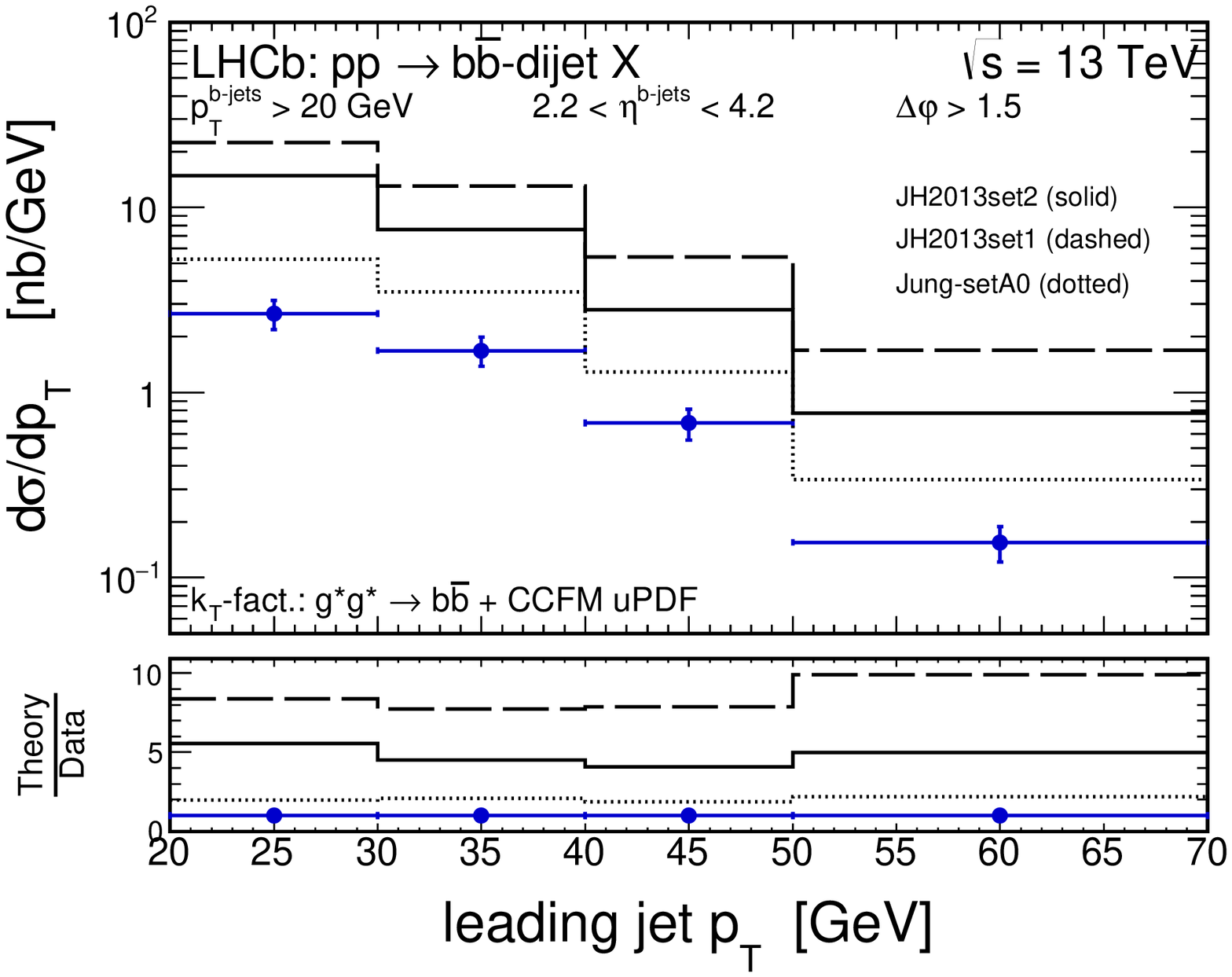}}
\end{minipage}
\begin{minipage}{0.47\textwidth}
  \centerline{\includegraphics[width=1.0\textwidth]{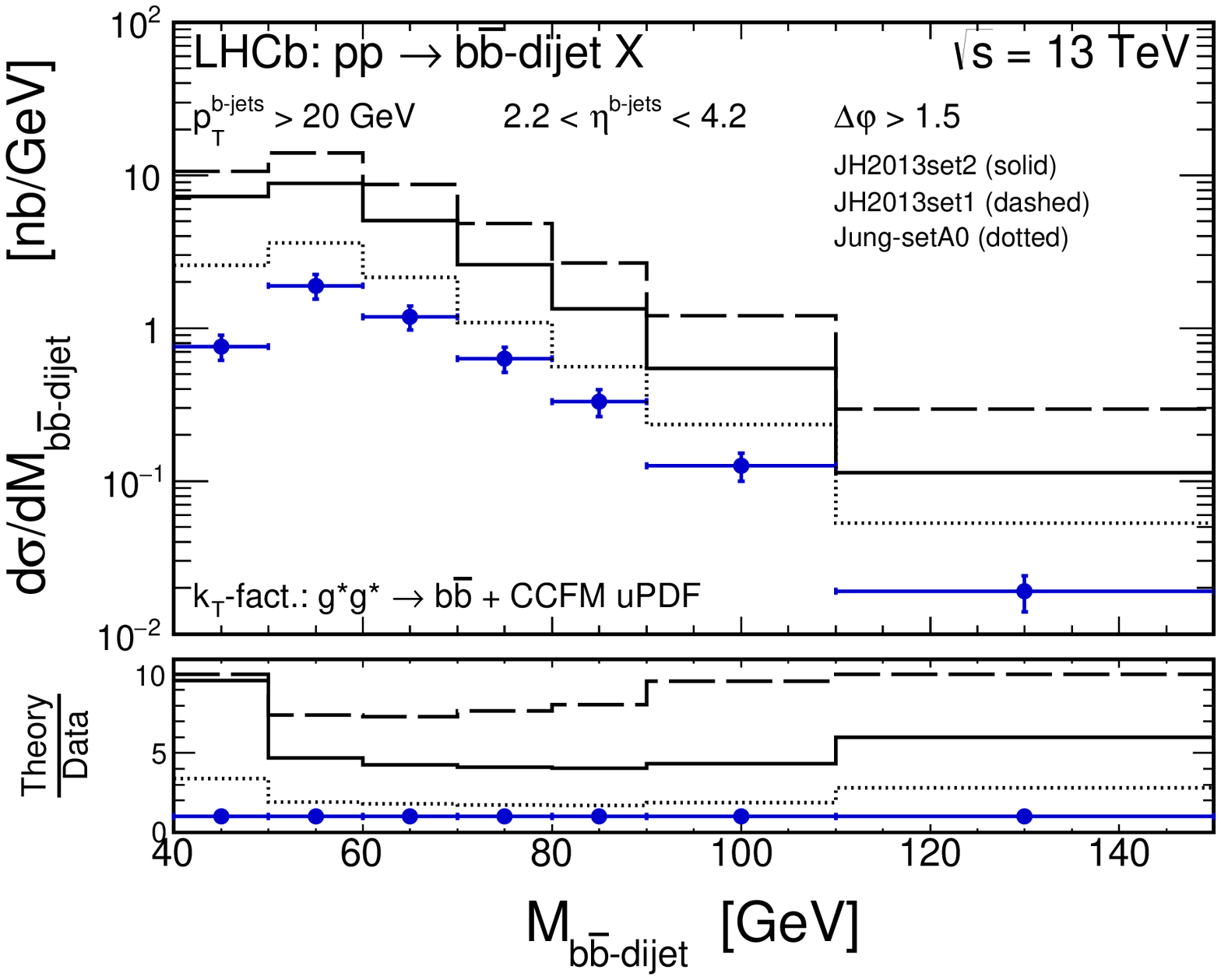}}
\end{minipage}
  \caption{The differential cross sections for forward production of $b\bar b$-dijets in $pp$-scattering at $\sqrt{s}=13$ TeV as a function of the leading jet $\eta$ (top left), the rapidity difference $\Delta y^{*}$ (top right), the leading jet $p_{T}$ (bottom left) and the dijet invariant mass $M_{b\bar b\text{-}\mathrm{dijet}}$ (bottom right). Here the dominant pQCD $g^*g^* \to b \bar b$ mechanism is taken into account. The theoretical histograms correspond to the $k_{T}$-factorization calculations obtained with the CCFM uPDFs.
\small 
}
\label{fig:6}
\end{figure}

\begin{figure}[!h]
\begin{minipage}{0.47\textwidth}
  \centerline{\includegraphics[width=1.0\textwidth]{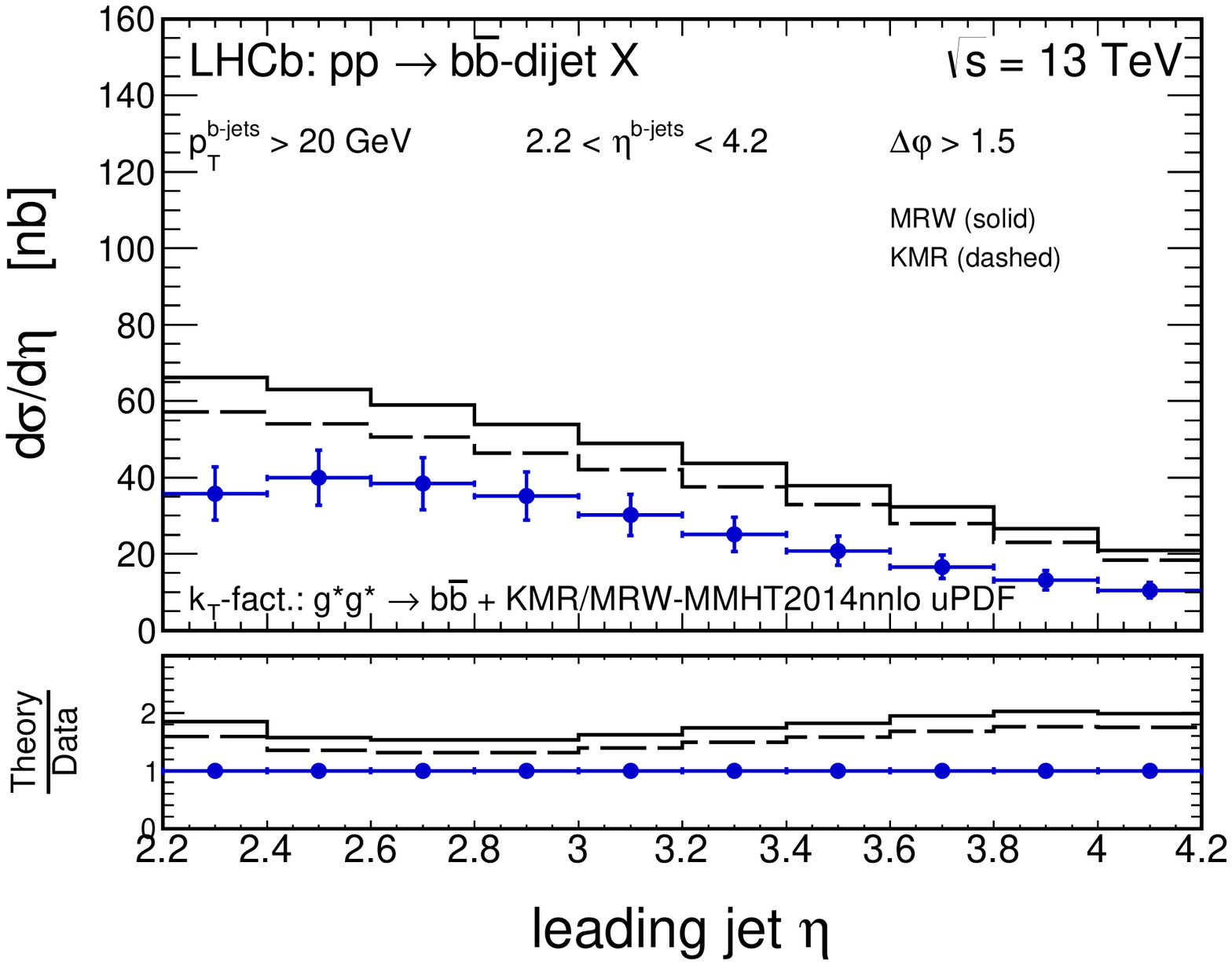}}
\end{minipage}
\begin{minipage}{0.47\textwidth}
  \centerline{\includegraphics[width=1.0\textwidth]{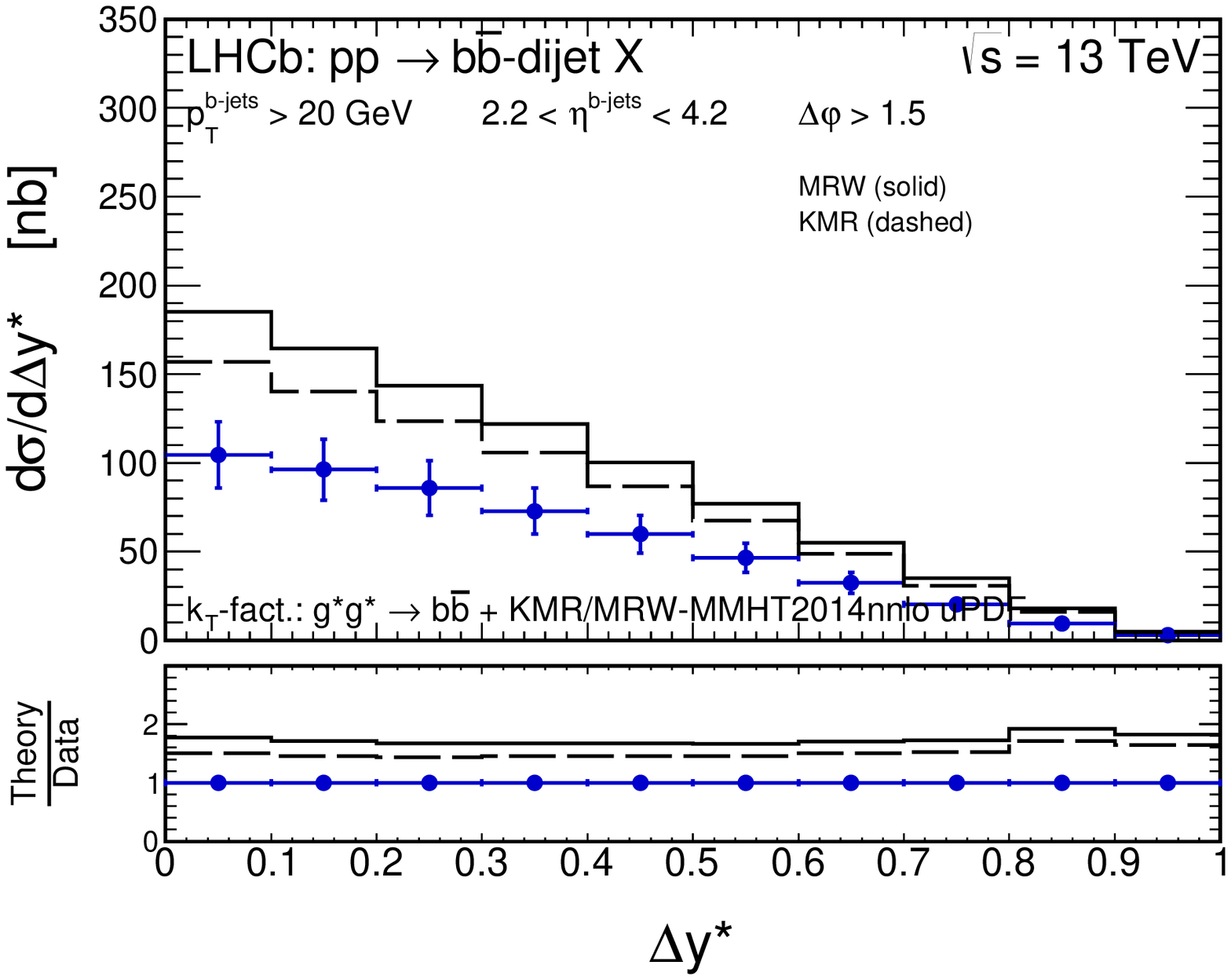}}
\end{minipage}
\begin{minipage}{0.47\textwidth}
  \centerline{\includegraphics[width=1.0\textwidth]{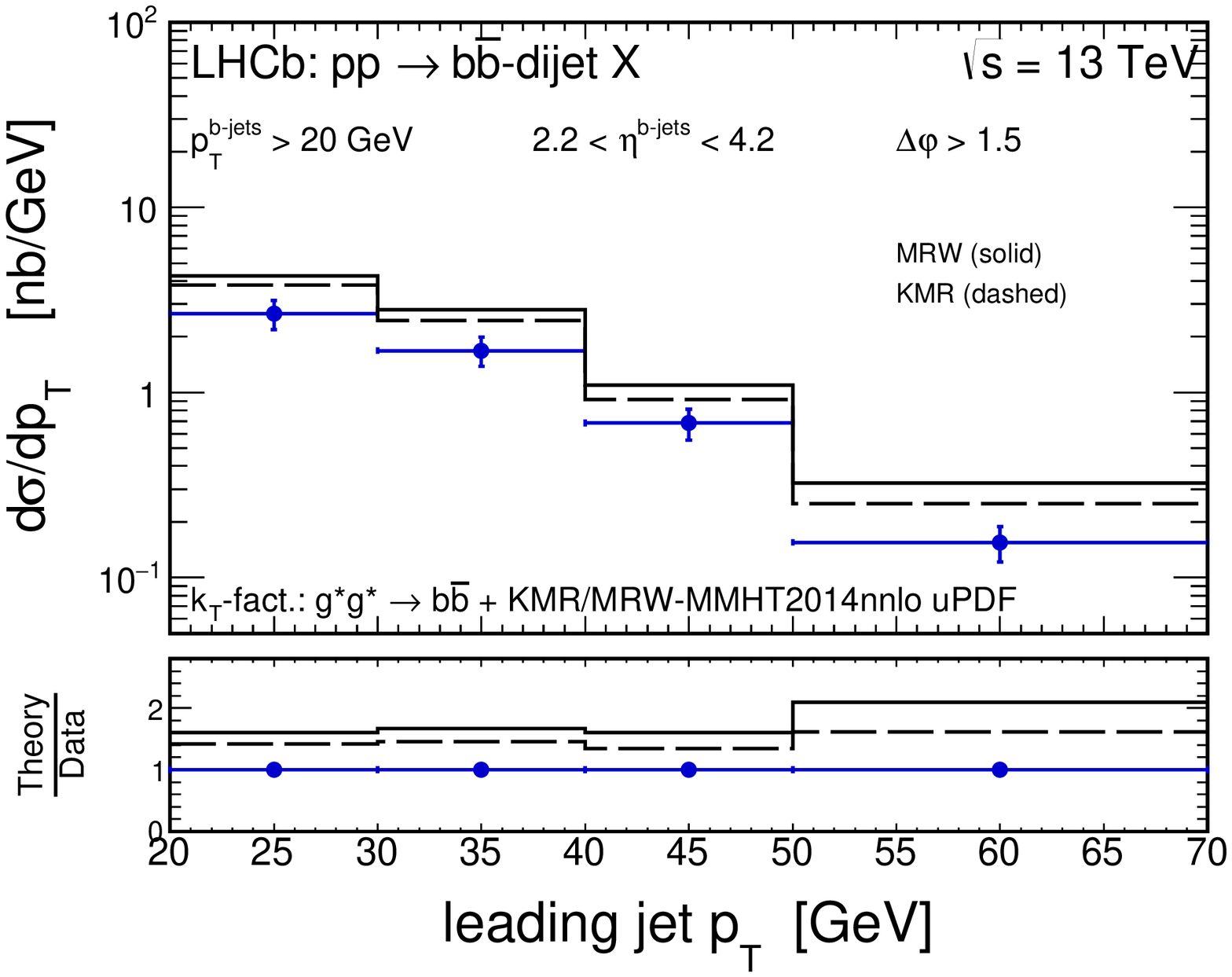}}
\end{minipage}
\begin{minipage}{0.47\textwidth}
  \centerline{\includegraphics[width=1.0\textwidth]{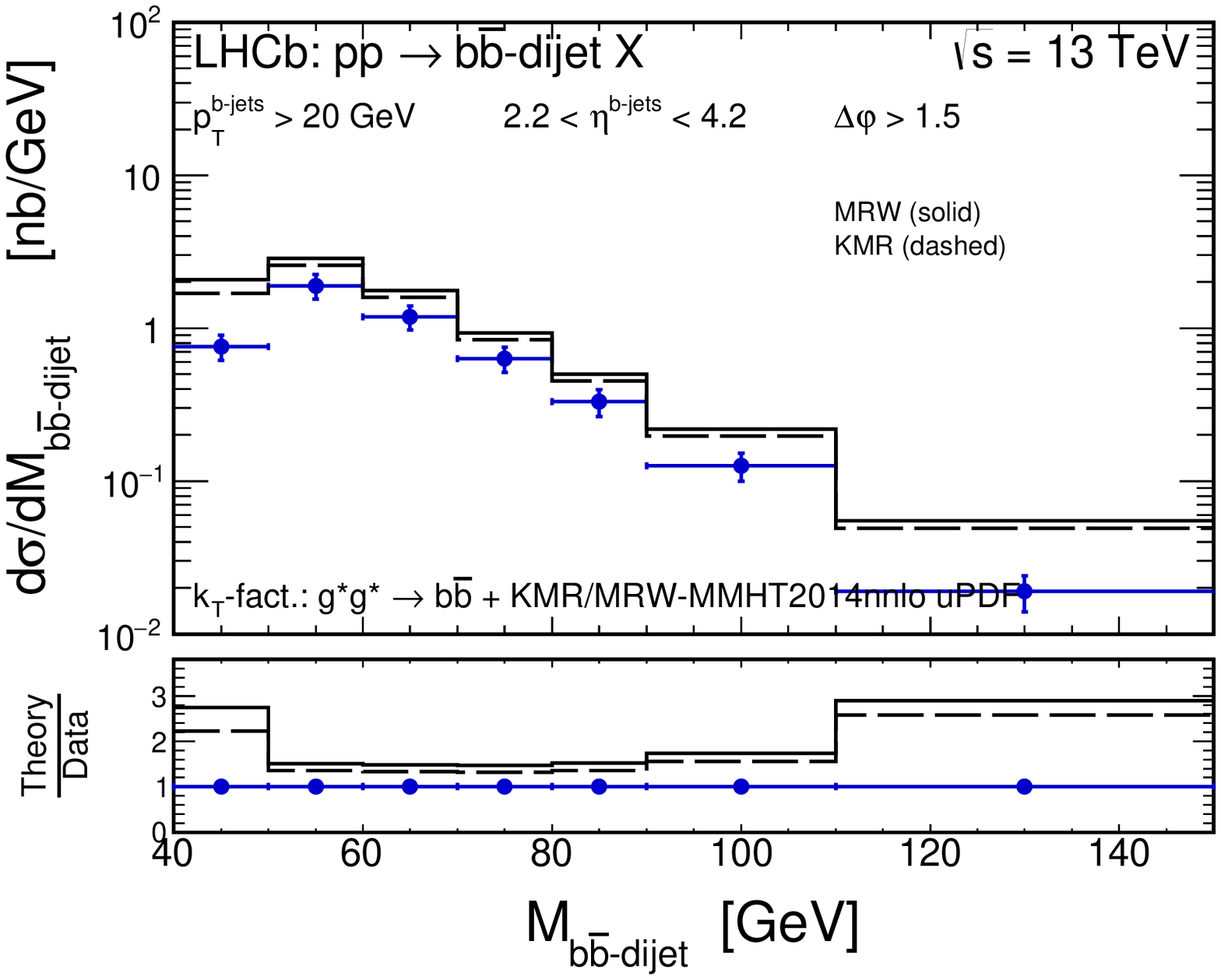}}
\end{minipage}
  \caption{The same as in Fig.~\ref{fig:6} but here the KMR and the MRW uPDFs are used.
\small 
}
\label{fig:7}
\end{figure}

\begin{figure}[!h]
\begin{minipage}{0.47\textwidth}
  \centerline{\includegraphics[width=1.0\textwidth]{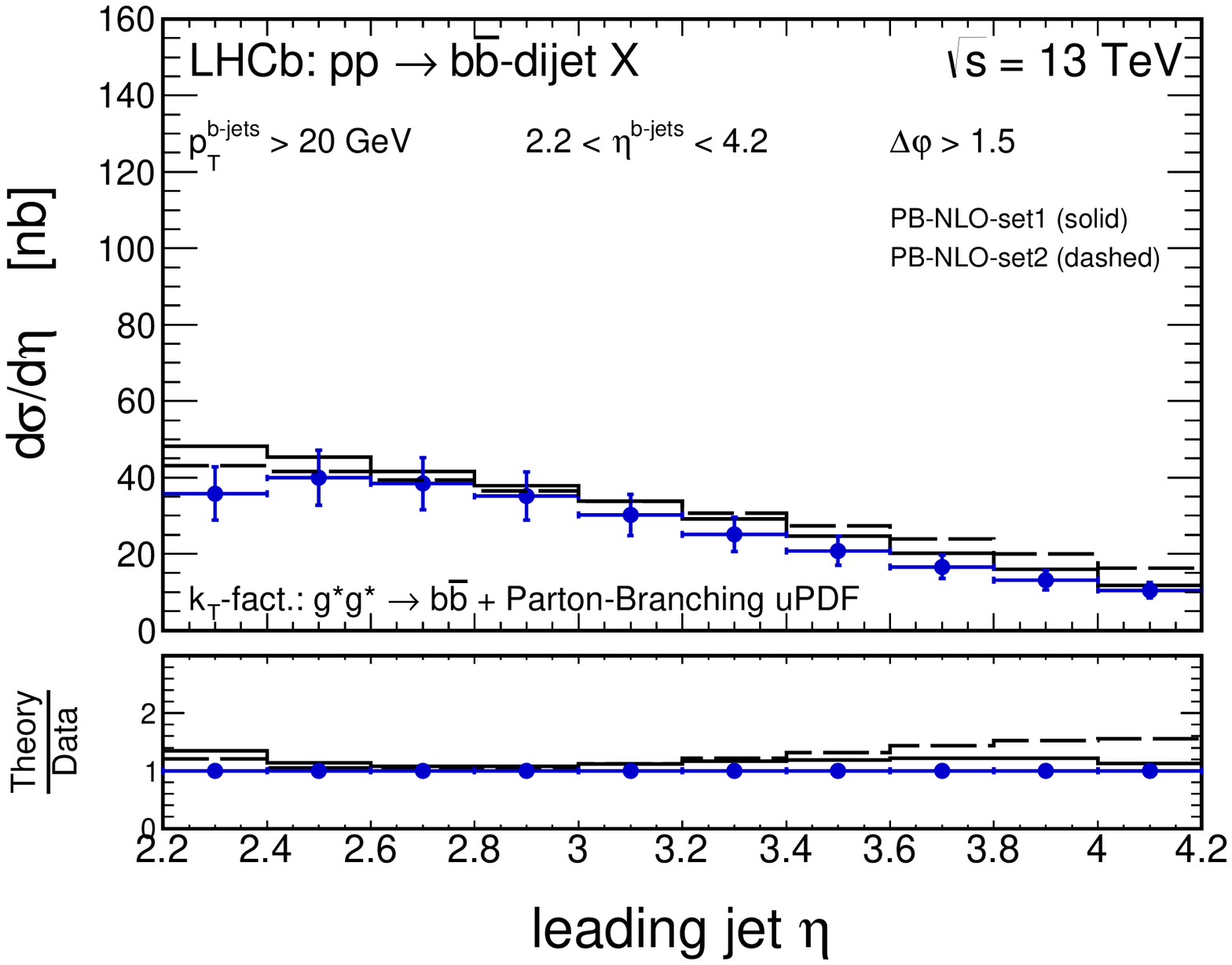}}
\end{minipage}
\begin{minipage}{0.47\textwidth}
  \centerline{\includegraphics[width=1.0\textwidth]{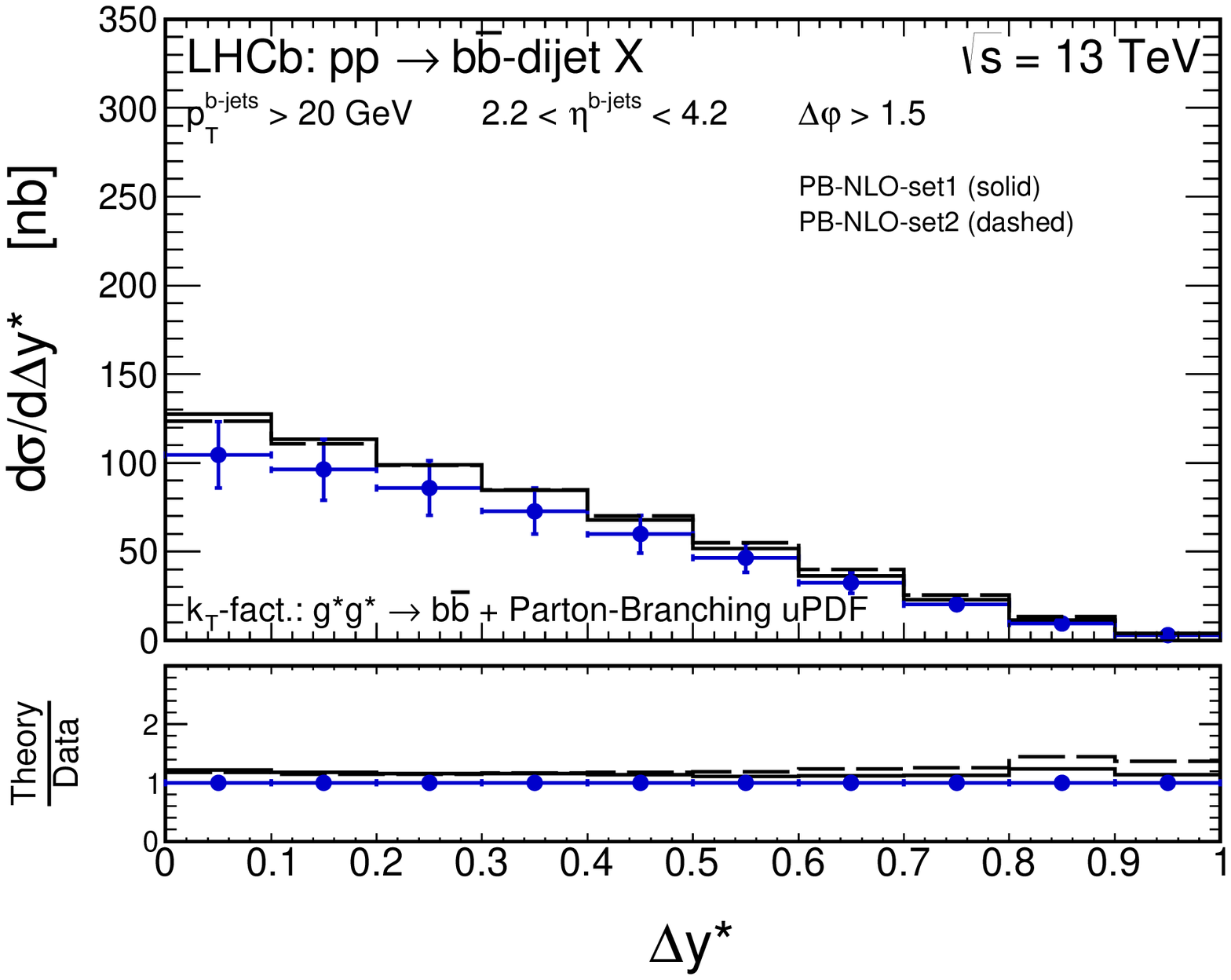}}
\end{minipage}
\begin{minipage}{0.47\textwidth}
  \centerline{\includegraphics[width=1.0\textwidth]{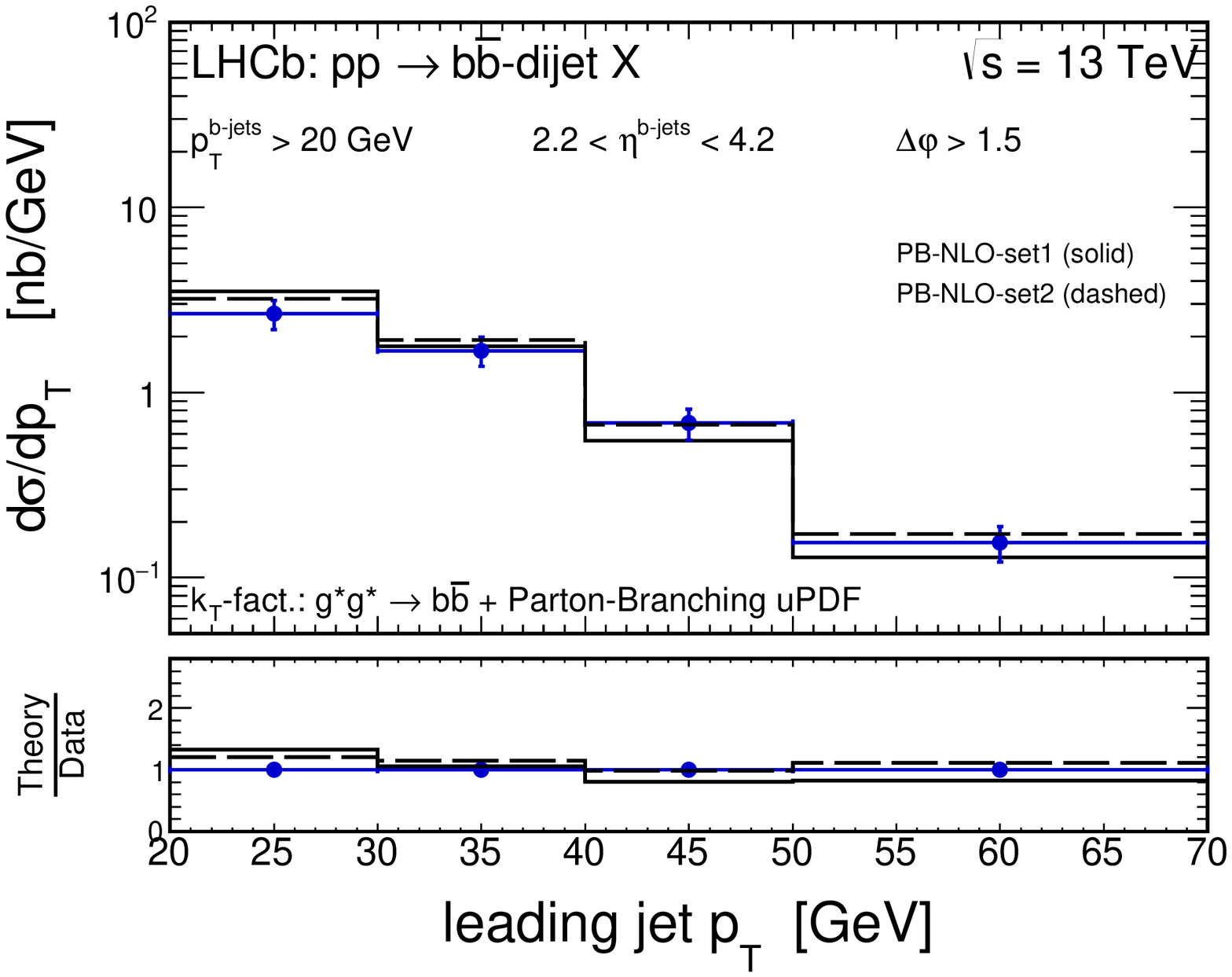}}
\end{minipage}
\begin{minipage}{0.47\textwidth}
  \centerline{\includegraphics[width=1.0\textwidth]{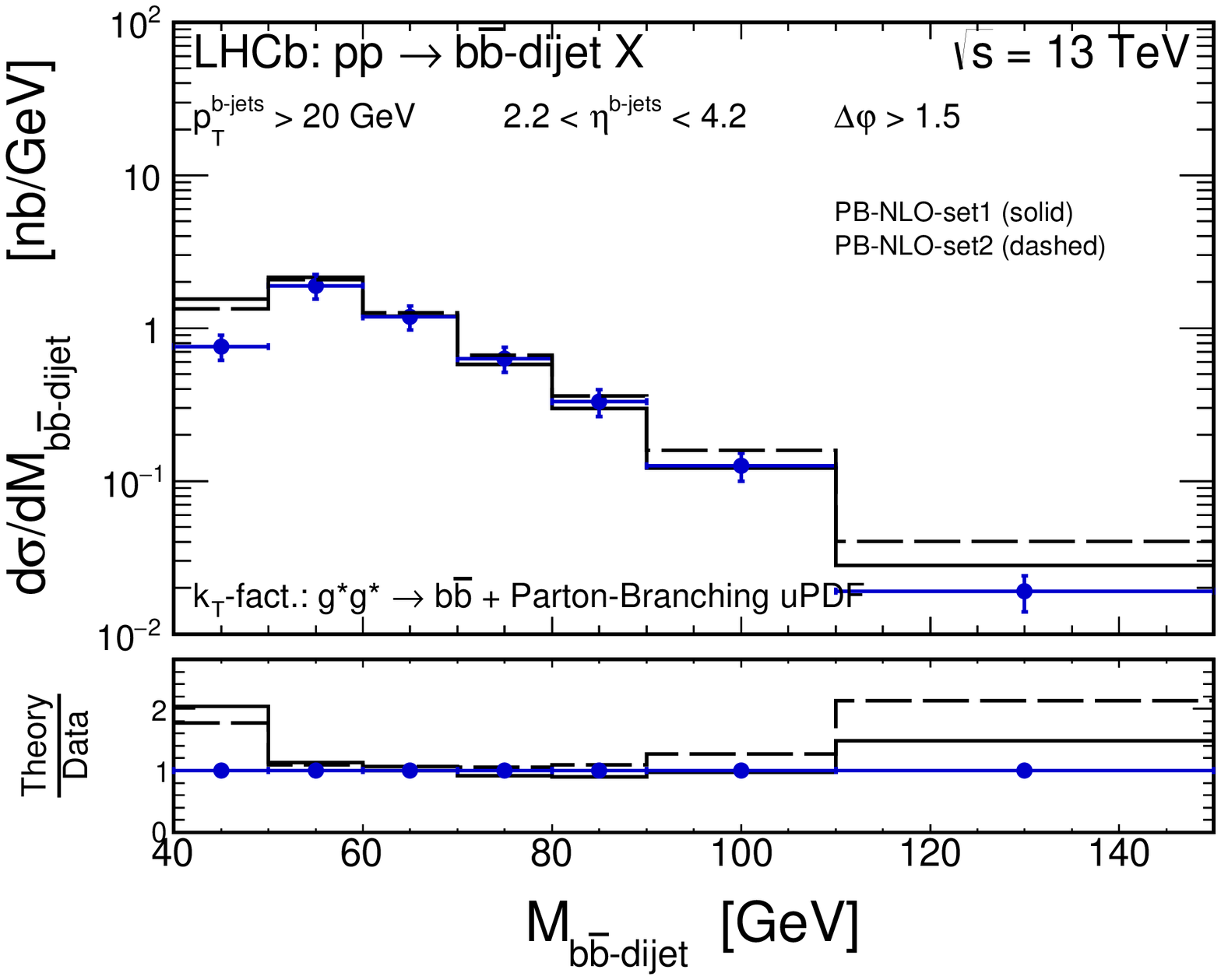}}
\end{minipage}
  \caption{The same as in Fig.~\ref{fig:6} but here the Parton Branching uPDFs are used.
\small 
}
\label{fig:8}
\end{figure}

Now let us go to the case of the $b\bar b$-dijet production. In Figs.~\ref{fig:6},~\ref{fig:7}, and~\ref{fig:8}, similarly as in previous subsection, we compare our predictions for the CCFM, KMR/MRW and the Parton-Branching gluon uPDFs to the LHCb experimental data on bottom dijets. For the LHCb experiment acceptance used here, one should not expect large effects related with the heavy quark mass. Although the experimental data show some small but visible difference of the cross sections for charm and bottom flavours, our theoretical predictions are almost insensitive to the heavy quark mass. The difference between our theoretical histograms for charm and bottom dijets is extremely small. We will discuss this issue in more detail in the following, when presenting results for the charm to bottom dijet ratio $R=\frac{c\bar c}{b\bar b}$ data.

Taking into account all the statements above, our basic findings about the application of different uPDF models in description of the LHCb $c\bar c$-dijet data applies also here. The main conclusion from a comparison of our predictions now with the $b\bar b$-dijet data is that the CCFM and the MRW/KMR gluon densities lead to a larger discrepancy between predictions and the data, overshooting the $b\bar b$-dijet data points even more stronger than in the $c\bar c$-dijet case. On the other hand the Parton-Branching densities describe the bottom data better than the charm one, where we have found a small missing strength, but here this is not the case.     

%----------------------------------------------------------------------------
%----------------------------------------------------------------------------

%----------------------------------------------------------------------------
%----------------------------------------------------------------------------

%----------------------------------------------------------------------------
%----------------------------------------------------------------------------

\subsection{Kinematics of the process probed by the LHCb experiment}

In the previous section we have shown that within the $k_{T}$-factorization approach some of popular models of the gluon uPDF
have difficulties with a reasonable description of the LHCb data on forward production of heavy flavoured dijets. In order to identify the possible reasons of the failure we wish to carefully illustrate the kinematics behind the considered process as probed by the LHCb experiment.

\begin{figure}[!h]
\begin{minipage}{0.33\textwidth}
  \centerline{\includegraphics[width=1.0\textwidth]{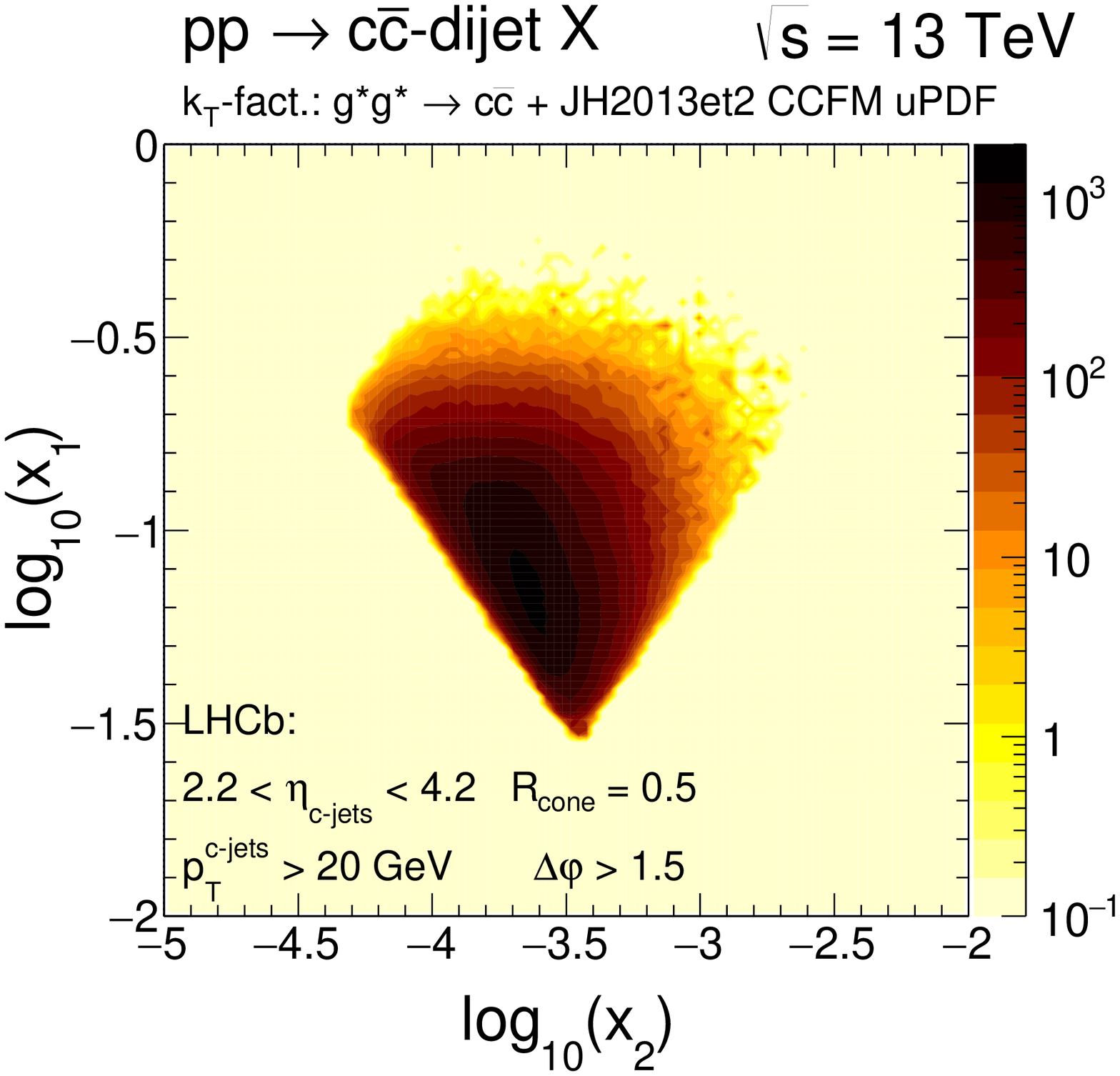}}
\end{minipage}
\begin{minipage}{0.33\textwidth}
  \centerline{\includegraphics[width=1.0\textwidth]{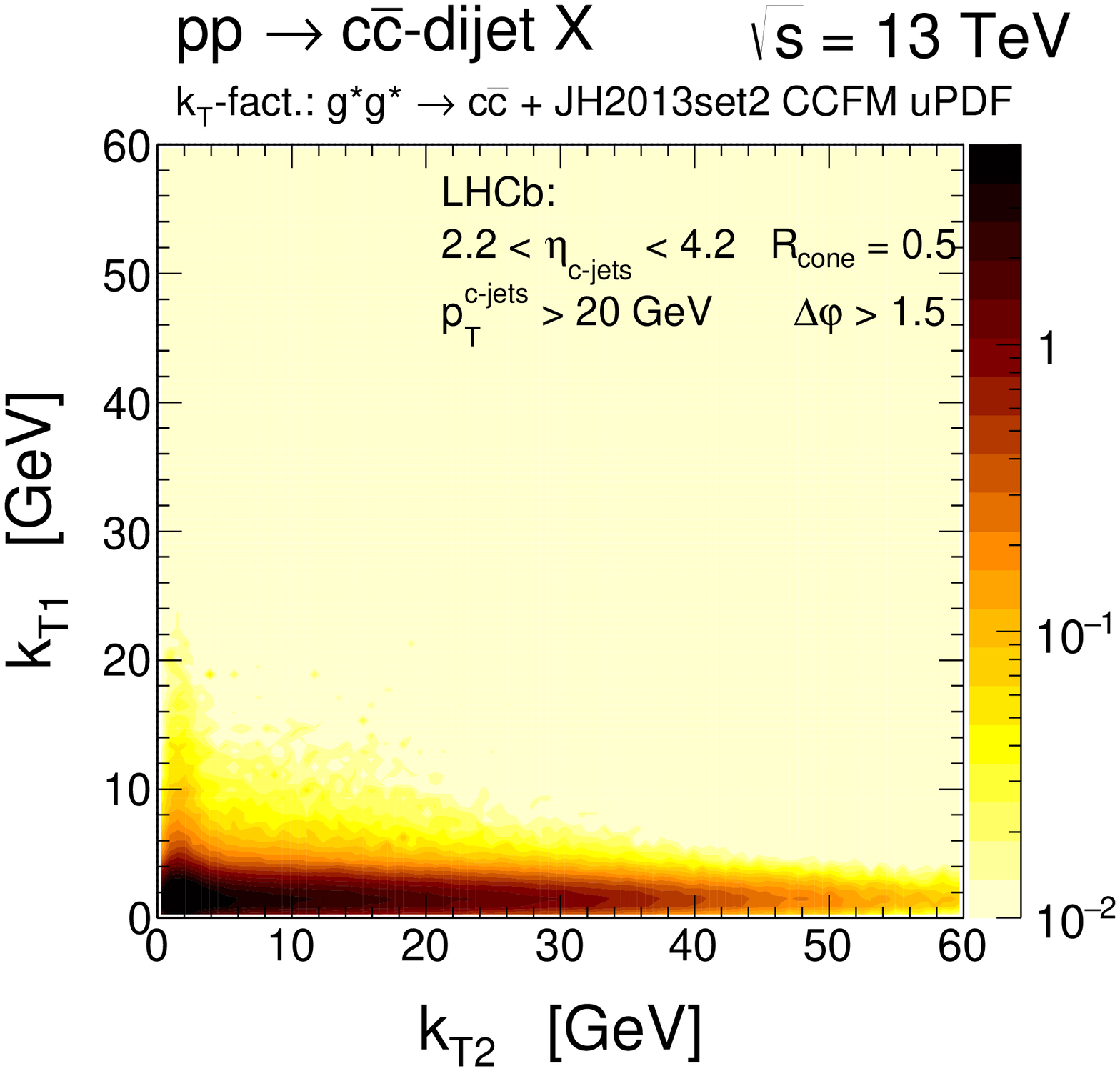}}
\end{minipage}
\begin{minipage}{0.33\textwidth}
  \centerline{\includegraphics[width=1.0\textwidth]{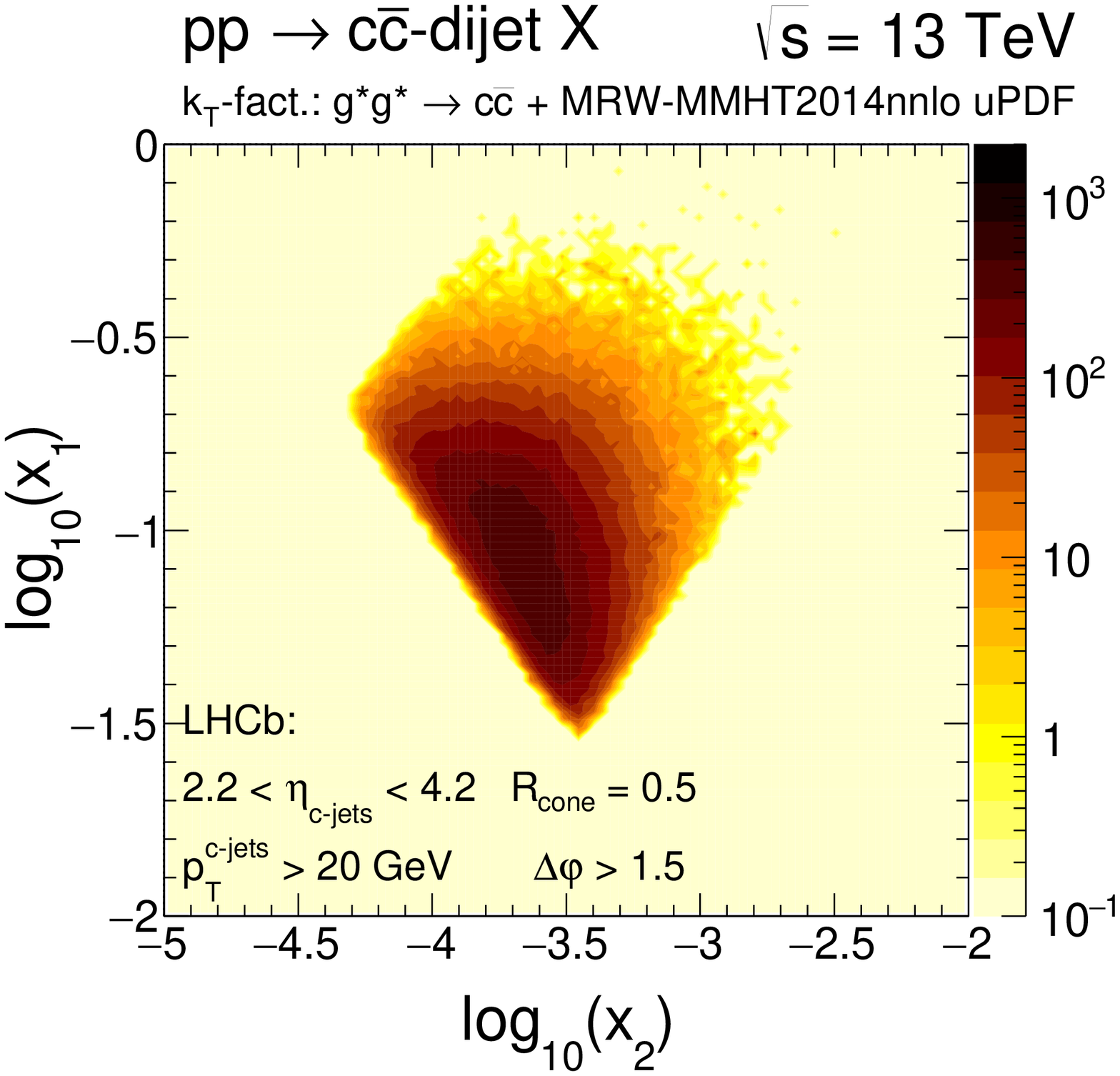}}
\end{minipage}
\begin{minipage}{0.33\textwidth}
  \centerline{\includegraphics[width=1.0\textwidth]{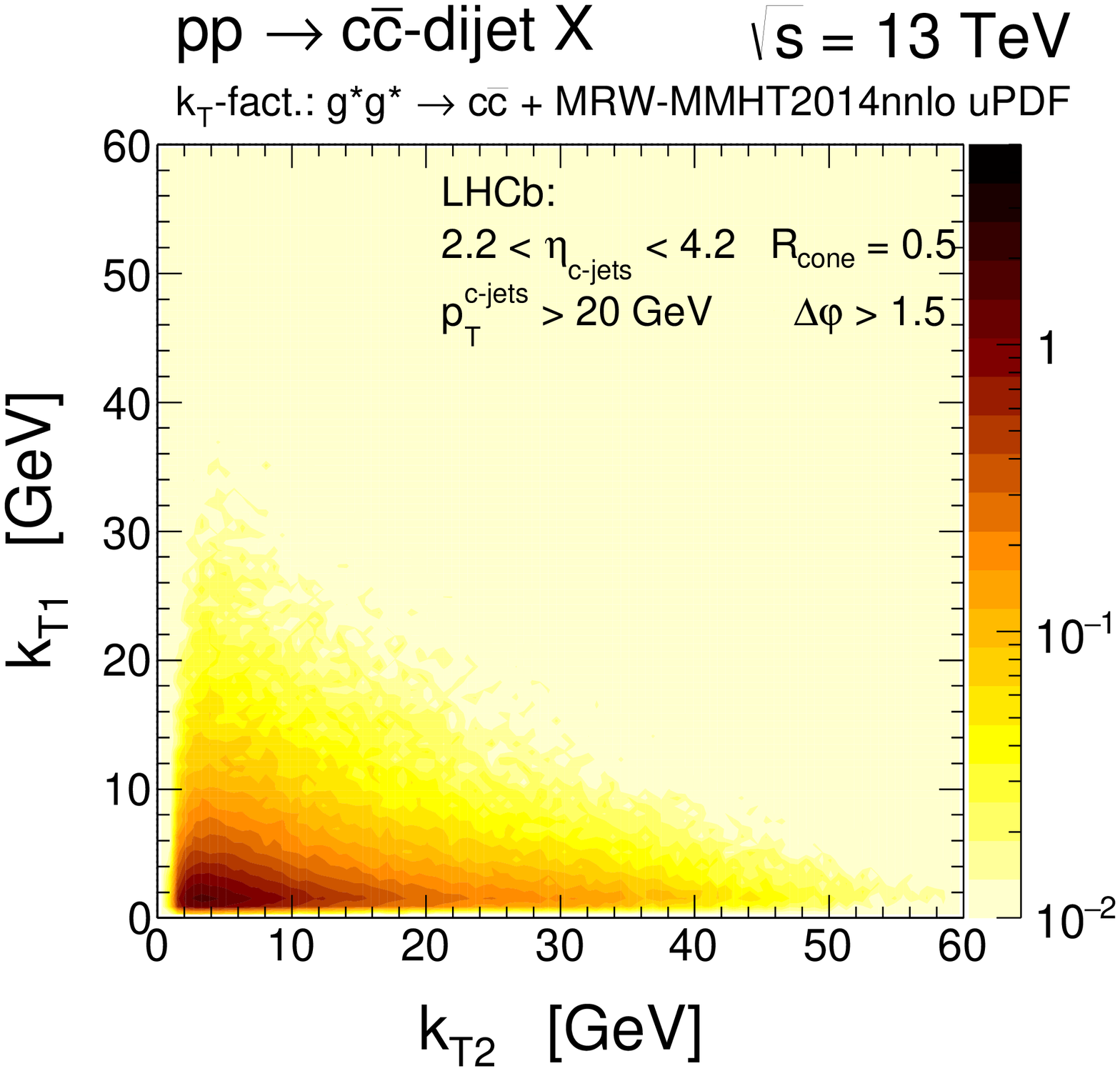}}
\end{minipage}
\begin{minipage}{0.33\textwidth}
  \centerline{\includegraphics[width=1.0\textwidth]{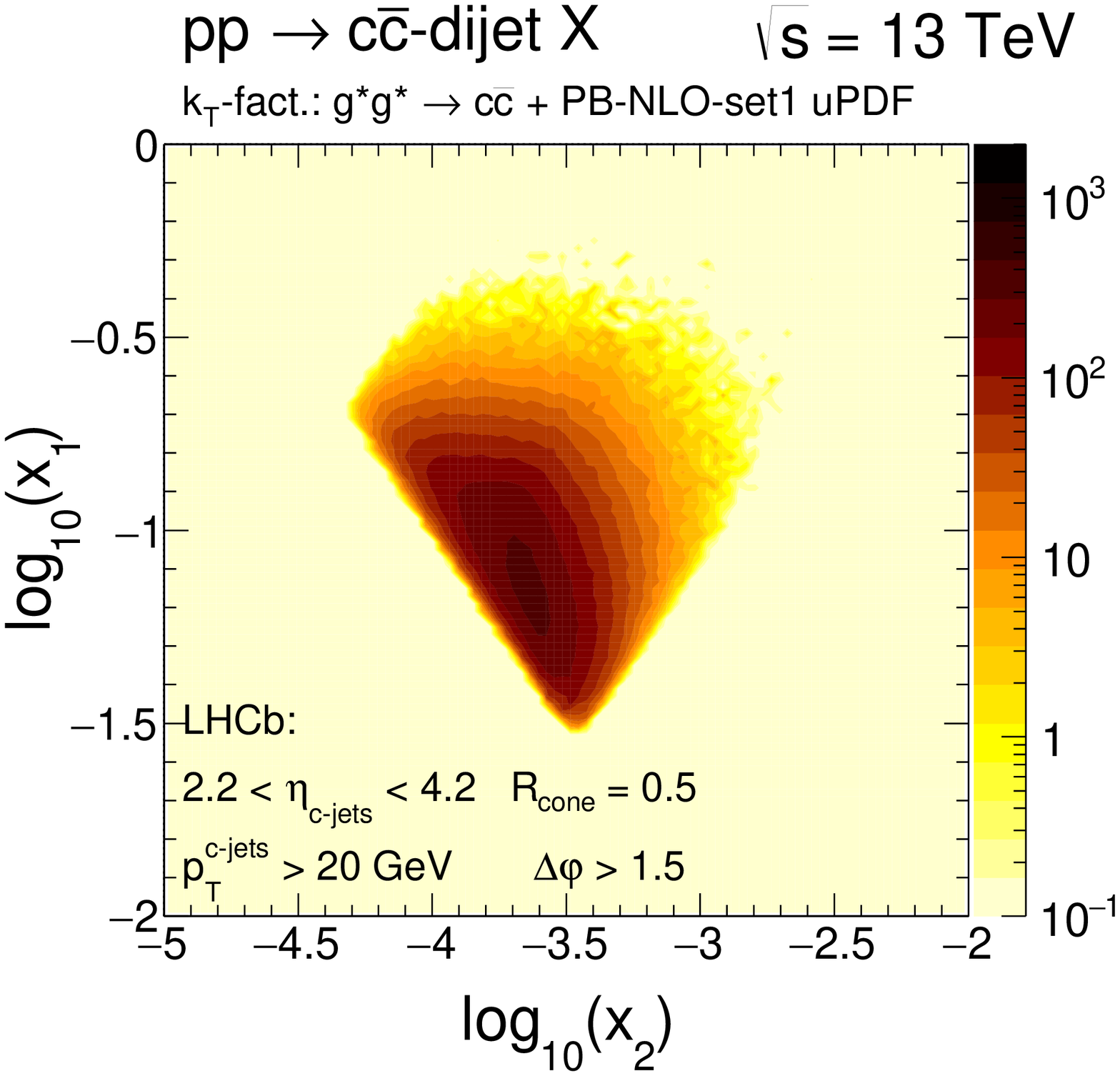}}
\end{minipage}
\begin{minipage}{0.33\textwidth}
  \centerline{\includegraphics[width=1.0\textwidth]{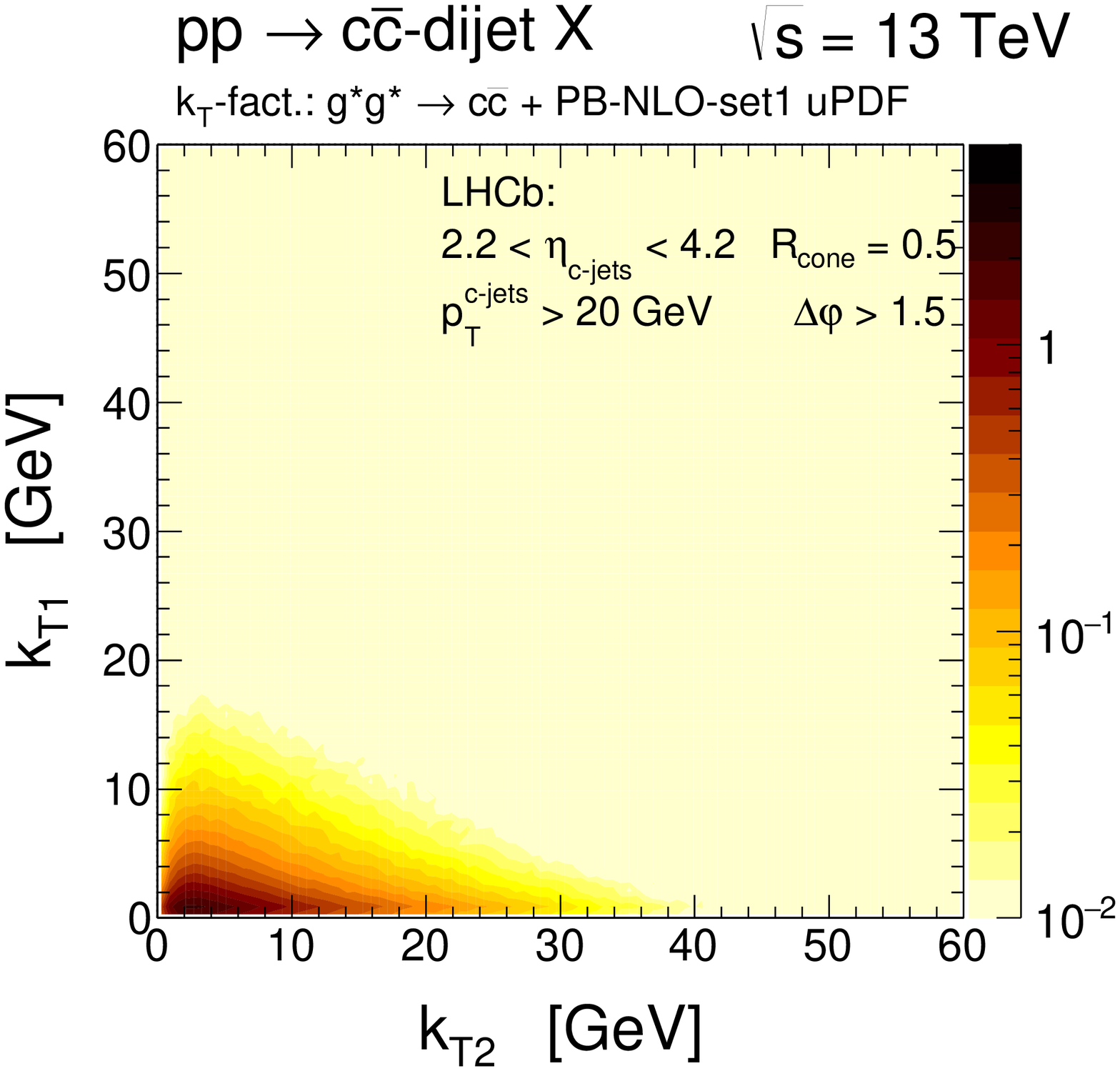}}
\end{minipage}
  \caption{
\small The double differential cross sections for forward production of $c\bar c$-dijets in $pp$-scattering at $\sqrt{s}=13$ TeV probed in the LHCb experiment \cite{LHCb:2020frr} as a function of the longitudinal momentum fractions $\log_{10}(x_{1}) \times \log_{10}(x_{2})$ (left panels) as well as the initial gluon transverse momenta $k_{t1} \times k_{t2}$ (right panels). The top, middle and bottom panels correspond to the JH2013set2 CCFM, MRW-MMHT2014nnlo, and the PB-NLO-set1 gluon uPDFs, respectively. 
}
\label{fig:9}
\end{figure}

\begin{figure}[!h]
\begin{minipage}{0.33\textwidth}
  \centerline{\includegraphics[width=1.0\textwidth]{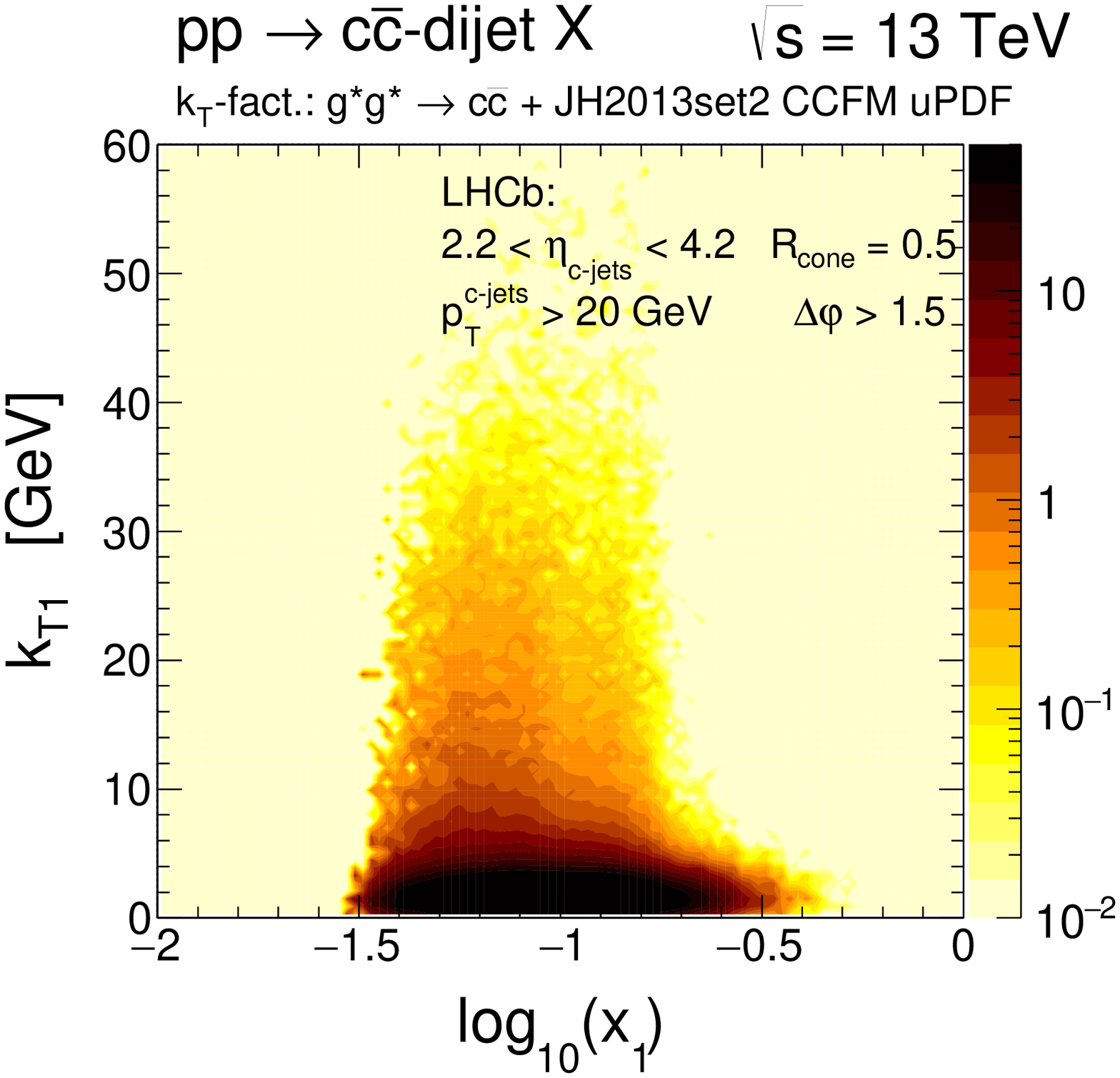}}
\end{minipage}
\begin{minipage}{0.33\textwidth}
  \centerline{\includegraphics[width=1.0\textwidth]{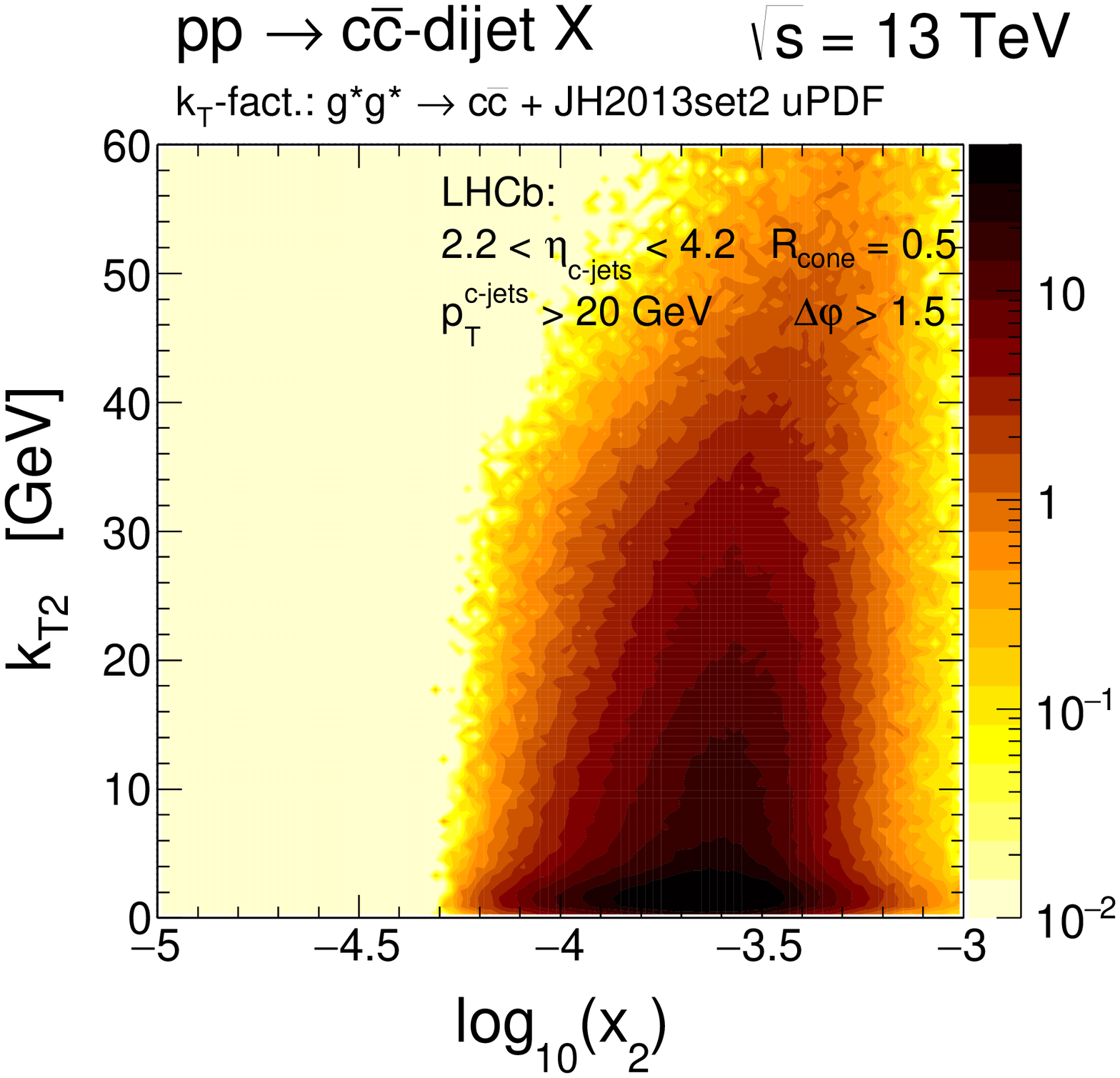}}
\end{minipage}
\begin{minipage}{0.33\textwidth}
  \centerline{\includegraphics[width=1.0\textwidth]{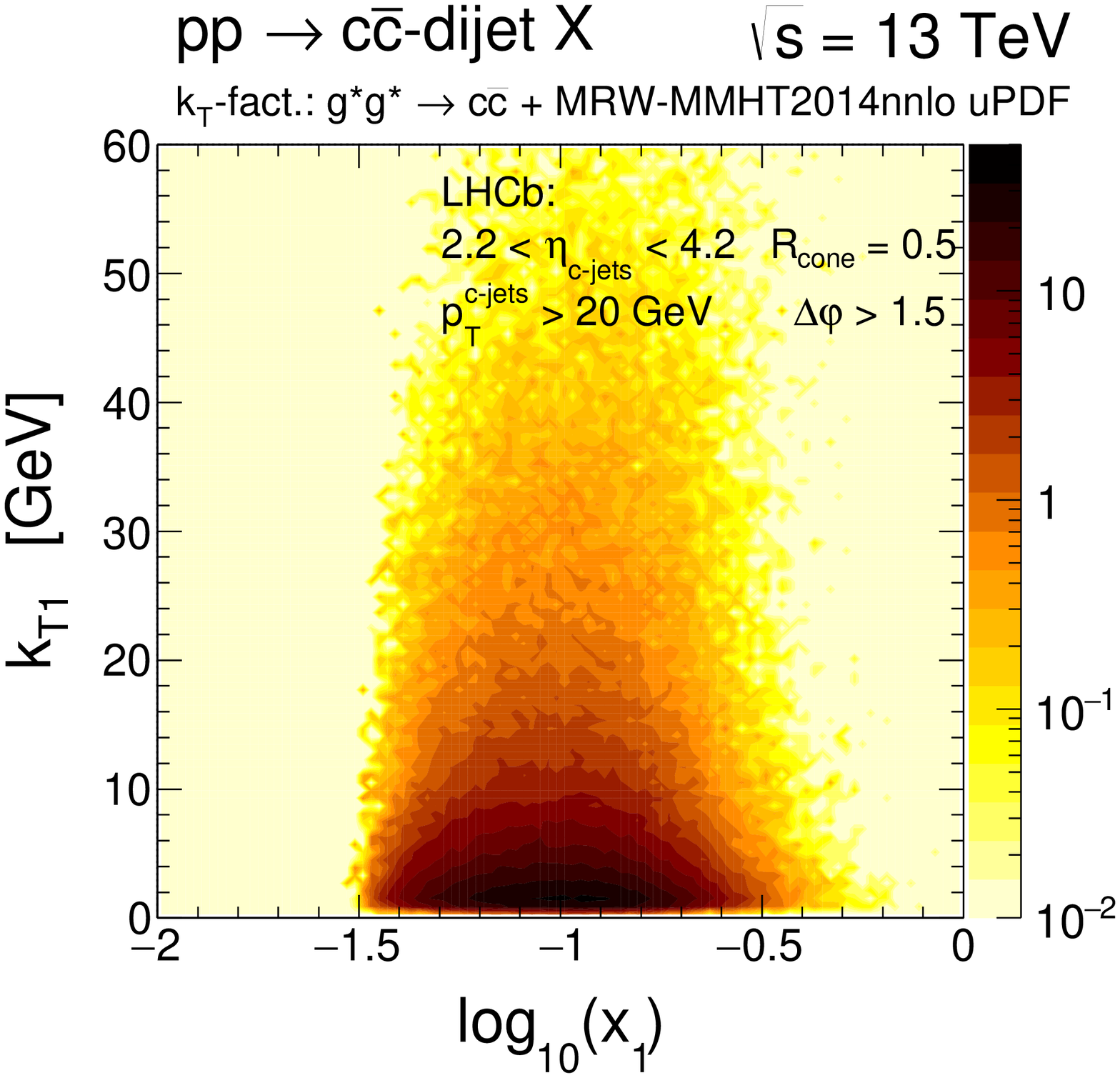}}
\end{minipage}
\begin{minipage}{0.33\textwidth}
  \centerline{\includegraphics[width=1.0\textwidth]{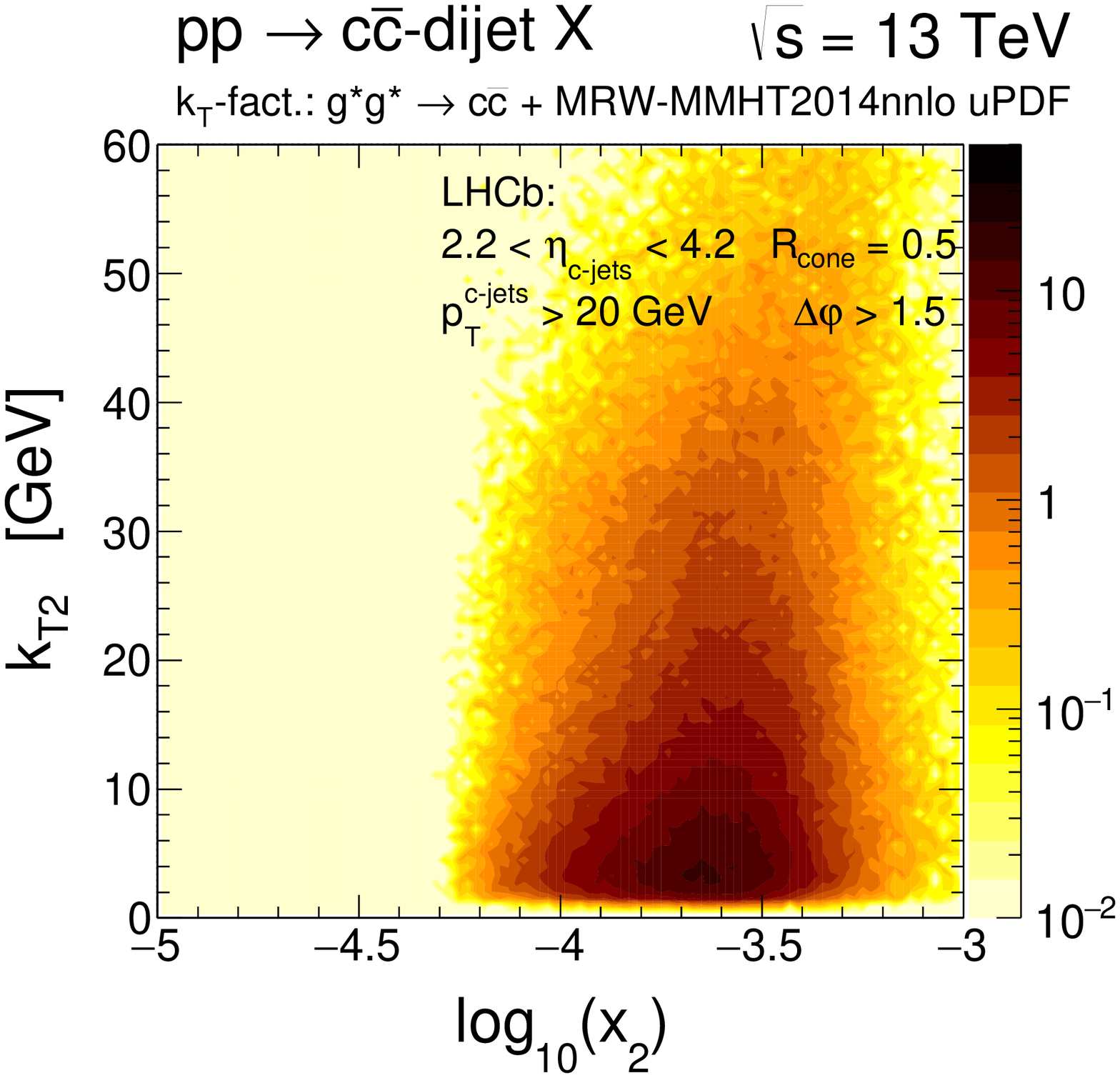}}
\end{minipage}
\begin{minipage}{0.33\textwidth}
  \centerline{\includegraphics[width=1.0\textwidth]{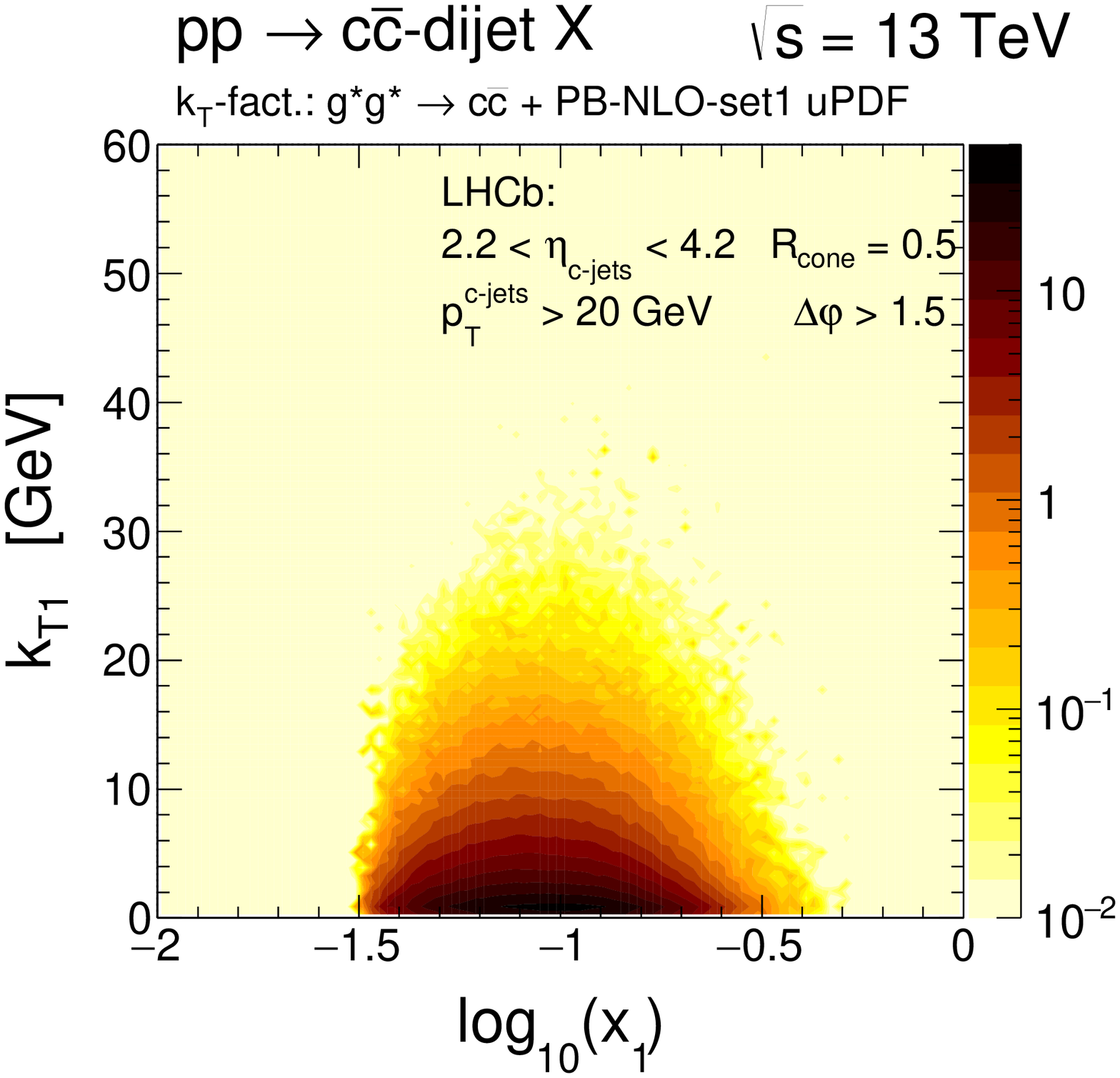}}
\end{minipage}
\begin{minipage}{0.33\textwidth}
  \centerline{\includegraphics[width=1.0\textwidth]{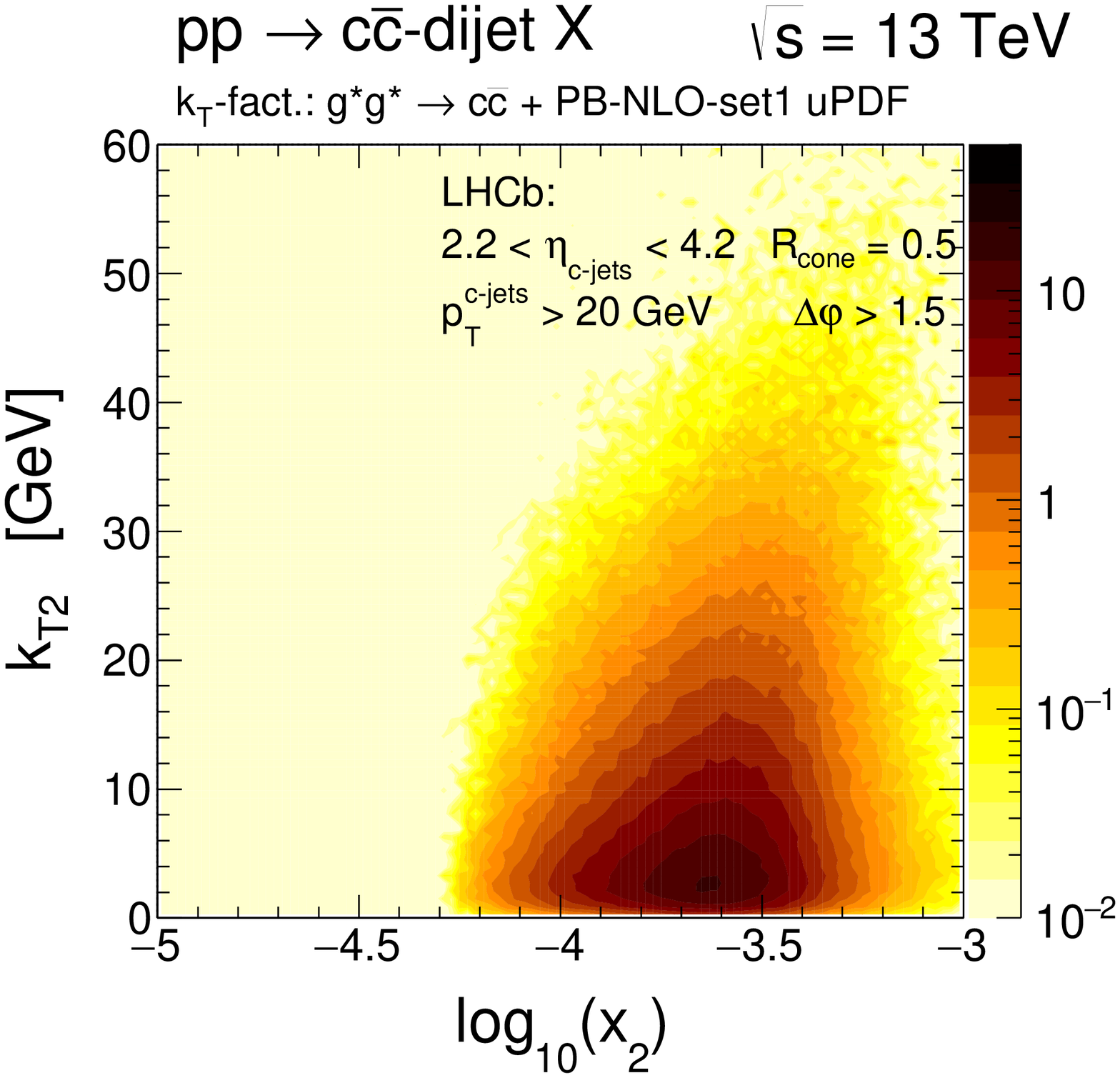}}
\end{minipage}
  \caption{
\small The same as in Fig.~\ref{fig:9} but here as a function of the initial gluon transverse momenta and the longitudinal momentum fractions - $k_{t1} \times \log_{10}(x_{1})$ (left panels) as well as $k_{t2} \times \log_{10}(x_{2})$ (right panels).
}
\label{fig:10}
\end{figure}

In Fig.~\ref{fig:9} we present the double differential cross sections for forward production of $c\bar c$-dijets in $pp$-scattering at $\sqrt{s}=13$ TeV probed in the LHCb experiment as a function of the longitudinal momentum fractions $\log_{10}(x_{1}) \times \log_{10}(x_{2})$ (left panels) as well as the initial gluon transverse momenta $k_{t1} \times k_{t2}$ (right panels). The top, middle and bottom panels correspond to the JH2013set2 CCFM, MRW-MMHT2014nnlo, and the PB-NLO-set1 gluon uPDFs, respectively. We see that we deal here with asymmetric configuration and probe the longitudinal momentum fractions in two different regions: $x_{1}$ is large with the maximum of the cross section around $10^{-1}$ and $x_{2}$ is much smaller with the maximum in the $10^{-4} < x_{2} < 10^{-3}$ range. At the same time, the large-$x$ gluon has rather small transverse momenta $k_{t1} \lesssim 5$ GeV, while the small-$x$ gluon takes much larger values of the $k_{t2}$ with the maximal range depending on the model of the gluon uPDF. The largest tail in $k_{t2}$ is observed for the case of the CCFM density, while the Parton-Branching model leads to the smallest one. The correlation between $x$ and $k_{t}$ is plotted explicitly in Fig.~\ref{fig:10}.
 
%----------------------------------------------------------------------------
%----------------------------------------------------------------------------

%----------------------------------------------------------------------------
%----------------------------------------------------------------------------

\begin{figure}[!h]
\begin{minipage}{0.33\textwidth}
  \centerline{\includegraphics[width=1.0\textwidth]{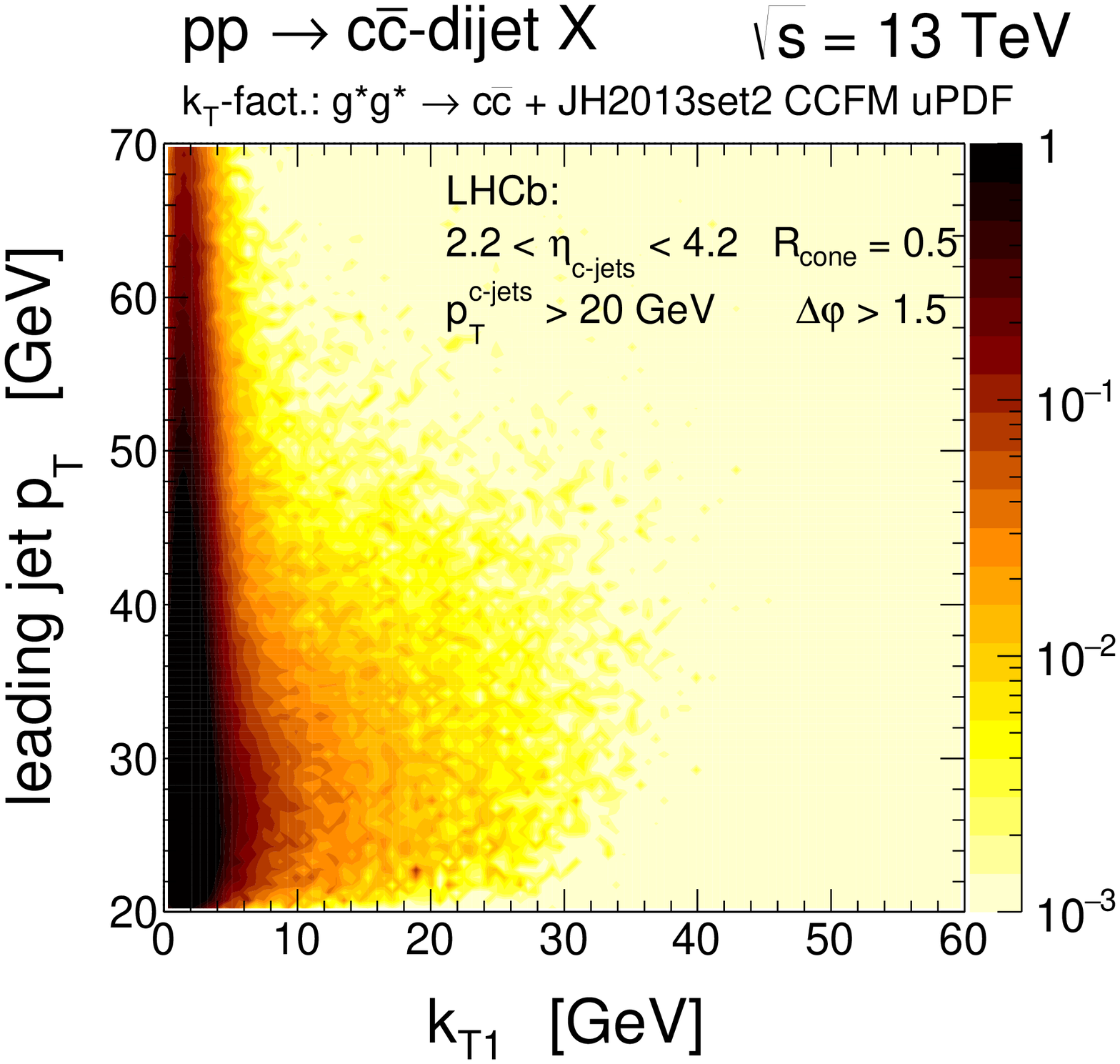}}
\end{minipage}
\begin{minipage}{0.33\textwidth}
  \centerline{\includegraphics[width=1.0\textwidth]{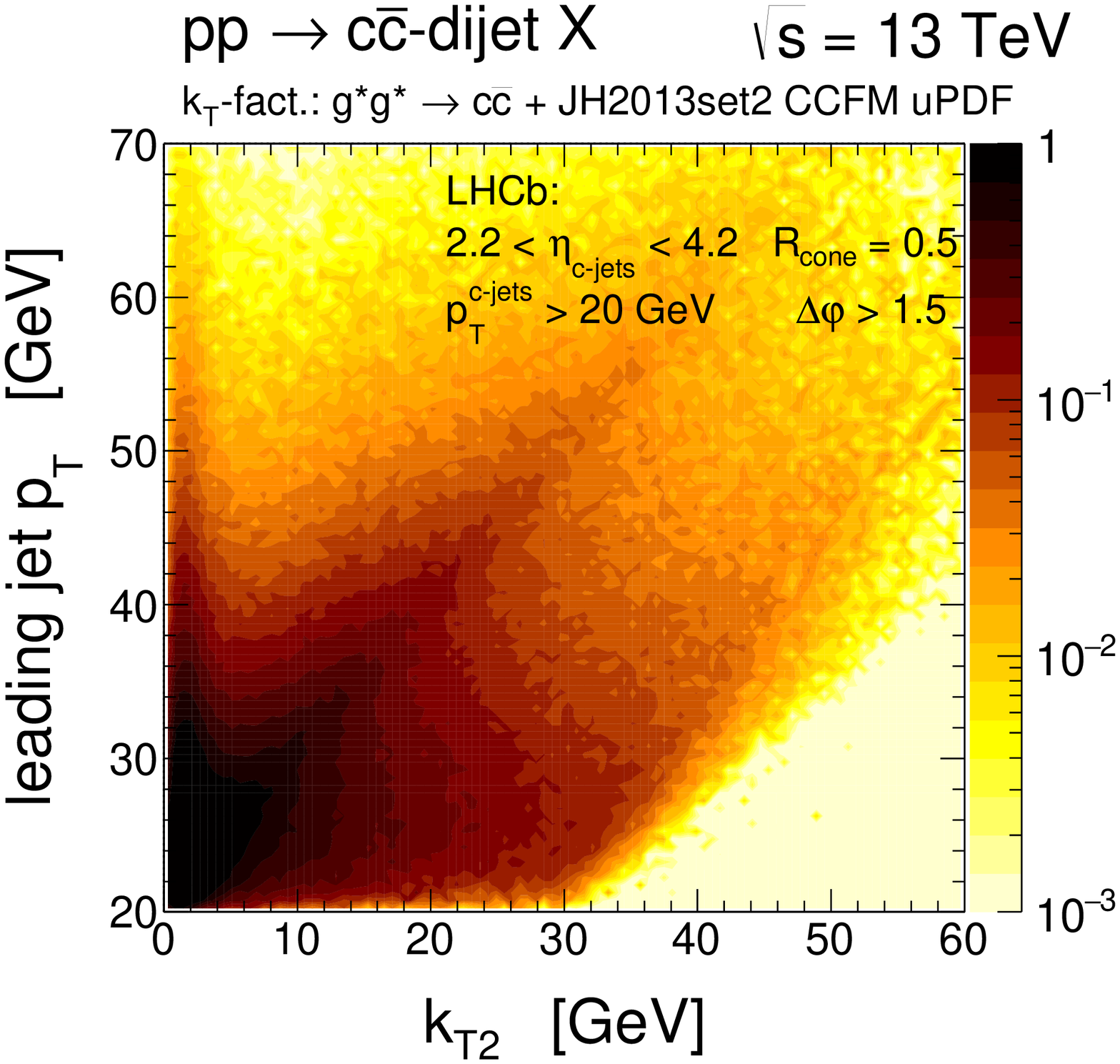}}
\end{minipage}
\begin{minipage}{0.33\textwidth}
  \centerline{\includegraphics[width=1.0\textwidth]{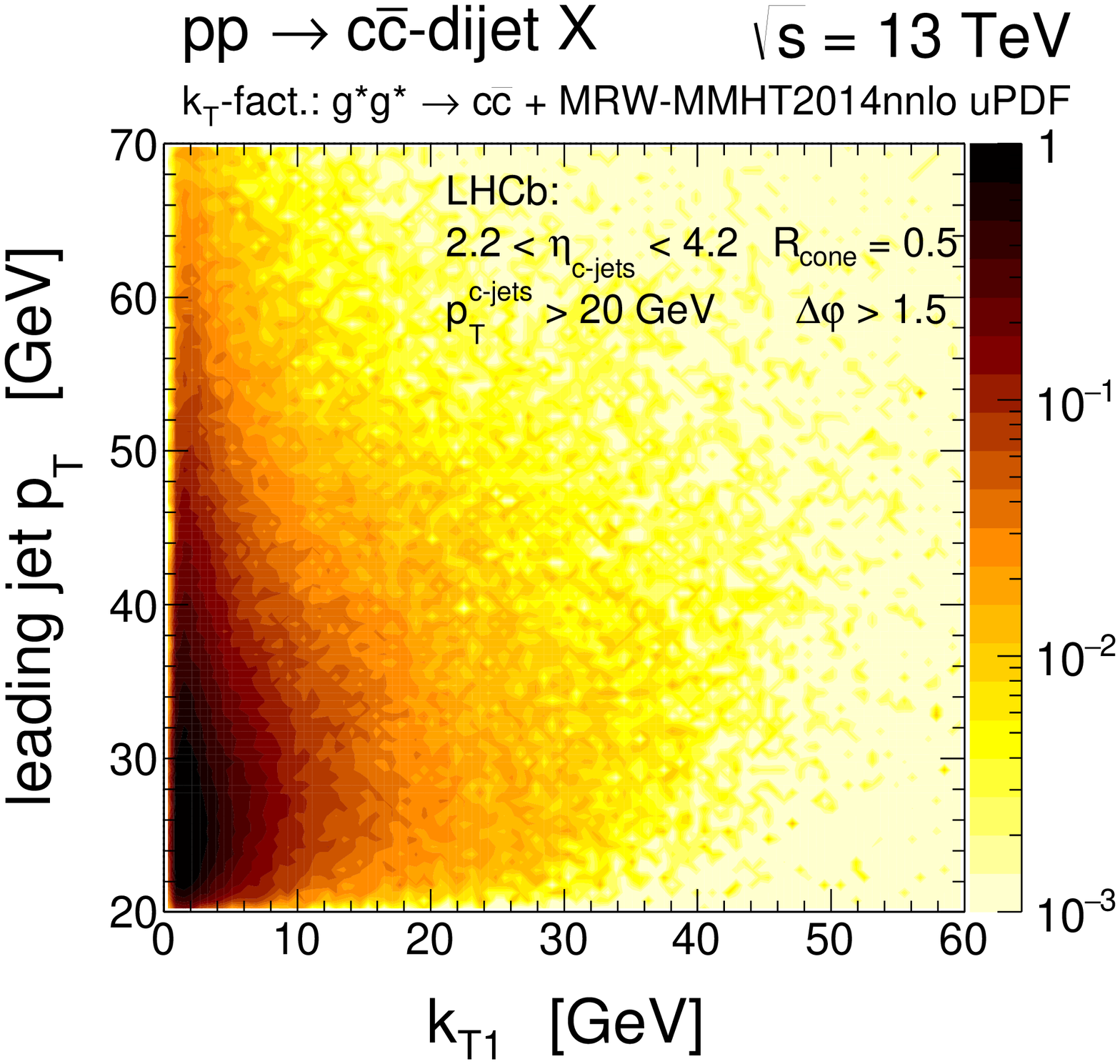}}
\end{minipage}
\begin{minipage}{0.33\textwidth}
  \centerline{\includegraphics[width=1.0\textwidth]{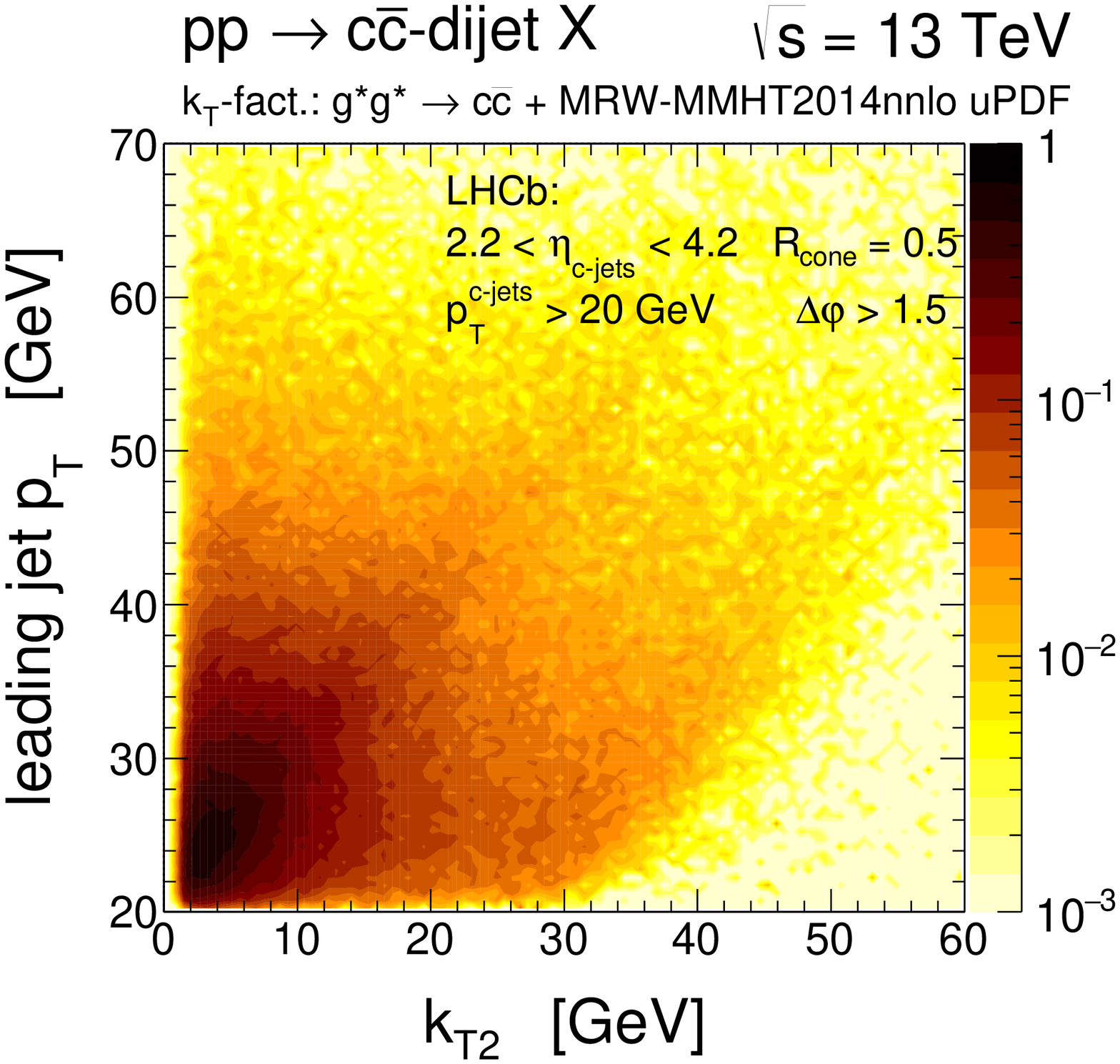}}
\end{minipage}
\begin{minipage}{0.33\textwidth}
  \centerline{\includegraphics[width=1.0\textwidth]{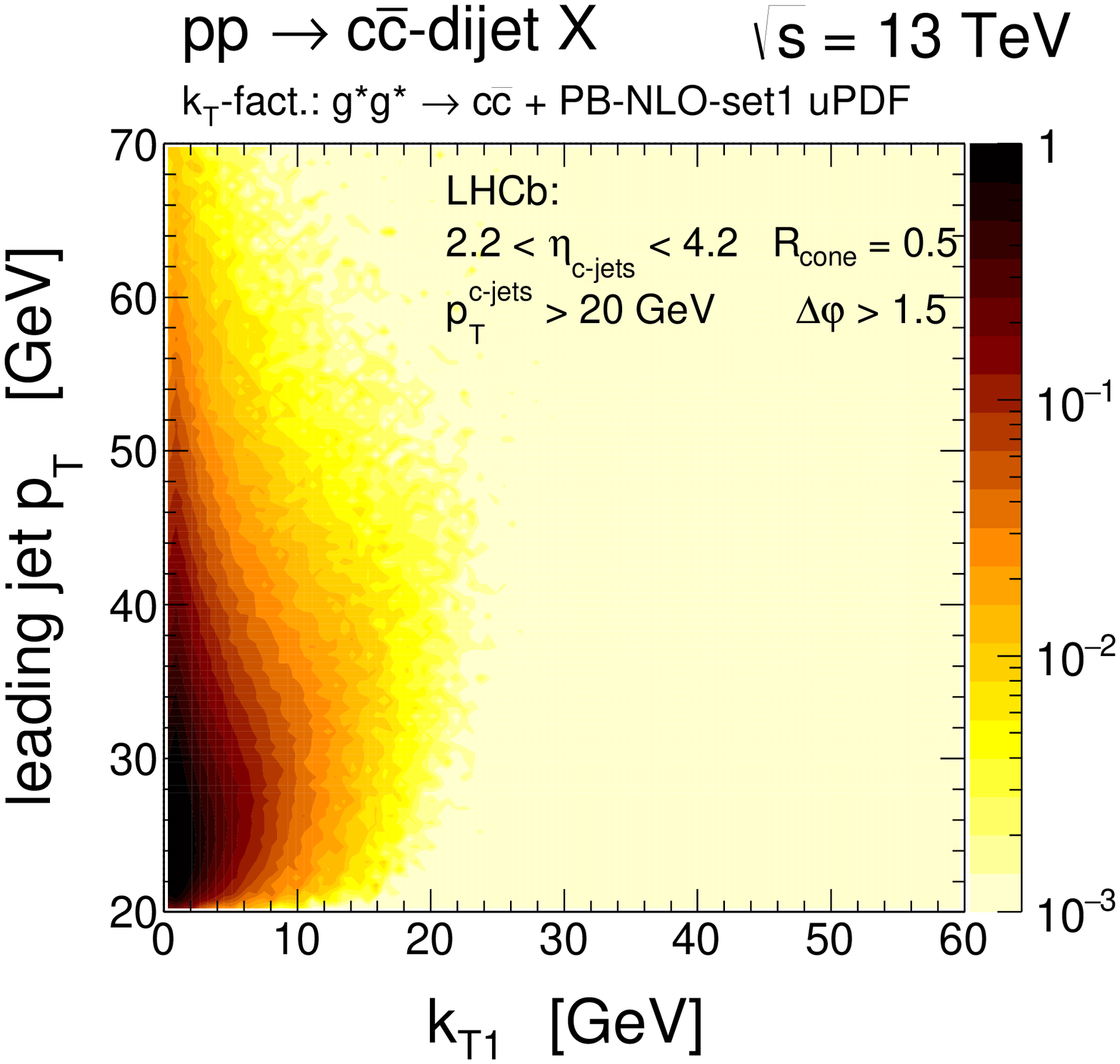}}
\end{minipage}
\begin{minipage}{0.33\textwidth}
  \centerline{\includegraphics[width=1.0\textwidth]{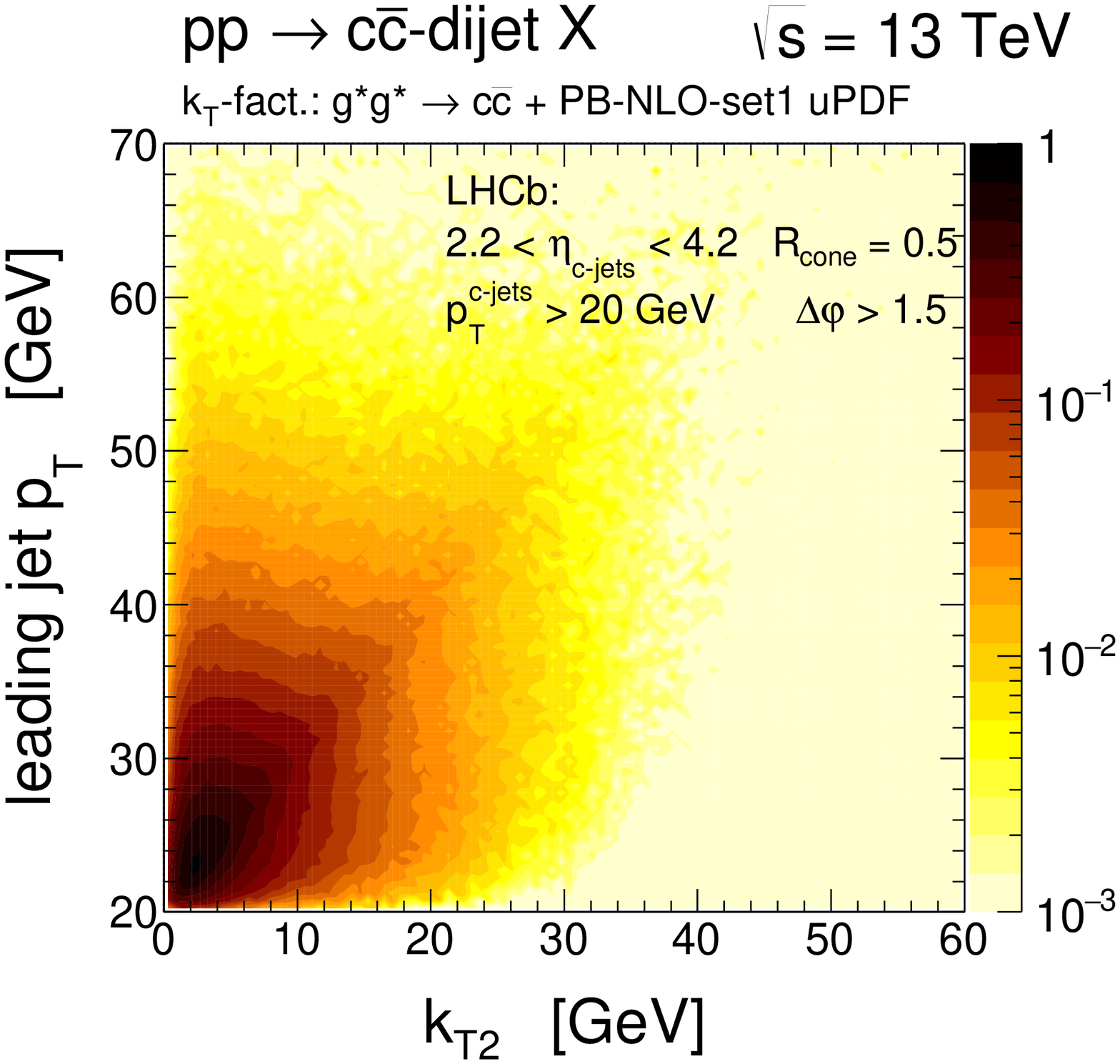}}
\end{minipage}
  \caption{
\small The same as in Fig.~\ref{fig:9} but here as a function of the leading $c$-jet transverse momentum and the initial gluon transverse momenta - $p_{T} \times k_{t1}$ (left panels) as well as $p_{T} \times k_{t2}$ (right panels).
}
\label{fig:11}
\end{figure}

It is also interesting to see how the initial transverse momenta of incoming gluons $k_{t}$ contribute to the leading jet $p_{T}$. The mutual relation is presented in Fig.~\ref{fig:11}. The leading jet $p_{T}$ distribution in the range of $20 < p_{T} < 50$ GeV is driven by both gluons incident transverse momenta. Above $50$ GeV mostly the large-$x$ gluon contributes to the spectrum.  

%----------------------------------------------------------------------------
%----------------------------------------------------------------------------

\subsection{The large-$\bm{x}$ behaviour of the gluon uPDFs}

As we have shown in the previous section within the present study one has to deal with gluon distributions for large- and small-$x$ simultaneously.
The unintegrated densities are usually devoted to the small-$x$ region where it can be safely used, however, the large-$x$ behaviour is challenging and might not be under full theoretical control. At large-$x$ one might expect that a given uPDF model should closely reproduce well known collinear PDFs when integrating out the $k_{t}$-dependence. Here we wish to take a closer look on this issue.

%----------------------------------------------------------------------------
\begin{figure}[!h]
\begin{minipage}{0.47\textwidth}
  \centerline{\includegraphics[width=1.0\textwidth]{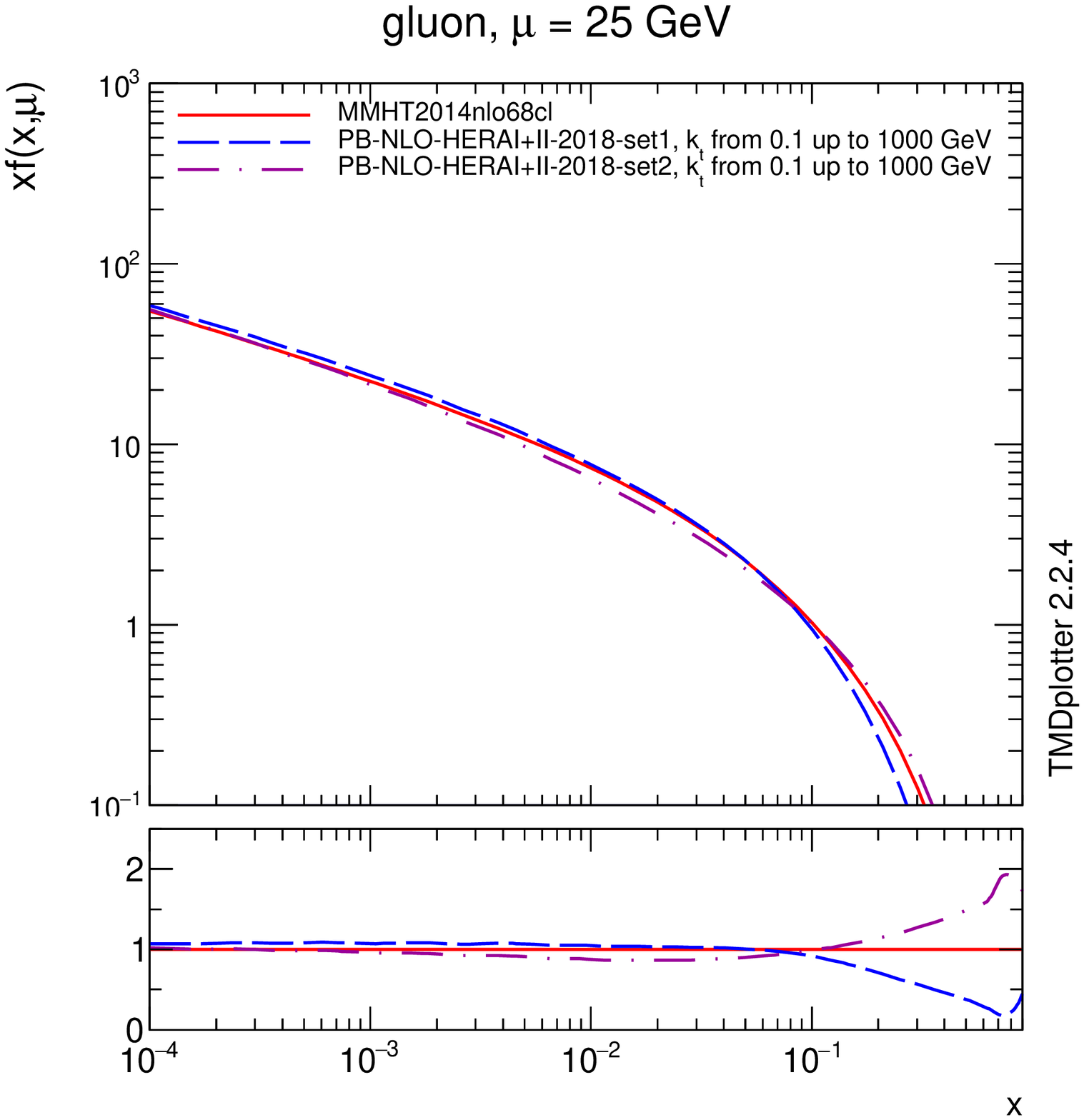}}
\end{minipage}
\begin{minipage}{0.47\textwidth}
  \centerline{\includegraphics[width=1.0\textwidth]{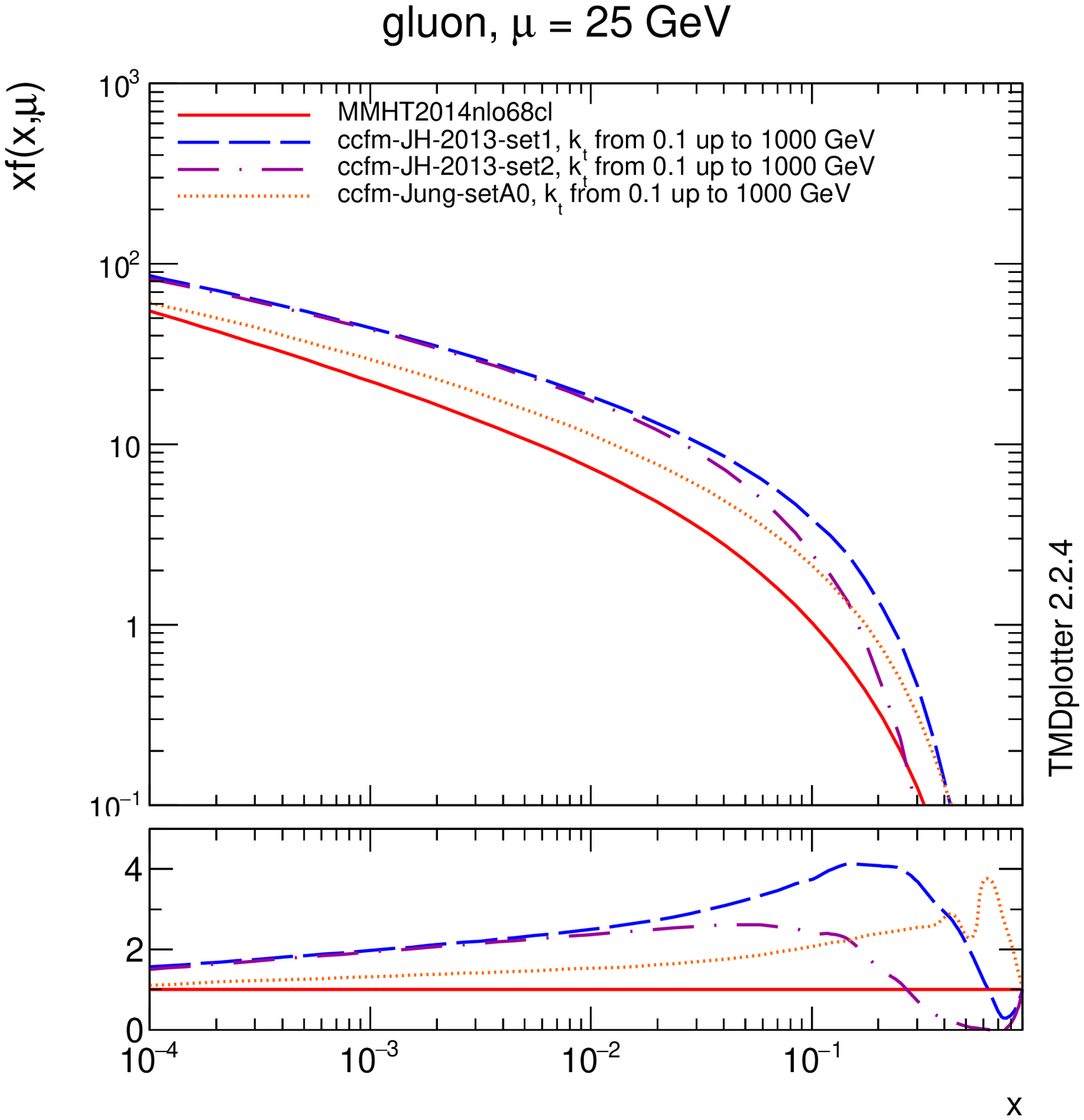}}
\end{minipage}
  \caption{The gluon uPDFs integrated over the transverse momentum $k_{t}$ as a function of the longitudinal momentum fraction $x$ at a given scale $\mu = 25$ GeV, relevant for the process under consideration. The left and right panels correspond to the Parton-Branching and the CCFM gluon uPDFs, respectively. As a reference the collinear MMHT2014nlo68cl gluon PDF is shown in addition.
\small 
}
\label{fig:12}
\end{figure}
%----------------------------------------------------------------------------

In Fig.~\ref{fig:12} we present the gluon uPDFs integrated over the transverse momentum $k_{t}$ as a function of the longitudinal momentum fraction $x$ at a given scale $\mu = 25$ GeV, relevant for the process under consideration. The left and right panels correspond to the Parton-Branching and the CCFM gluon uPDFs, respectively. As a reference the collinear MMHT2014nlo68cl gluon PDF is shown. We found that the Parton-Branching uPDF reproduces very strictly the collinear PDF, while for the CCFM uPDFs this is not the case. 

%----------------------------------------------------------------------------
\begin{figure}[!h]
\begin{minipage}{0.47\textwidth}
  \centerline{\includegraphics[width=1.0\textwidth]{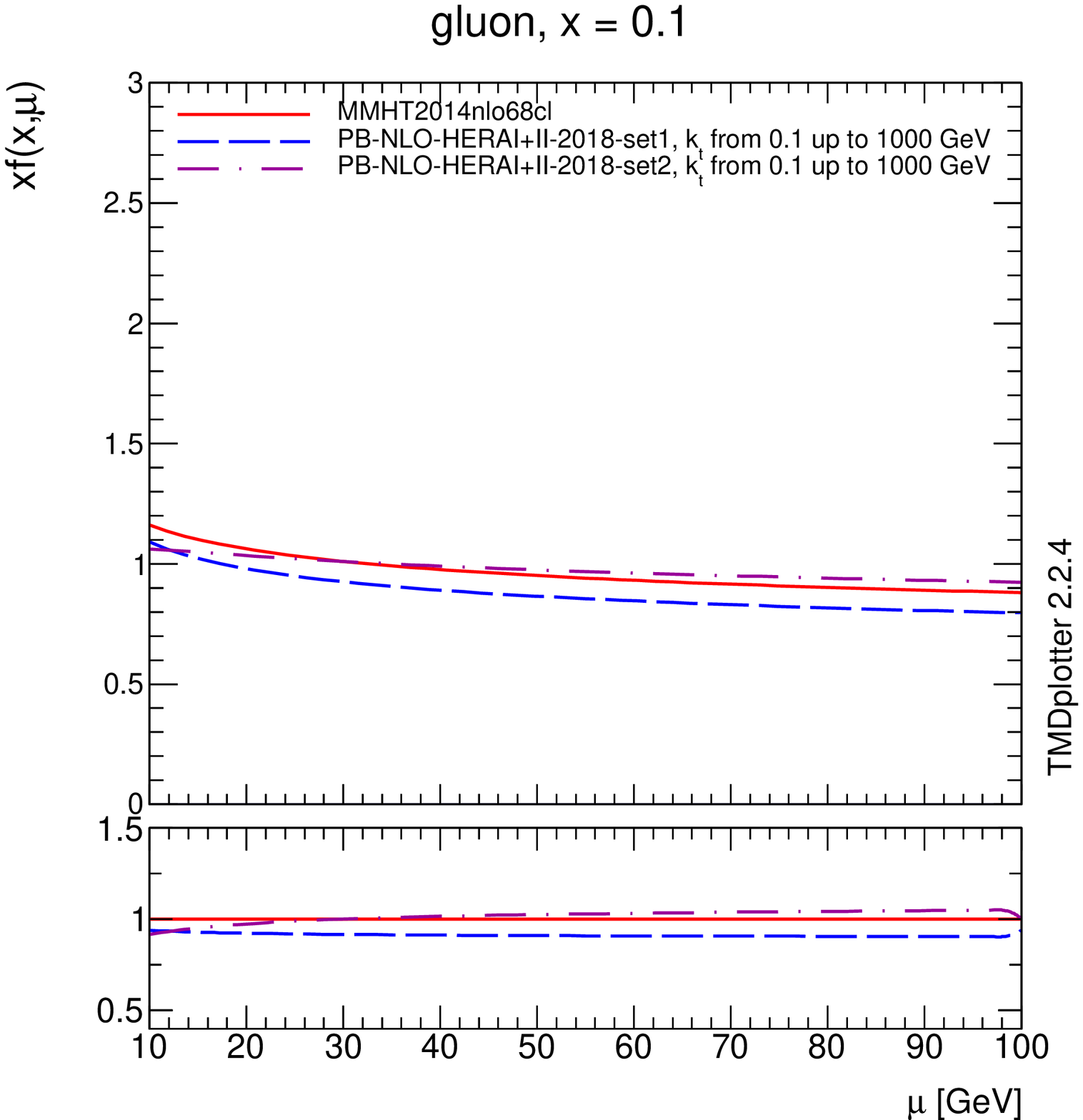}}
\end{minipage}
\begin{minipage}{0.47\textwidth}
  \centerline{\includegraphics[width=1.0\textwidth]{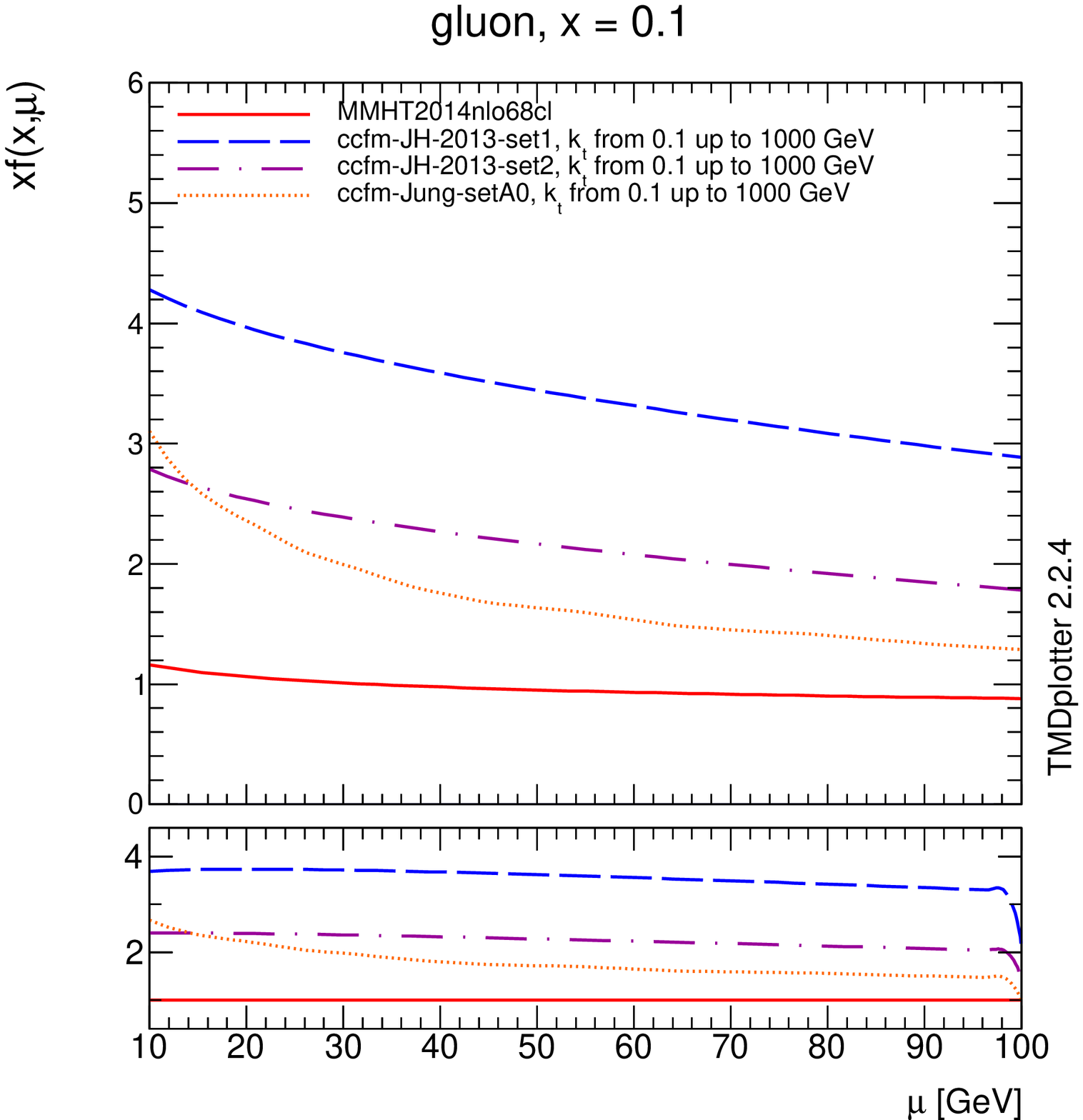}}
\end{minipage}
\begin{minipage}{0.47\textwidth}
  \centerline{\includegraphics[width=1.0\textwidth]{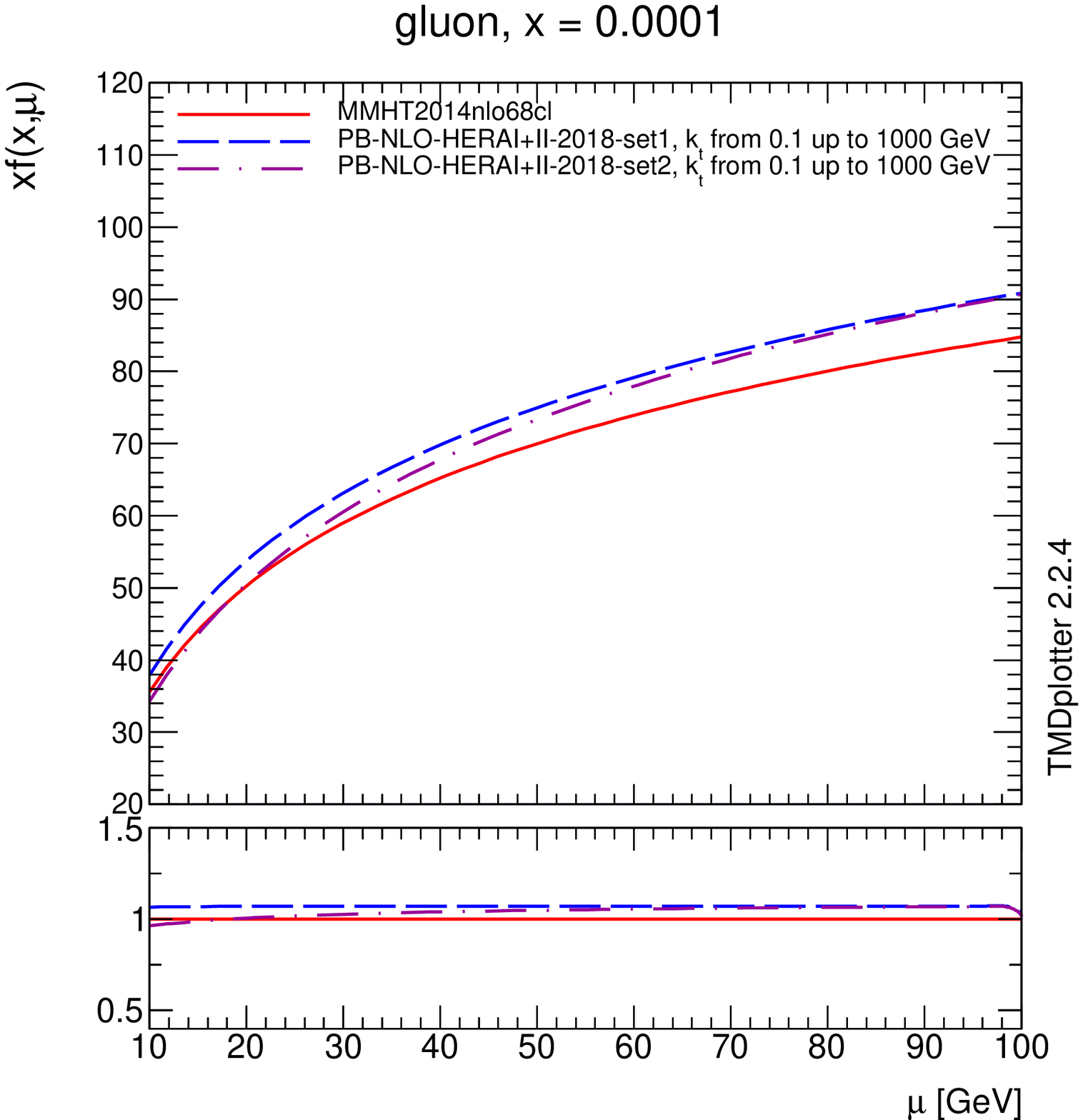}}
\end{minipage}
\begin{minipage}{0.47\textwidth}
  \centerline{\includegraphics[width=1.0\textwidth]{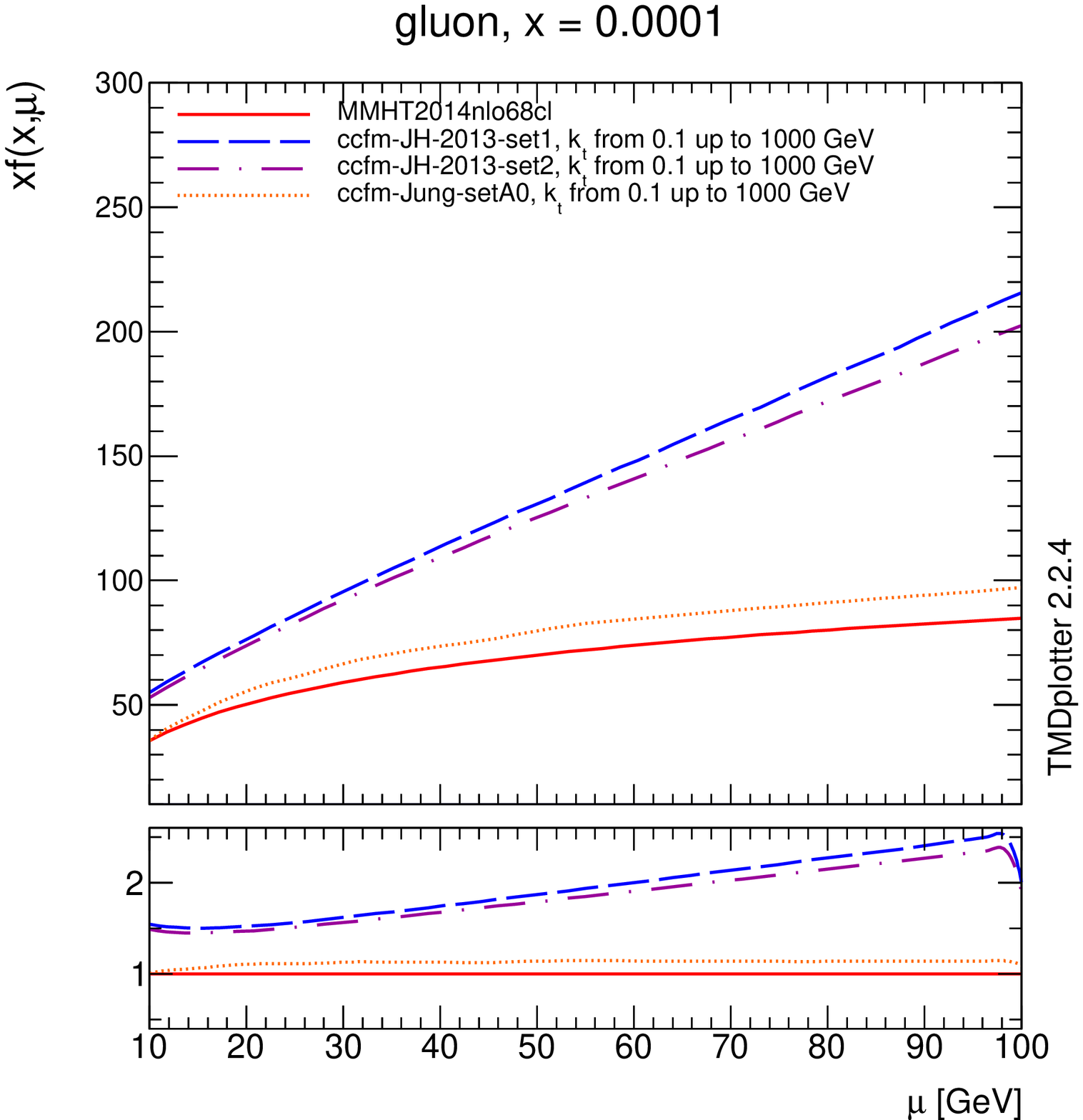}}
\end{minipage}
  \caption{The gluon uPDFs integrated over the transverse momentum $k_{t}$ as a function of the factorization scale $\mu$ for a given longitudinal momentum fraction $x$. The left and right panels correspond to the Parton-Branching and the CCFM gluon uPDFs, respectively. The top panels are obtained for $x=10^{-1}$ and the bottom ones for $x=10^{-4}$. 
As a reference the collinear MMHT2014nlo68cl gluon PDF is shown in addition.
\small 
}
\label{fig:13}
\end{figure}
%----------------------------------------------------------------------------

In Fig.~\ref{fig:13} we show in addition the gluon uPDFs integrated over the transverse momentum $k_{t}$ as a function of the factorization scale $\mu$ for a given longitudinal momentum fraction $x$. The left and right panels correspond to the Parton-Branching and the CCFM gluon uPDFs, respectively. The top panels are obtained for $x=10^{-1}$ and the bottom ones for $x=10^{-4}$. 
As a reference the collinear MMHT2014nlo68cl gluon PDF is also shown. Here we see that for large-$x$ gluon the relation between the integrated uPDFs and the collinear PDF does not change with the scale $\mu$. On the other hand, the integrated gluon uPDFs grow faster with the scale than their collinear counterparts. This effect is stronger especially in the case of the JH2013 CCFM uPDF sets.

Having shown somewhat technical problems within full $k_{T}$-factorization approach in the following we shall consider also the hybrid factorization approach.

\subsection{Predictions of the hybrid factorization framework}

All the calculations below preformed within the hybrid model are done using the collinear MMHT2014nlo68cl gluon PDF on large-$x$ side.
We have checked numerically that at large-$x$ there is no significant difference between different sets of the gluon PDF at NLO, given in the literature by different PDF groups. 

\subsubsection{Charm dijets}

\begin{figure}[!h]
\begin{minipage}{0.47\textwidth}
  \centerline{\includegraphics[width=1.0\textwidth]{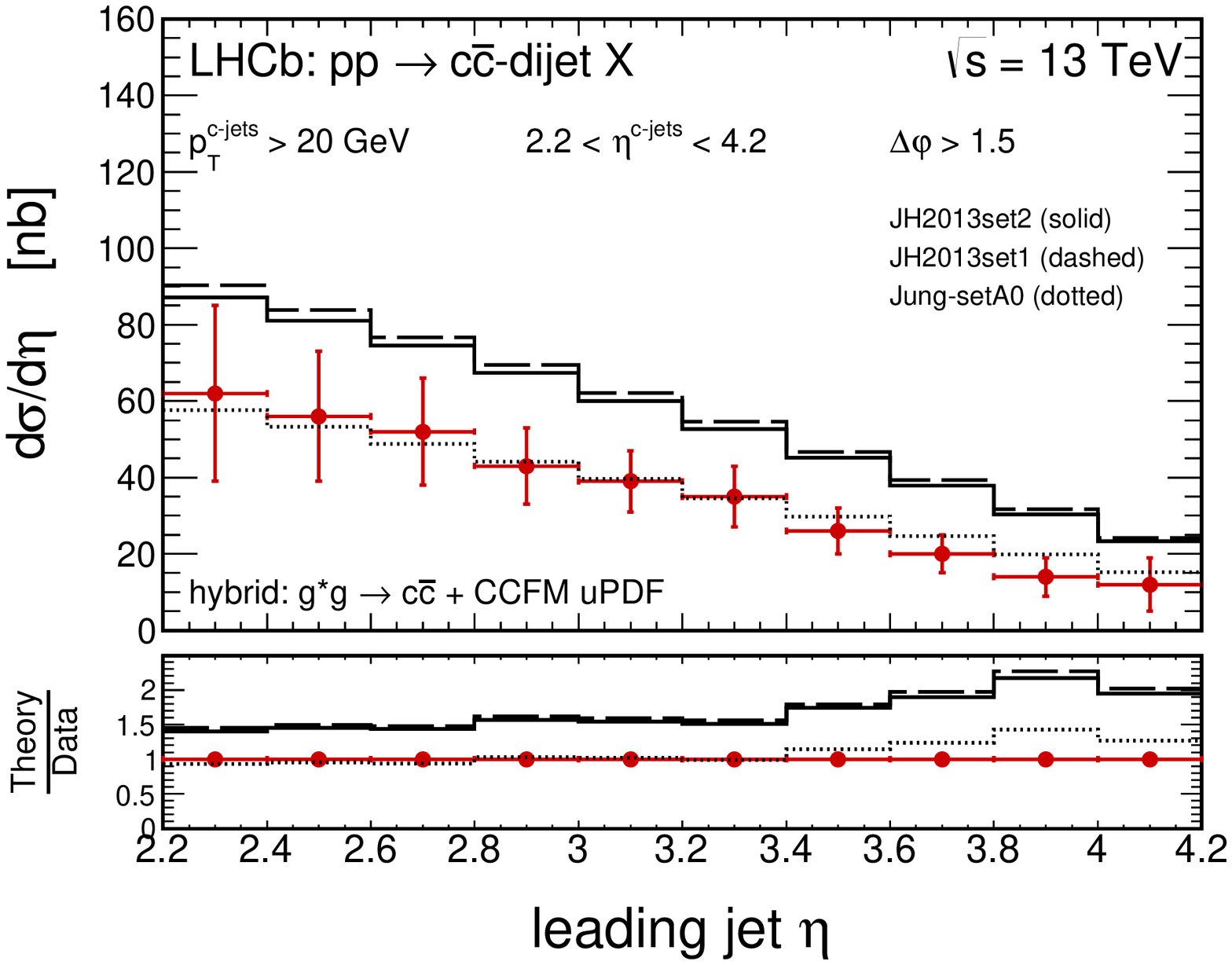}}
\end{minipage}
\begin{minipage}{0.47\textwidth}
  \centerline{\includegraphics[width=1.0\textwidth]{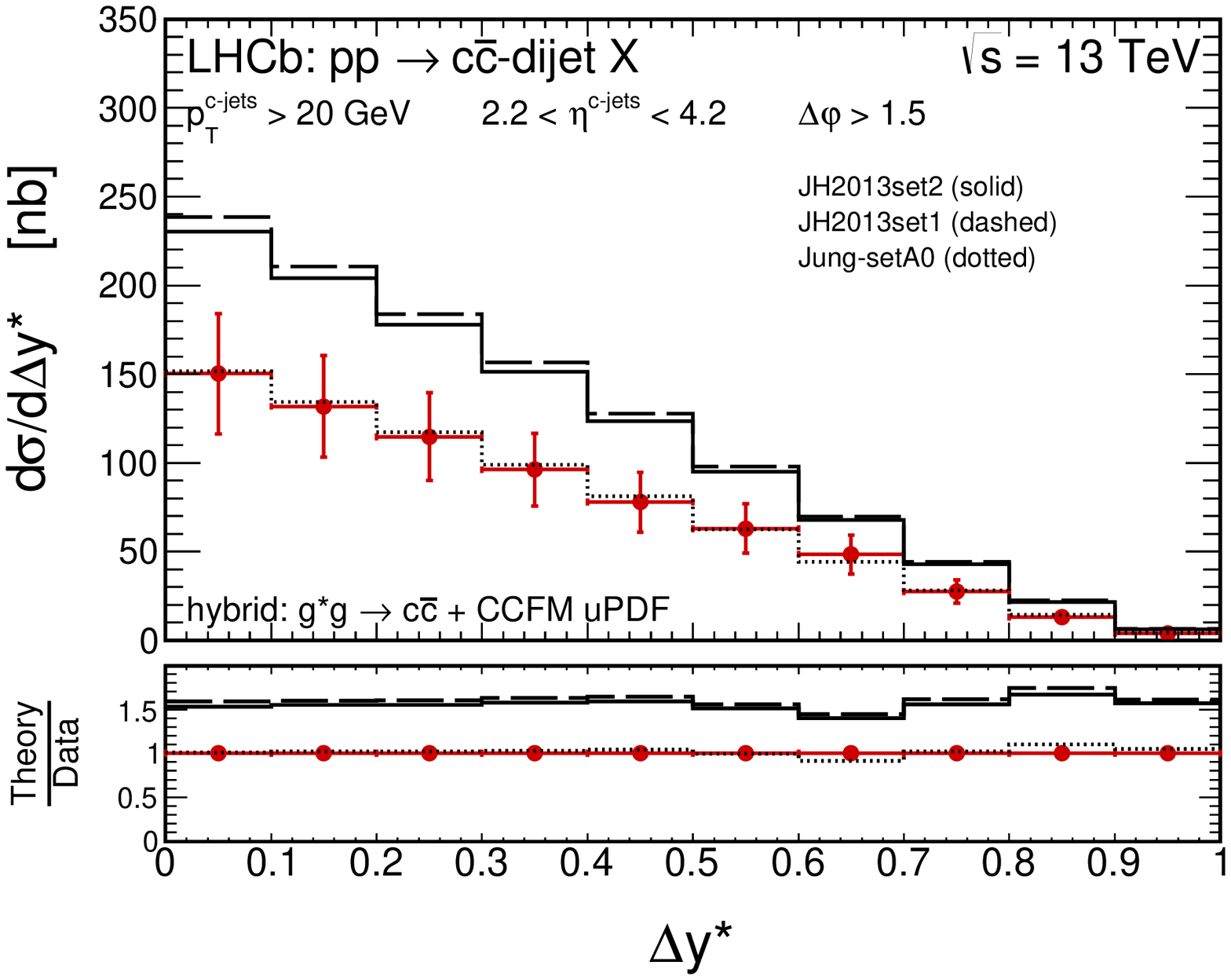}}
\end{minipage}
\begin{minipage}{0.47\textwidth}
  \centerline{\includegraphics[width=1.0\textwidth]{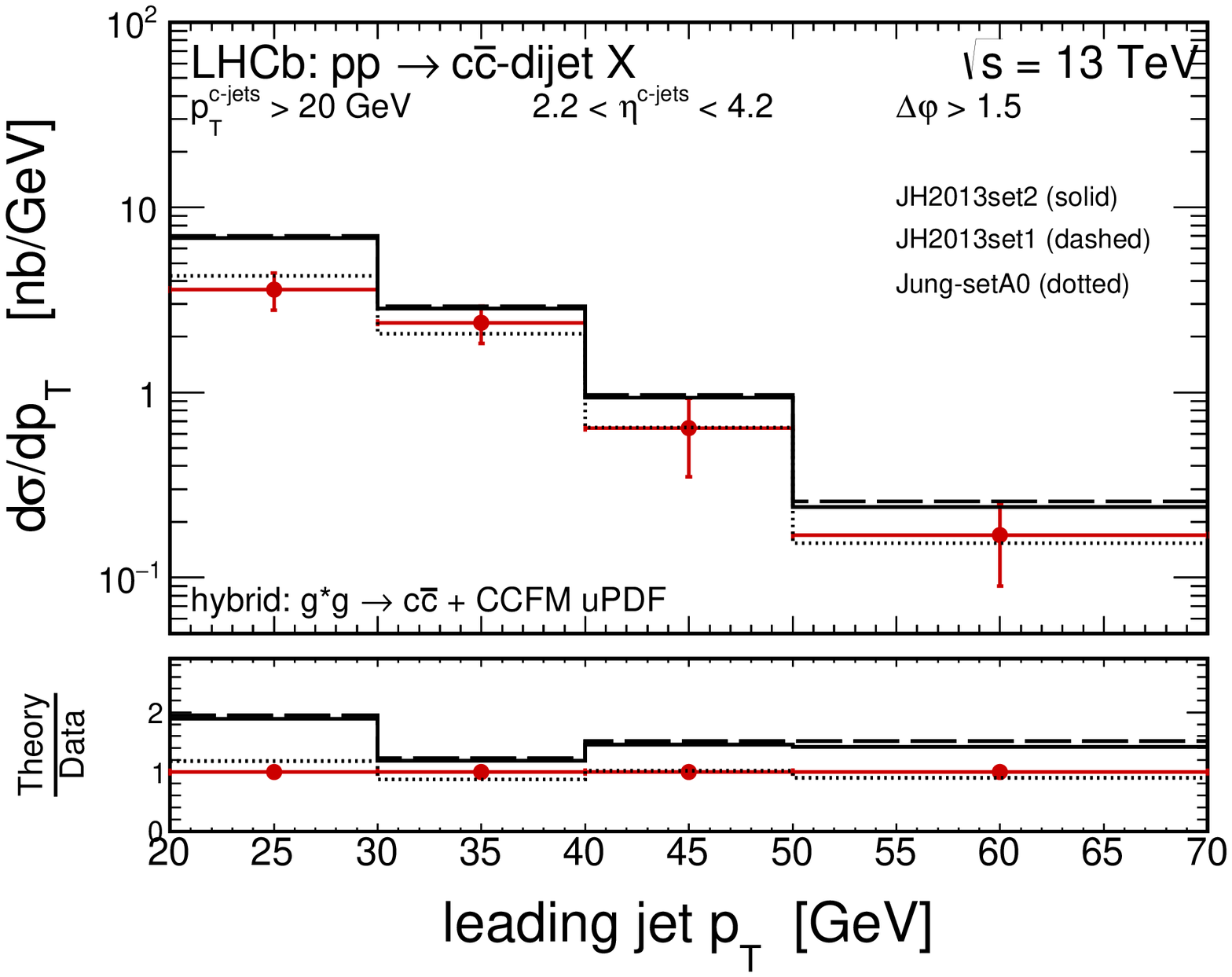}}
\end{minipage}
\begin{minipage}{0.47\textwidth}
  \centerline{\includegraphics[width=1.0\textwidth]{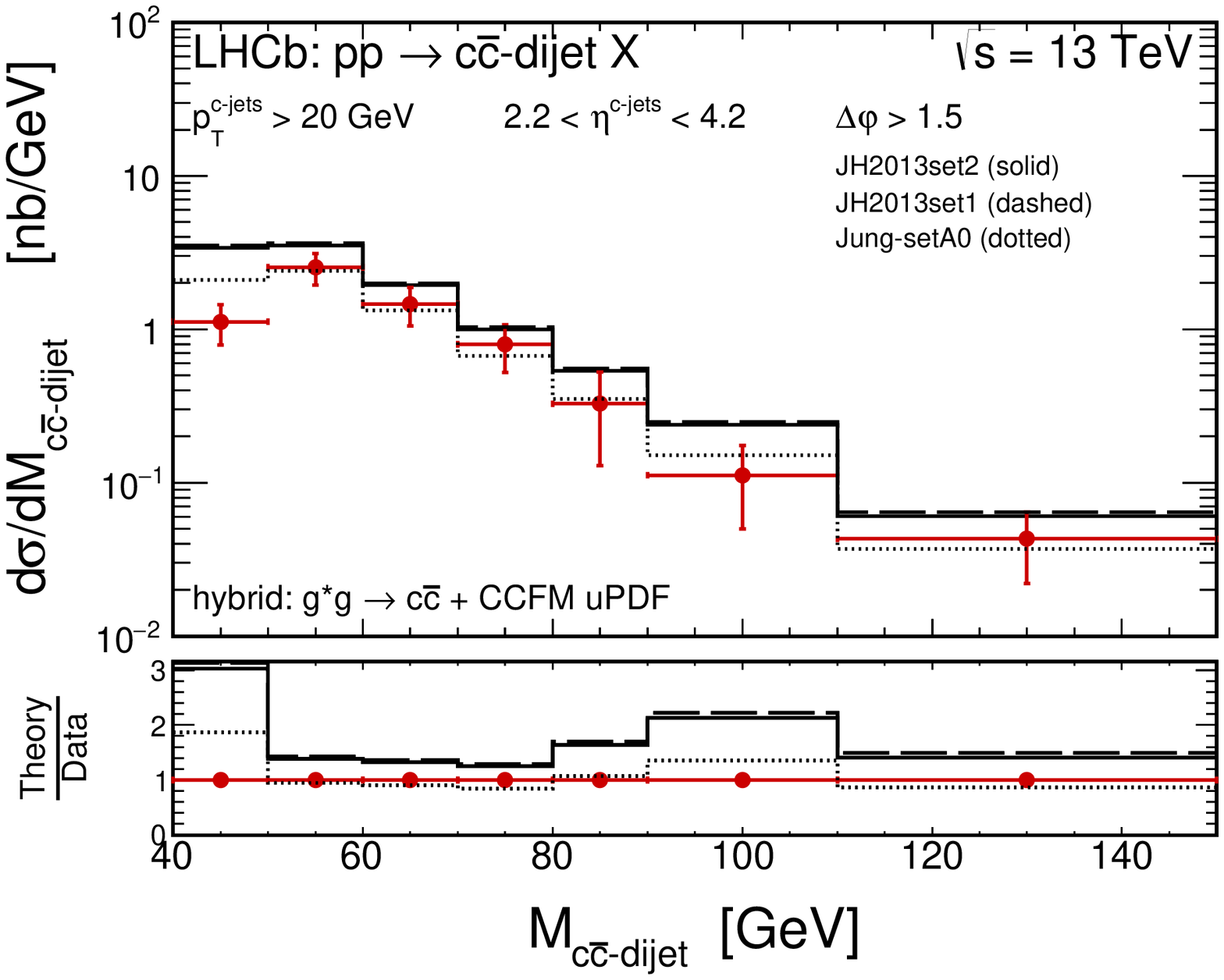}}
\end{minipage}
  \caption{The differential cross sections for forward production of $c\bar c$-dijets in $pp$-scattering at $\sqrt{s}=13$ TeV as a function of the leading jet $\eta$ (top left), the rapidity difference $\Delta y^{*}$ (top right), the leading jet $p_{T}$ (bottom left) and the dijet invariant mass $M_{c\bar c\text{-}\mathrm{dijet}}$ (bottom right). Here the dominant pQCD $g^*g \to c \bar c$ mechanism is taken into account. The theoretical histograms correspond to the hybrid model calculations obtained with the CCFM uPDFs.
\small 
}
\label{fig:14}
\end{figure}

The results of the hybrid model will be shown in an analogous manner as for the full $k_{T}$-factorization. 
We start with forward production of the $c\bar c$-dijets in $pp$-scattering at $\sqrt{s}=13$ TeV. In Fig.~\ref{fig:14} we show the corresponding differential cross sections as a function of the leading jet $\eta$ (top left panels), the rapidity difference $\Delta y^{*}$ (top right panels), the leading jet $p_{T}$ (bottom left panels) and the dijet invariant mass $M_{c\bar c\text{-}\mathrm{dijet}}$ (bottom right panels). The theoretical histograms are obtained for the interaction of off-shell small-$x$ and on-shell large-$x$ gluon distributions using three different sets of the CCFM uPDFs: JH2013set1 (dashed), JH2013set2 (solid), and Jung-setA0 (dotted). We observe that within the hybrid approach the discrepancies between predictions and the experimental data decrease with respect to the case of the full $k_{T}$-factorization calculations. Within the hybrid factorization and the Jung setA0 gluon uPDF we get an excellent description of the experimental distributions. Also here the JH2013 sets of the CCFM gluon uPDFs overestimate the data points by less than a factor of 2 only, which is also a significant improvement. 

%----------------------------------------------------------------------------
%----------------------------------------------------------------------------

\begin{figure}[!h]
\begin{minipage}{0.47\textwidth}
  \centerline{\includegraphics[width=1.0\textwidth]{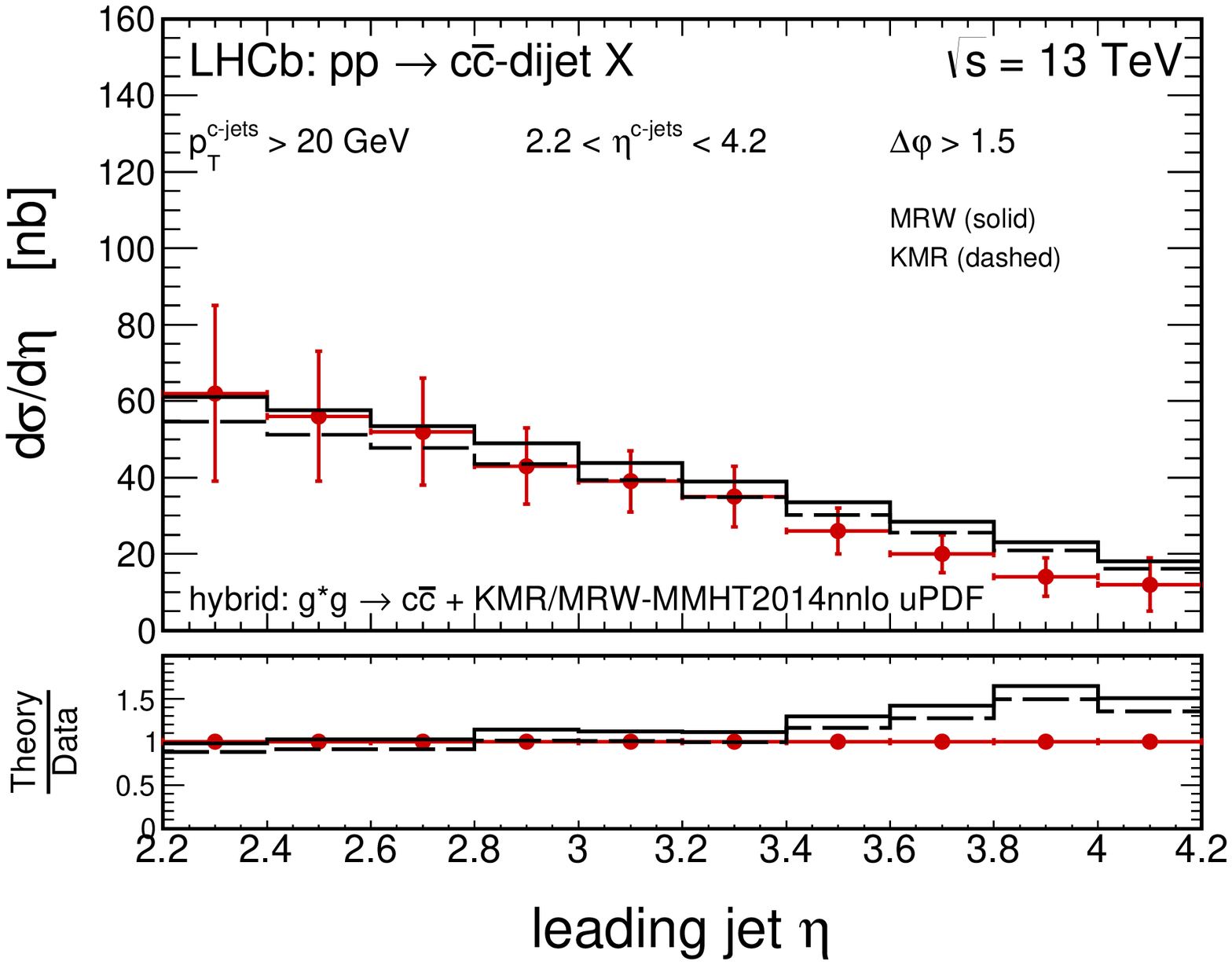}}
\end{minipage}
\begin{minipage}{0.47\textwidth}
  \centerline{\includegraphics[width=1.0\textwidth]{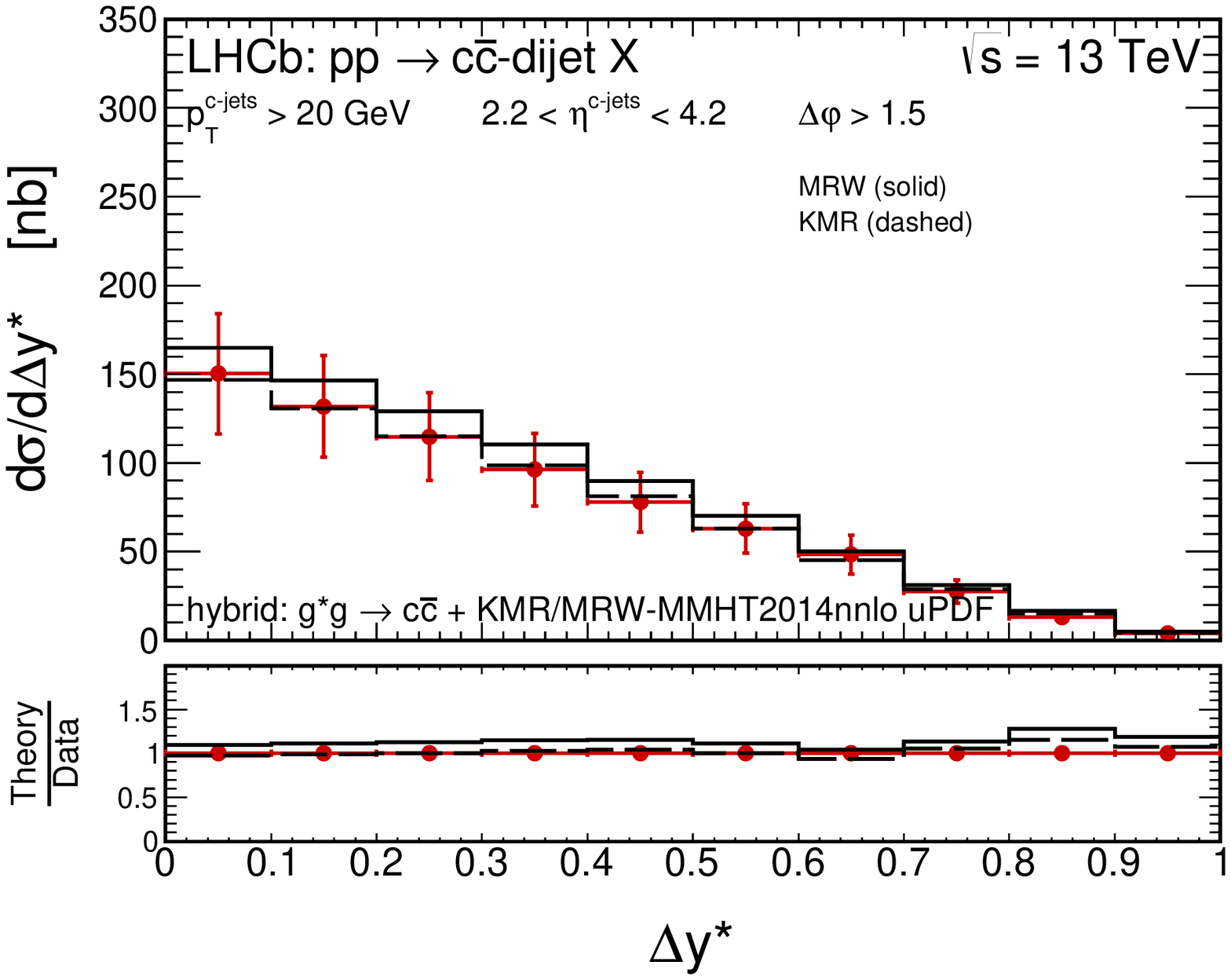}}
\end{minipage}
\begin{minipage}{0.47\textwidth}
  \centerline{\includegraphics[width=1.0\textwidth]{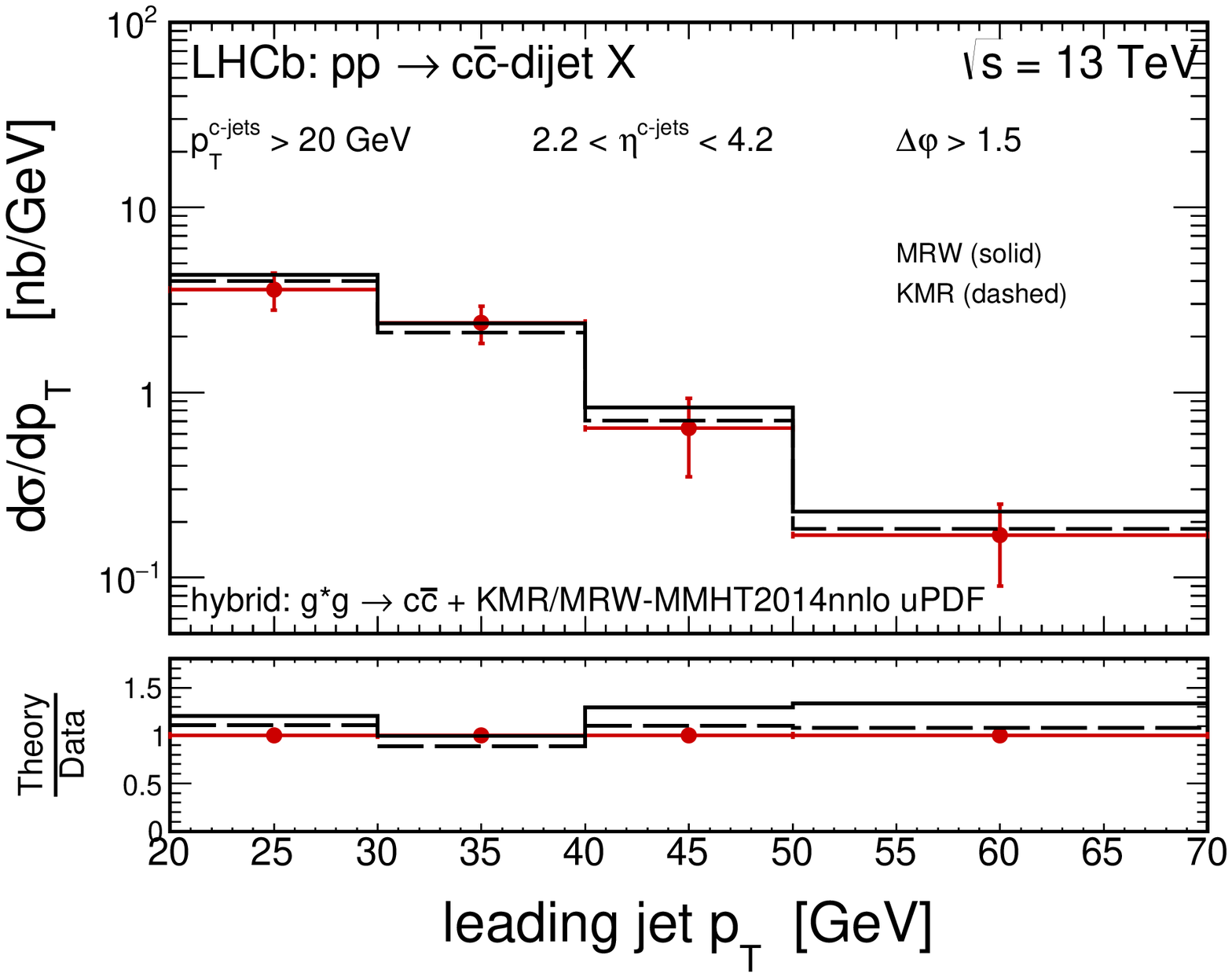}}
\end{minipage}
\begin{minipage}{0.47\textwidth}
  \centerline{\includegraphics[width=1.0\textwidth]{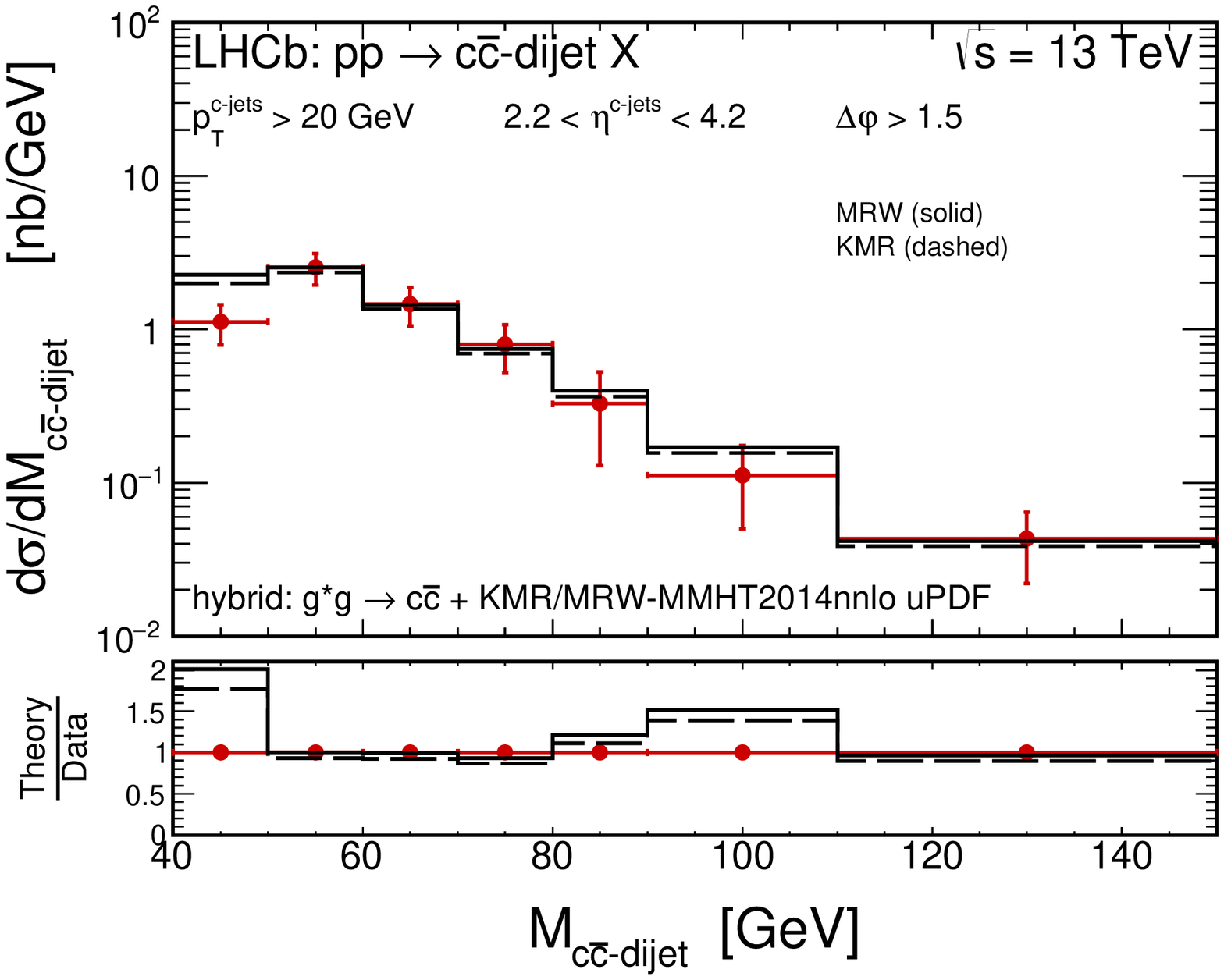}}
\end{minipage}
  \caption{The same as in Fig.~\ref{fig:14} but here the KMR and the MRW uPDFs are used.
\small 
}
\label{fig:15}
\end{figure}

A similar effect is observed also in the case of the KMR/MRW uPDF. As can be seen in Fig.~\ref{fig:15}, using the hybrid approach leads to better agreement with the data and removes a slight tendency to overshoot the data, reported previously in the full $k_{T}$-factorization case. 

\begin{figure}[!h]
\begin{minipage}{0.47\textwidth}
  \centerline{\includegraphics[width=1.0\textwidth]{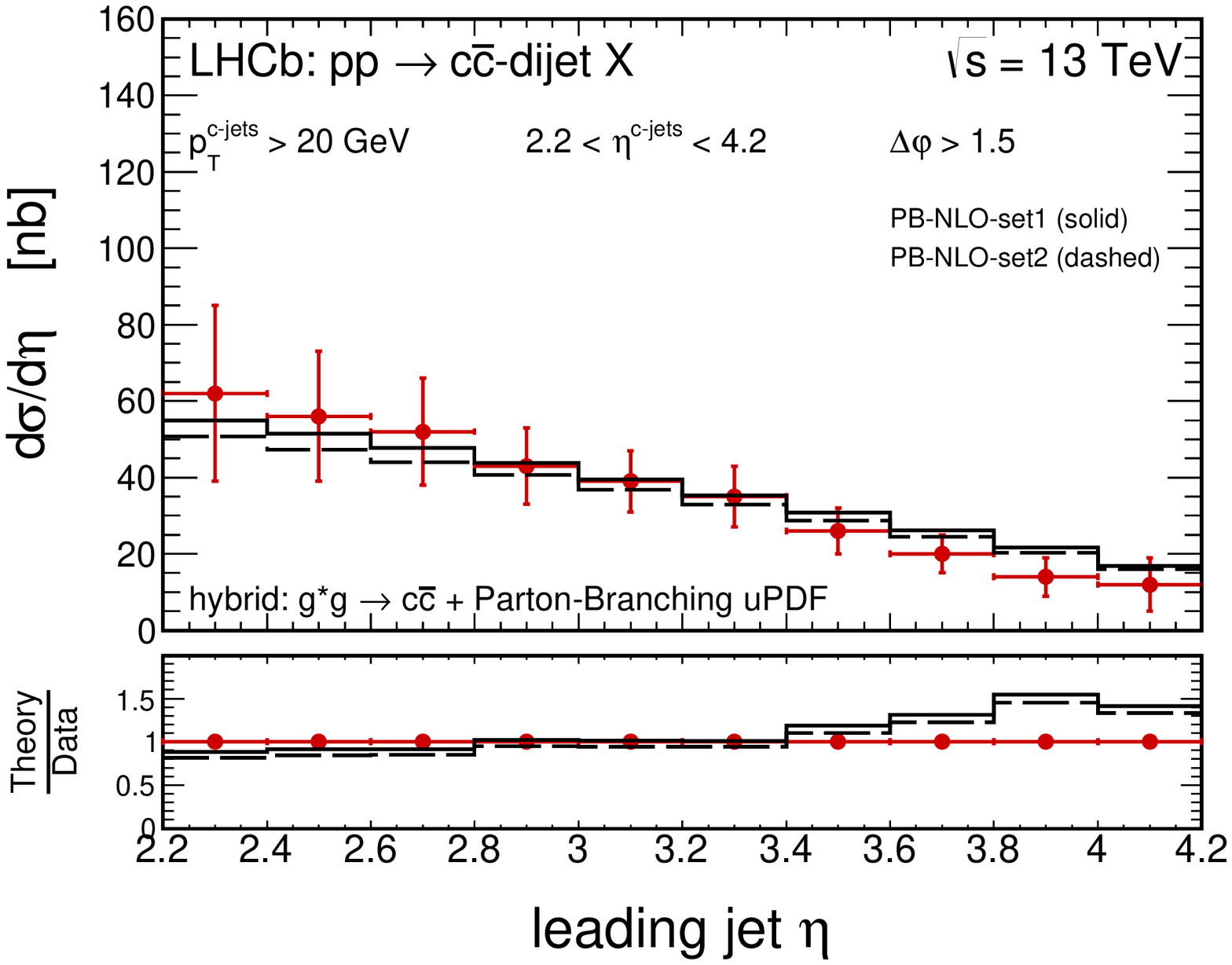}}
\end{minipage}
\begin{minipage}{0.47\textwidth}
  \centerline{\includegraphics[width=1.0\textwidth]{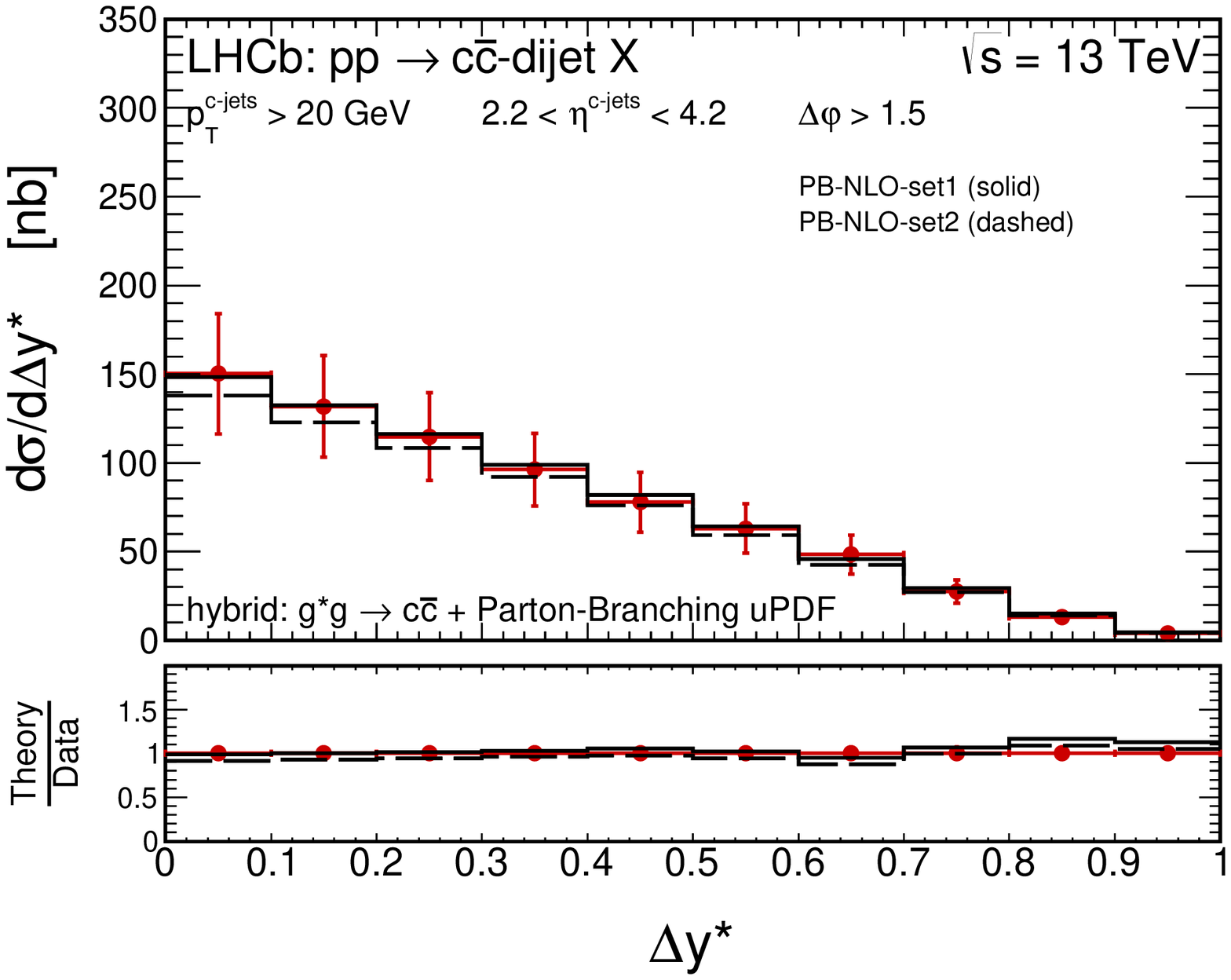}}
\end{minipage}
\begin{minipage}{0.47\textwidth}
  \centerline{\includegraphics[width=1.0\textwidth]{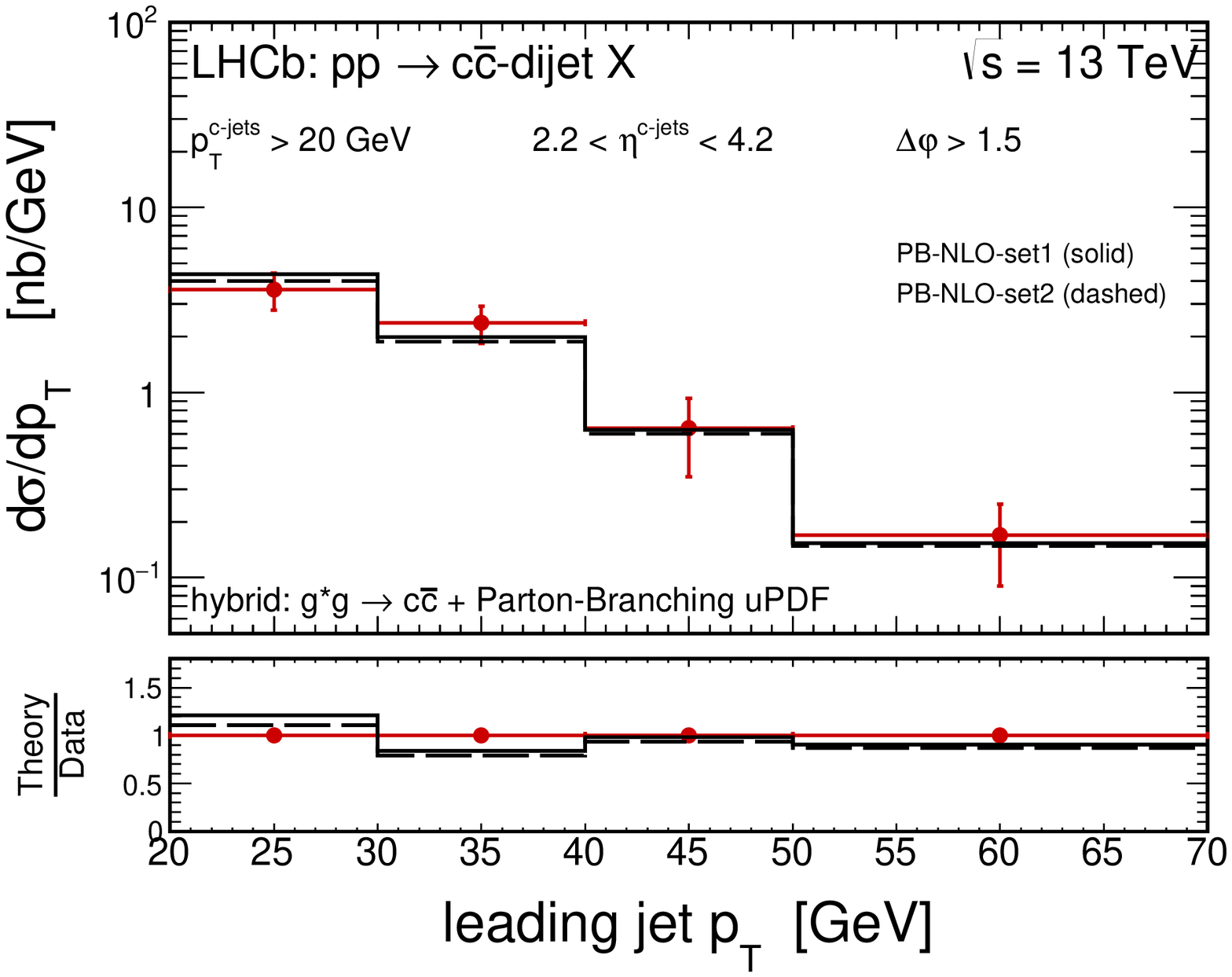}}
\end{minipage}
\begin{minipage}{0.47\textwidth}
  \centerline{\includegraphics[width=1.0\textwidth]{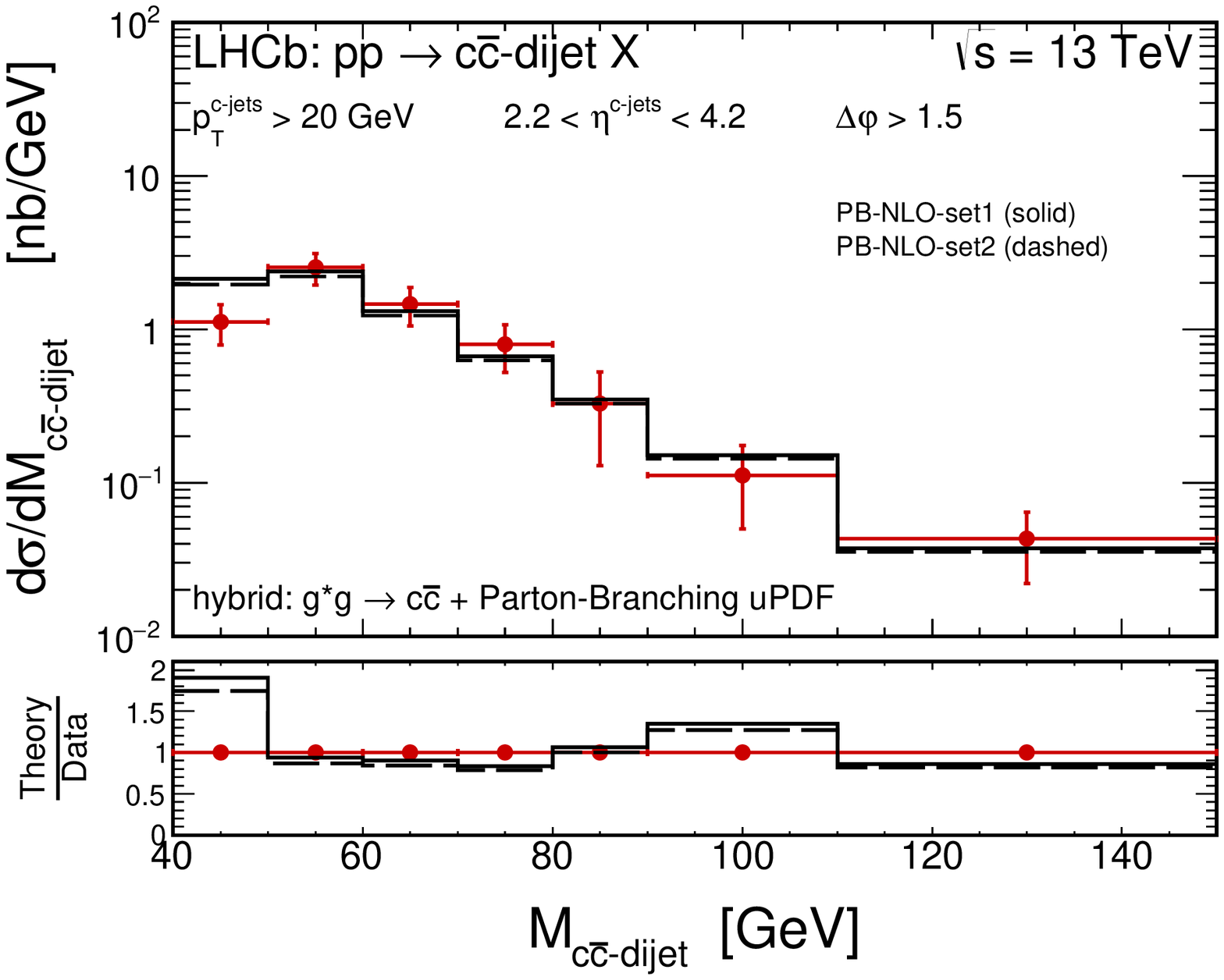}}
\end{minipage}
  \caption{The same as in Fig.~\ref{fig:14} but here the Parton Branching uPDFs are used.
\small 
}
\label{fig:16}
\end{figure}

On the other hand, the hybrid model results obtained with the Parton-Branching uPDFs show an opposite effect (see Fig.~\ref{fig:16}).
Here we observe a small enhancement of the cross section but it works in a good direction and improves the agreement with the experimental data, resulting in an excellent description of all four measured differential distributions.  

%----------------------------------------------------------------------------
%----------------------------------------------------------------------------

%----------------------------------------------------------------------------
%----------------------------------------------------------------------------

%----------------------------------------------------------------------------
\begin{figure}[!h]
\begin{minipage}{0.47\textwidth}
  \centerline{\includegraphics[width=1.0\textwidth]{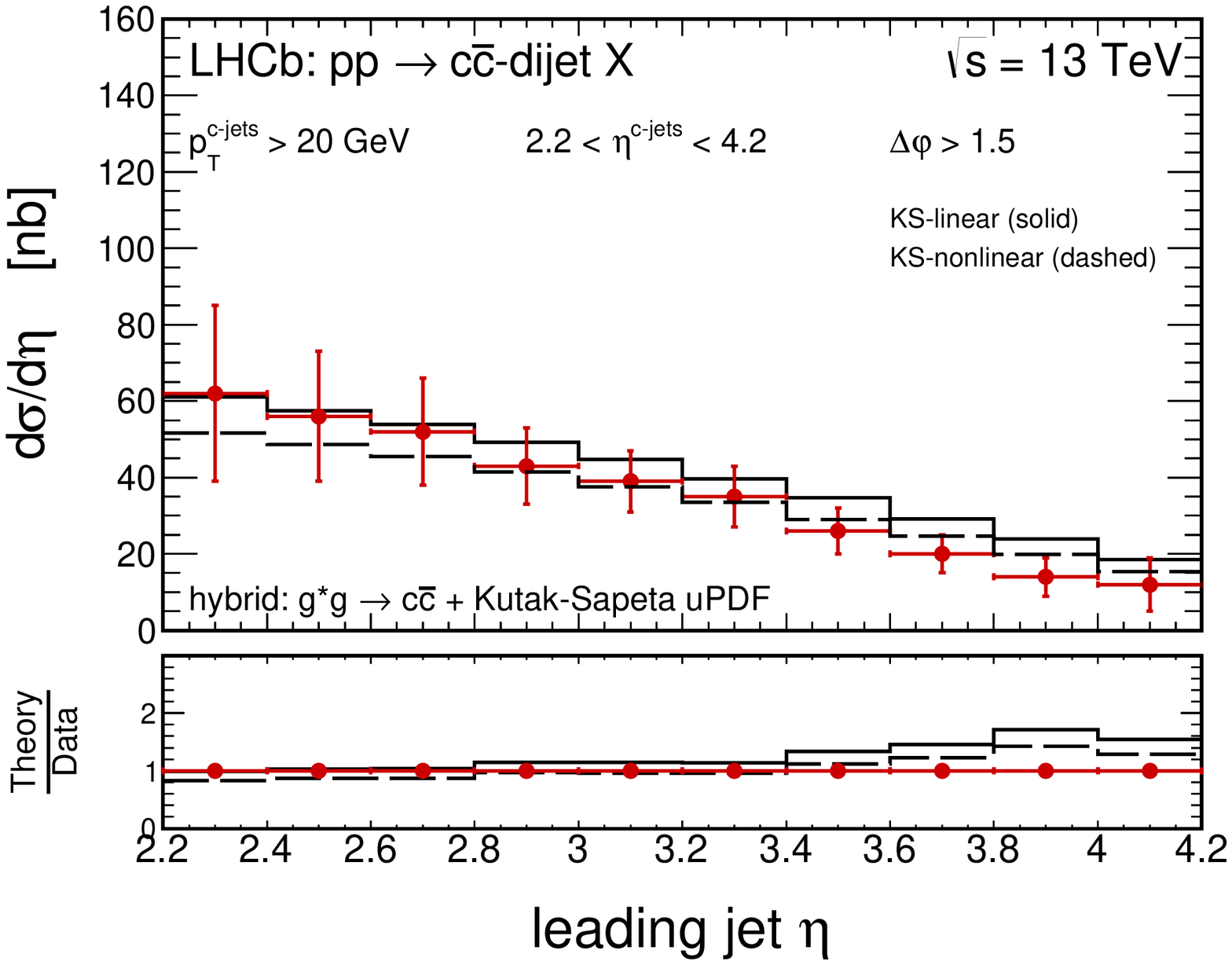}}
\end{minipage}
\begin{minipage}{0.47\textwidth}
  \centerline{\includegraphics[width=1.0\textwidth]{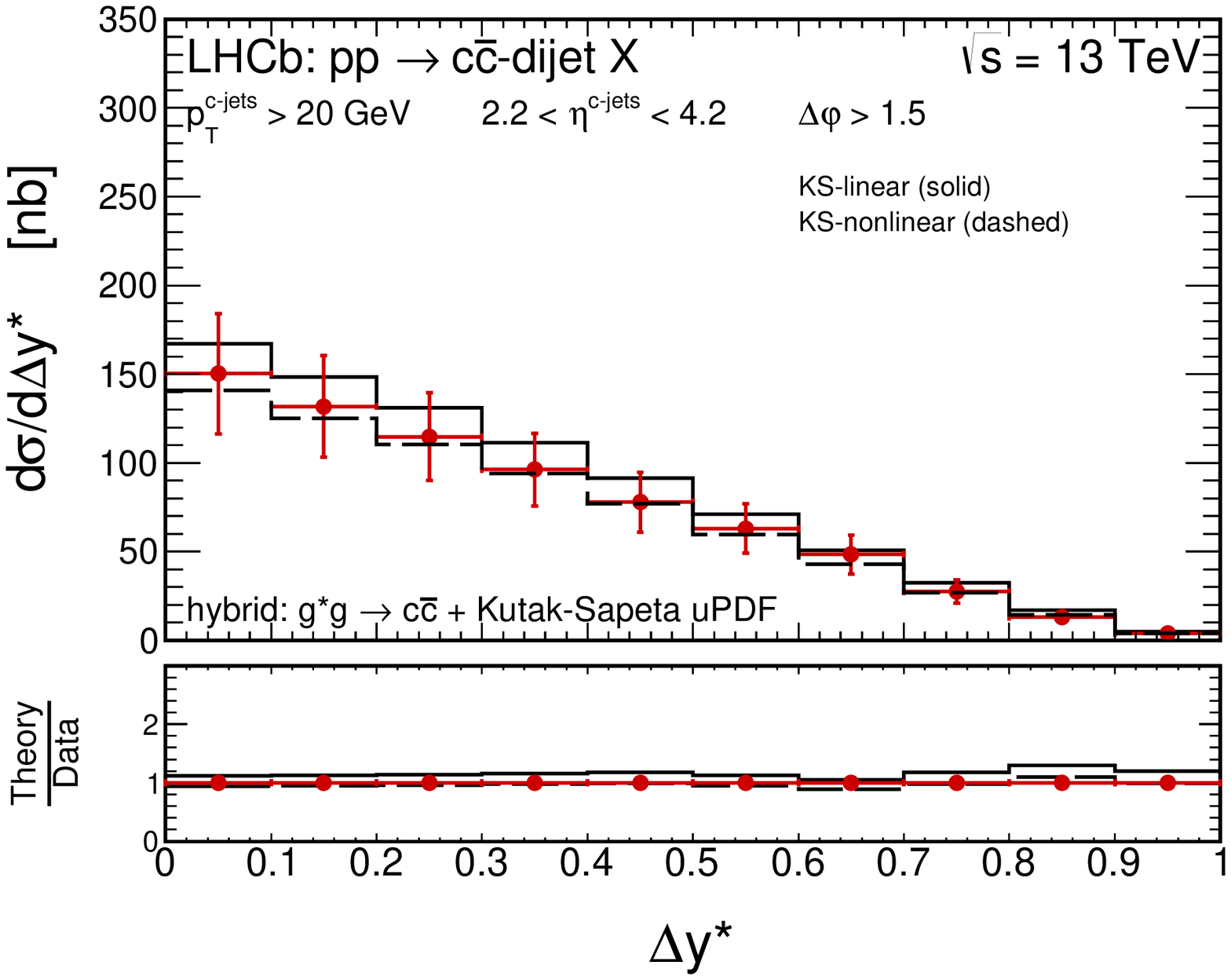}}
\end{minipage}
\begin{minipage}{0.47\textwidth}
  \centerline{\includegraphics[width=1.0\textwidth]{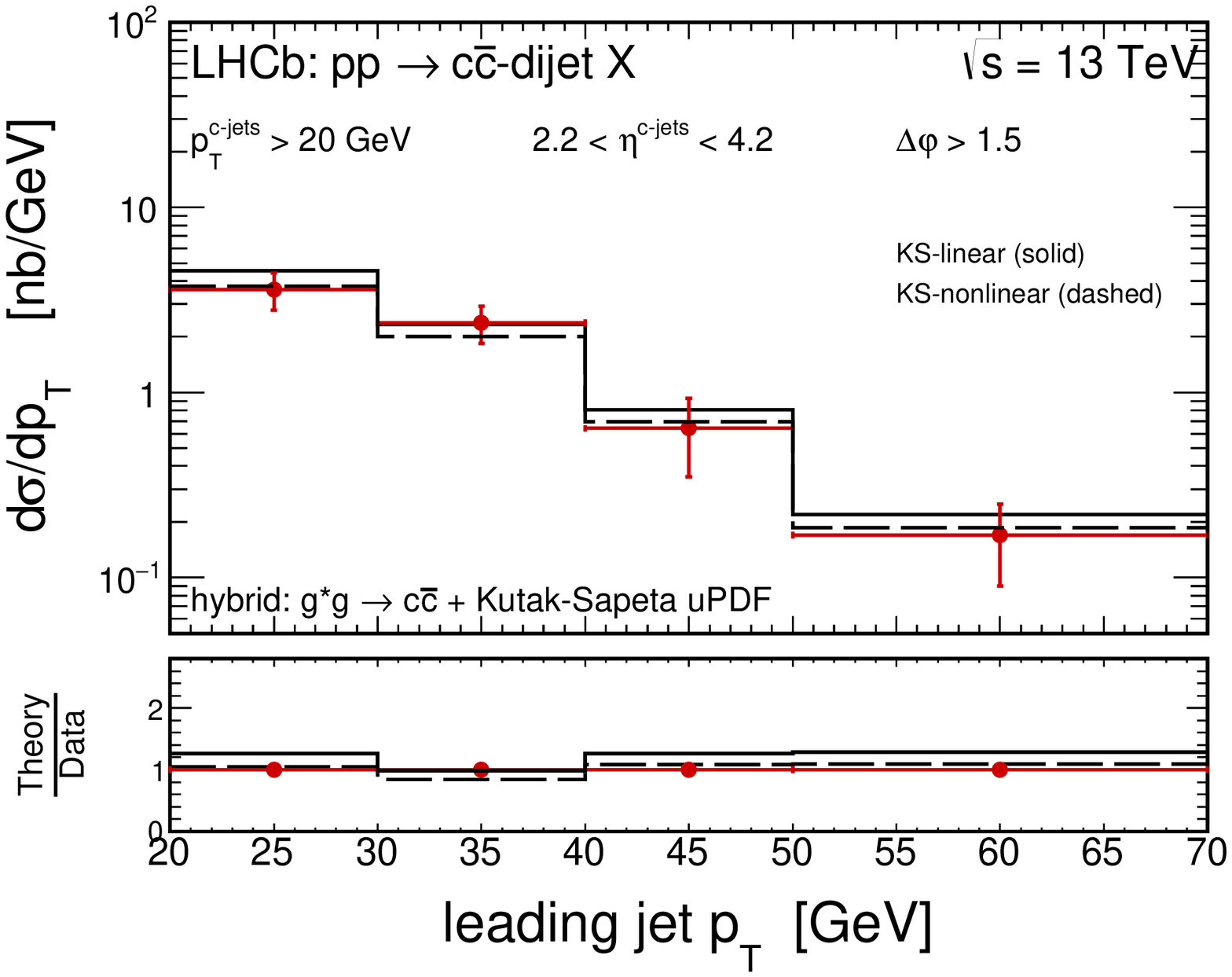}}
\end{minipage}
\begin{minipage}{0.47\textwidth}
  \centerline{\includegraphics[width=1.0\textwidth]{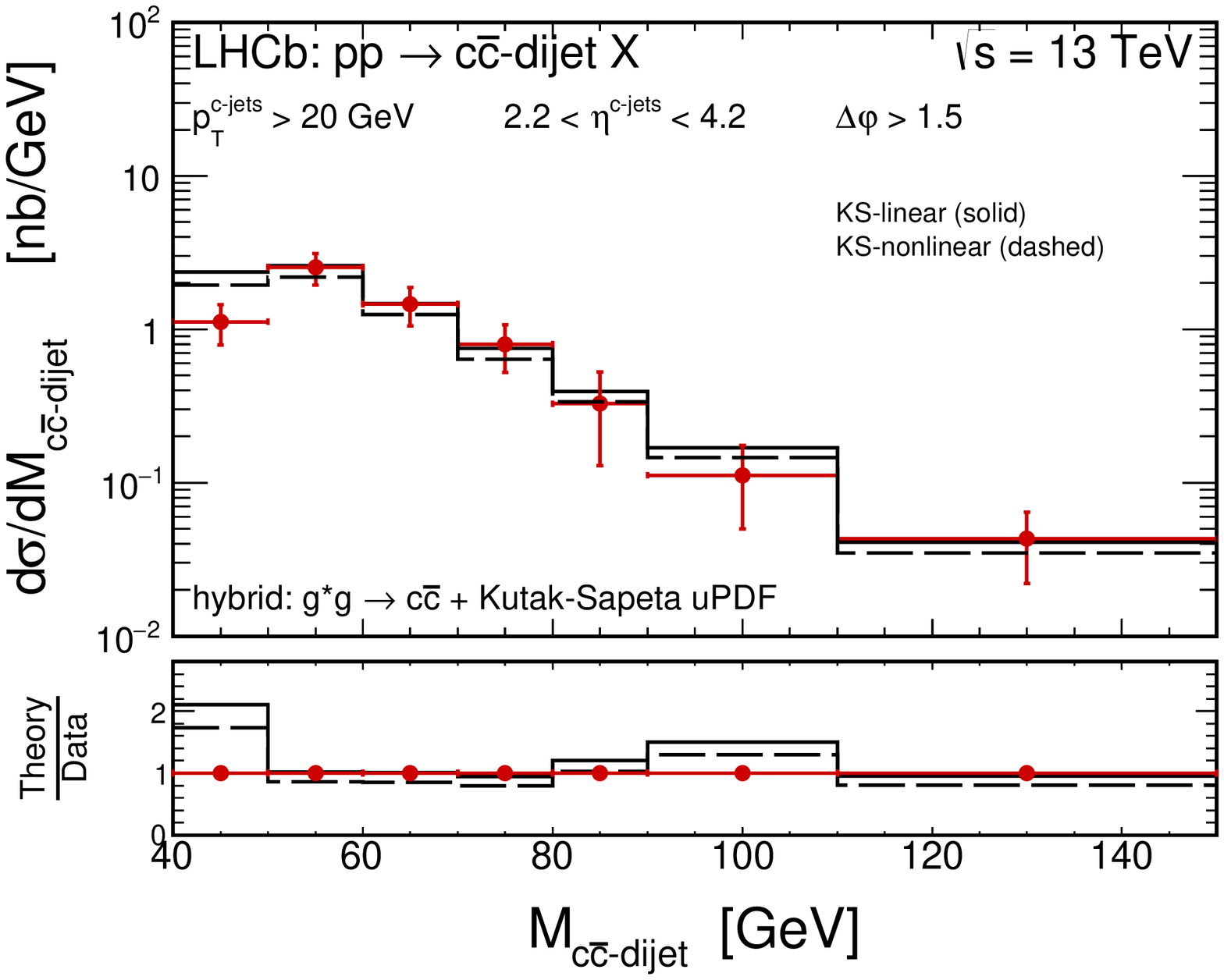}}
\end{minipage}
  \caption{The same as in Fig.~\ref{fig:14} but here the Kutak-Sapeta uPDFs are used.
\small 
}
\label{fig:17}
\end{figure}
%----------------------------------------------------------------------------
\clearpage
\subsubsection{Bottom dijets}

\begin{figure}[!h]
\begin{minipage}{0.47\textwidth}
  \centerline{\includegraphics[width=1.0\textwidth]{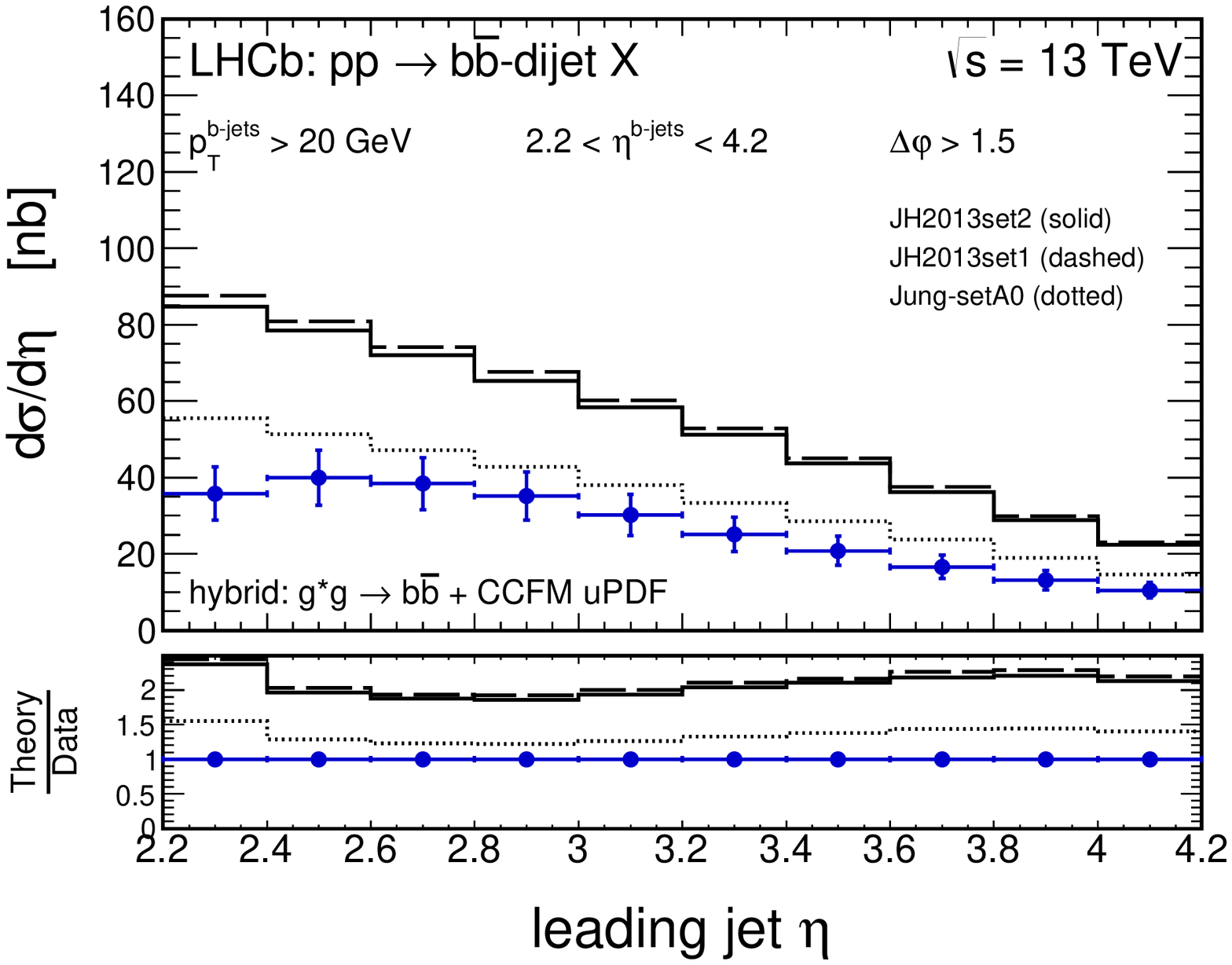}}
\end{minipage}
\begin{minipage}{0.47\textwidth}
  \centerline{\includegraphics[width=1.0\textwidth]{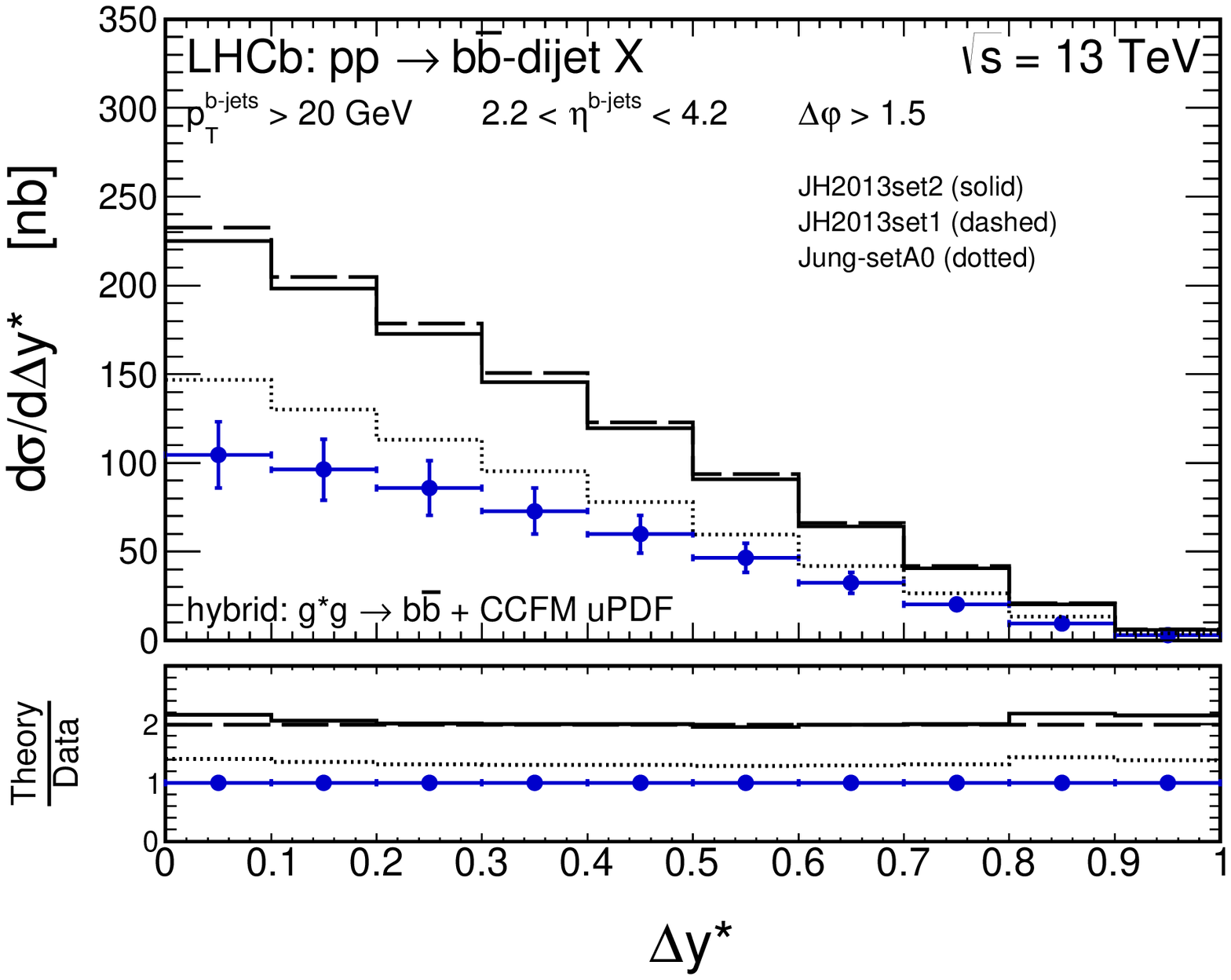}}
\end{minipage}
\begin{minipage}{0.47\textwidth}
  \centerline{\includegraphics[width=1.0\textwidth]{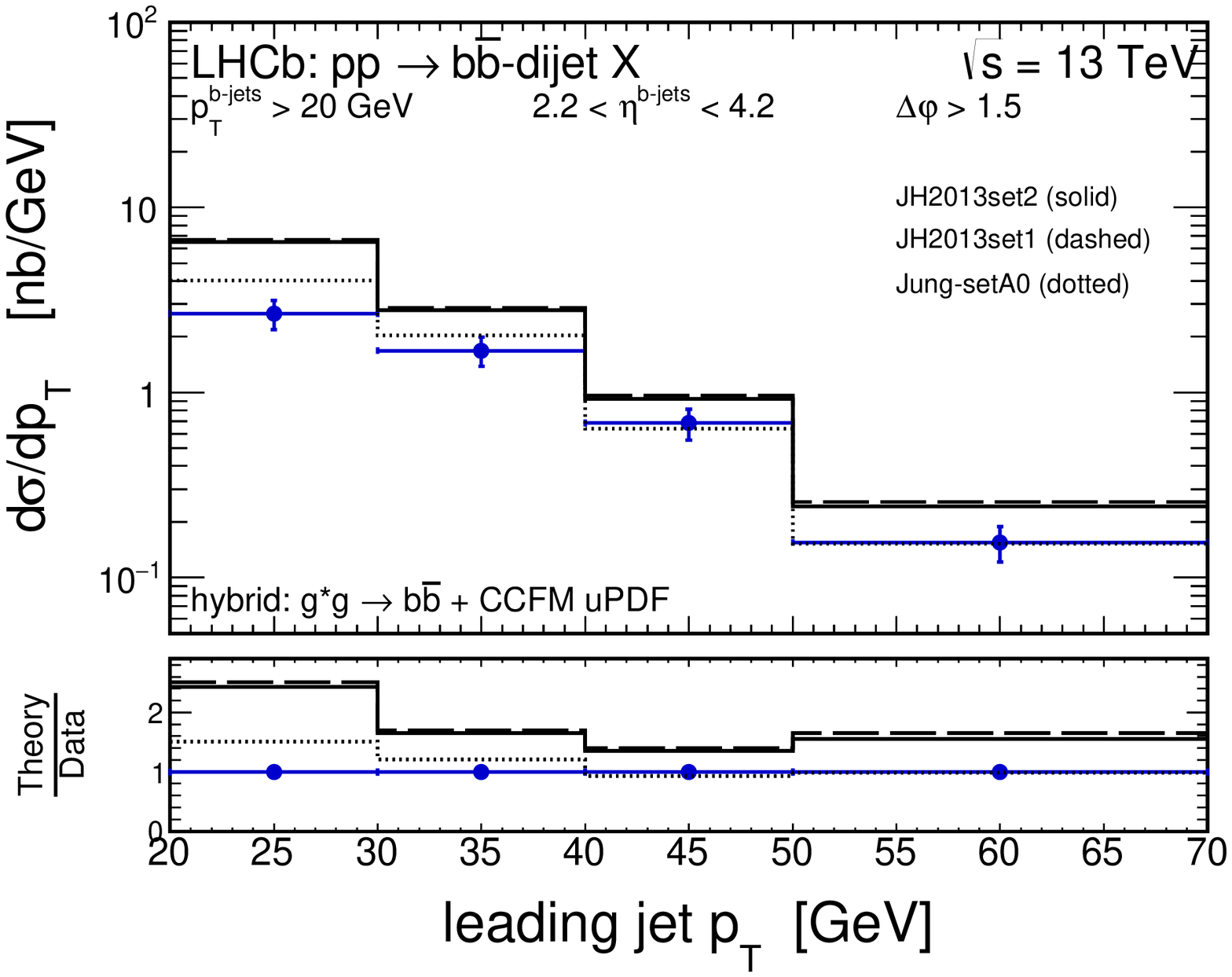}}
\end{minipage}
\begin{minipage}{0.47\textwidth}
  \centerline{\includegraphics[width=1.0\textwidth]{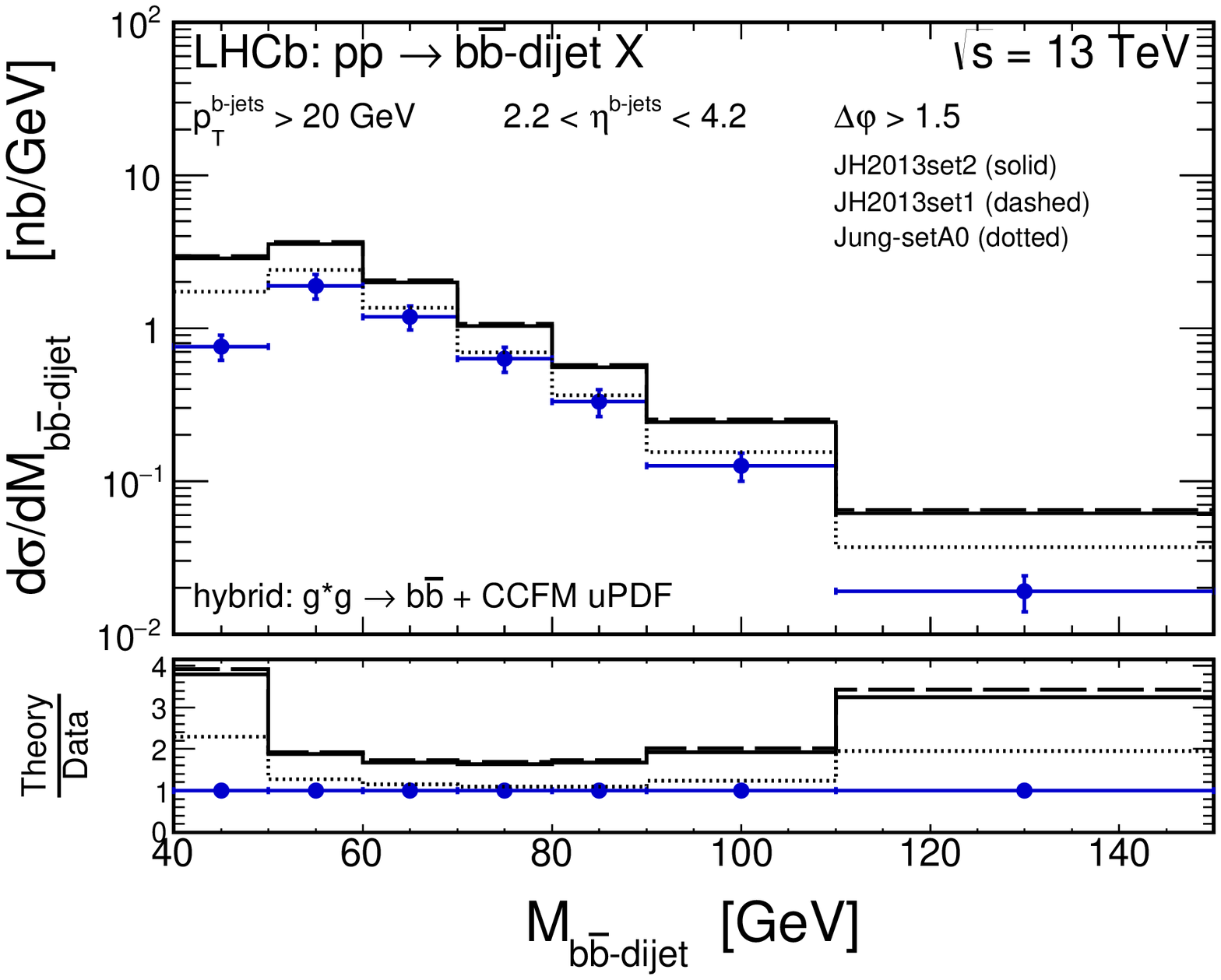}}
\end{minipage}
  \caption{The differential cross sections for forward production of $b\bar b$-dijets in $pp$-scattering at $\sqrt{s}=13$ TeV as a function of the leading jet $\eta$ (top left), the rapidity difference $\Delta y^{*}$ (top right), the leading jet $p_{T}$ (bottom left) and the dijet invariant mass $M_{b\bar b\text{-}\mathrm{dijet}}$ (bottom right). Here the dominant pQCD $g^*g \to b \bar b$ mechanism is taken into account. The theoretical histograms correspond to the hybrid model calculations obtained with the CCFM uPDFs.
\small 
}
\label{fig:18}
\end{figure}

\begin{figure}[!h]
\begin{minipage}{0.47\textwidth}
  \centerline{\includegraphics[width=1.0\textwidth]{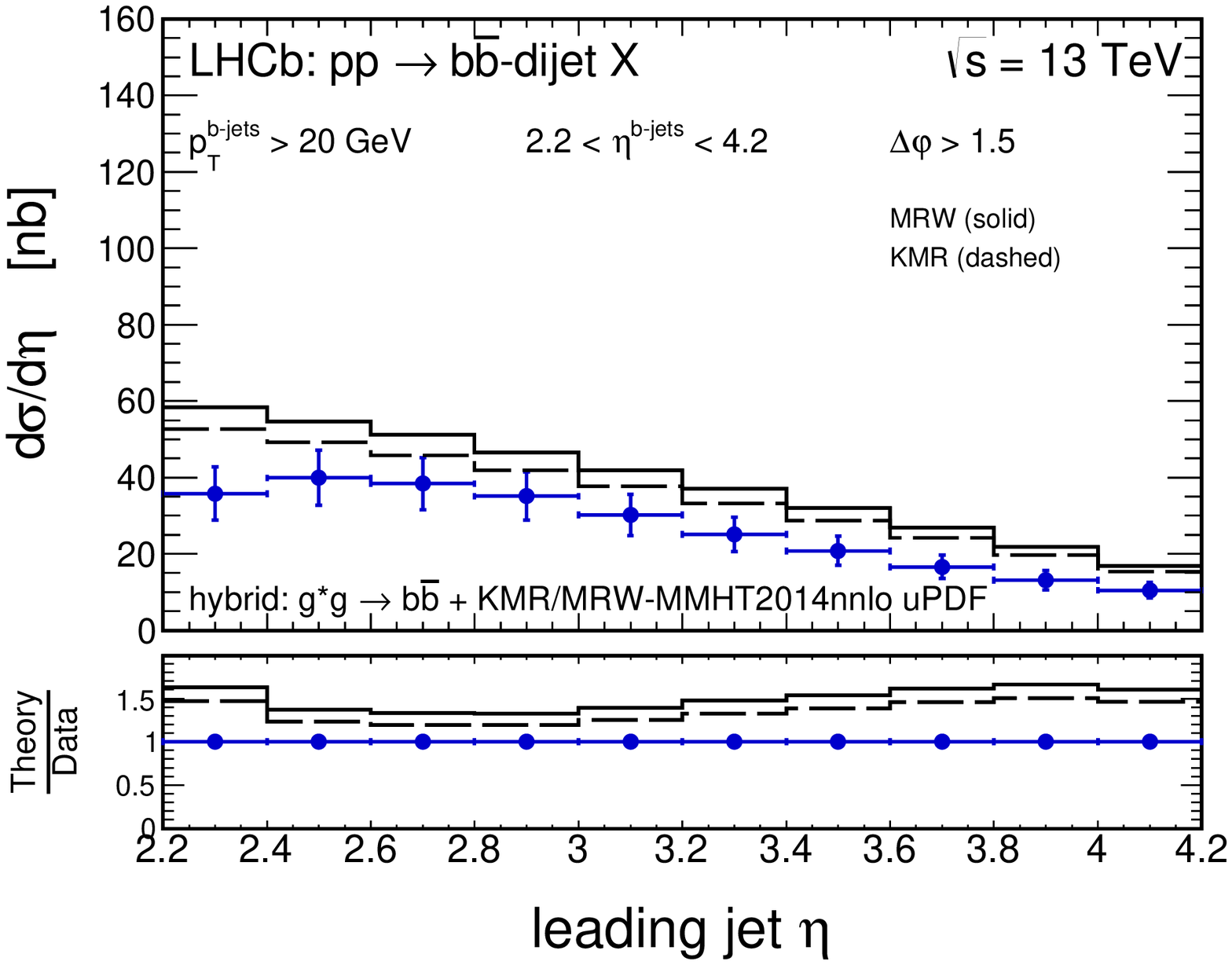}}
\end{minipage}
\begin{minipage}{0.47\textwidth}
  \centerline{\includegraphics[width=1.0\textwidth]{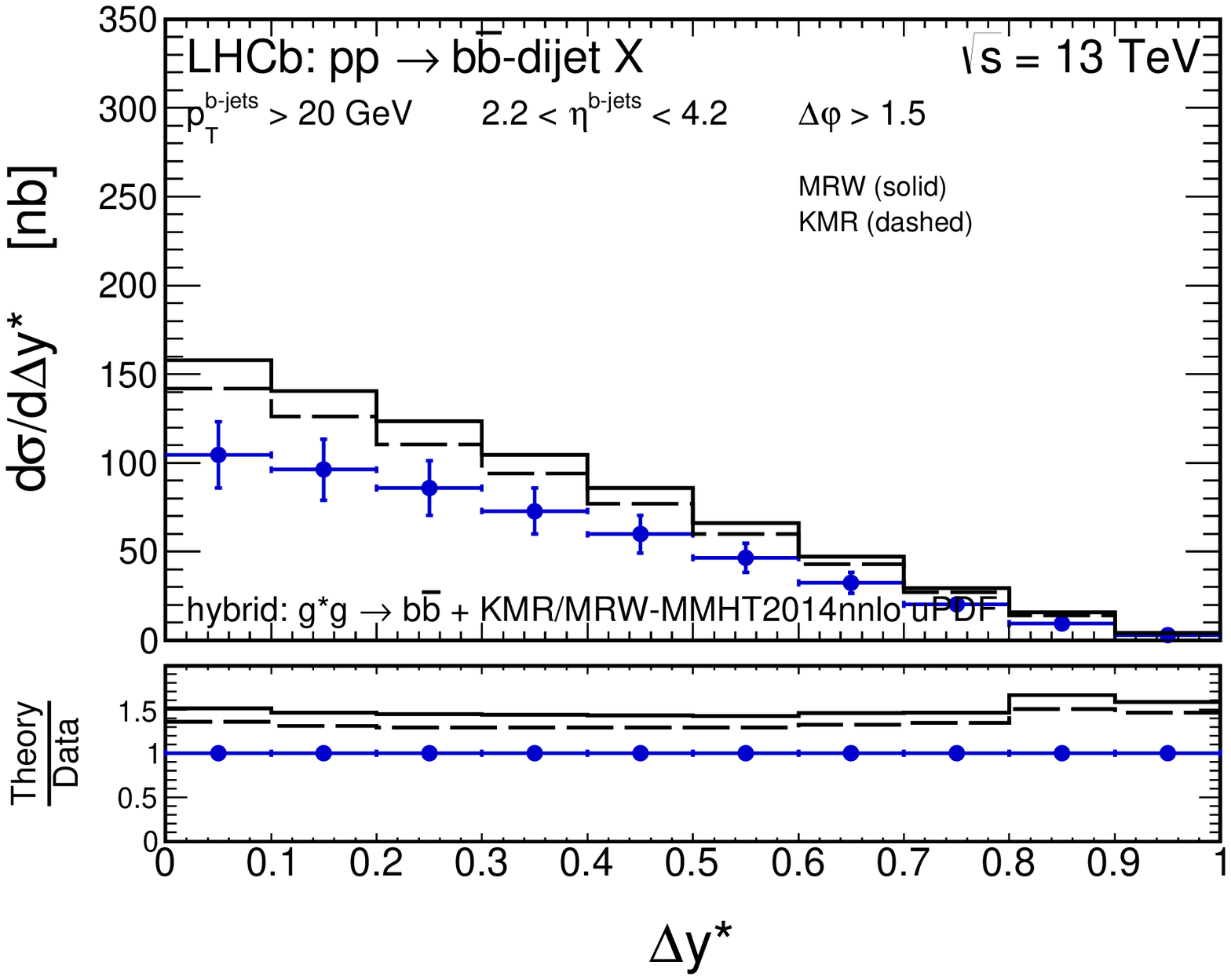}}
\end{minipage}
\begin{minipage}{0.47\textwidth}
  \centerline{\includegraphics[width=1.0\textwidth]{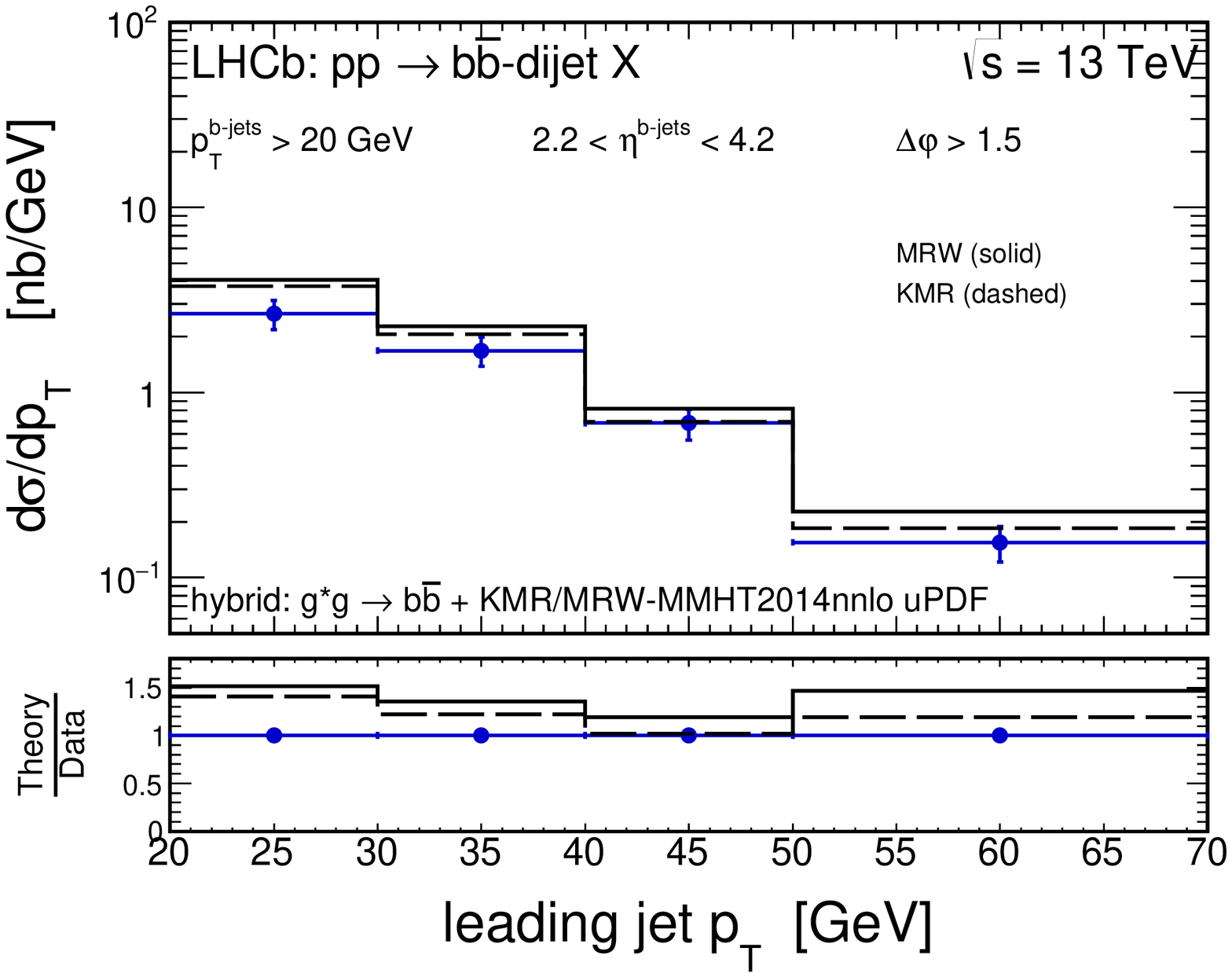}}
\end{minipage}
\begin{minipage}{0.47\textwidth}
  \centerline{\includegraphics[width=1.0\textwidth]{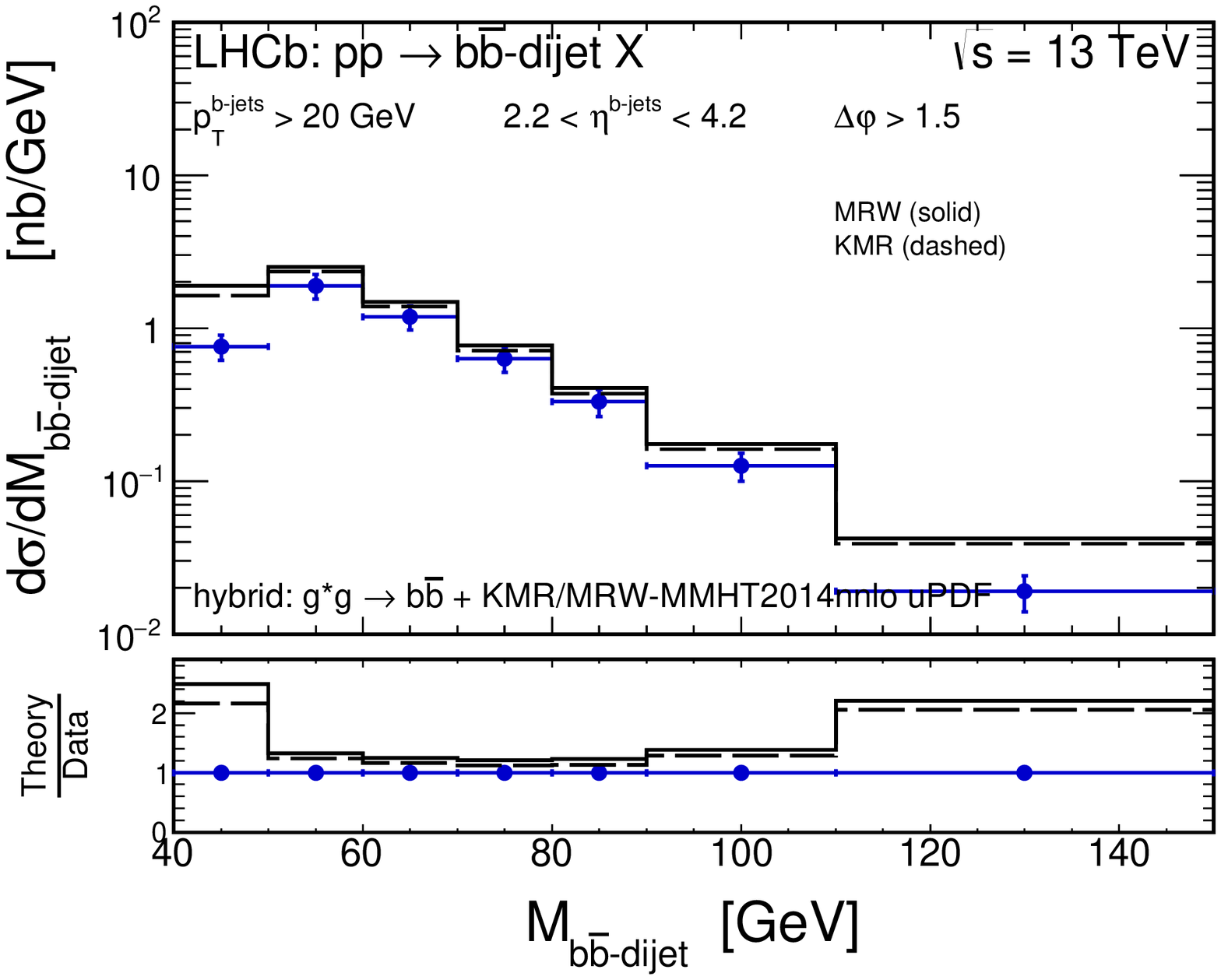}}
\end{minipage}
  \caption{The same as in Fig.~\ref{fig:18} but here the KMR and the MRW uPDFs are used.
\small 
}
\label{fig:19}
\end{figure}

\begin{figure}[!h]
\begin{minipage}{0.47\textwidth}
  \centerline{\includegraphics[width=1.0\textwidth]{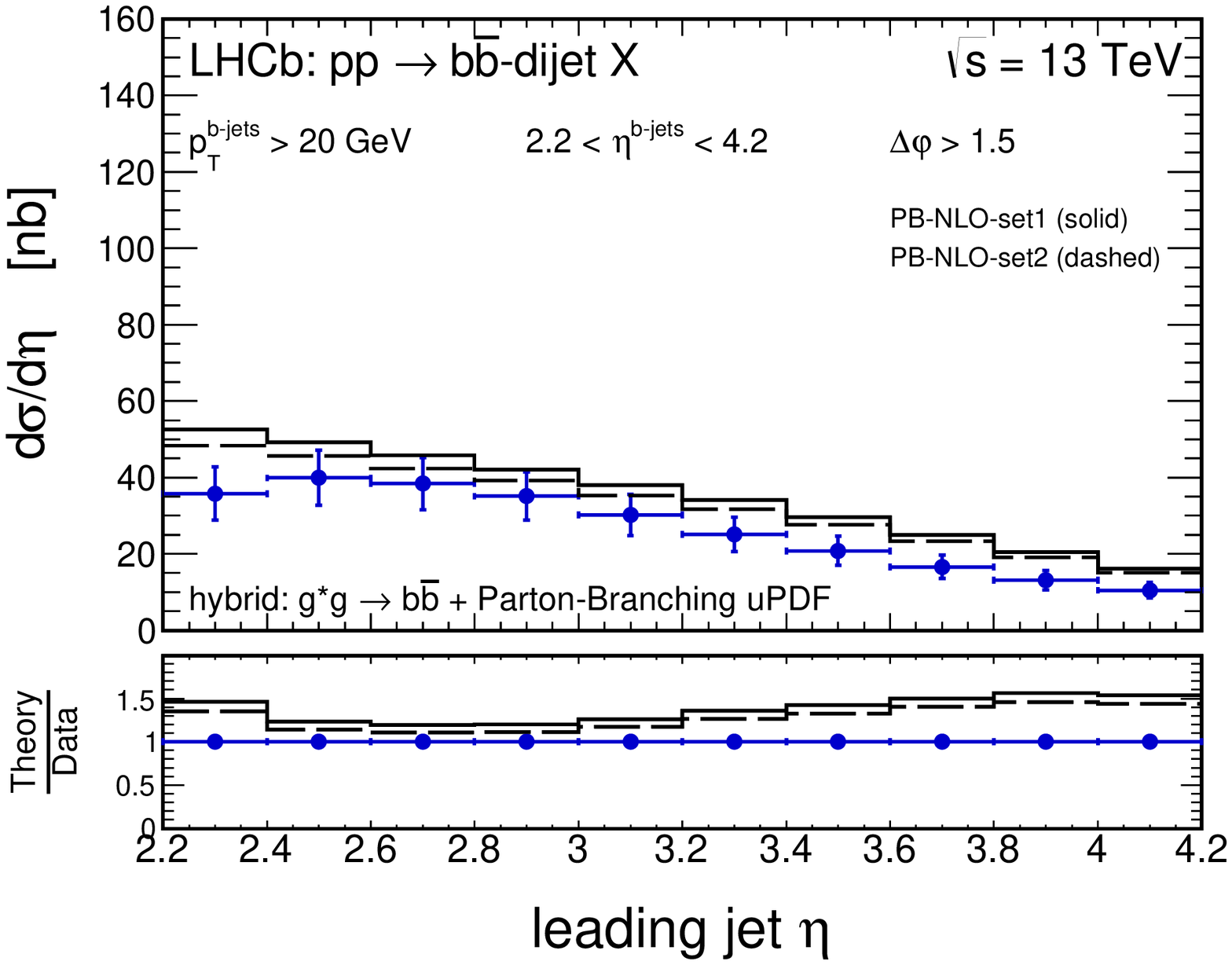}}
\end{minipage}
\begin{minipage}{0.47\textwidth}
  \centerline{\includegraphics[width=1.0\textwidth]{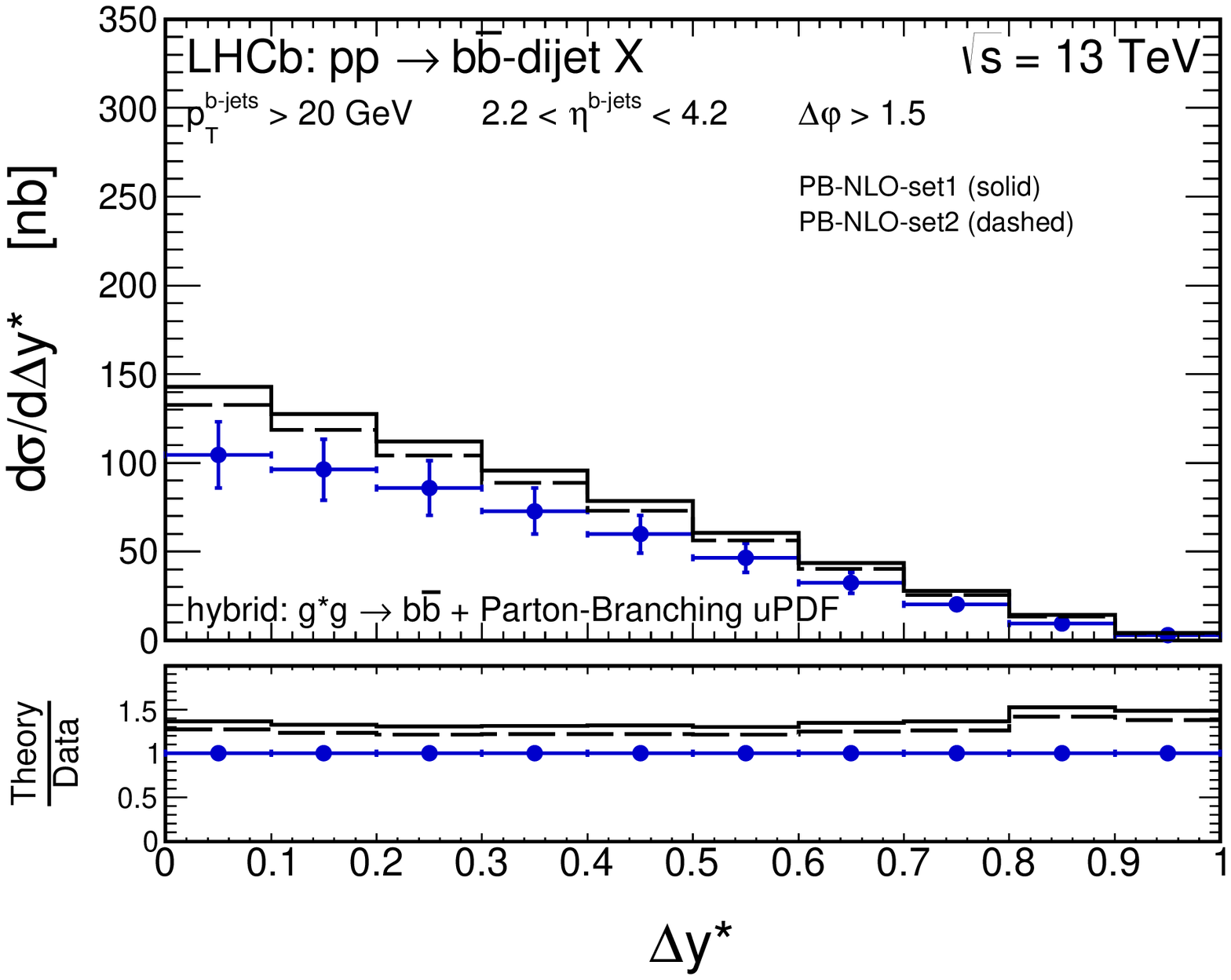}}
\end{minipage}
\begin{minipage}{0.47\textwidth}
  \centerline{\includegraphics[width=1.0\textwidth]{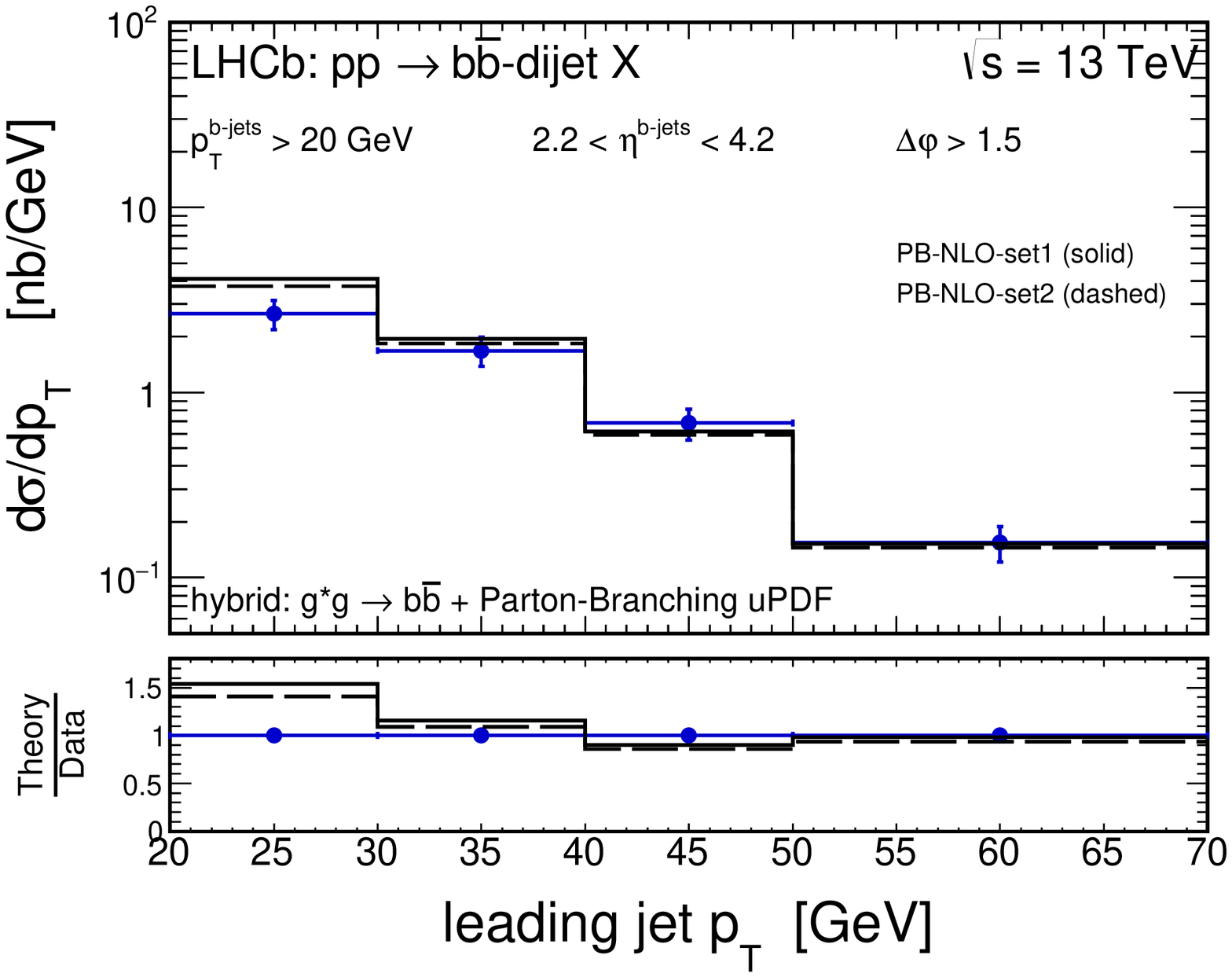}}
\end{minipage}
\begin{minipage}{0.47\textwidth}
  \centerline{\includegraphics[width=1.0\textwidth]{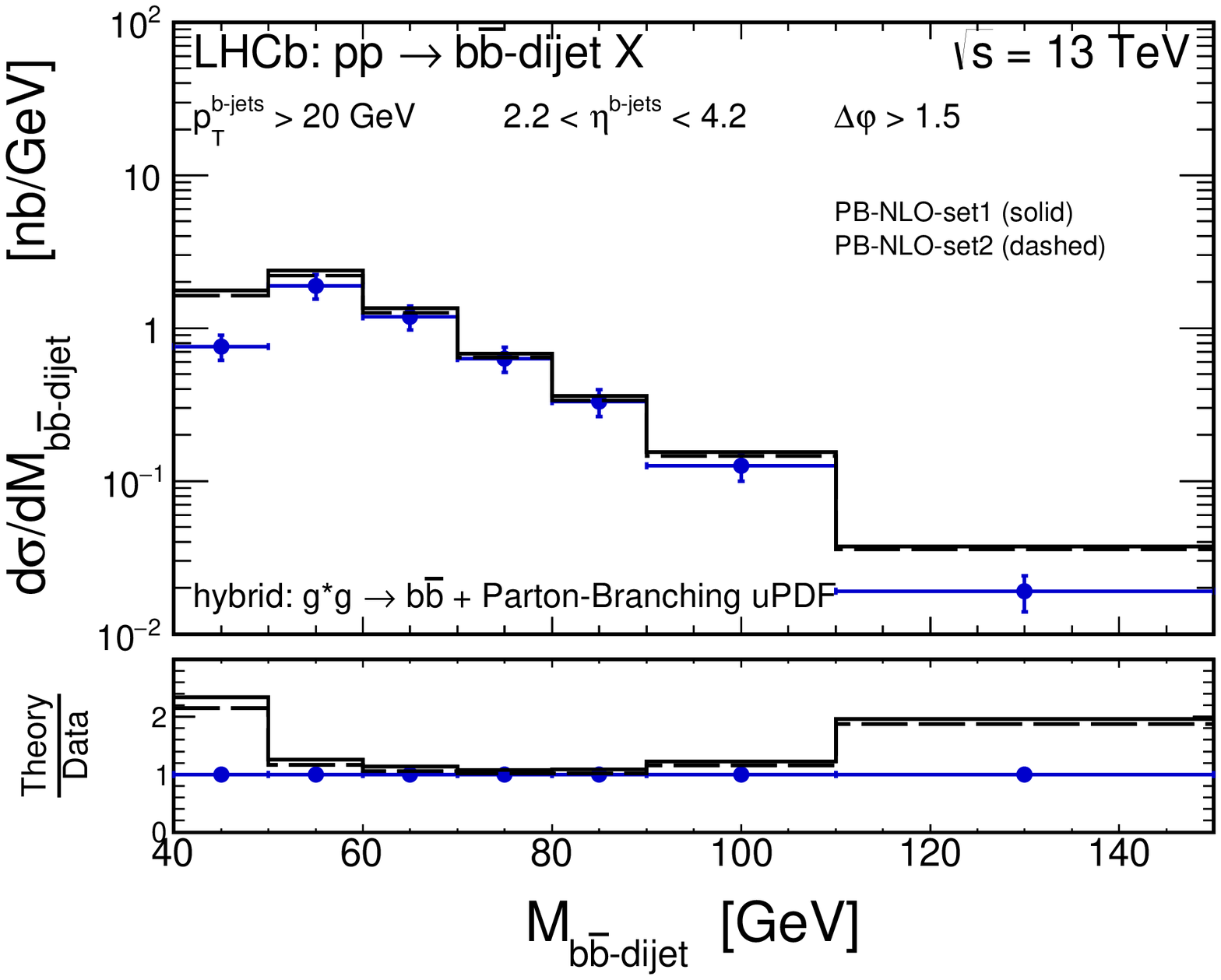}}
\end{minipage}
  \caption{The same as in Fig.~\ref{fig:18} but here the Parton-Branching uPDFs are used.
\small 
}
\label{fig:20}
\end{figure}

\begin{figure}[!h]
\begin{minipage}{0.47\textwidth}
  \centerline{\includegraphics[width=1.0\textwidth]{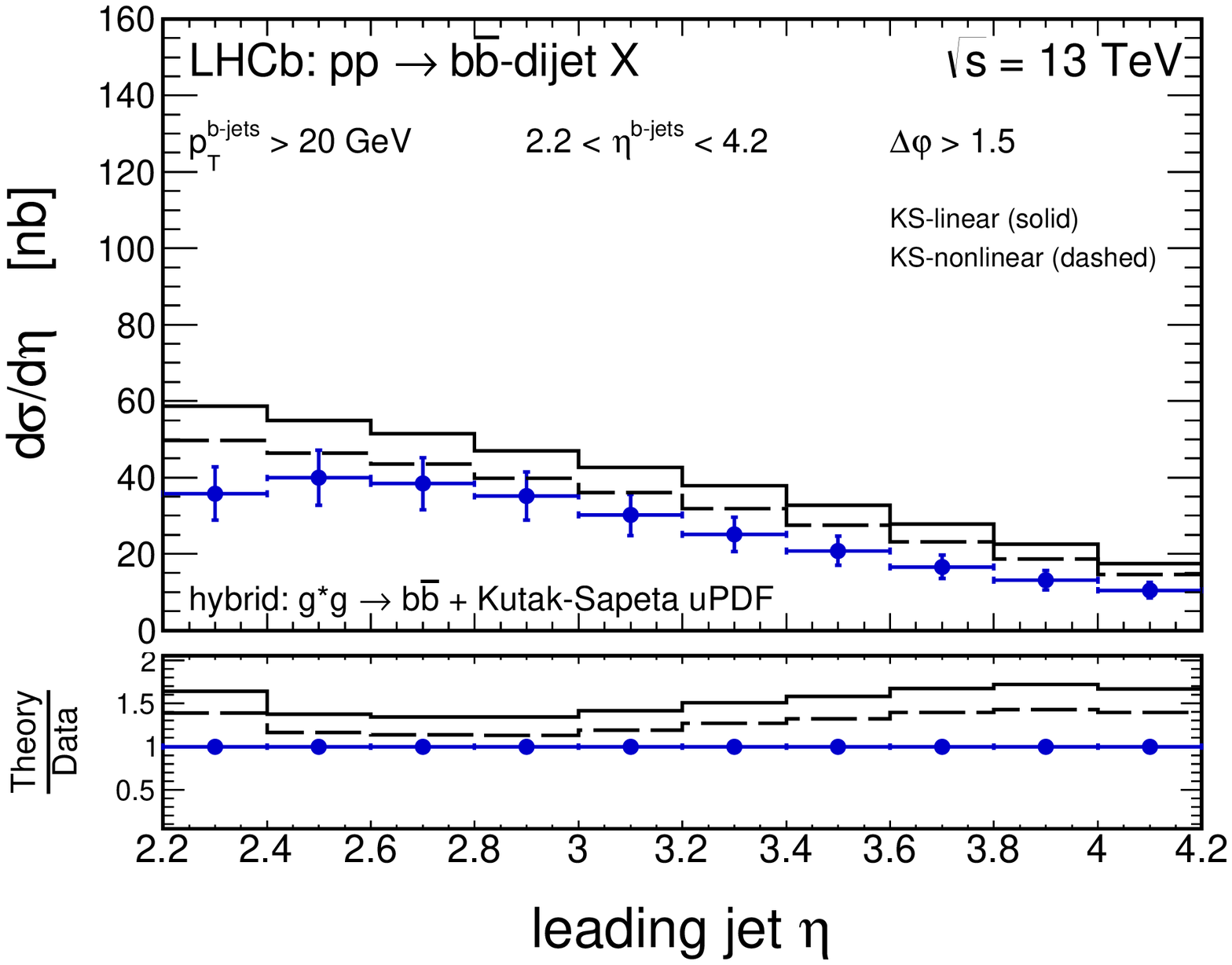}}
\end{minipage}
\begin{minipage}{0.47\textwidth}
  \centerline{\includegraphics[width=1.0\textwidth]{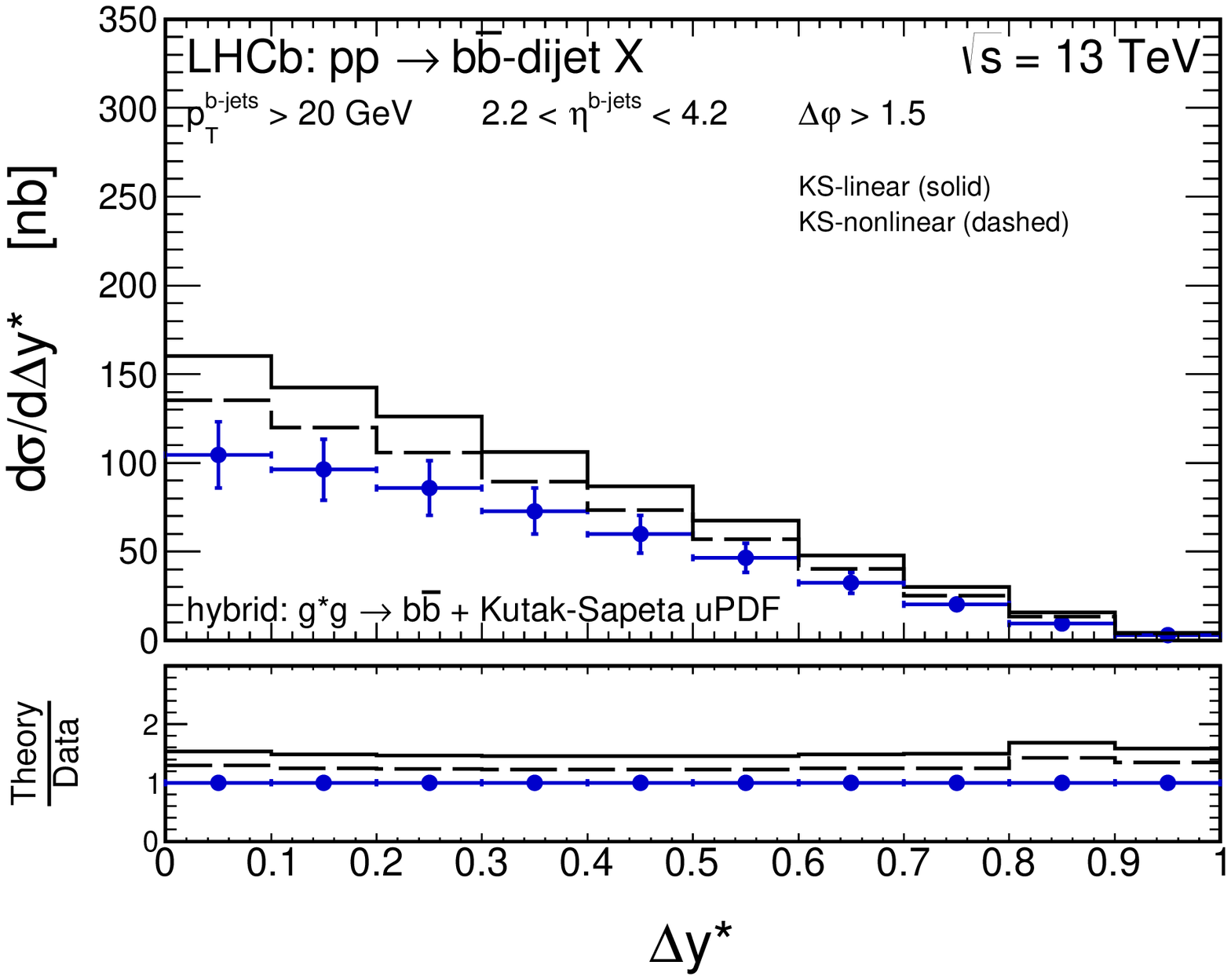}}
\end{minipage}
\begin{minipage}{0.47\textwidth}
  \centerline{\includegraphics[width=1.0\textwidth]{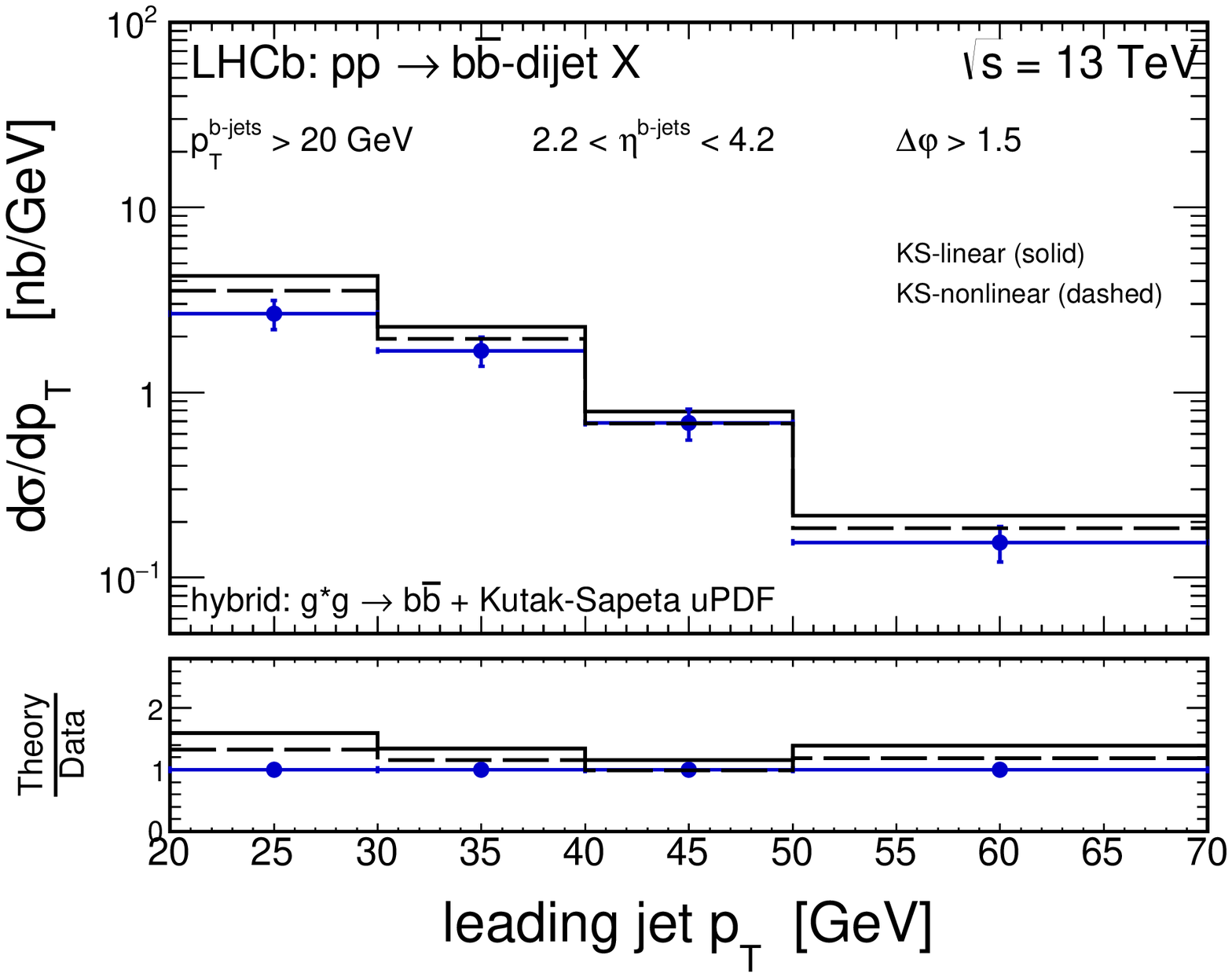}}
\end{minipage}
\begin{minipage}{0.47\textwidth}
  \centerline{\includegraphics[width=1.0\textwidth]{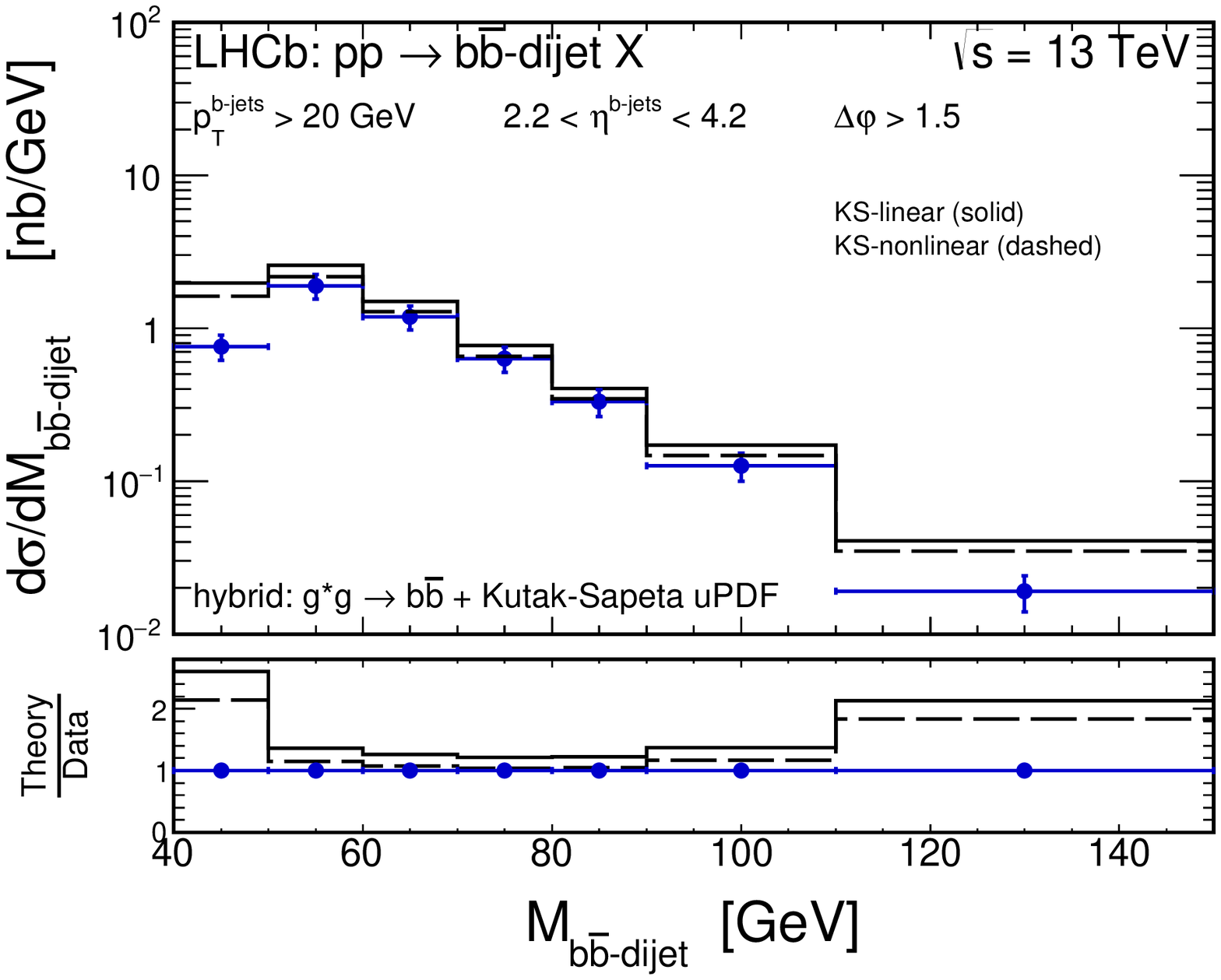}}
\end{minipage}
  \caption{The same as in Fig.~\ref{fig:18} but here the Kutak-Sapeta uPDFs are used.
\small 
}
\label{fig:21}
\end{figure}

Now let us go to bottom dijet production within the hybrid approach. Here we repeat the analysis from the previous subsection but for the $b\bar b$-dijet LHCb data (see Figs.~\ref{fig:18}, ~\ref{fig:19}, ~\ref{fig:20}, and ~\ref{fig:21}). As was already mentioned our predictions are almost independent of the heavy quark mass so the theoretical distributions are almost identical to those from the previous subsection, however, the experimental data points are changed. The Jung setA0, the KMR, and the KS-linear gluon uPDFs provide a very good description of the charm dijet data within the hybrid model, here showing a visible tendency to overshoot the bottom dijet data. Again the PB-NLO-set1 uPDF leads to the best agreement with the data, however, here some problem appears for the very first bin in the leading jet transverse momentum where the theoretical cross section slightly overestimates the experimental point. The discrepancy affects also the leading jet $\eta$ and the rapidity difference $\Delta y^{*}$ distributions. 

To summarize this section, we conclude that the best agreement between the LHCb heavy flavoured dijet data and the predictions of the hybrid model can be obtained for the PB-NLO-set1 uPDF. However, the results of the model show a small tendency to underestimate and overestimate the experimental distributions for the $c\bar c$- and the $b\bar b$-dijets, respectively. It may suggest, that within our model the production cross section ratio $R=\frac{c\bar c}{b\bar b}$ is not well reproduced.

%----------------------------------------------------------------------------
%----------------------------------------------------------------------------

%----------------------------------------------------------------------------
%----------------------------------------------------------------------------

%----------------------------------------------------------------------------
%----------------------------------------------------------------------------
\clearpage
%----------------------------------------------------------------------------
%----------------------------------------------------------------------------

\subsection{The charm/bottom dijet ratio}

\begin{figure}[!h]
\begin{minipage}{0.47\textwidth}
  \centerline{\includegraphics[width=1.0\textwidth]{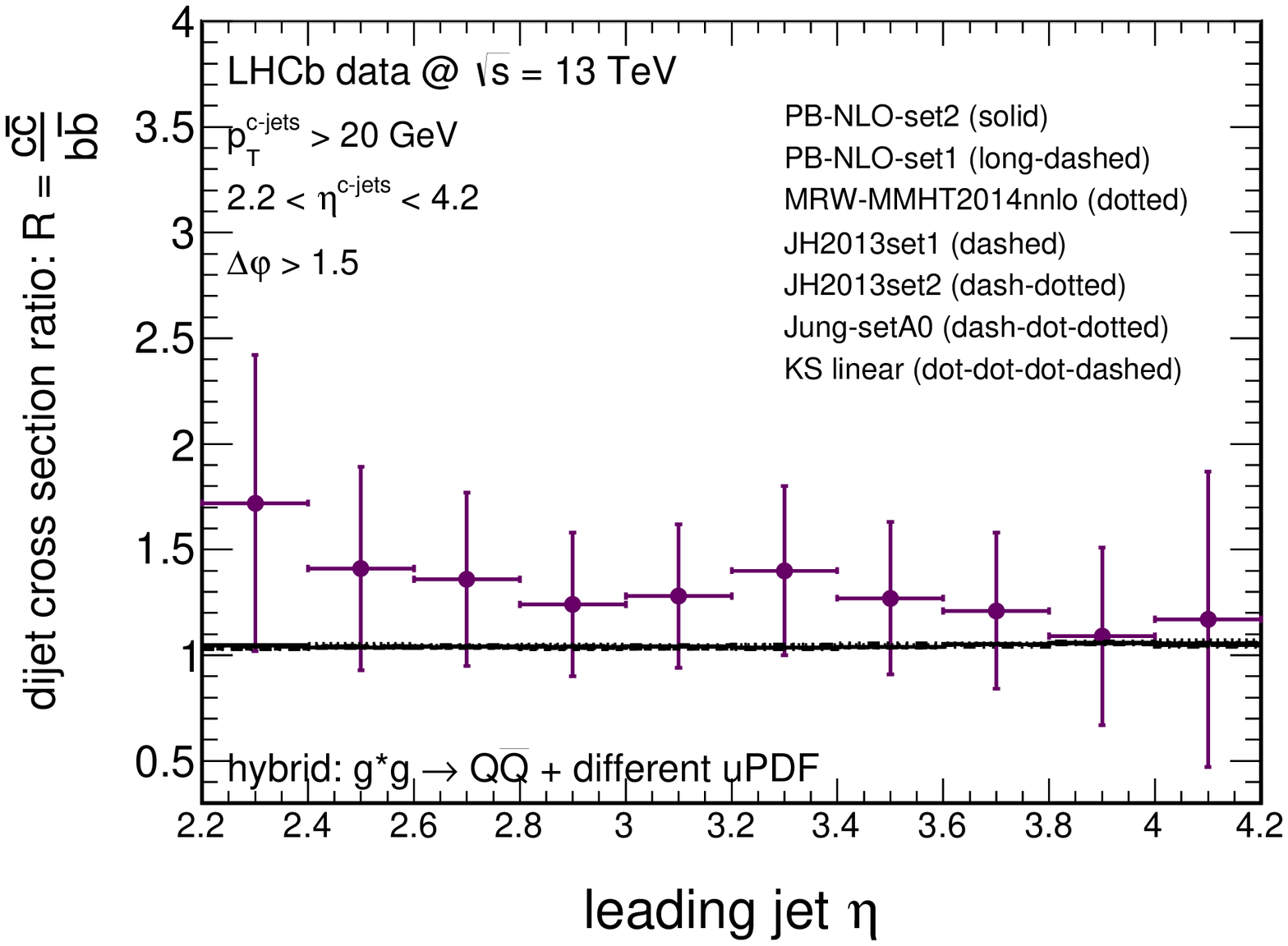}}
\end{minipage}
\begin{minipage}{0.47\textwidth}
  \centerline{\includegraphics[width=1.0\textwidth]{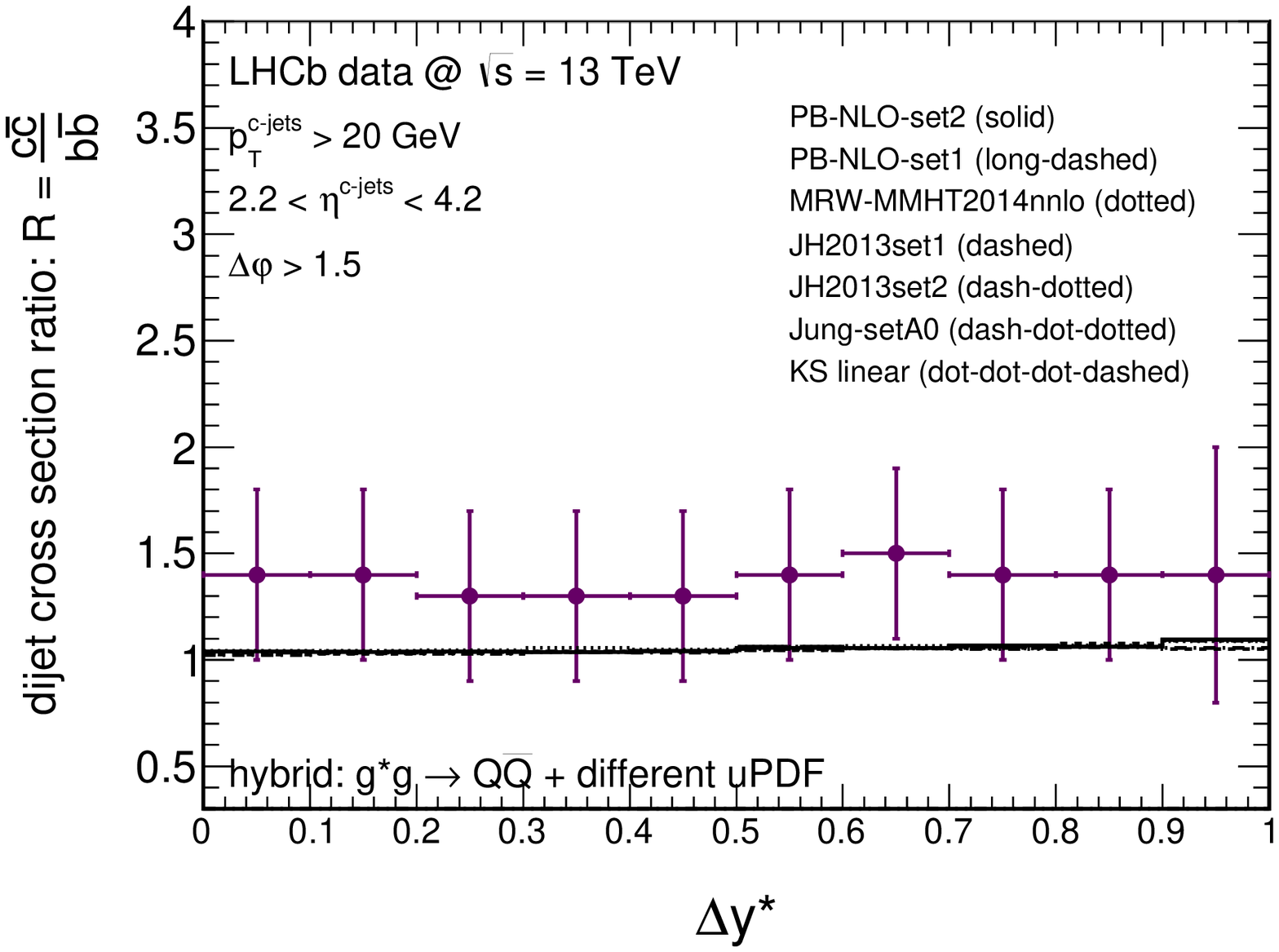}}
\end{minipage}
\begin{minipage}{0.47\textwidth}
  \centerline{\includegraphics[width=1.0\textwidth]{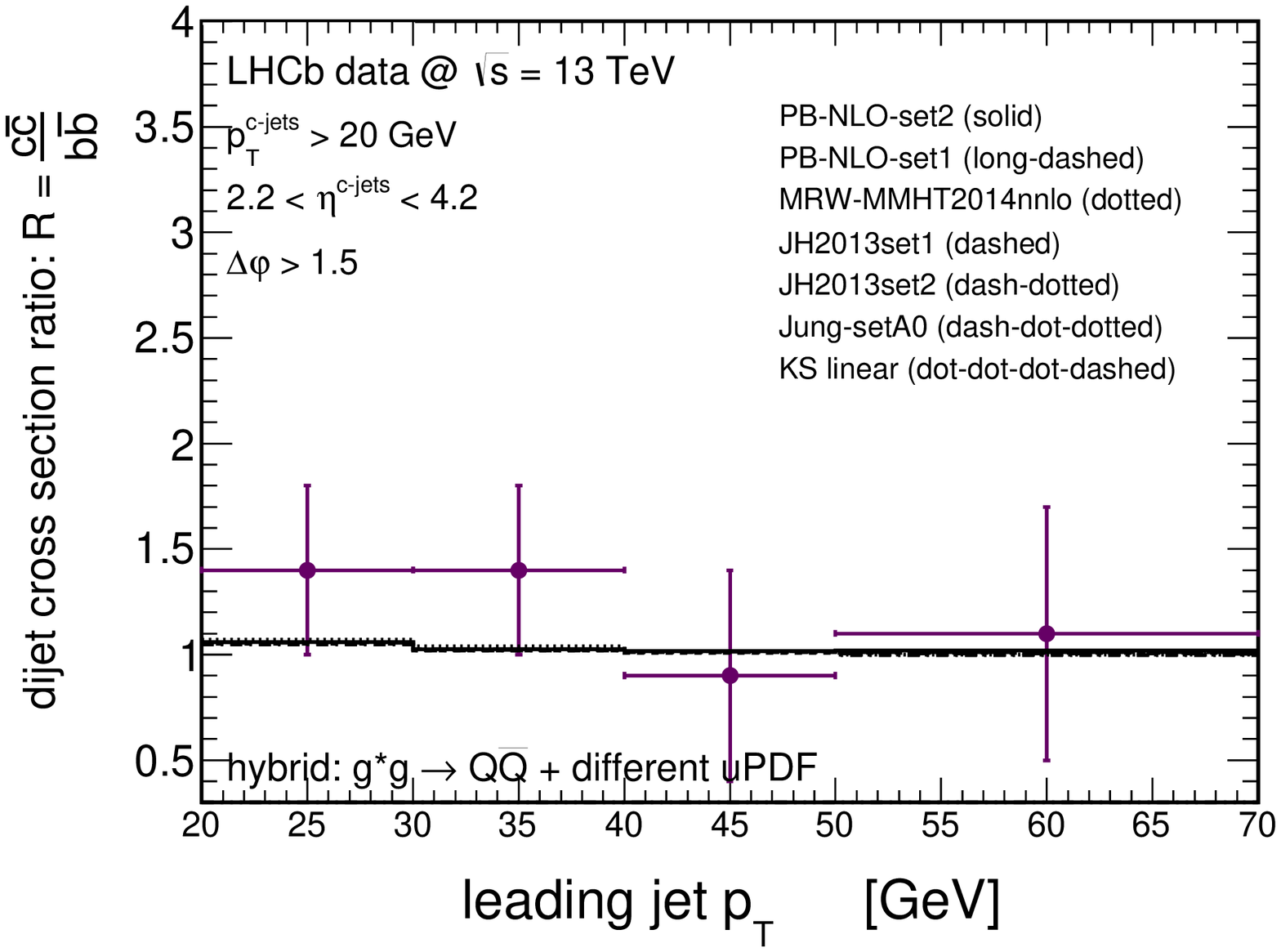}}
\end{minipage}
\begin{minipage}{0.47\textwidth}
  \centerline{\includegraphics[width=1.0\textwidth]{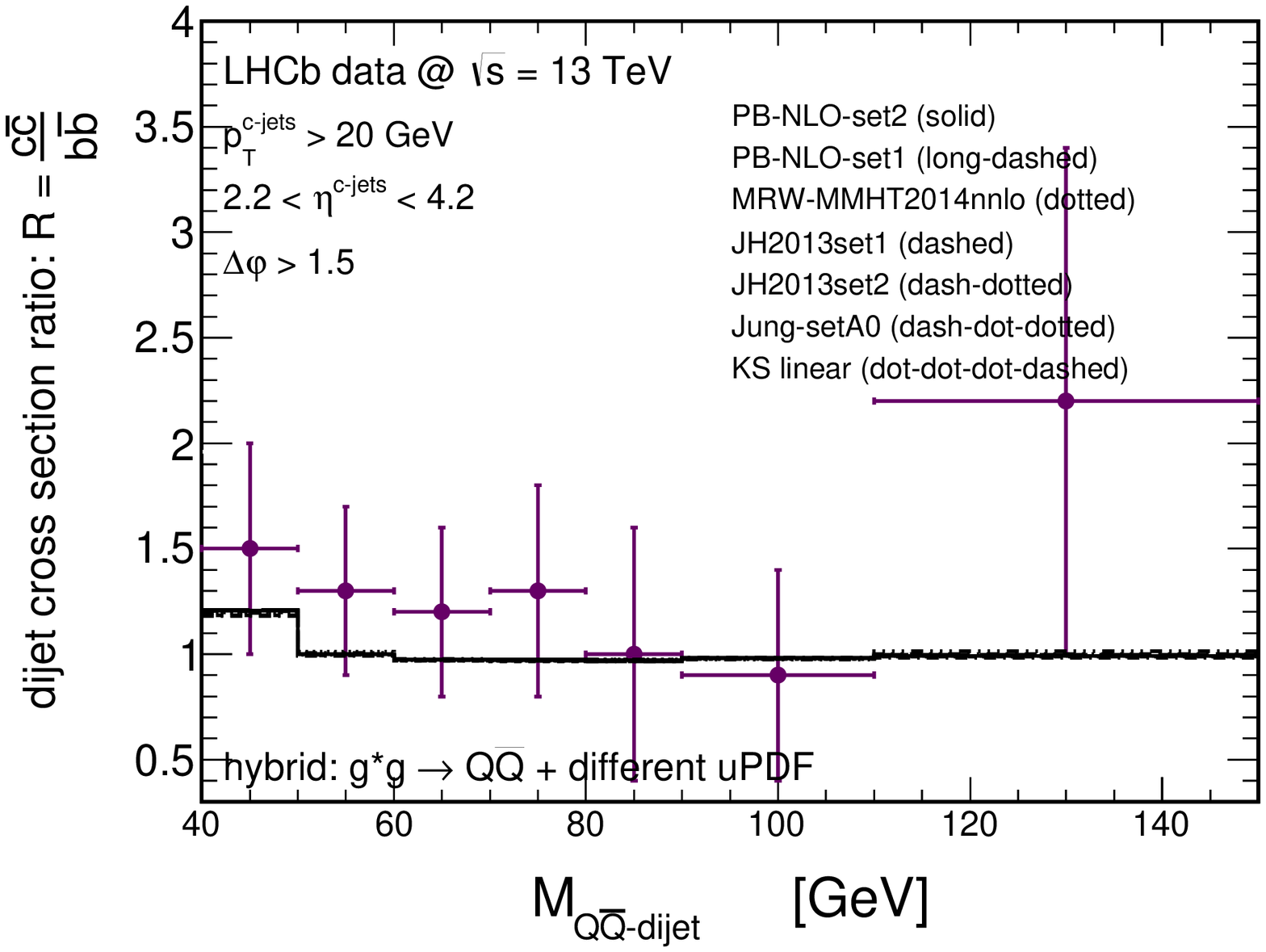}}
\end{minipage}
  \caption{The ratio $R=\frac{c\bar c}{b\bar b}$ of the dijet differential cross sections in $pp$-scattering at $\sqrt{s}=13$ TeV as a function of the leading jet $\eta$ (top left), the rapidity difference $\Delta y^{*}$ (top right), the leading jet $p_{T}$ (bottom left) and the dijet invariant mass $M_{Q\bar Q\text{-}\mathrm{dijet}}$ (bottom right). The theoretical histograms correspond to the hybrid model calculations obtained with different gluon uPDFs. Here the default set $\mu^{2}= m_{T}^2$ of the renormalization and factorization scales is used. 
\small 
}
\label{fig:28}
\end{figure}

The last conclusion above can be directly examined beacuse the LHCb collaboration has also presented the
corresponding distributions of the ratio $R=\frac{c\bar c}{b\bar b}$ as a function of the leading jet $\eta$, the rapidity difference $\Delta y^{*}$, the leading jet $p_{T}$ and the dijet invariant mass $M_{Q\bar Q\text{-}\mathrm{dijet}}$. A comparison
of our predictions to the experimental distributions is shown in Fig.~\ref{fig:28}. The theoretical ratio is identical for all models of the gluon uPDF used here. The predicted ratios are slightly above 1 and are almost independent of the considered kinematic variables. The experimental results correspond to a rather larger values. In general, we obtain a theory/data agreement, however, we reached only the lower experimental limits defined by large total experimental uncertainties. We will continue the discussion of the charm to bottom ratio in the following.      

%----------------------------------------------------------------------------
%----------------------------------------------------------------------------

\clearpage
\subsection{The scale dependence of the predictions}

Now let us discuss main uncertainties of our predictions. As we have already shown an important source of uncertainty in the present study is modelling of the unintegrated gluon densities in the proton. Here we wish to focus on the second important source of uncertainties that is the scale dependence of our predictions.

\begin{figure}[!h]
\begin{minipage}{0.47\textwidth}
  \centerline{\includegraphics[width=1.0\textwidth]{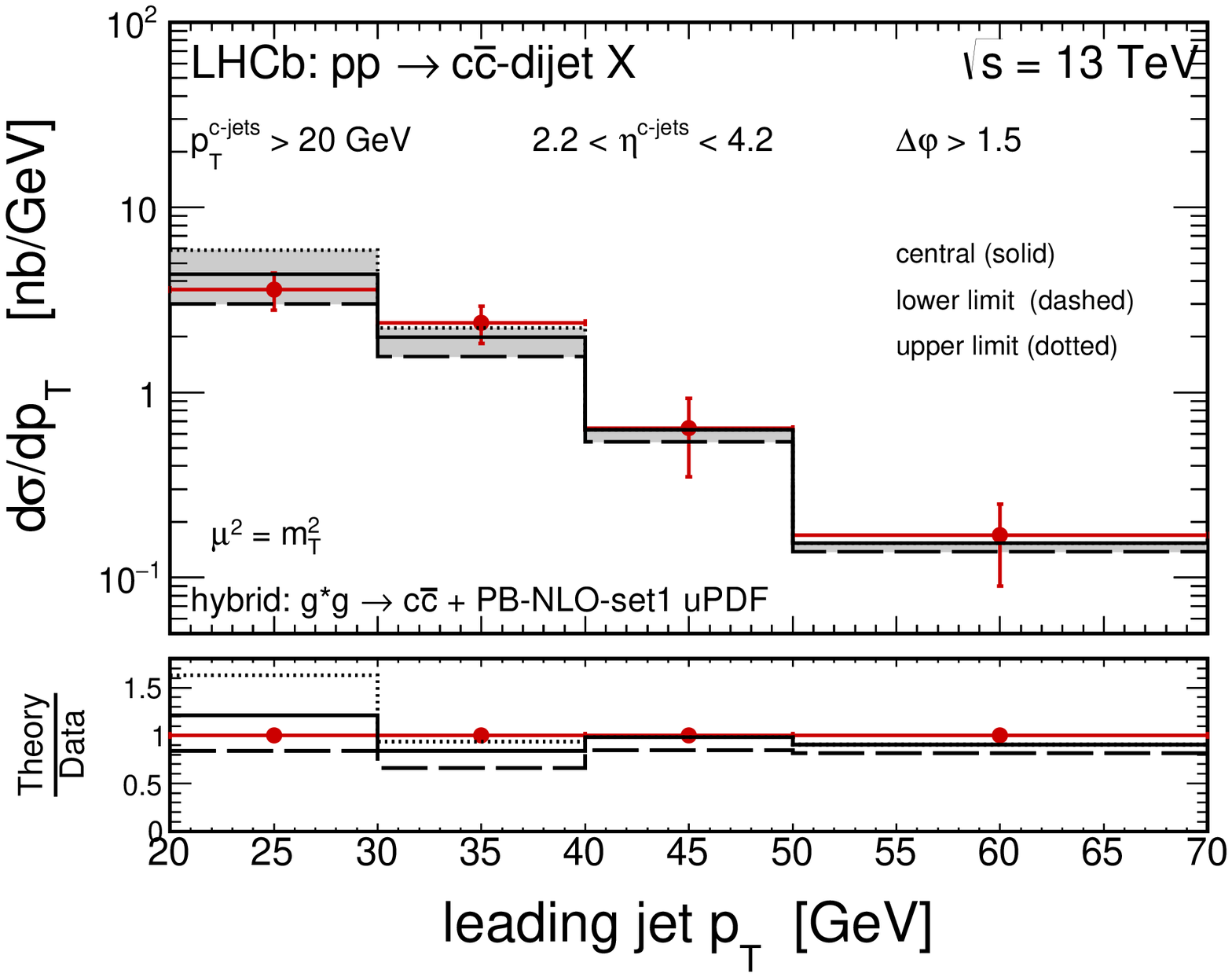}}
\end{minipage}
\begin{minipage}{0.47\textwidth}
  \centerline{\includegraphics[width=1.0\textwidth]{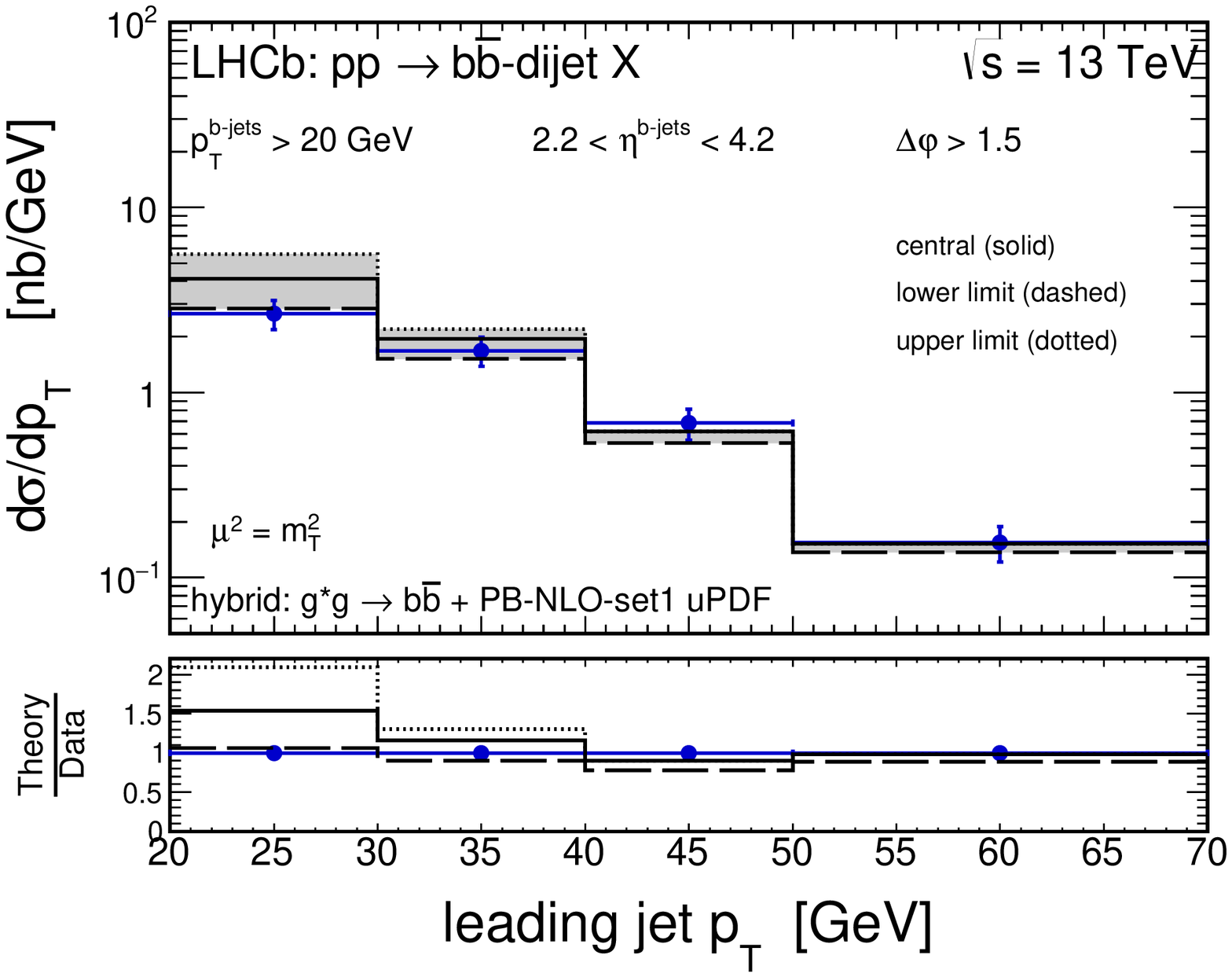}}
\end{minipage}
\begin{minipage}{0.47\textwidth}
  \centerline{\includegraphics[width=1.0\textwidth]{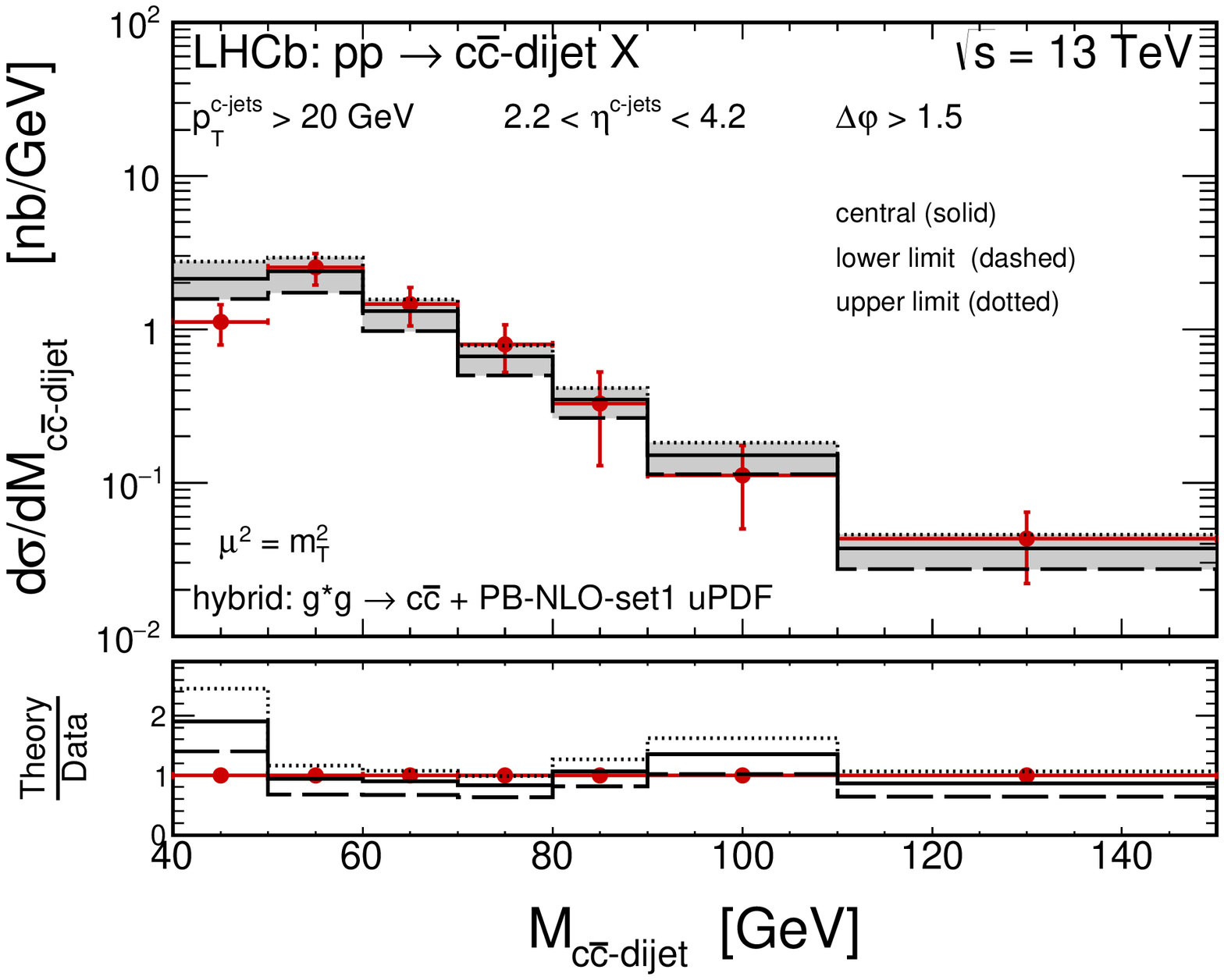}}
\end{minipage}
\begin{minipage}{0.47\textwidth}
  \centerline{\includegraphics[width=1.0\textwidth]{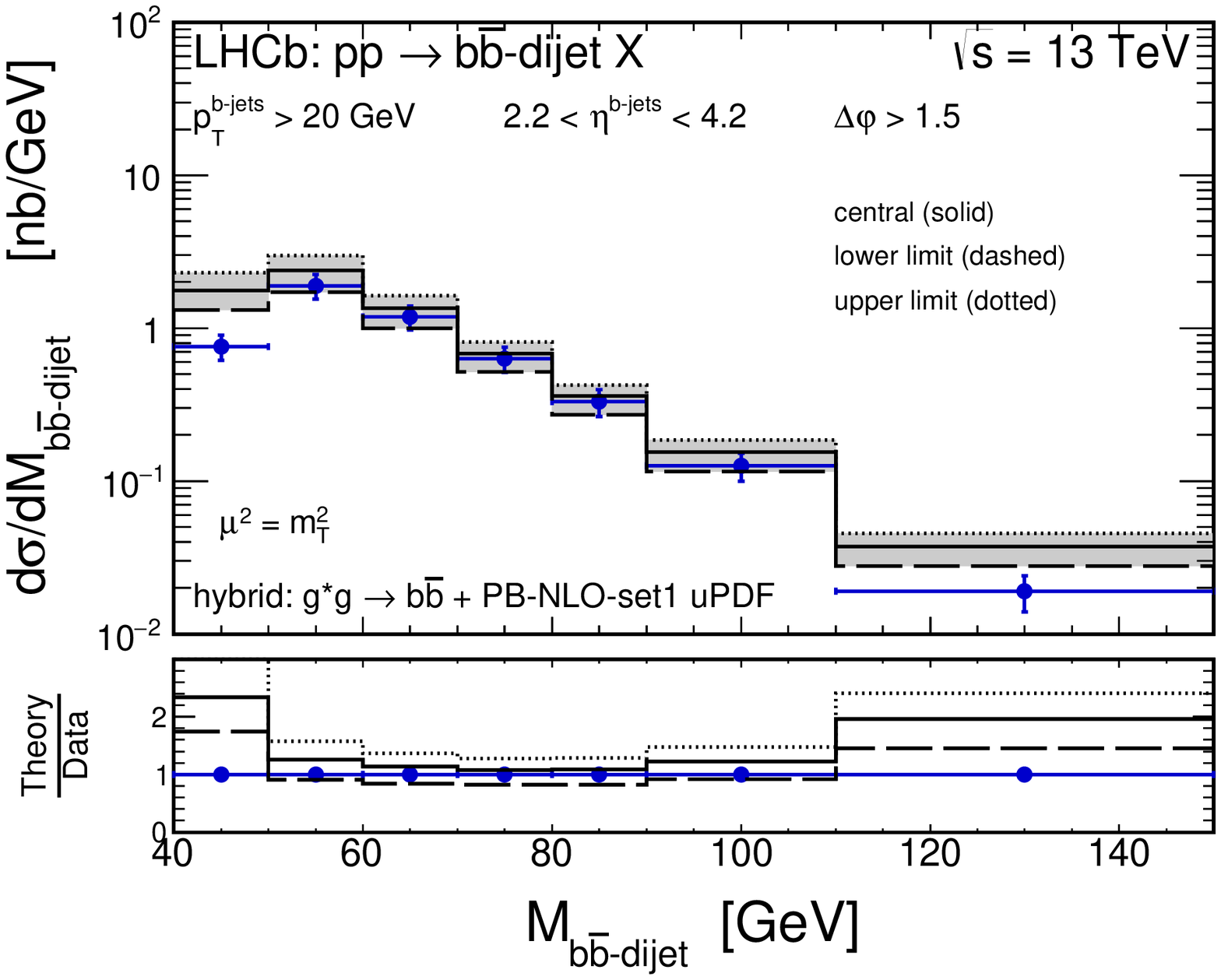}}
\end{minipage}
  \caption{The differential cross sections for forward production of $c\bar c$- (left panels) and $b\bar b$-dijets (right panels) in $pp$-scattering at $\sqrt{s}=13$ TeV as a function of the leading jet $p_{T}$ (top panels) and the dijet invariant mass $M_{Q\bar Q\text{-}\mathrm{dijet}}$ (bottom panels).  The shaded bands represent the scale uncertainties calculated by varying the central set of the renormalizaton and factorization scales independently by a factor 2 up and down. Here the hybrid model with the PB-NLO-set1 gluon uPDF is used. 
\small 
}
\label{fig:22}
\end{figure}

\begin{figure}[!h]
\begin{minipage}{0.47\textwidth}
  \centerline{\includegraphics[width=1.0\textwidth]{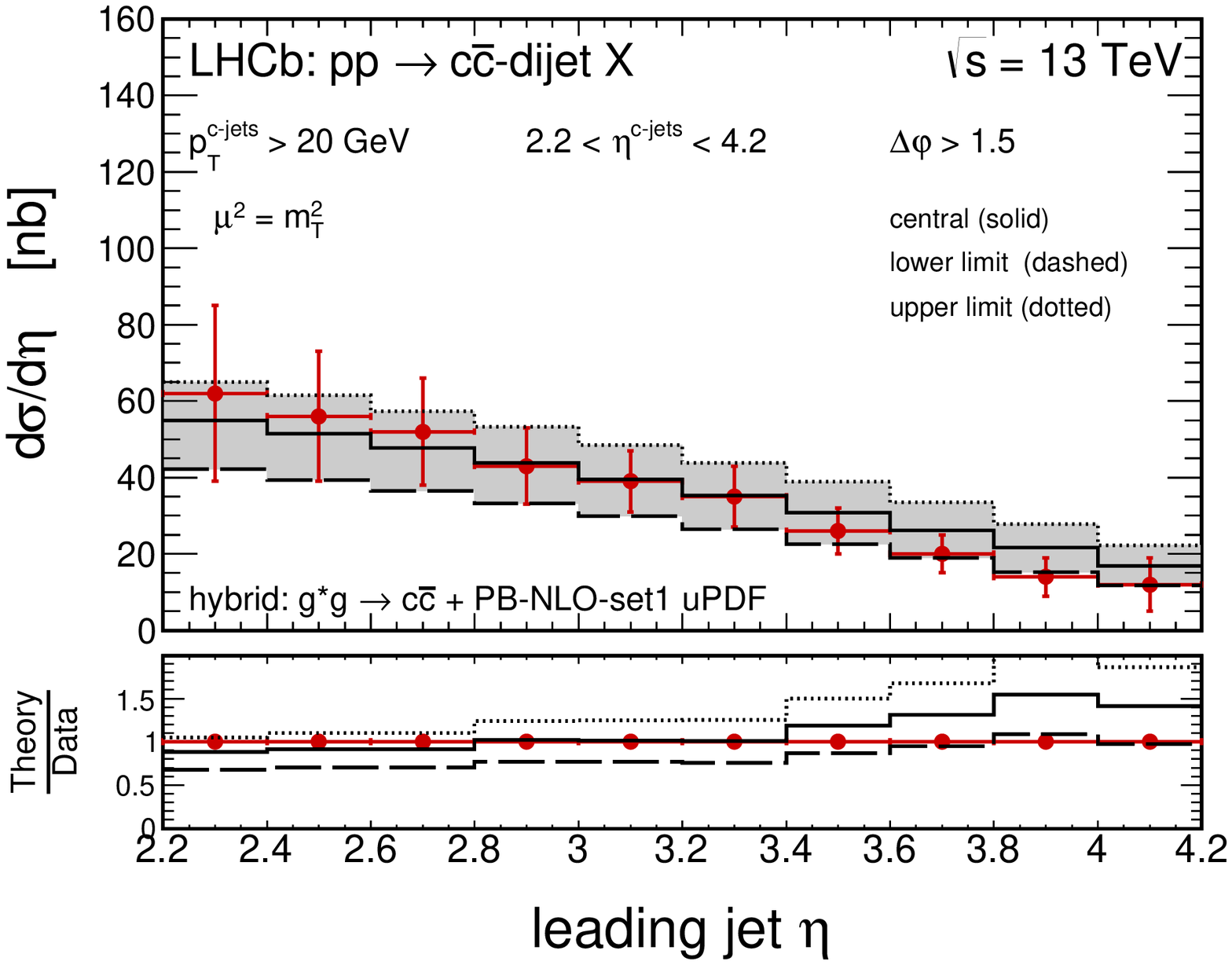}}
\end{minipage}
\begin{minipage}{0.47\textwidth}
  \centerline{\includegraphics[width=1.0\textwidth]{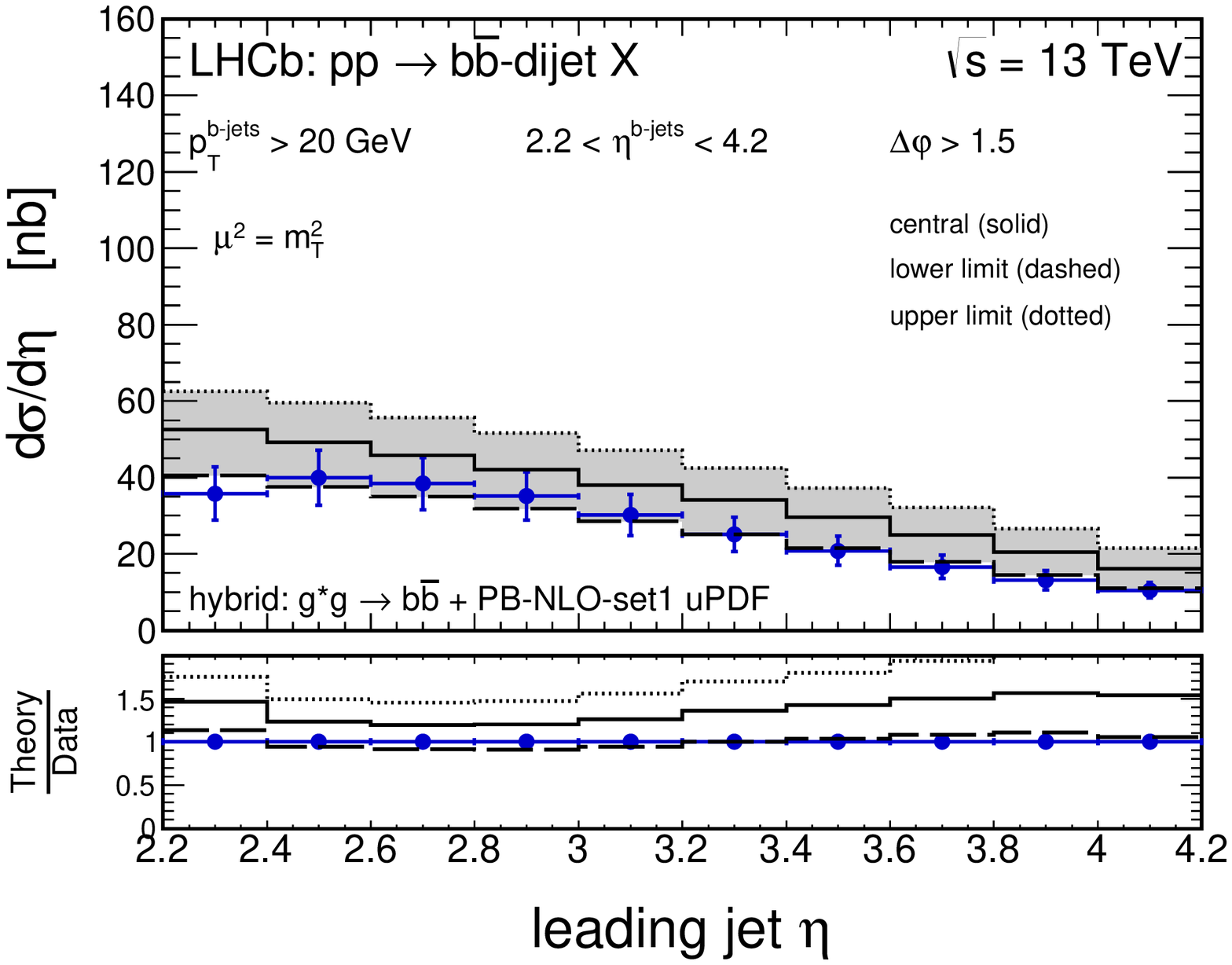}}
\end{minipage}
\begin{minipage}{0.47\textwidth}
  \centerline{\includegraphics[width=1.0\textwidth]{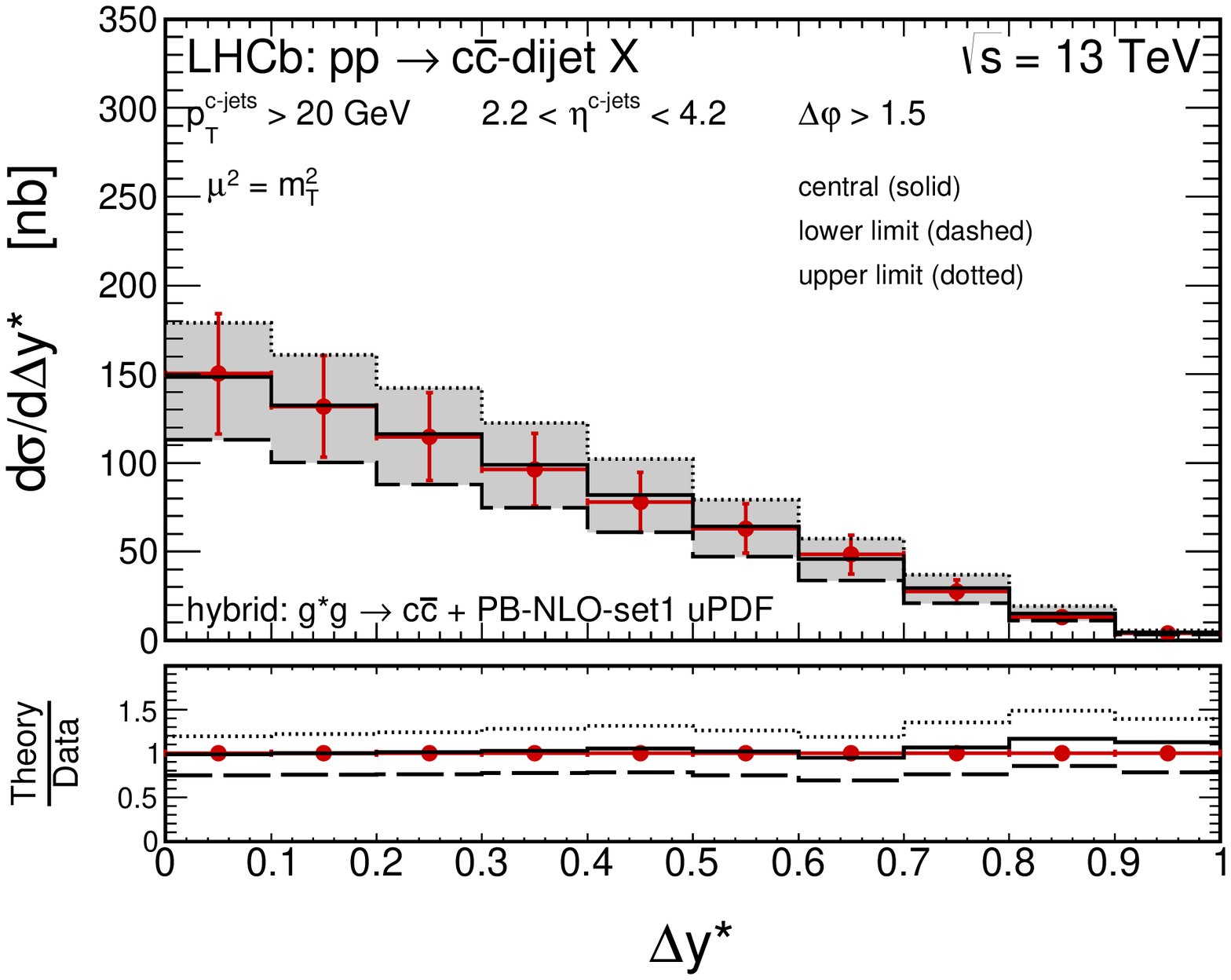}}
\end{minipage}
\begin{minipage}{0.47\textwidth}
  \centerline{\includegraphics[width=1.0\textwidth]{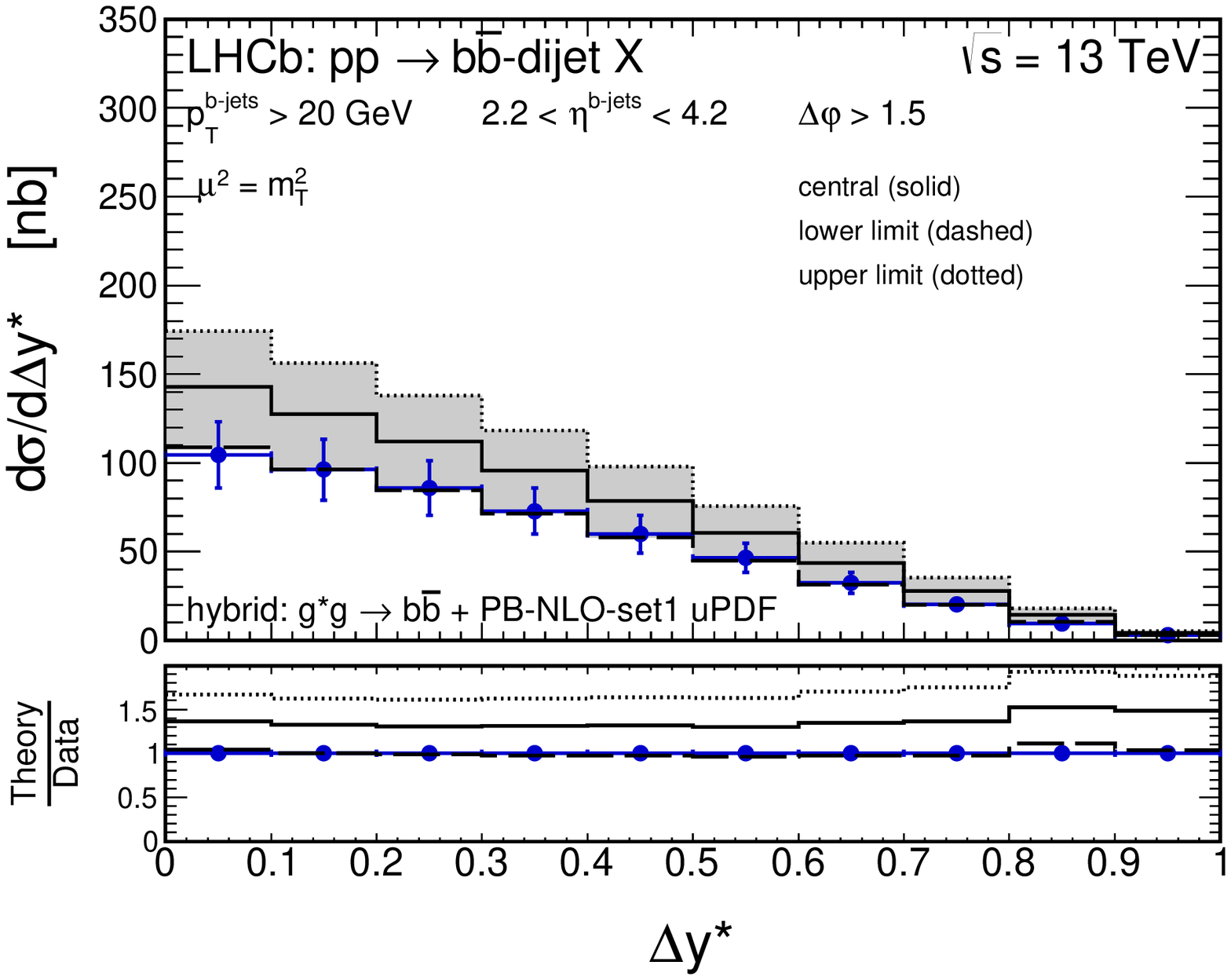}}
\end{minipage}
  \caption{The same as in Fig.~\ref{fig:22} but here the leading jet $\eta$ (top panels) and the rapidity difference $\Delta y^*$ (bottom panels) distributions are shown.
\small 
}
\label{fig:23}
\end{figure}

In Figs.~\ref{fig:22} and ~\ref{fig:23} we plot the hybrid model results for the PB-NLO-set1 gluon uPDF with the shaded bands that represent the scale uncertainties calculated by varying the central set of the renormalizaton and factorization scales independently by a factor 2 up and down. Here the central set of the scales corresponds to the $\mu^{2}=m_{t}^{2}$ case. As we can see the sensitivity is sizeable but not huge and seems to be decreasing with the leading jet transverse momentum.

%----------------------------------------------------------------------------
%----------------------------------------------------------------------------

\begin{figure}[!h]
\begin{minipage}{0.47\textwidth}
  \centerline{\includegraphics[width=1.0\textwidth]{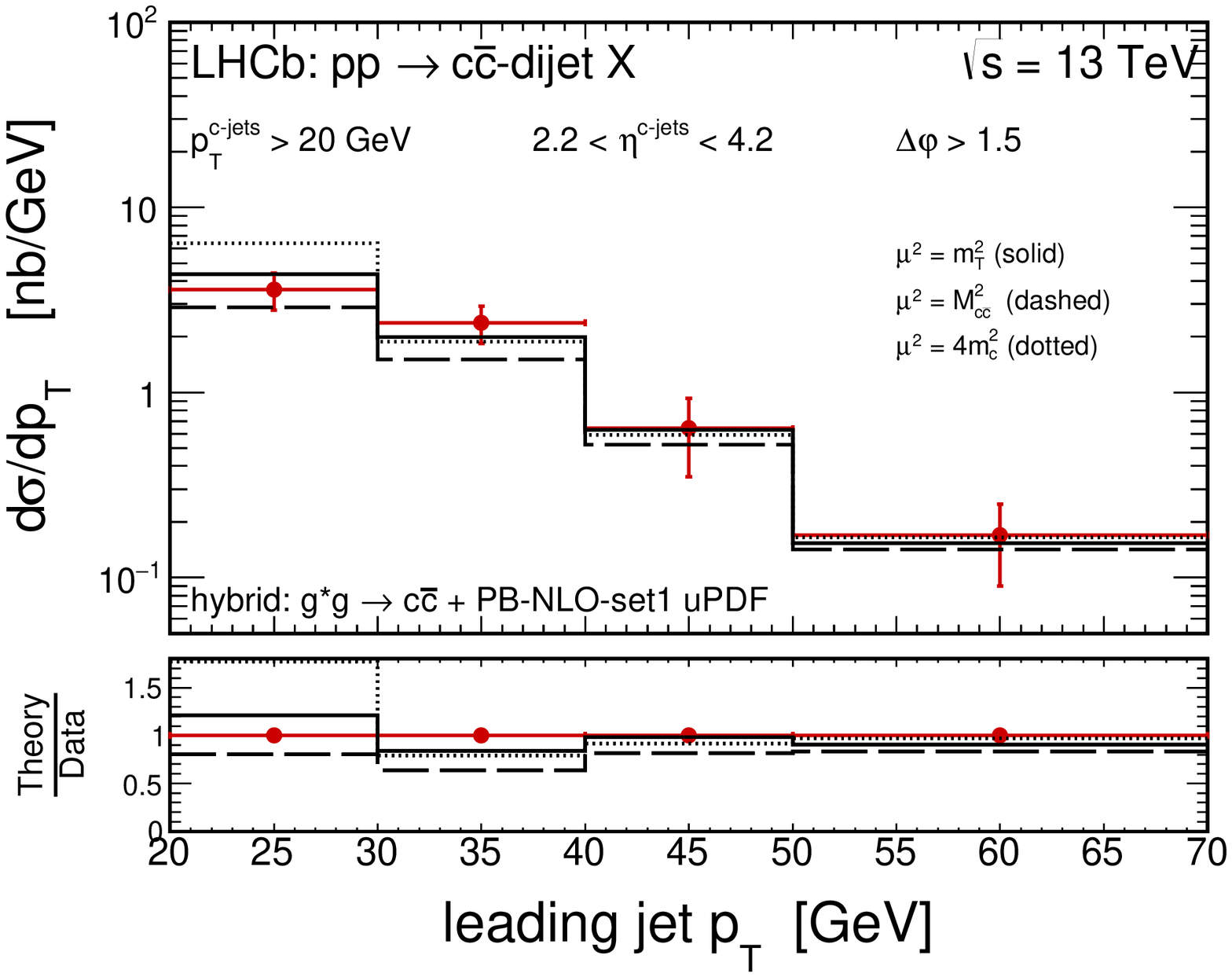}}
\end{minipage}
\begin{minipage}{0.47\textwidth}
  \centerline{\includegraphics[width=1.0\textwidth]{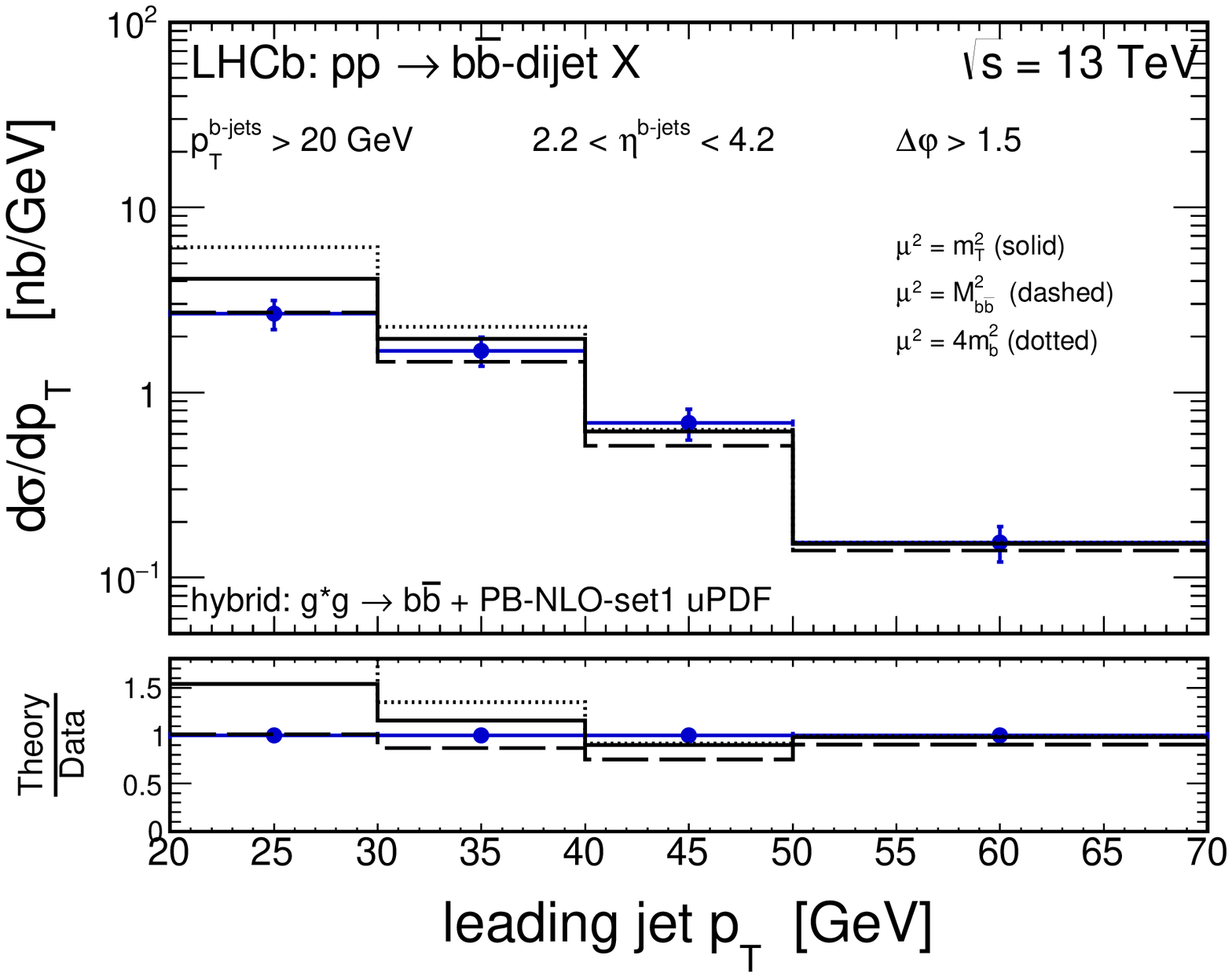}}
\end{minipage}
\begin{minipage}{0.47\textwidth}
  \centerline{\includegraphics[width=1.0\textwidth]{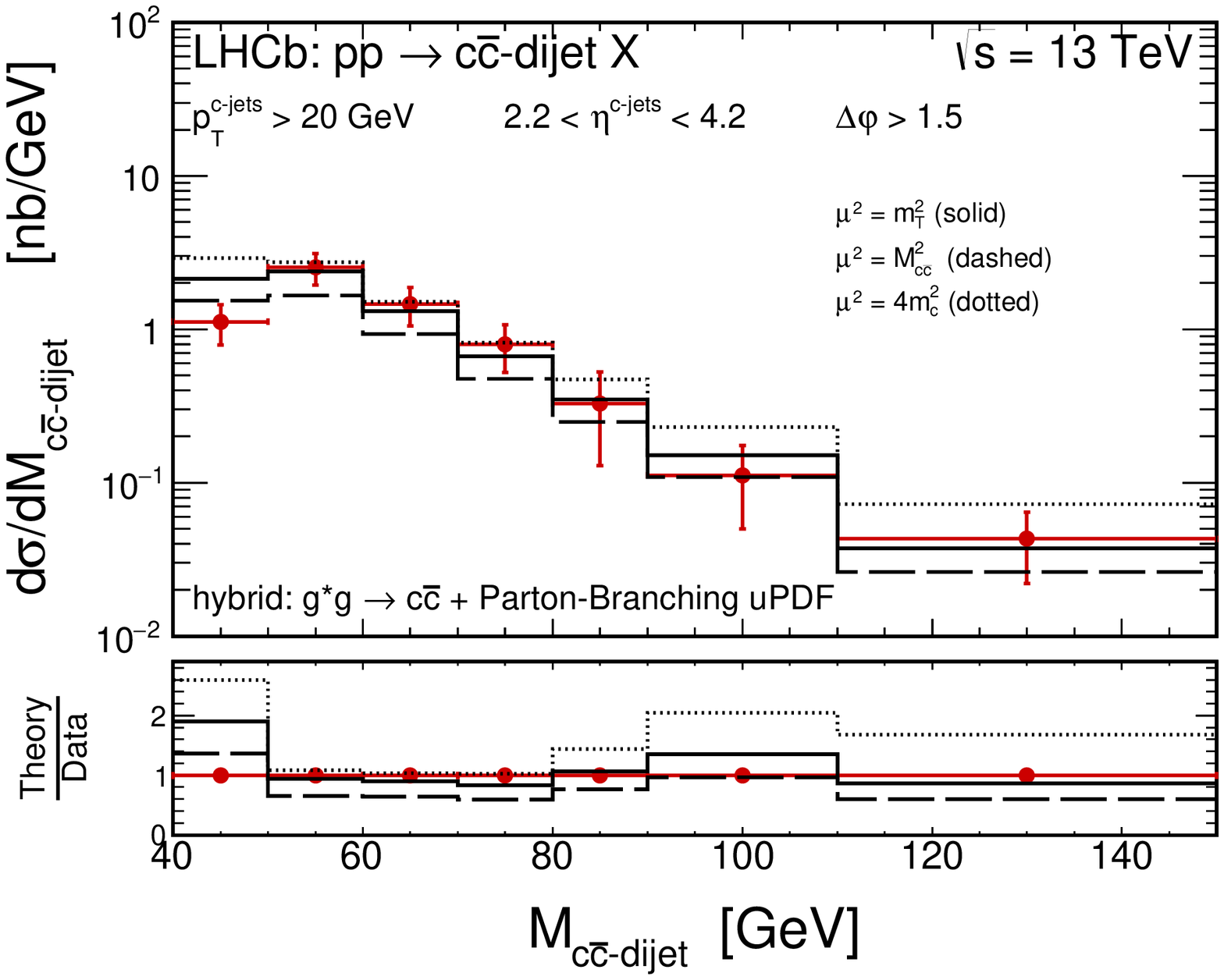}}
\end{minipage}
\begin{minipage}{0.47\textwidth}
  \centerline{\includegraphics[width=1.0\textwidth]{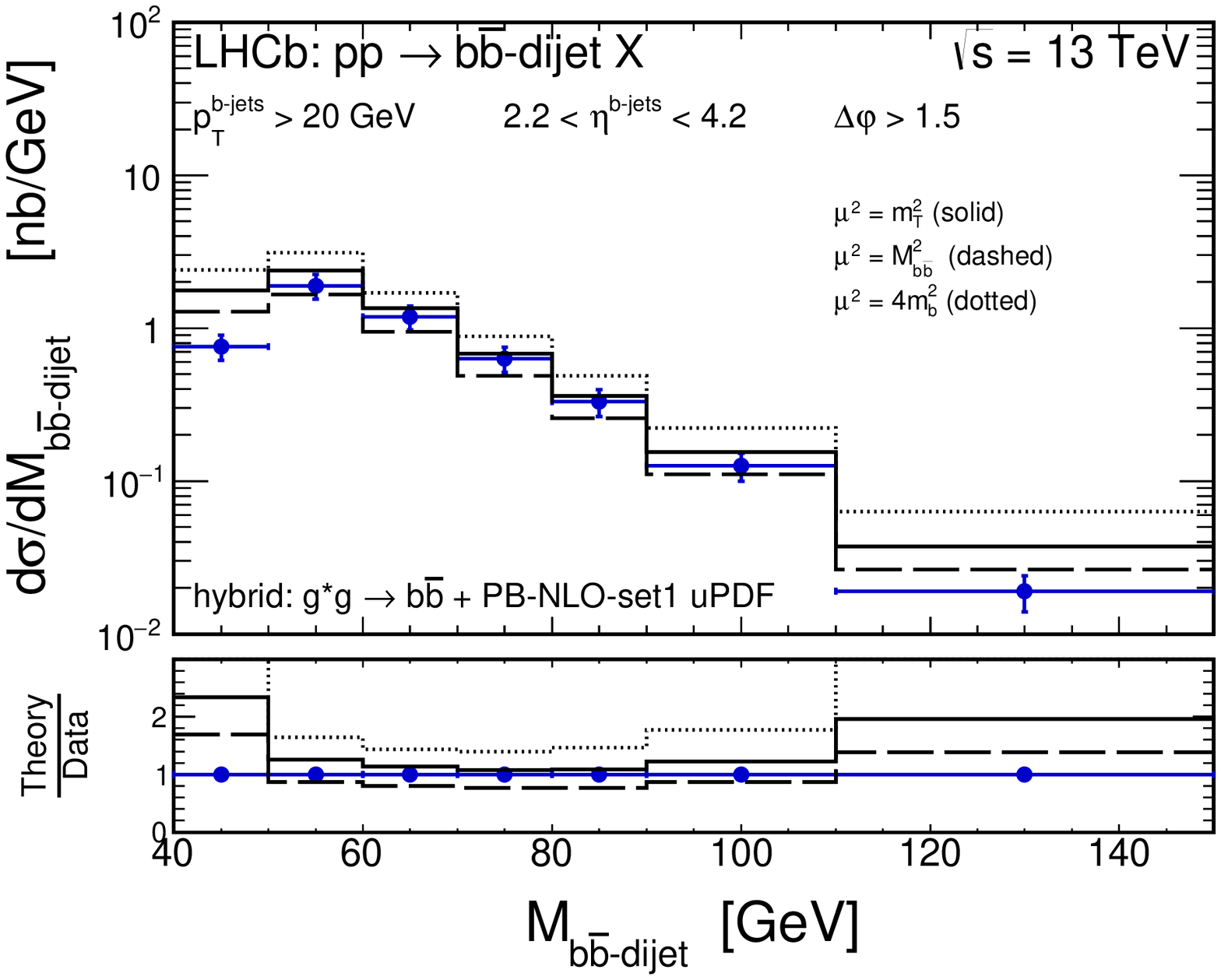}}
\end{minipage}
  \caption{The differential cross sections for forward production of $c\bar c$- (left panels) and $b\bar b$-dijets (right panels) in $pp$-scattering at $\sqrt{s}=13$ TeV as a function of the leading jet $p_{T}$ (top panels) and the dijet invariant mass $M_{Q\bar Q\text{-}\mathrm{dijet}}$ (bottom panels).  The three different lines correspond to the different choices for the central set of the renormalization and factorization scale. More details can be found in the figure. Here the hybrid model with the PB-NLO-set1 gluon uPDF is used. 
\small 
}
\label{fig:24}
\end{figure}

\begin{figure}[!h]
\begin{minipage}{0.47\textwidth}
  \centerline{\includegraphics[width=1.0\textwidth]{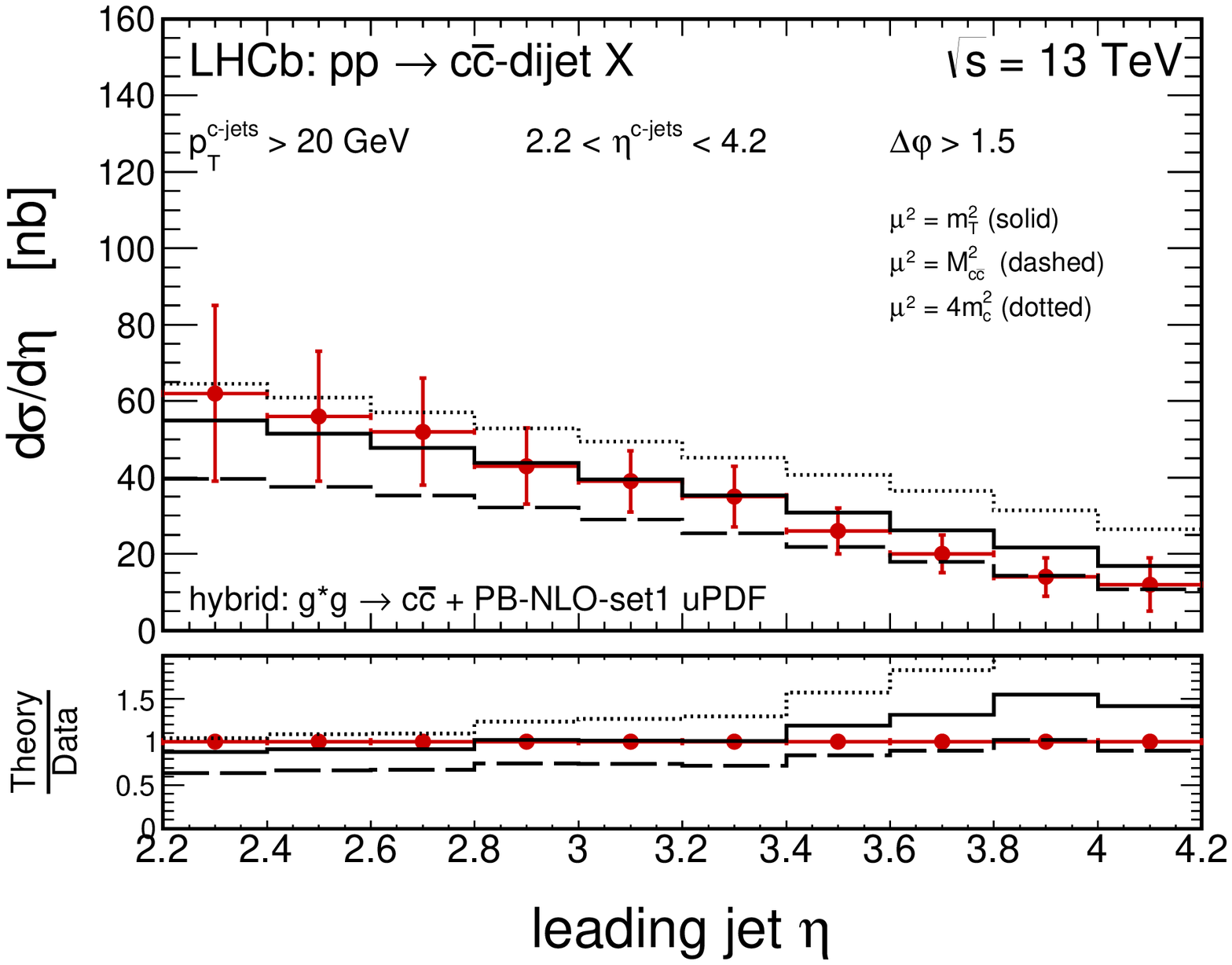}}
\end{minipage}
\begin{minipage}{0.47\textwidth}
  \centerline{\includegraphics[width=1.0\textwidth]{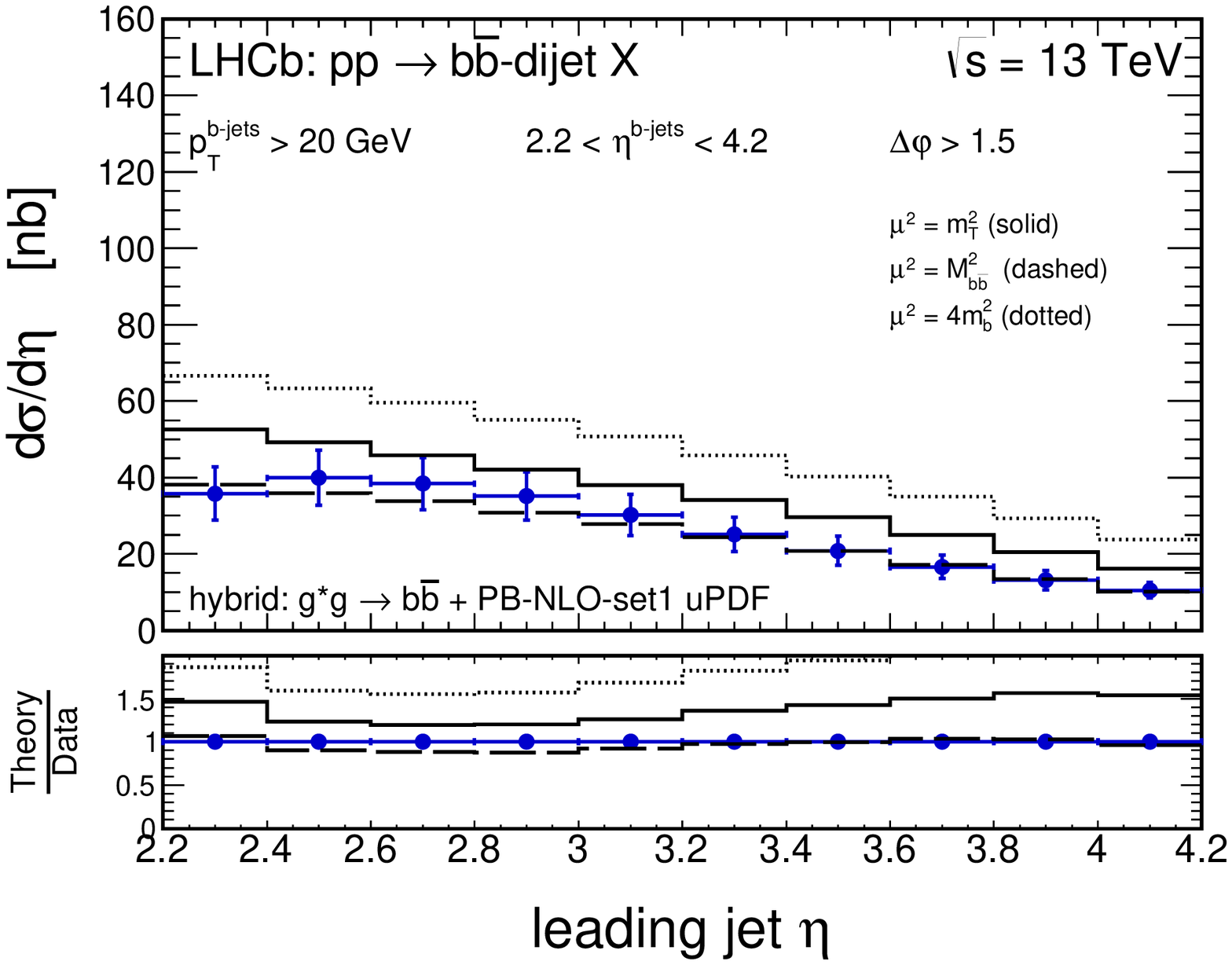}}
\end{minipage}
\begin{minipage}{0.47\textwidth}
  \centerline{\includegraphics[width=1.0\textwidth]{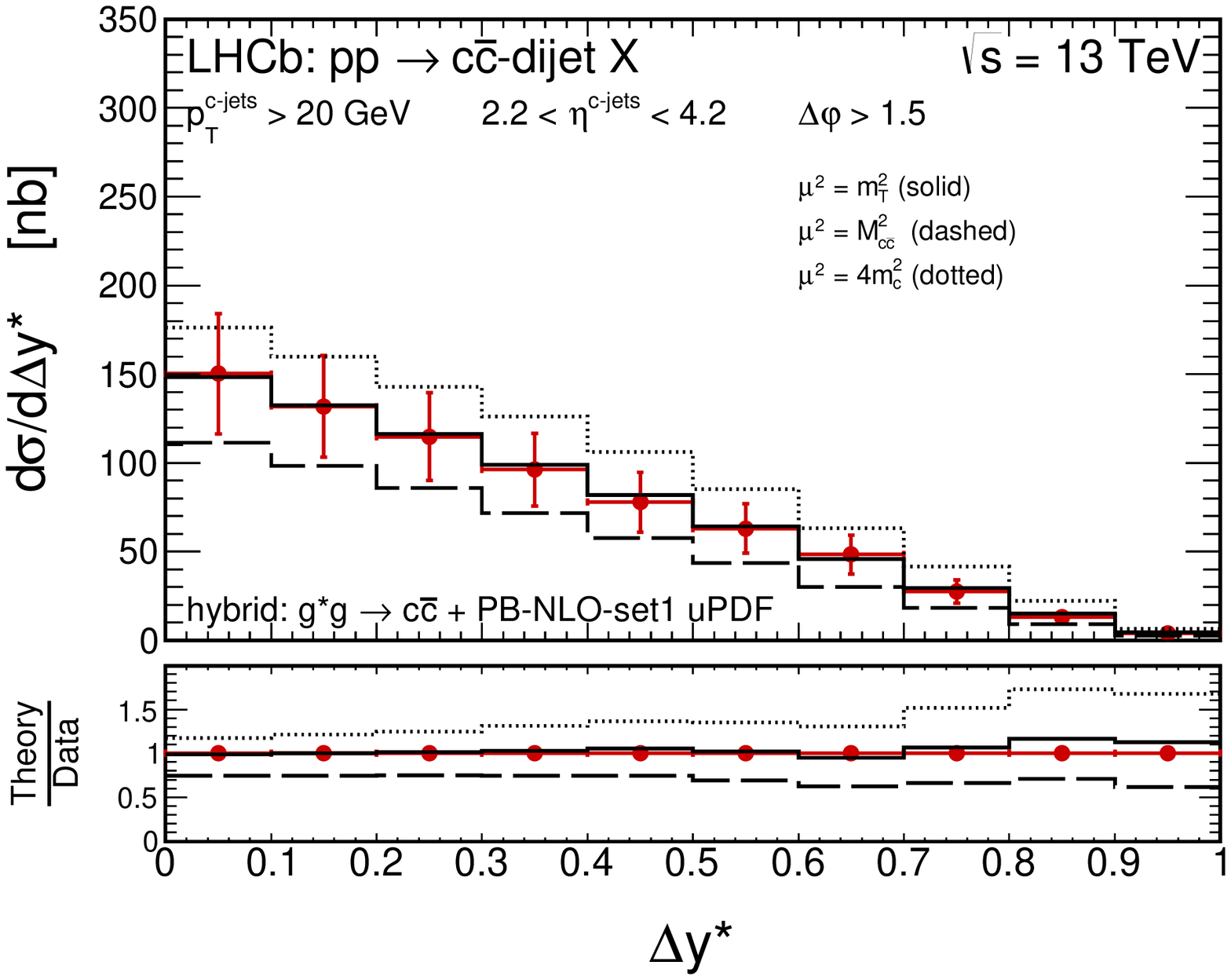}}
\end{minipage}
\begin{minipage}{0.47\textwidth}
  \centerline{\includegraphics[width=1.0\textwidth]{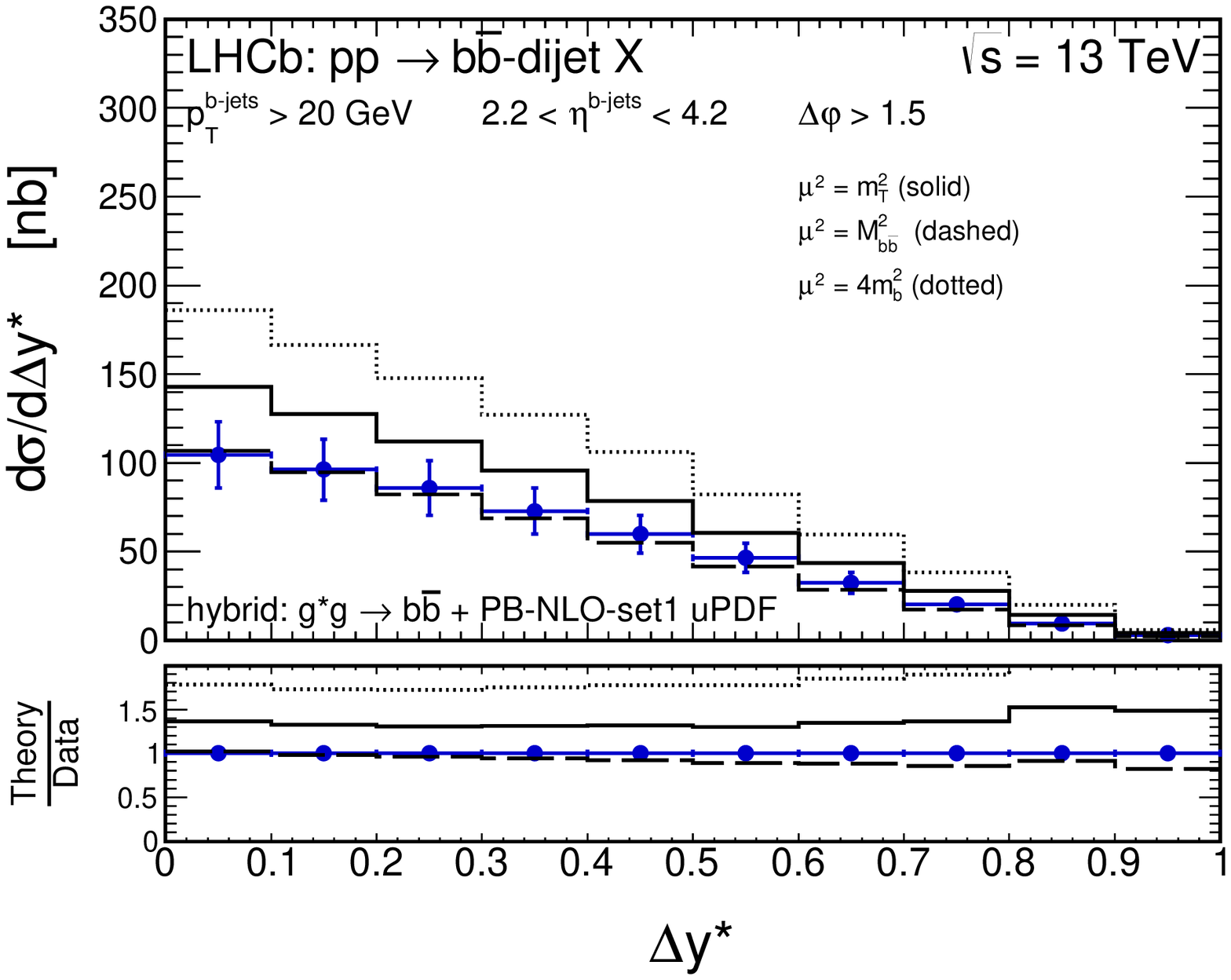}}
\end{minipage}
  \caption{The same as in Fig.~\ref{fig:24} but here the leading jet $\eta$ (top panels) and the rapidity difference $\Delta y^*$ (bottom panels) distributions are shown.
\small 
}
\label{fig:25}
\end{figure}

In Figs.~\ref{fig:24} and ~\ref{fig:25} we present in addition how our results depend on the definition of the central set for the scales. The sensitivity has a similar size as above and one has to keep in mind that different choices of the central sets might slightly modify the overall picture. For example, using the $\mu^{2} = M_{Q\bar Q}^2$ we get slightly smaller cross sections which means that within this choice we get better agreement with the data for bottom  than for charm production -- in opposite to conclusions from our default scale set.

\begin{figure}[!h]
\begin{minipage}{0.47\textwidth}
  \centerline{\includegraphics[width=1.0\textwidth]{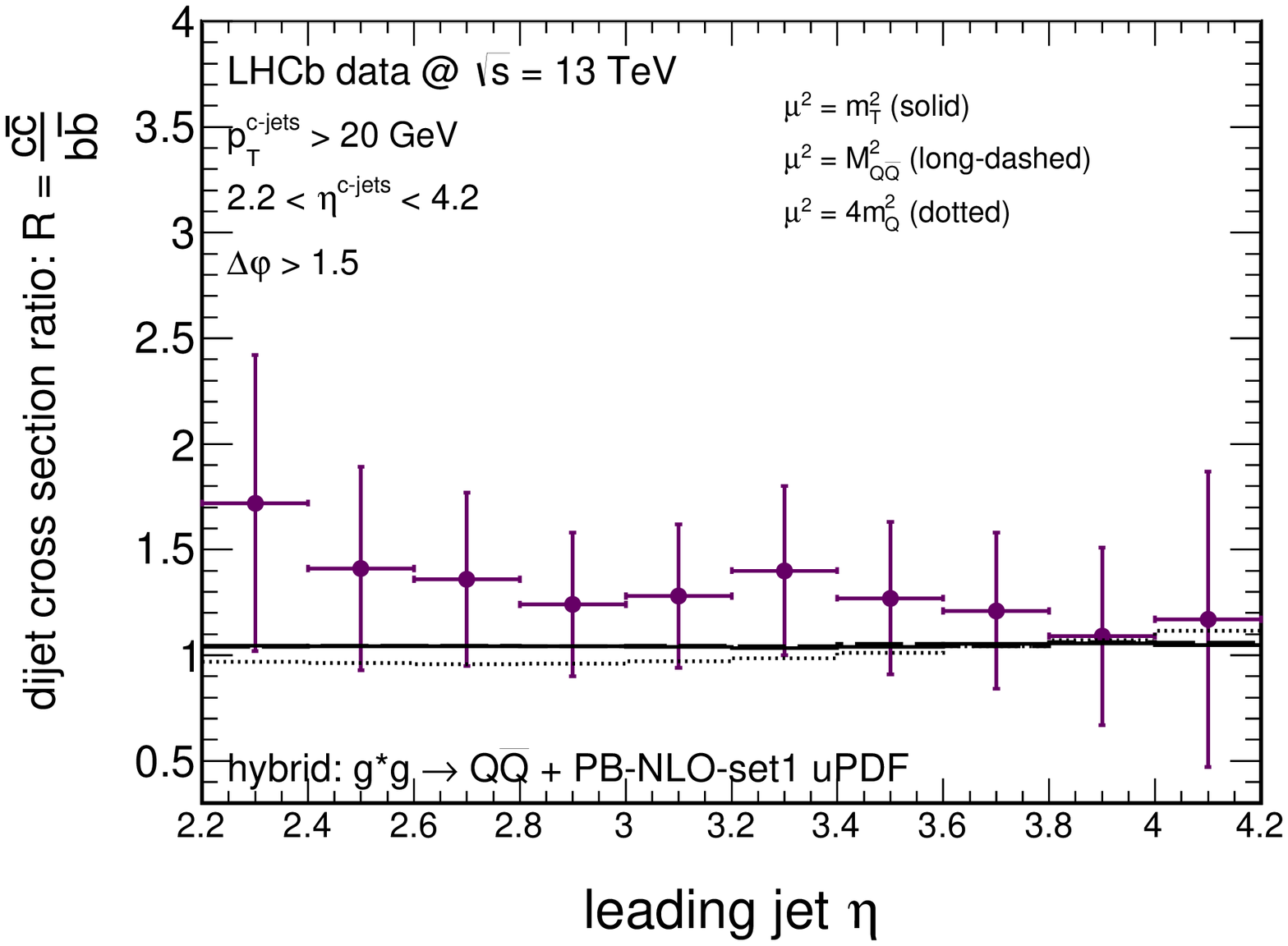}}
\end{minipage}
\begin{minipage}{0.47\textwidth}
  \centerline{\includegraphics[width=1.0\textwidth]{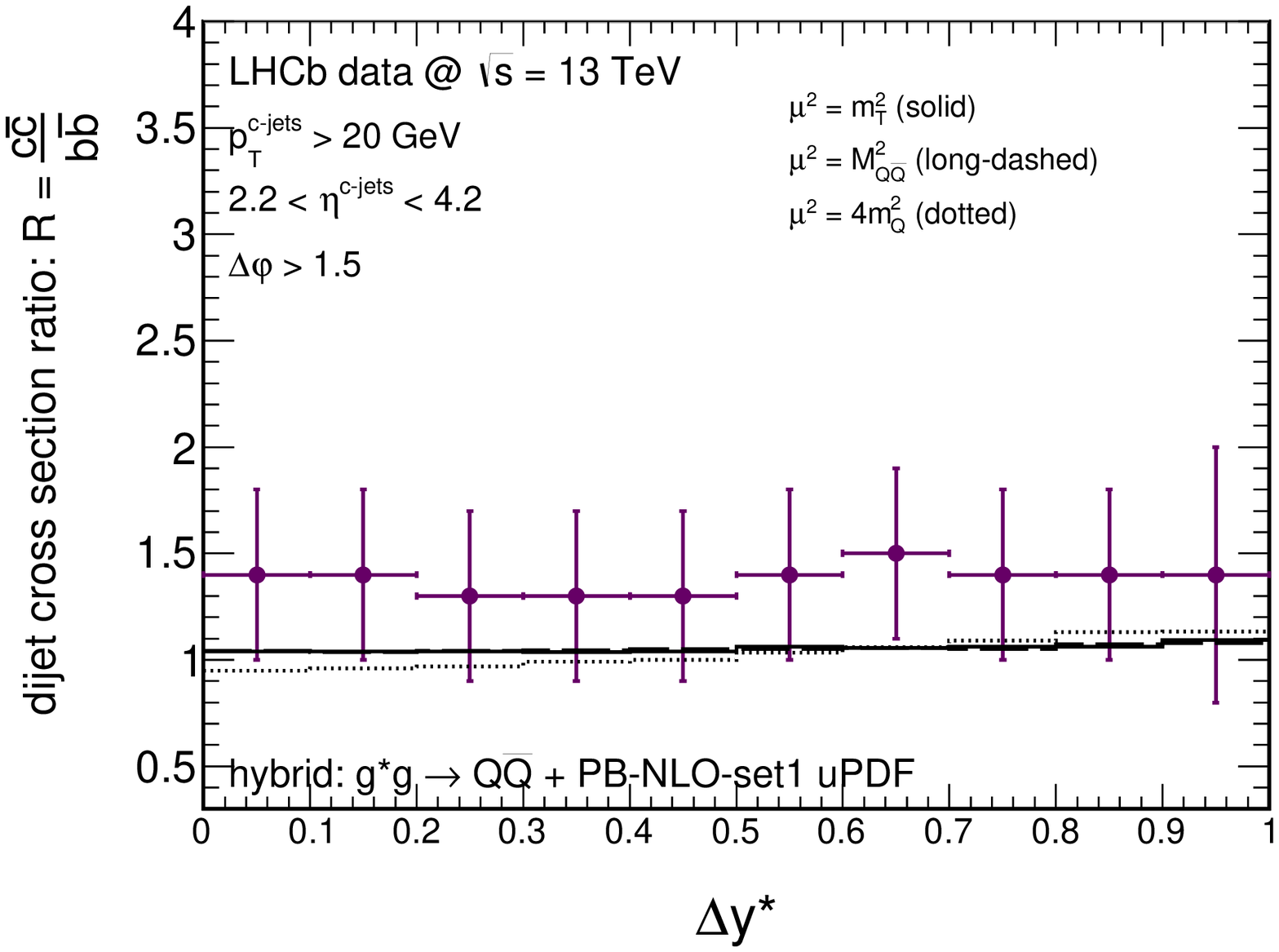}}
\end{minipage}
\begin{minipage}{0.47\textwidth}
  \centerline{\includegraphics[width=1.0\textwidth]{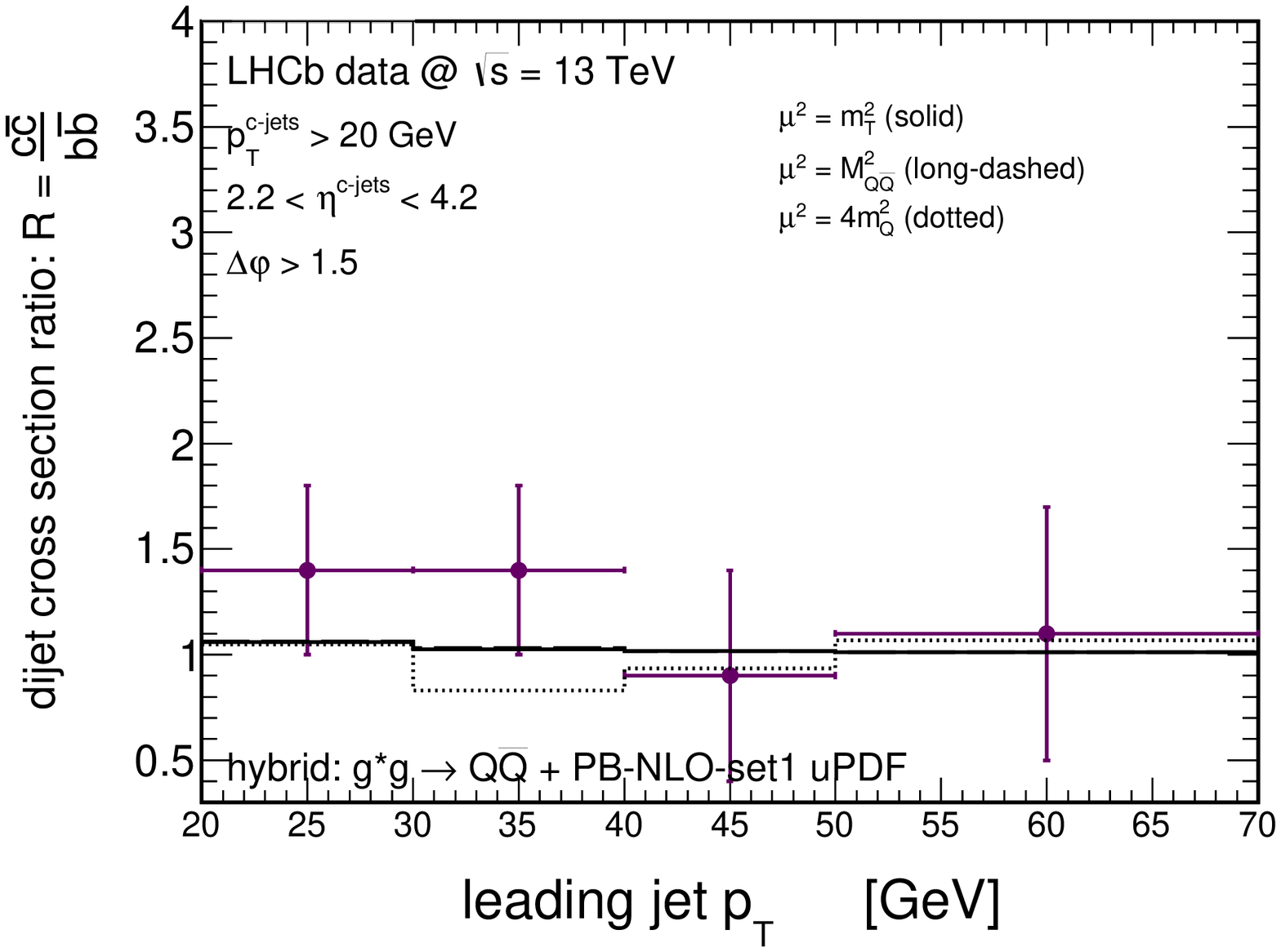}}
\end{minipage}
\begin{minipage}{0.47\textwidth}
  \centerline{\includegraphics[width=1.0\textwidth]{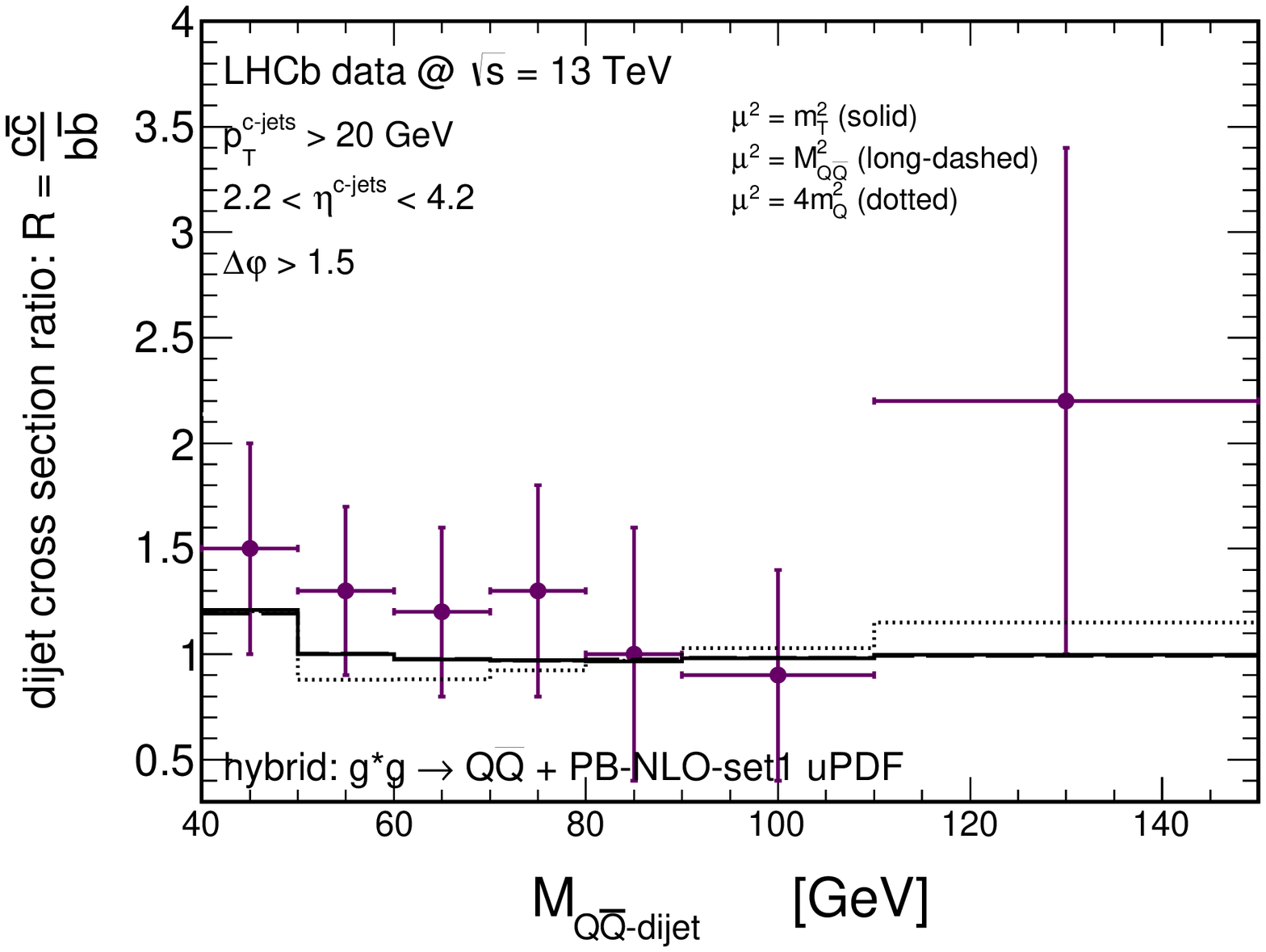}}
\end{minipage}
  \caption{The same as in Fig.~\ref{fig:28} but here the three different lines correspond to the different choices for the central set of the renormalization and factorization scale variables. More details can be found in the figure. Here the hybrid model with the PB-NLO-set1 gluon uPDF is used.
\small 
}
\label{fig:29}
\end{figure}

What we find interesting here, is that playing with the definition of the central scale does not improve our description of the charm to bottom production cross section ratios (see Fig.~\ref{fig:29}).   

%----------------------------------------------------------------------------
%----------------------------------------------------------------------------

%----------------------------------------------------------------------------
%----------------------------------------------------------------------------

%----------------------------------------------------------------------------
%----------------------------------------------------------------------------

%----------------------------------------------------------------------------
%----------------------------------------------------------------------------

\subsection{A comparison with the results of the collinear approach}

\begin{figure}[!h]
\begin{minipage}{0.47\textwidth}
  \centerline{\includegraphics[width=1.0\textwidth]{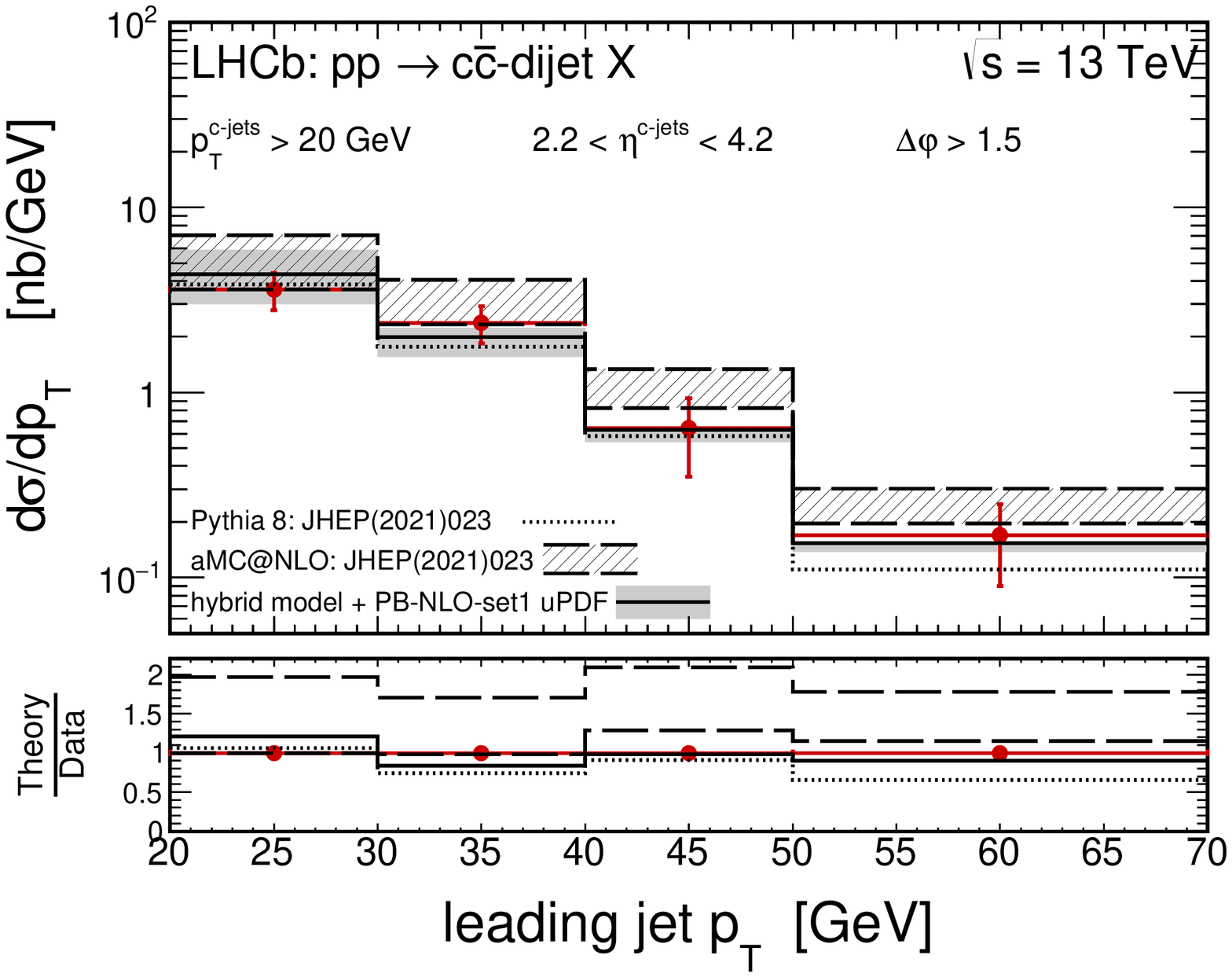}}
\end{minipage}
\begin{minipage}{0.47\textwidth}
  \centerline{\includegraphics[width=1.0\textwidth]{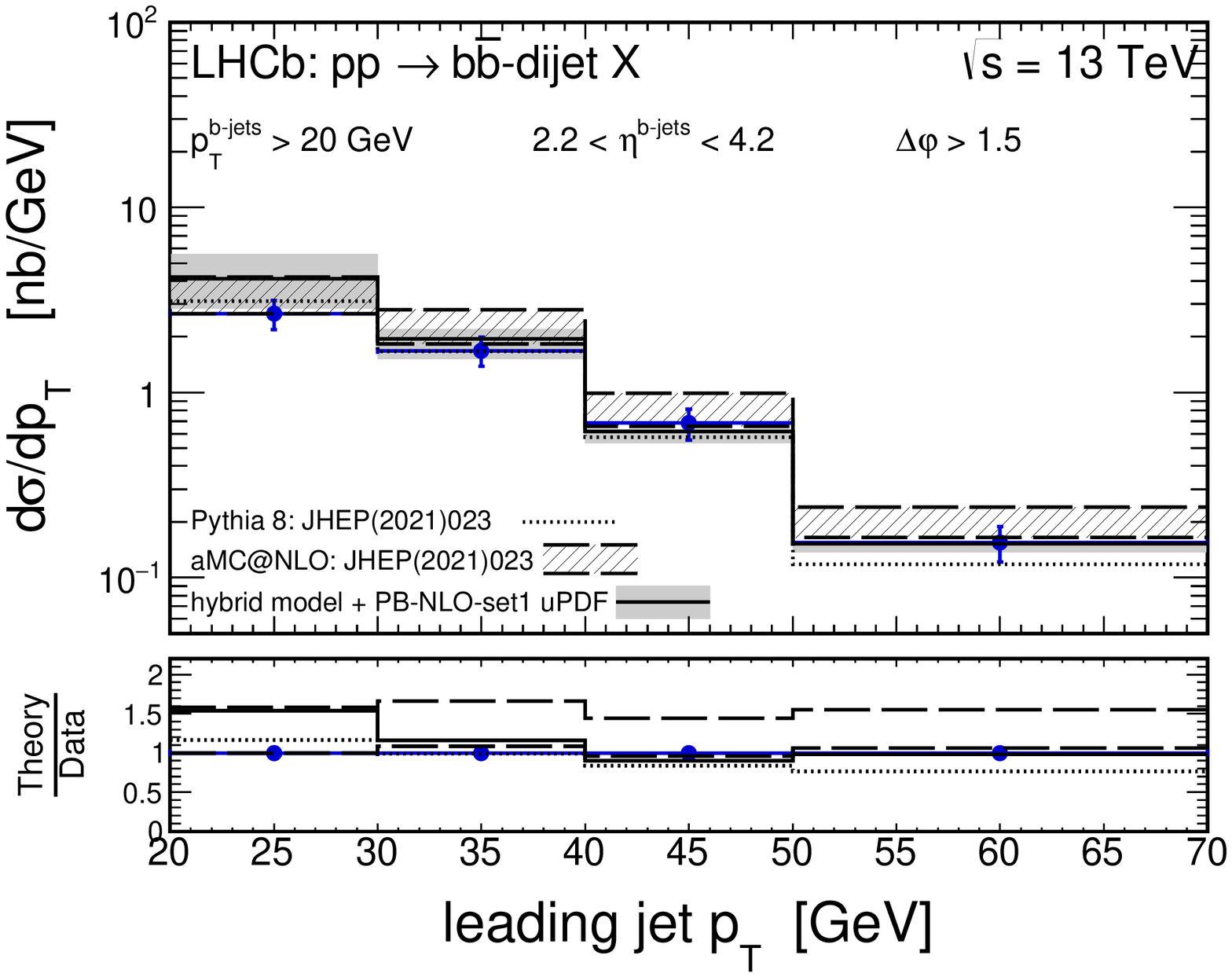}}
\end{minipage}
\begin{minipage}{0.47\textwidth}
  \centerline{\includegraphics[width=1.0\textwidth]{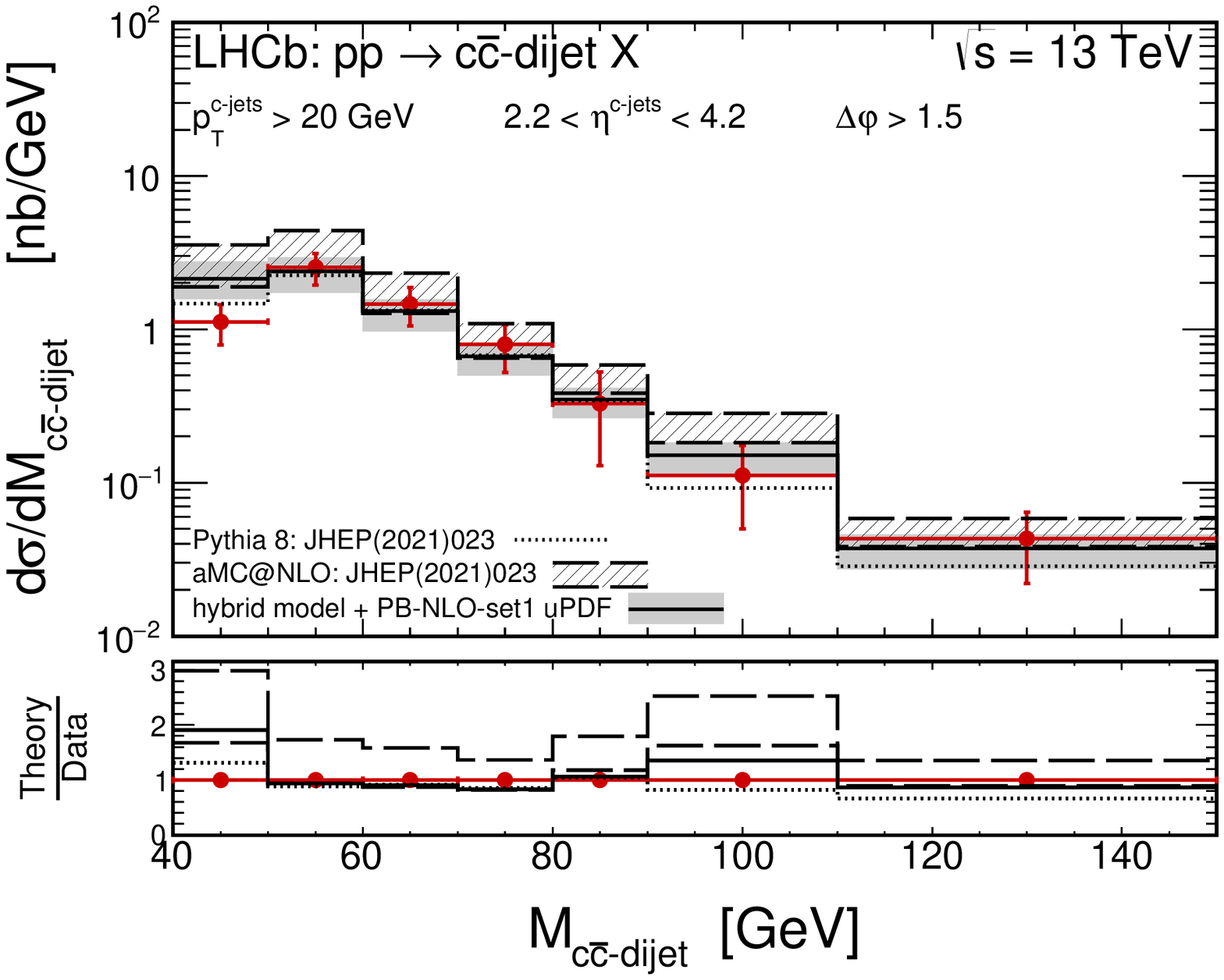}}
\end{minipage}
\begin{minipage}{0.47\textwidth}
  \centerline{\includegraphics[width=1.0\textwidth]{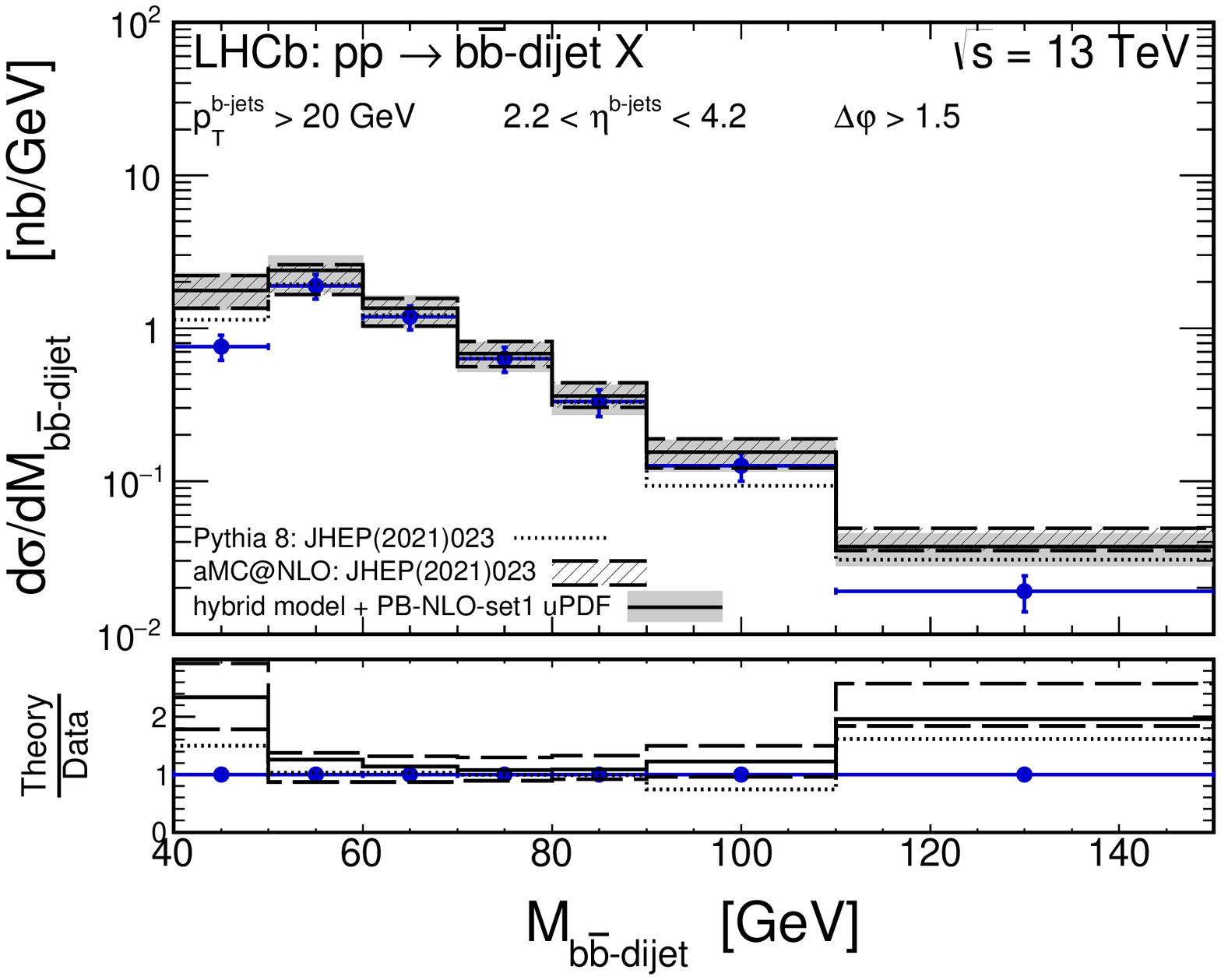}}
\end{minipage}
  \caption{The differential cross sections for forward production of $c\bar c$- (left panels) and $b\bar b$-dijets (right panels) in $pp$-scattering at $\sqrt{s}=13$ TeV as a function of the leading jet $p_{T}$ (top panels) and the dijet invariant mass $M_{Q\bar Q\text{-}\mathrm{dijet}}$ (bottom panels).
Here we compare our predictions of the hybrid model obtained with the PB-NLO-set1 gluon uPDF (solid bands) with two different collinear approach calculations: aMC@NLO (dashed bands) and \textsc{Pythia8} (dotted histograms), both taken from Ref.~\cite{LHCb:2020frr}. More details can be found in the figure.
\small 
}
\label{fig:30}
\end{figure}

\begin{figure}[!h]
\begin{minipage}{0.47\textwidth}
  \centerline{\includegraphics[width=1.0\textwidth]{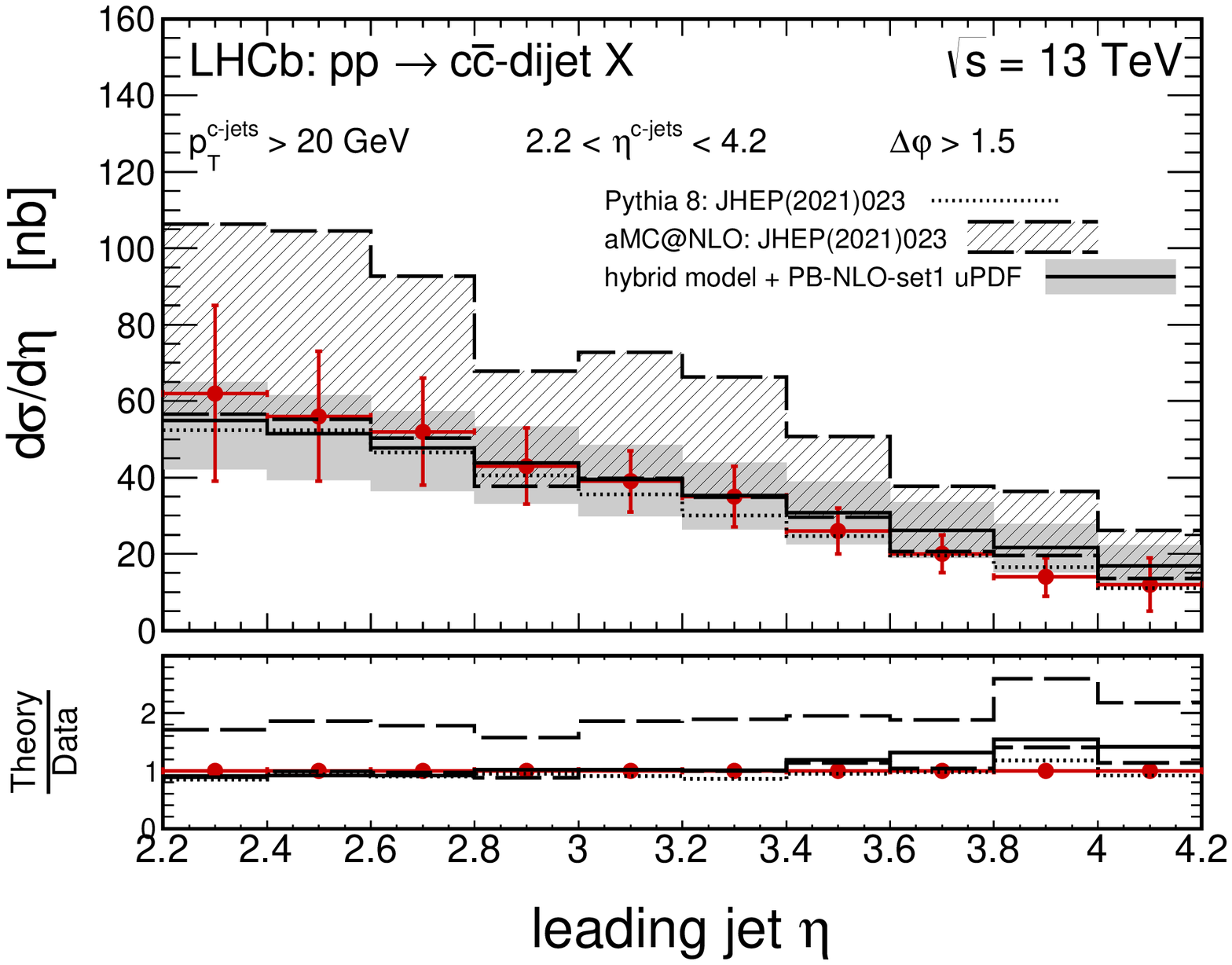}}
\end{minipage}
\begin{minipage}{0.47\textwidth}
  \centerline{\includegraphics[width=1.0\textwidth]{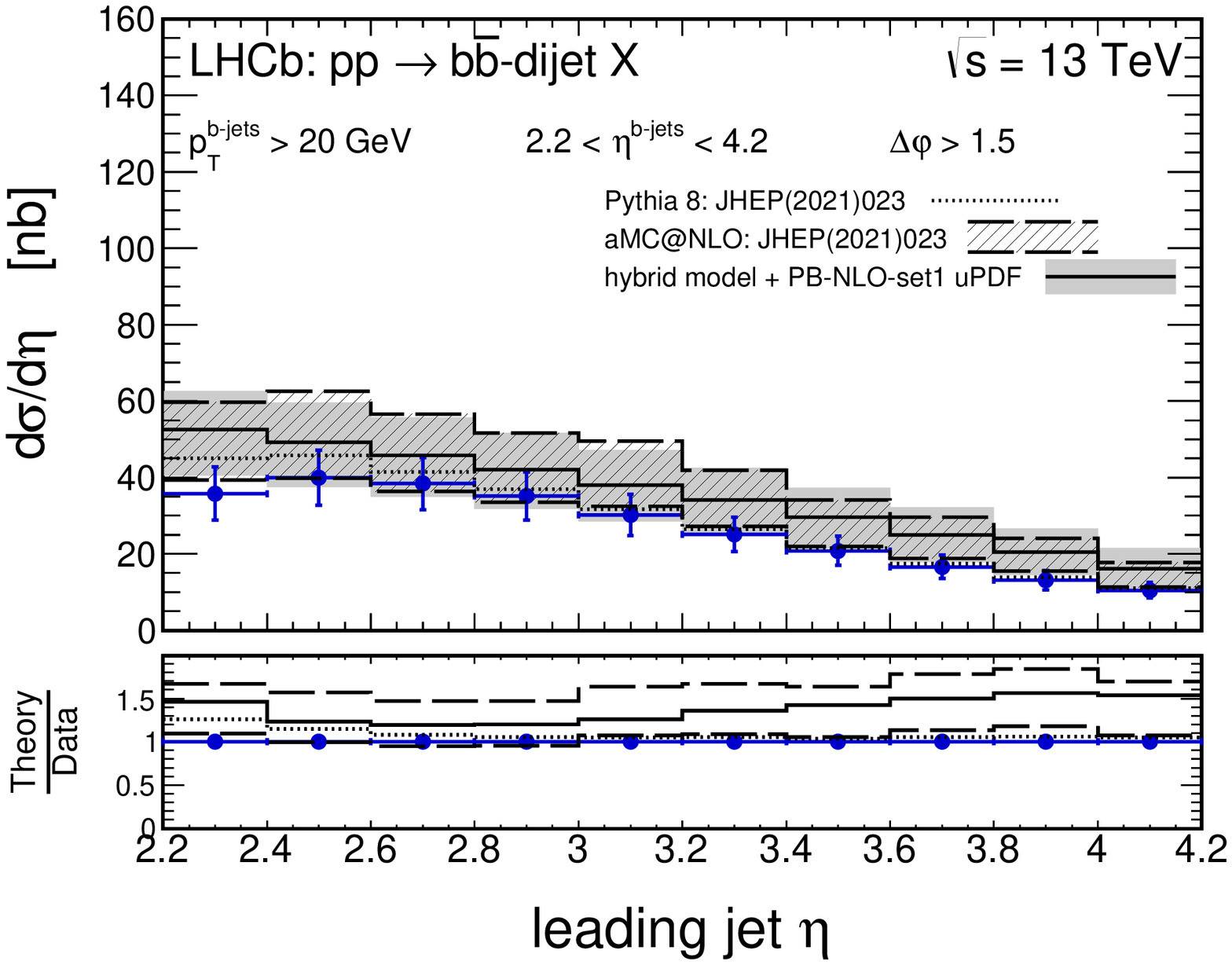}}
\end{minipage}
\begin{minipage}{0.47\textwidth}
  \centerline{\includegraphics[width=1.0\textwidth]{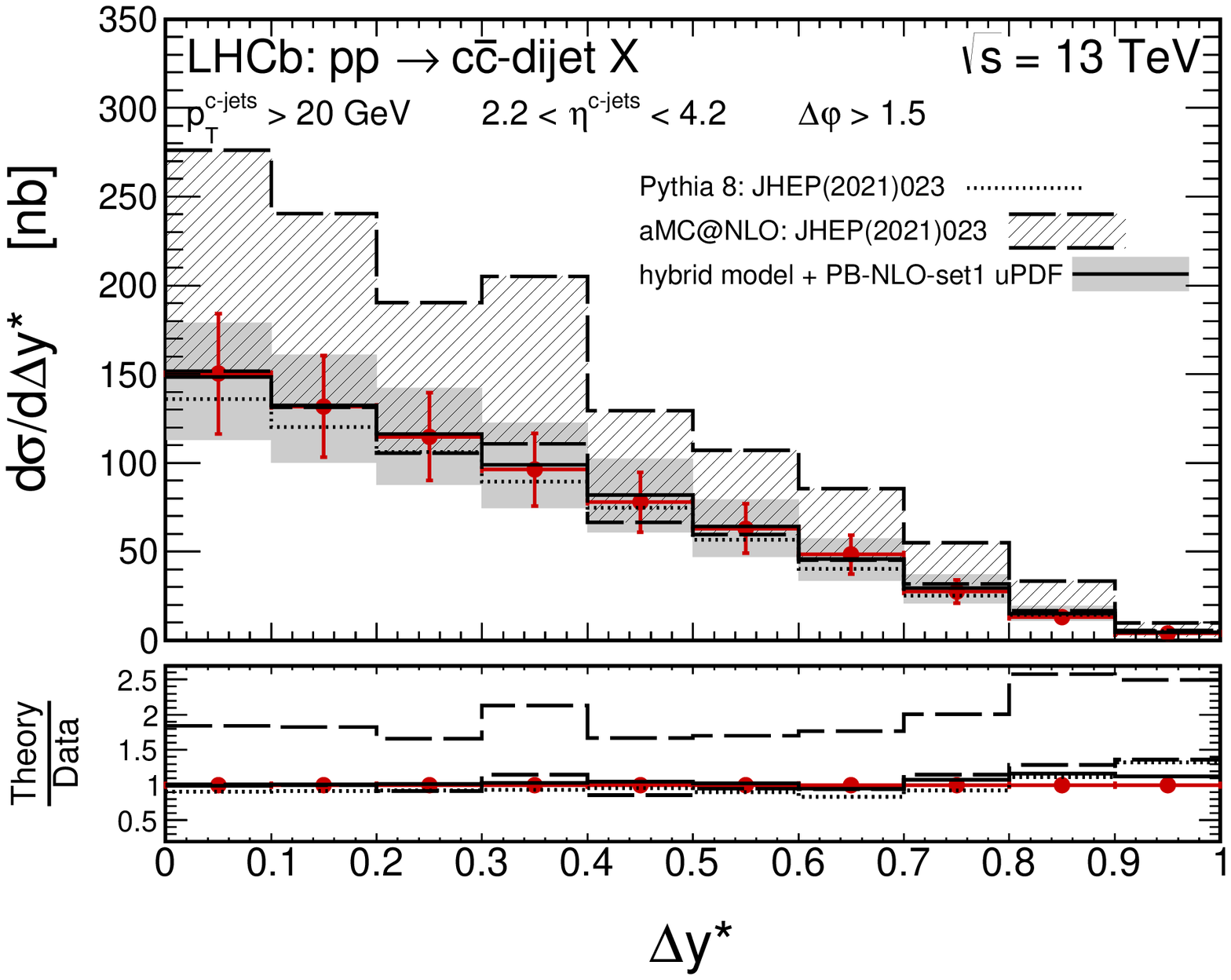}}
\end{minipage}
\begin{minipage}{0.47\textwidth}
  \centerline{\includegraphics[width=1.0\textwidth]{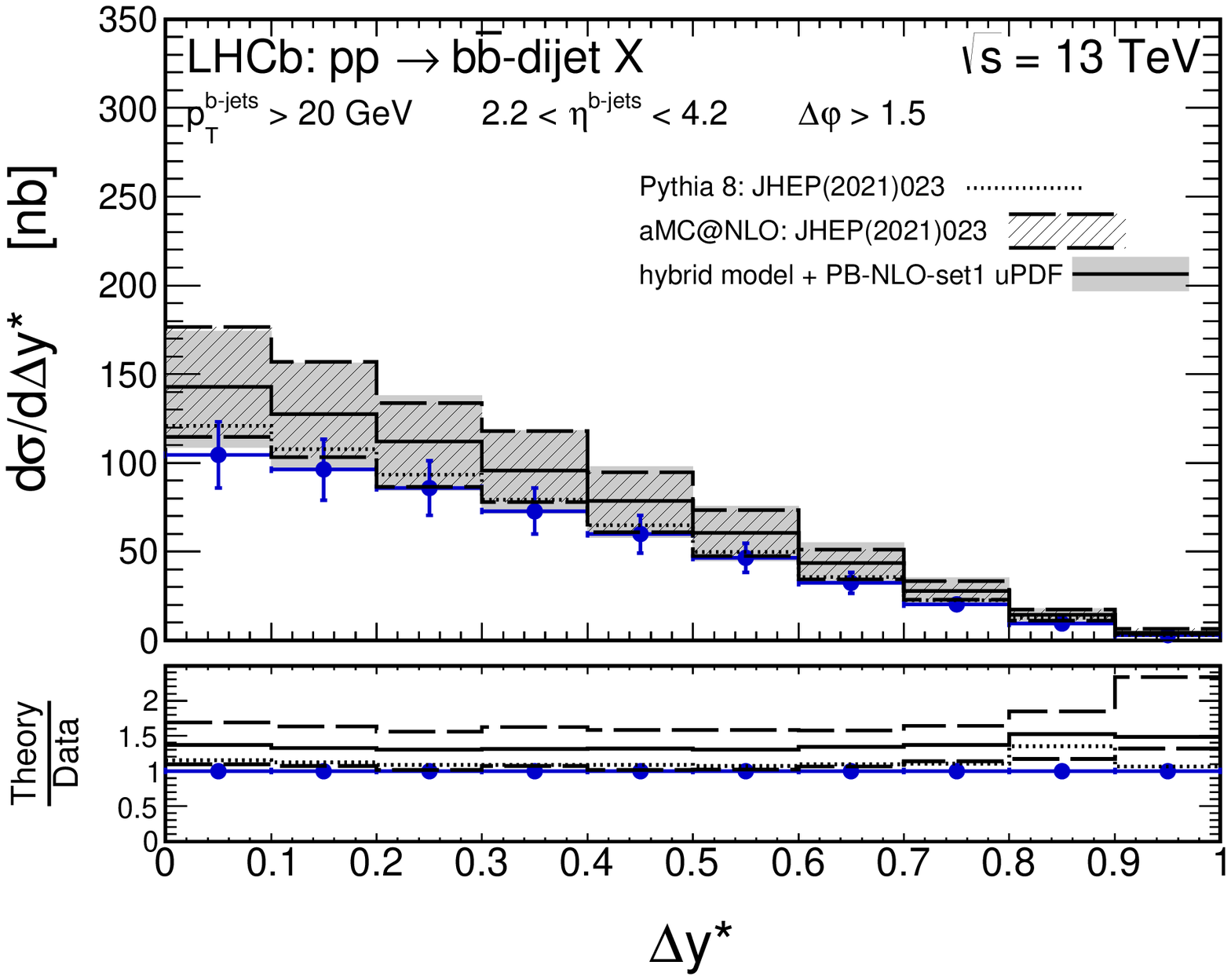}}
\end{minipage}
  \caption{The same as in Fig.~\ref{fig:30} but here the leading jet $\eta$ (top panels) and the rapidity difference $\Delta y^*$ (bottom panels) distributions are shown.
\small 
}
\label{fig:31}
\end{figure}

\begin{figure}[!h]
\begin{minipage}{0.47\textwidth}
  \centerline{\includegraphics[width=1.0\textwidth]{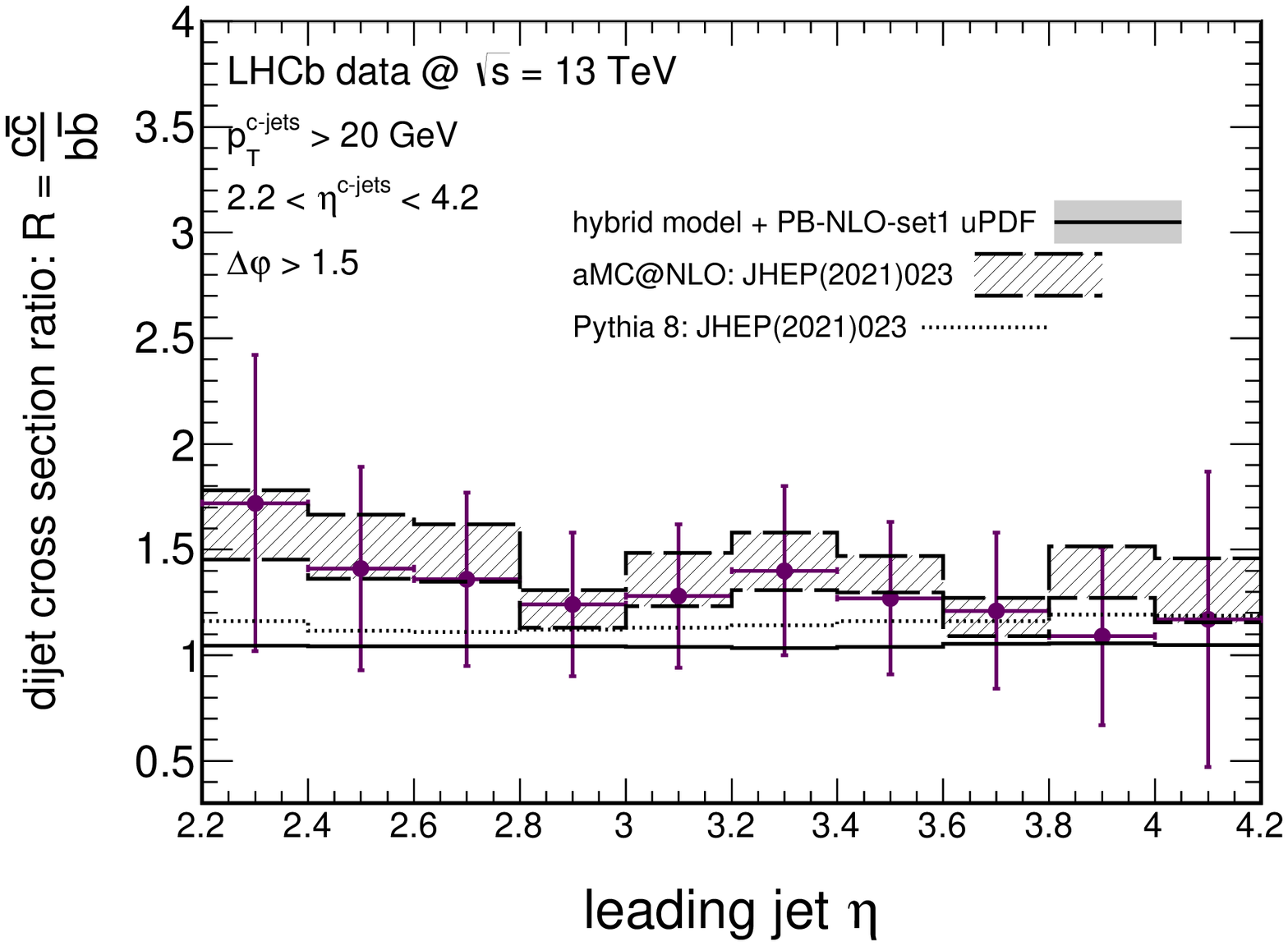}}
\end{minipage}
\begin{minipage}{0.47\textwidth}
  \centerline{\includegraphics[width=1.0\textwidth]{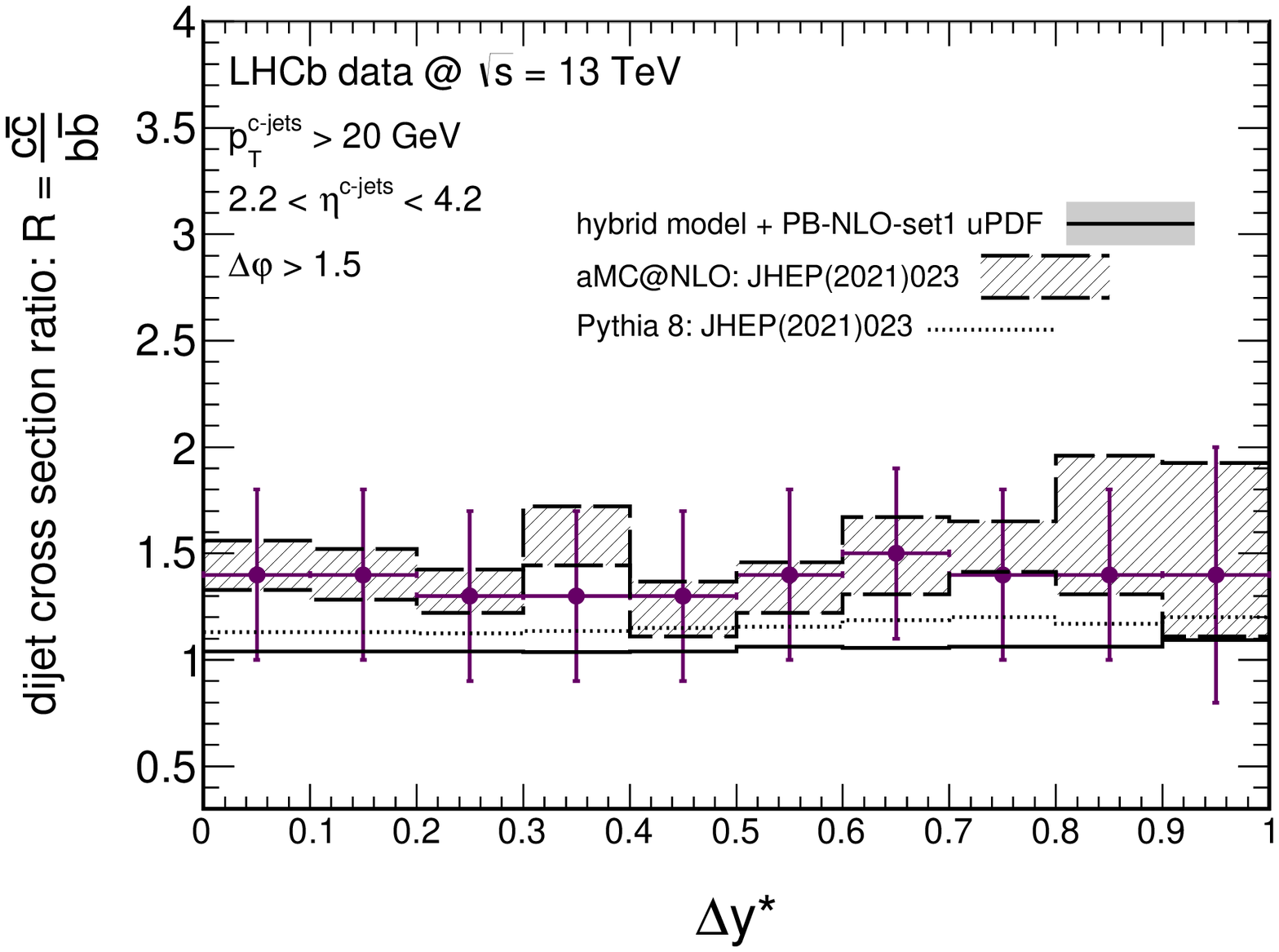}}
\end{minipage}
\begin{minipage}{0.47\textwidth}
  \centerline{\includegraphics[width=1.0\textwidth]{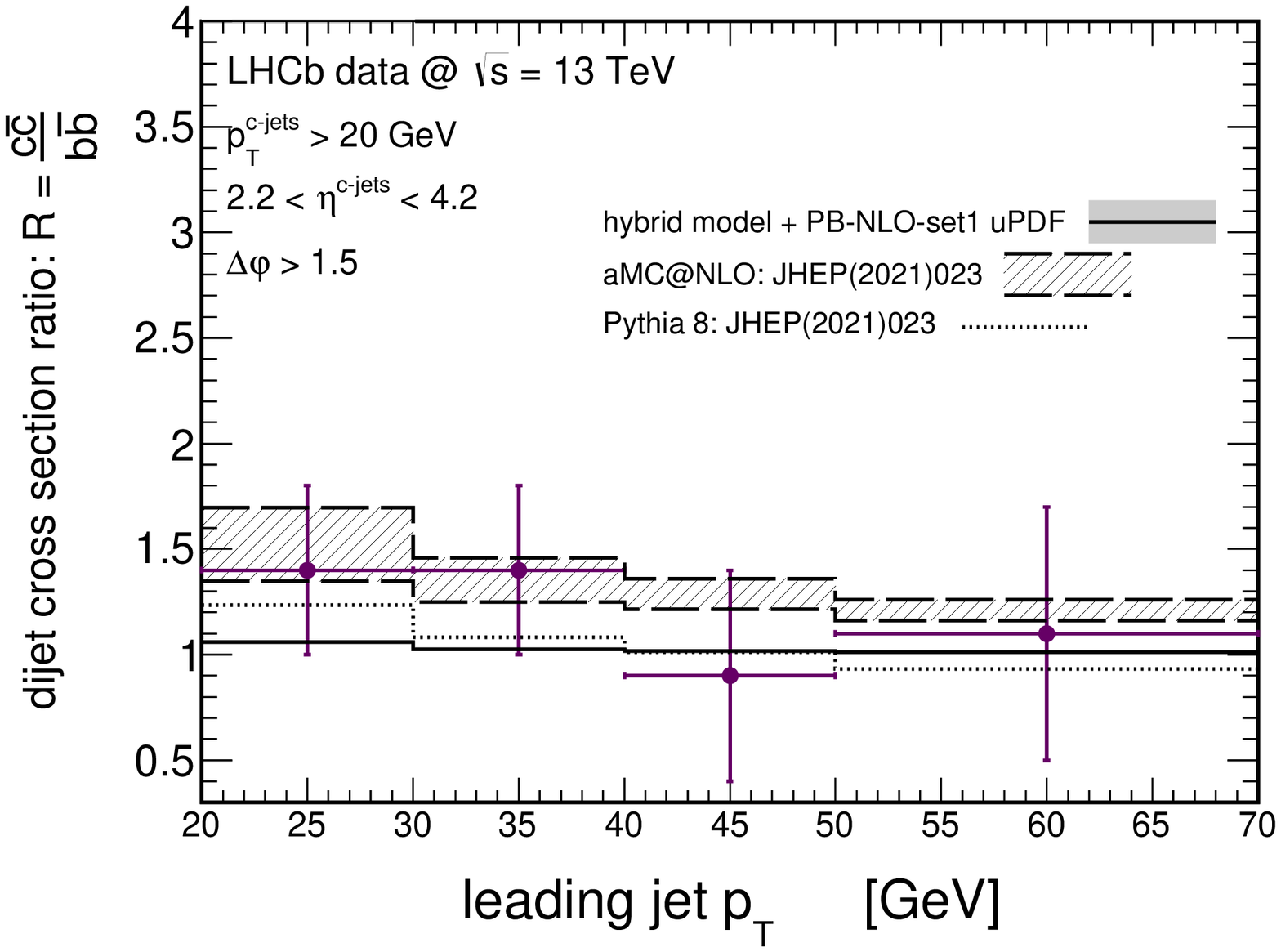}}
\end{minipage}
\begin{minipage}{0.47\textwidth}
  \centerline{\includegraphics[width=1.0\textwidth]{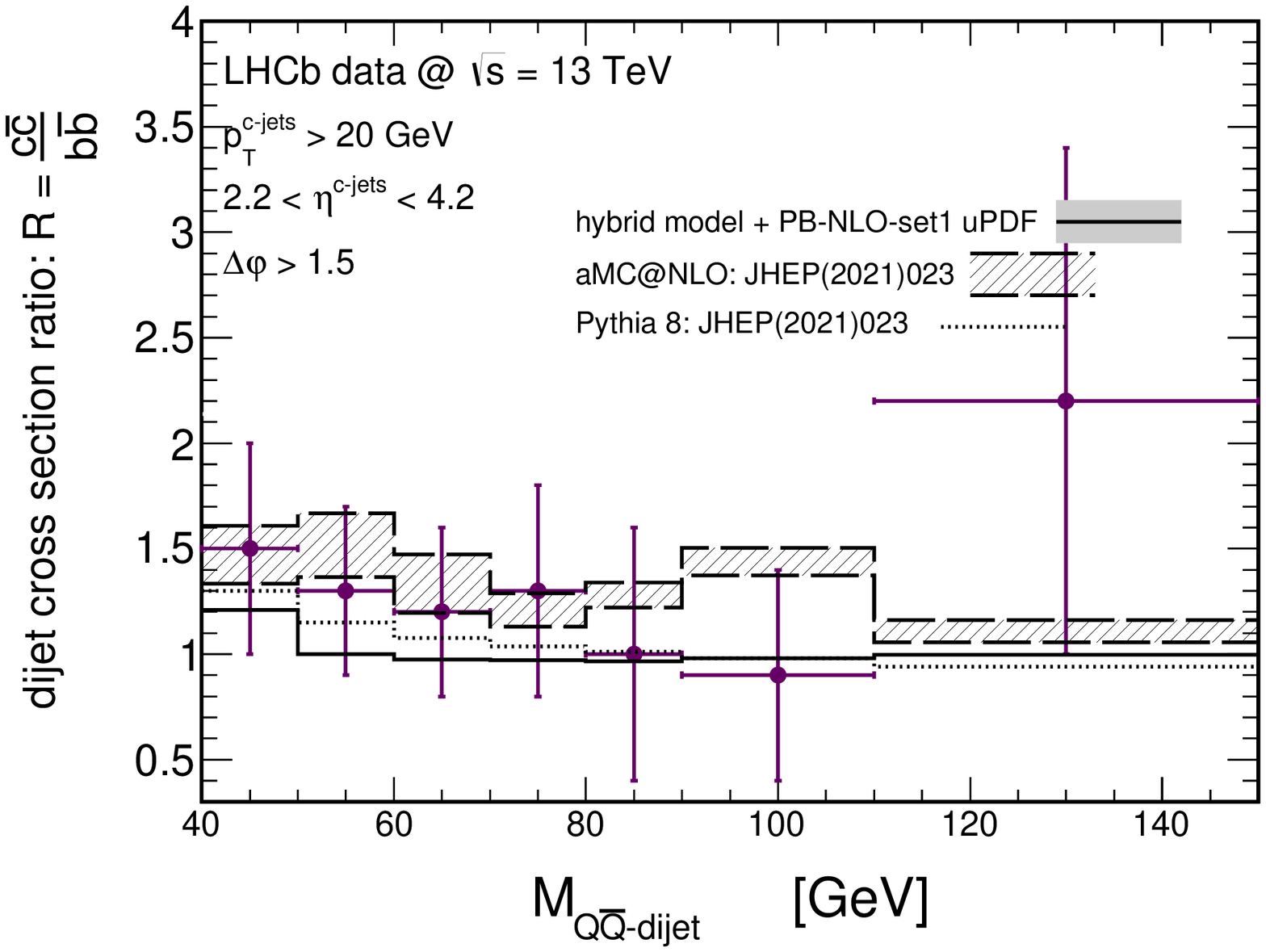}}
\end{minipage}
  \caption{The ratio $R=\frac{c\bar c}{b\bar b}$ of the dijet differential cross sections in $pp$-scattering at $\sqrt{s}=13$ TeV as a function of the leading jet $\eta$ (top left), the rapidity difference $\Delta y^{*}$ (top right), the leading jet $p_{T}$ (bottom left) and the dijet invariant mass $M_{b\bar b\text{-}\mathrm{dijet}}$ (bottom right). Here we compare our predictions of the hybrid model obtained with the PB-NLO-set1 gluon uPDF (solid histograms) with two different collinear approach calculations: aMC@NLO (dashed bands) and \textsc{Pythia8} (dotted histograms), both taken from Ref.~\cite{LHCb:2020frr}. More details can be found in the figure.
\small 
}
\label{fig:32}
\end{figure}

Finally, we wish to compare our predictions based on the $k_{T}$-factorization framework with those obtained according to the collinear approximation. In Figs.~\ref{fig:30}, ~\ref{fig:31}, and ~\ref{fig:32} we plot our hybrid model results for the PB-NLO-set1 uPDF (solid bands and histograms) and results of two different collinear approach calculations: \textsc{Madgraph}5 aMC@NLO \cite{Alwall:2014hca} (dashed bands) and \textsc{Pythia8} \cite{Sjostrand:2007gs} (dotted histograms), both taken from Ref.~\cite{LHCb:2020frr}. The two models correspond to NLO and LO matrix element calculations, respectively, supplemented further with dedicated parton-showers.

%----------------------------------------------------------------------------
%----------------------------------------------------------------------------

%----------------------------------------------------------------------------
%----------------------------------------------------------------------------

The main conclusion is that all three models presented here lead to a rather consistent results. The \textsc{Pythia8} predictions are very similar to the hybrid model results for both, $c\bar c$- and $b\bar b$-dijets. For the case of $b\bar b$ production also the aMC@NLO framework gives pretty much the same results. Here, all three approaches show some small tendency to overshoot the LHCb data. For the $c\bar c$-dijets we observe that the aMC@NLO leads to a slightly larger cross sections with respect to the two other models, showing some tendency to overshoot the LHCb data within the central prediction. For this model only the lower limit seems to be compatible with the experimental data and with the central predictions of the hybrid model and \textsc{Pythia8} as well. The situation has a direct consequence for the $R=\frac{c\bar c}{b\bar b}$ distributions. The hybrid model and the \textsc{Pythia8} results give the ratio close to 1, while the aMC@NLO leads to a slightly larger values, which are more preferred by the LHCb data.
Definite conclusions here are quite limited because of large theoretical and experimental uncertainties. As was already mentioned the hybrid model and the \textsc{Pythia8} calculations are based on the LO matrix element with real higher-order corrections taken into account by uPDFs or parton-shower, respectively. Both of them do not include virtual corrections which might be responsible for observed differences with respect to the aMC@NLO. Unfortunately, a full NLO framework within the hybrid (or full) $k_{T}$-factorization approach for heavy quark dijets is quite challenging and still not established.         

%----------------------------------------------------------------------------
%----------------------------------------------------------------------------

%----------------------------------------------------------------
\section{A comment on finite jet size effects}
%----------------------------------------------------------------

So far we have calculated distributions at the parton level.
The agreement with experimental data is, however, quite good,
both for $c \bar c$ and $b \bar b$ dijets.
In principle, one could worry about jet size effects.
The LHCb collaboration chosen the jet cone size $R_{\textrm{cone}} = 0.5$ in their
analysis \cite{LHCb:2020frr}.
Could {this affect} our partonic results?

In general, a thorough answer to this question requires modelling of $c$ and $b$
jets which is rather complicated and goes beyond the scope of 
the present paper. Instead of following this path, we will try to estimate a
possible effect as was done for one jet case in \cite{GKNPP2017}.

\begin{figure}[!h]
\begin{minipage}{0.33\textwidth}
  \centerline{\includegraphics[width=1.0\textwidth]{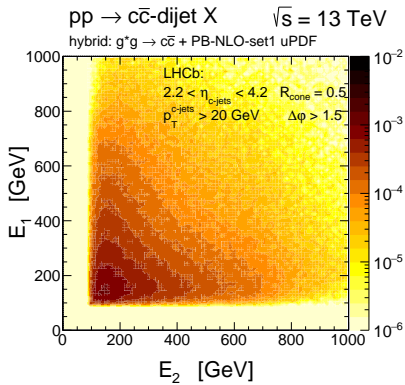}}
\end{minipage}
  \caption{
\small The double differential cross sections in energies of both jets for forward production of $c\bar c$-dijets in $pp$-scattering at $\sqrt{s}=13$ TeV probed in the LHCb experiment \cite{LHCb:2020frr} as a function of energies of both jets. 
}
\label{fig:dsig_dE1dE2}
\end{figure}

It is energy of the jet which decides about the shape of the jet
\cite{GKNPP2017}. The larger energy of the jet the smaller finite cone
size effects.
In Fig.~\ref{fig:dsig_dE1dE2} we show distribution in jet energies
for the LHCb kinematics.

%----------------------------------------------------------------------------
%----------------------------------------------------------------------------

The typical jet energies are relatively large
$E_1, E_2 >$ 200 GeV.
In the approach from \cite{GKNPP2017} the basic ingredient
is $d E_g / dx dk^2$ where $E_g$ is energy of the emitted gluon,
$x$ is its longitudinal momentum fraction
and $k^2$ is transverse momentum squared of the gluon with respect
to the heavy quark ($c$ or $b$ in our case). This quantity depends
on the mass of the emitting quark. In principle, one can expect
even a spectacular effect known under the name dead cone \cite{DKT91}.
The dead cone effect was observed experimentally recently for 
the first time by the ALICE collaboration \cite{ALICE_dead_cone}.
Energy loss due to emission outside of the jet cone can be then written as
\begin{equation}
\Delta E_g = \int \frac{d E_g}{ d x d k^2} 
\theta \left(\theta > \theta_{\mathrm{open}}(R_{\mathrm{cone}}) \right) \;,
\label{Delta_E}
\end{equation}
where $\theta_{\mathrm{open}}(R_{\mathrm{cone}})$ is jet openning angle corresponding to a given
jet radius $R_{\mathrm{cone}}$. The distribution $\frac{d E_g}{ d x d k^2}$ is obtained  \cite{GKNPP2017} from a
generalization of the Gunion-Bertsch formula \cite{GB82} 
for the gluon number distribution $\frac{d n_g}{d x d  k^2}$.

At high energies, as in our case, such effects are completely
unimportant as dead cone angle is of the order of a small fraction
of one degree. In our case, it is the energy (transverse momentum) which
escapes from experimentally defined jet cones.
We have estimated that in our case the relative energy loss $\Delta E / E$, 
including range of jet energies and $R_{\mathrm{cone}} = 0.5$, is of the order of a few percent.
It is reasonable to expect the same is true for $\Delta p_T / p_T$.
The main effect is due to the fact that the cuts are imposed on the
measured transverse momenta, not momenta of heavy quarks/antiquarks 
which are of course bigger than the measured ones. 
In our case we have to apply this procedure to both measured jets. 
This leads to a damping of the cross section by approximately a few percent.
We have checked that the damping is practically the same for 
$c/{\bar c}$ and $b / {\bar b}$ quarks/antiquarks. In our energy range
the quark mass effect is negligible.
We conclude that jet cone size effect is most probably not responsible for
the charm-to-bottom ratios discussed in our paper.
We think that the effect of the shape of the heavy quark jets
requires further studies, also in the present context.

%----------------------------------------------------------------------------
\begin{figure}[!h]
\begin{minipage}{0.47\textwidth}
  \centerline{\includegraphics[width=1.0\textwidth]{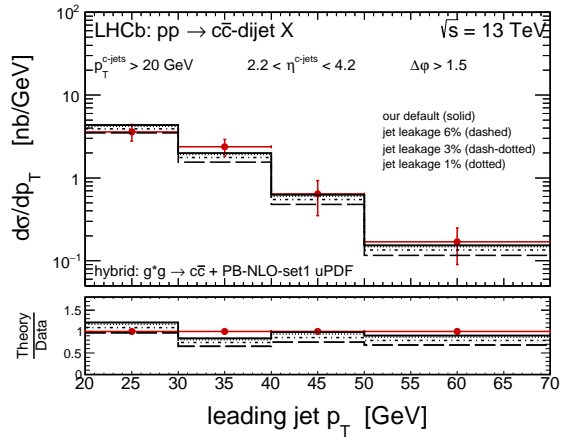}}
\end{minipage}
  \caption{
\small The differential cross sections for forward production of $c\bar c$-dijets in $pp$-scattering at $\sqrt{s}=13$ TeV as a function of the leading jet $p_{T}$ for three different values of the jet leakage effect.
}
\label{fig:leakage}
\end{figure}
%----------------------------------------------------------------------------

In Fig.~\ref{fig:leakage} we show results for different (independent off transverse momentum and pseudorapidity)
values of the leakage of the energy outside of the jet cone radius $R_{\mathrm{cone}} = 0.5$. Here we have assumed in addition $\frac{\Delta p_{T}}{p_{T}} \approx \frac{\Delta E}{E}$ for each of the two jets. Although the assumed leakage (1\%, 3\%, 6\%) is rather small the resulting effect on the cross section normalization is rather sizeable. 

%--------------------------
\section{Conclusions}
%--------------------------

In the present paper we have studied the $c \bar c$- and $b \bar b$-dijet production in $pp$-scattering for $\sqrt{s}$ = 13 TeV and LHCb acceptance
limitations on jet pseudorapidities 2.2 $< \eta_{jet1,2} <$ 4.2, transverse momenta $p_{T}^{jet1,2} >$ 20 GeV, jet cone $R_{\mathrm{cone}} =$ 0.5 and on the azimuthal angle between the jets $\Delta \varphi > 1.5$.
The differential cross sections were calculated within
$k_T$-factorization and hybrid approaches. We have used different 
unintegrated parton distribution functions for gluon, including the PB, the KMR/MRW and various CCFM models.
We have calculated distributions in leading-jet transverse momentum $p_{T}$, leading-jet pseudorapidity $\eta$, rapidity difference bewteen jets $\Delta y^*$ and in dijet invariant mass $M_{Q\bar Q\text{-}\mathrm{dijet}}$.
We found that an agreement between the predictions and the data within the full $k_T$-factorization is strongly related to the modelling of the large-$x$ behaviour of the gluon uPDFs which is usually not well constrained. For this case only the PB-NLO-set1 gluon uPDF leads to a satisfactory description of the LHCb data. Rest of the models seem to visibly overestimate the experimental data points. This was understood as due to incorrect effective integrated gluon distribution obtained from the unintegrated one for large-$x$.

The problem may be avoided by following the hybrid factorization. Then a good description of the measured distributions is obtained with the PB-NLO-set1, the KMR-MMHT2014nnlo, the Kutak-Sapeta and the Jung setA0 CCFM gluon uPDFs for both $c\bar c$ and $b\bar b$ production taking into account experiemental uncertainties. Only the most recent JH-2013-set1 and JH-2013-set2 CCFM gluon uPDFs seems to overestimate the data sets even within the hybrid approach. In any case we observe a small tendency of our predictions to slightly overestimate the data for $b\bar b$-dijets while the corresponding data for $c\bar c$-dijets can be perfectly described within the same gluon uPDFs. In general, taking into account theoretical uncertainties our hybrid model predictions are consistent with collinear results of the \textsc{Pythia8} and the aMC@NLO frameworks. 

We have presented also the ratios of $c \bar c$ to $b \bar b$ cross sections. We get cross sections ratio very close to 1. This could be
understood by the fact that for large transverse momenta the effect
of quark mass is rather small. Slightly larger, but with large
systematic error bars, effect was obtained in the experimental extraction.
It is difficult to understand in the moment the possible disagreement.
In principle, it can be due to inappropriate extraction of
$c / \bar c$ jets (e.g. by admixture of light quarks and/or 
$b / {\bar b}$ jets). We have also discussed the effect of damping
of the cross section within a simple model
due to finite jet radius. It may be expected that statistically a part of 
the parton (gluon) energy escapes outside of the jet cones.
Within this, a bit naive approach, one may expect sizeable corrections.
A much better approach would be to take into account internal structure of $c$- and $b$-jets due to parton shower.
This definitely goes beyond the scope of the present paper.

In our $k_T$-factorization approach the production of $c$ and $\bar c$ as well as $b$ and $\bar b$ is identical (symmetric).
Asymmetry effects ($\frac{d\sigma(c)}{d\xi} \ne \frac{d\sigma(\bar{c})}{d\xi}$ and $\frac{d\sigma(b)}{d\xi} \ne \frac{d\sigma(\bar{b})}{d\xi}$, where $\xi$ represents schematically $y$ and $p_{T}$) were discussed in Ref.~\cite{Gauld:2019doc}. The asymmetry effects found there are rather small, of the order of 1 \%. Also the contribution of electroweak processes, as discussed in Ref.~\cite{Gauld:2019doc}, is rather small. So we think these interesting effects are not crucial for our studies.

\vskip+5mm
{\bf Acknowledgments}\

A.S. is indebted to Marcin Kucharczyk for a discussion about jet size and jet definition.
This work and the stay of Rafa{\l} Maciu{\l}a in Lund was supported by the Bekker Program of the Polish National Agency for Academic Exchange under Contract No. PPN/BEK/2020/1/00187/U/00001.
This study was also partially supported by the Polish National Science Center grant UMO-2018/31/B/ST2/03537
and by the Center for Innovation and Transfer of Natural Sciences and Engineering Knowledge in Rzesz{\'o}w. R.P.~is supported in part by the Swedish Research Council grants, contract numbers 621-2013-4287 and 2016-05996, as well as by the European Research Council (ERC) under the European Union's Horizon 2020 research and innovation programme (grant agreement No 668679).

%-------------------------------------------------------------------------------------

%-------------------------------------------------------------------------------------

\end{document}